Universidade de Aveiro
2022

Jorge Filipe
Mónico Delgado

# Buracos Negros com Rotação e Cabelo Escalar e Objetos Compactos sem Horizonte dentro e além da Relatividade Geral

# Spinning Black Holes with Scalar Hair and Horizonless Compact Objects within and beyond General Relativity

Universidade de Aveiro
2022

**Jorge Filipe Mónico Delgado**

# Buracos Negros com Rotação e Cabelo Escalar e Objetos Compactos sem Horizonte dentro e além da Relatividade Geral

# Spinning Black Holes with Scalar Hair and Horizonless Compact Objects within and beyond General Relativity

*"I want to know God's thoughts. The rest are details."*

— Albert Einstein

Universidade de Aveiro
2022

**Jorge Filipe Mónico Delgado**

# Buracos Negros com Rotação e Cabelo Escalar e Objetos Compactos sem Horizonte dentro e além da Relatividade Geral

# Spinning Black Holes with Scalar Hair and Horizonless Compact Objects within and beyond General Relativity

Tese apresentada à Universidade de Aveiro para cumprimento dos requisitos necessários à obtenção do grau de Doutor em Física, realizada sob a orientação científica do Doutor Carlos Alberto Ruivo Herdeiro, Investigador Coordenador do Departamento de Matemática da Universidade de Aveiro, e do Doutor Eugen Radu, Investigador Doutorado do Departamento de Matemática da Universidade de Aveiro.

*To my dearly beloved girlfriend and partner Ana Pedro and my parents Jorge and Otília.*

**o júri / the jury**

| | |
|---|---|
| presidente / president | Doutor José Carlos Esteves Duarte Pedro<br>Professor Catedrático, Universidade de Aveiro |
| vogais / examiners committee | Doutora Betti Hartmann<br>Professora, University College London |
| | Doutor Yakov Mihajlovic Shnir<br>Leading Research Scientist, Joint Institute for Nuclear Research |
| | Doutor Luis Carlos Bassalo Crispino<br>Professor Titular, Universidade Federal do Pará |
| | Doutora Lara Rodrigues Da Costa Gomes De Sousa<br>Investigadora Júnior, Universidade do Porto |
| | Doutor Carlos Alberto Ruivo Herdeiro<br>Investigador Coordenador, Universidade de Aveiro |

**agradecimentos / acknowledgements**

First and foremost, I offer my gratitude to my supervisor, Prof. Dr. Carlos Alberto Ruivo Herdeiro and my co-supervisor, Dr. Eugen Radu, whose support was one of the best. This thesis was only possible due to their guidance, encouragement, effort, and patience to answer my questions. I hope to continue to work with them in the future.

To my dearly beloved Ana Pedro Paracana, I would like to give my most significant, sincere and passionate thank you for everything she taught me and for all the memories we built together. She has always brightened my life with her smile and presence. Her support, help, care, and love are undoubtedly the reasons why I can finish this thesis and why I grew to become a better person overall. After all that we have been through, she is the person that I want to spend the rest of my days with.

To Alexandre Pombo and João Oliveira, for the countless hours discussing physics and all other sorts of topics, working and travelling together, I want to express my deepest thanks. Friends with different opinions and with whom one can have a friendly competition always helps oneself to grow and develop new skills and perspectives. For that, I am very thankful.

I want to thank several close friends of mine, mainly Rui Lopes, Maria João, Ana Oliveira, Ana Rita Paracana, Mário Redondo and João Simões, for all the time spent together, building everlasting friendships and memories I shall not forget. I also want to thank João Peixoto, André Ferreira e André Sá, for the companionship and all the good memories when we lived together. To Rodrigo Antunes, my academic godson, no words can express my gratitude. He always helped me through words of encouragement and time spent together, having healthy discussions and drinking coffee and beer. Thank you for allowing me to be your academic godfather and for all the help and good memories you provided.

For all the sympathy, patience and help to deal with the bureaucracies and documentation regarding all type of topics, I want to present my thank you to Cristina Grosso. Without her, I would still be stuck dealing with paperwork regarding the first year of my PhD.

Last but definitely not least, I want to express my most tremendous gratitude to my family for always supporting me, allowing me to pursue my dreams, and for everything they taught me. All that I am and have is mainly due to my family, and for that, I can not thank them enough. I am incredibly grateful for having them all by my side.

This thesis received financial support from the Fundação para a Ciência e Tecnologia (FCT) through the grants PD/BI/129008/2017 and SFRH/BD/130784/2017.

**Resumo**

Os últimos anos trouxeram-nos uma era de ouro da física gravitacional observacional. Os vários trabalhos realizados pela colaboração LIGO/Virgo/KAGRA sobre ondas gravitacionais e pela colaboração EHT sobre a sombra e deflexão da luz ao redor do buraco negro supermassivo no centro de M87 apontarão a comunidade científica na direção correta para encontrar uma resposta para a hipótese de Kerr. Para seguir essa direção, é necessária a construção e análise sistemática das propriedades físicas de soluções dentro da Relatividade Geral com campos adicionais ou dentro de teorias da gravidade modificadas. Nesta tese, forneceremos tal construção e análise de objetos compactos dentro da teoria (complexa-)Einstein-Klein-Gordon com vários potenciais escalares e dentro de uma teoria escalar-tensorial em particular - a teoria de Horndesky com simetria de deslocamento.
Após uma breve introdução a alguns tópicos-chave que serão úteis ao longo desta tese, apresentamos uma discussão sobre a geometria do horizonte de buracos negros de Kerr com e sem cabelo escalar. Seguimos com a construção e estudo das mesmas soluções cabeludas discutidas no capítulo anterior, mas com índices harmónicos azimutais mais elevados. Nos dois capítulos seguintes, introduzimos um potencial escalar diferente baseado no potencial do axião da Cromodinâmica Quântica e obtivemos e estudamos objetos compactos sem horizonte e buracos negros. Passamos então para a teoria de Horndeski com simetria de deslocamento, onde realizamos construções e análises semelhantes às já mencionadas. Por fim, derivamos uma relação entre a estabilidade radial dos anéis de luz e as órbitas circulares do tipo tempo em torno deles. Seguimos com um estudo sobre o quão eficiente é a conversão de energia gravitacional em radiação quando uma partícula do tipo tempo cai em direção a todos os objetos compactos estudados nesta tese. Terminamos com algumas conclusões e observações.

**Abstract**

The last years have brought upon us a golden age of observational gravitational physics. The several observations by the LIGO/Virgo/KAGRA collaboration about gravitational waves and by the EHT collaboration about the shadow and lensing of light around the supermassive black hole in the centre of M87 will point the scientific community in the correct direction to find an answer to the Kerr hypothesis. In order to follow that direction, the systematic construction and analysis of the physical properties of solutions within General Relativity with additional fields or within modified theories of gravity is necessary. In this thesis, we shall provide such construction and analysis for compact objects within (complex-)Einstein-Klein-Gordon theory with various scalar potentials and within a particular scalar-tensor theory – the shift-symmetric Horndesky theory.

After a brief introduction to some key topics that shall be useful throughout this thesis, we present a discussion about the horizon geometry of Kerr black holes with and without scalar hair. We follow up with the construction and study of the same hairy solutions discussed in the previous chapter but with higher azimuthal harmonic indexes. In the following two chapters, we introduce a different scalar potential based on the Quantum Chromodynamic axion potential and obtain and study both horizonless compact objects and black holes. We then go to the shift-symmetric Horndeski theory, where we perform similar constructions and analyses to the ones already mentioned. Lastly, we derive a relation between the radial stability of light-rings and timelike circular orbits around them. We follow up with a study on how efficient it is the conversion of gravitational energy in radiation as a timelike particle falls towards all compact objects studied throughout this thesis. We end with some conclusions and remarks.

# Contents

# List of Figures

# List of Tables

# Glossary

| | |
|---|---|
| **ADM** | Arnowitt, Deser and Misner |
| **ALP** | Axion like Particles |
| **BH** | Black Hole |
| **BS** | Boson Star |
| **BL** | Boyer-Lindquist |
| **CO** | Circular Orbit |
| **DEC** | Dominant Energy Condition |
| **EsGB** | Einstein-scalar-Gauss-Bonnet |
| **GB** | Gauss-Bonnet |
| **GR** | General Relativity |
| **ISCO** | Innermost Stable Circular Orbit |
| **KBHsAH** | Kerr Black Holes with Axionic Hair |
| **KBHsSH** | Kerr Black Holes with Scalar Hair |
| **KG** | Klein-Gordon |
| **LR** | Light-Ring |
| **MSCO** | Marginally Stable Circular Orbit |
| **NHEK** | Near Horizon Extremal Kerr |
| **PDE** | Partial Differential Equation |
| **QCD** | Quantum Chromodynamics |
| **RABS** | Rotating Axion Boson Star |
| **SZ** | Sotiriou and Zhou |
| **SCO** | Stable Circular Orbit |
| **STCO** | Stable Timelike Circular Orbit |
| **SEC** | Strong Energy Condition |
| **TCO** | Timelike Circular Orbit |
| **UCO** | Unstable Circular Orbit |
| **UTCO** | Unstable Timelike Circular Orbit |
| **WEC** | Weak Energy Condition |
| **ZAMO** | Zero Angular Momentum Observer |

CHAPTER 1

# Introduction

General Relativity (GR) is the best and most well-tested gravity theory to date. Albert Einstein developed it more than one hundred years ago, and it revolutionised how we think about gravity. Instead of being purely a force, as Newton has envisioned it, it is the manifestation of the geometry of a four-dimensional spacetime, where both space and time are connected. The well-known Einstein field equations characterise such manifestation,

$$R_{\mu\nu} - \frac{1}{2}g_{\mu\nu}R = \frac{8\pi G}{c^4}T_{\mu\nu} \ , \tag{1.1}$$

where $R_{\mu\nu}$ and $R$ are the Ricci tensor and scalar, respectively, $g_{\mu\nu}$ is the metric associated with the four-dimensional spacetime, and $T_{\mu\nu}$ is the energy-momentum tensor that encapsulates the information about the energy and matter present on the spacetime[1].

In general, the Einstein field equations yield a set of coupled non-linear differential equations that are highly arduous to solve analytically. However, a few weeks after Einstein published his theory, Schwarzschild obtained the first nontrivial solution of the vacuum Einstein equations, *i.e.* with a vanishing energy-momentum tensor, which is known as the *Schwarzschild* solution [1], [2]. This solution encodes the gravitational field outside of a static, asymptotically flat and spherically symmetric object (that later would be identified as a black hole), and its total mass entirely characterises it. Shorty after, Reissner and Nordström obtained, independently, the electromagnetic generalisation of the previous solution by solving the Einstein equations together with the Maxwell equations [3], [4]. Such a solution became known as the *Reissner-Nordström* black hole (BH). It has the same symmetries as the Schwarzschild BH. Still, since the problem now includes the Maxwell equations, there will be an electromagnetic field on the spacetime and a non-trivial electromagnetic energy-momentum tensor. Thus, the Reissner-Nordström solution is characterise by its total mass and an electric charge, that contains the information regarding the electromagnetic field.

The rotating generalisation of the Schwarzschild solution would only appear almost fifty years later (showing how strenuous the task of obtaining rotating solutions of Einstein equations is), in 1963, when Kerr successfully derived what is now known as the *Kerr* BH [5]. This solution encodes the gravitational field outside a stationary, asymptotically flat and axially symmetric object in vacuum. Since the spacetime is rotating, a new physical quantity characterised the solution beside its total mass – the total angular momentum. Two years later, Newman and collaborators found the charged generalisation of the Kerr BH – the *Kerr-Newman* BH [6]. The physical quantities that characterise this solution are the same as the ones found for Kerr, but, since there is an electromagnetic field, the Kerr-Newman BH is also characterised by its electric charge.

[1] After this point, unless stated otherwise, we shall use the signature $(-, +, +, +)$ and natural units such that $G = c = 1$.

Both the Kerr and Kerr-Newman solutions provided a breakthrough for BH Physics, showing that rotating solutions to the Einstein field equations were possible to obtain. But, they also helped the construction of powerful *uniqueness theorems* proposed by Werner Israel [7], [8], Brandon Carter [9] and David Robinson [10], [11] at the end of the 60s and beginning of the 70s. Israel proved that the only static, regular outside of the event horizon, asymptotically flat solution of the vacuum (electrovacuum) Einstein equations is the Schwarzschild [7] (Reissner-Nordström [8]) solution. Carter and Robinson then generalised Israel's theorem for the rotating case [9]–[11]. They show that the Kerr (Kerr-Newman) solution is the only stationary, axially symmetry, regular outside of the event horizon, asymptotically flat solution of the vacuum (electrovacuum) Einstein equations. In the end, such powerful theorems simply state that all possible BH solutions of the Einstein-Maxwell equations, under the assumptions/symmetries above, are entirely identical up to three physical quantities: the total mass, angular momentum and electric charge.

The existence of these uniqueness theorems led to the construction of a conjecture that states that the outcome of gravitational collapse in the presence of any energy-matter fields is a Kerr-Newman BH utterly described by its mass, angular momentum and electric charge, which can be measured asymptotically through a Gauss law. All remaining information about the energy-matter fields, commonly denoted as "hair", is lost behind the event horizon. Such conjecture is known as the *no-hair conjecture* [12]. Even though this conjecture involves the dynamical endpoint of gravitational collapse and not simply the existence of a stationary BH with a generic energy-matter field, many researchers found motivation in the above conjecture to try to find stationary BH solutions with either new global charges (*primary* hair) or new nontrivial fields – that can be related to standard global charges (*secondary* hair) – which are not measured through a Gauss law. With such motivation, throughout the 80s and 90s, the scientific community found several hairy BH solutions, most of them with nonlinear energy-matter fields [13]–[15]. However, it appeared impossible to find physically well-motivated hairy BHs for one of the simplest energy-matter sources: the scalar field – a more complete review than the one present here can be found in [16].

The study of scalar fields is well motivated. In Cosmology, several dark energy/matter models use scalar fields. In Particle Physics, researchers at the Large Hadron Collider in CERN found, in 2012, a fundamental scalar field in Nature by experimental observation of a scalar particle known as the Higgs boson [17], [18]. Furthermore, since it is possible to model the canonical scalar field as perfect fluid with some equation of state, they can act as a proxy of real matter [19]. Given the motivations for studying scalar fields, it makes sense to ask if one can obtain stationary, axially symmetric, regular on and outside the horizon and asymptotically flat hairy BHs with a scalar field. As mentioned in the previous paragraph, the answer appears to be no, and the main reason for the negative answer is related to the no-hair theorem developed by Jacob Bekenstein in 1972 – *Bekenstein's theorem* [20], [21].

## 1.1 Bekenstein's Theorem

Bekenstein's theorem states that, under some generic assumptions, a BH cannot support a surrounding scalar field. This can be shown by considering a stationary, axially symmetric, regular on and outside an event horizon and asymptotically flat BH. Due to the stationarity and axially symmetry, the BH possesses two Killing vectors fields that can be written, in a spherical type coordinate system, as $(t, r, \theta, \varphi)$, $t^\mu\partial_\mu = \partial_t$ and $\varphi^\mu\partial_\mu = \partial_\varphi$. Let us also consider the following assumptions [16],

1. *The scalar field is minimally coupled with Einstein's gravity.* The corresponding action reads,
$$\mathcal{S} = \int d^4x\sqrt{-g}\left(\frac{R}{16\pi} - \frac{1}{2}\partial_\mu\Psi\partial^\mu\Psi - V(\Psi)\right) , \tag{1.2}$$
where $\Psi$ is the scalar field, and $V(\Psi)$ is the scalar field potential. The field equations can be obtained by varying the action with respect to the metric, $g_{\mu\nu}$, and the scalar field, $\Psi$. The former yields Equation 1.1 with an energy-momentum tensor associated with the scalar field, and the latter yields the Klein-Gordon (KG) equation,
$$\Box\Psi - V'(\Psi) = 0 , \tag{1.3}$$
where $\Box \equiv \frac{1}{\sqrt{-g}}\partial_\mu\left[\sqrt{-g}g^{\mu\nu}\partial_\nu\right]$ is the d'Alembert operator, and the prime denotes the first derivative with respect to the argument.
2. *The scalar field inherits the spacetime symmetries.* Such assumption implies that $\partial_t\Psi = \partial_\varphi\Psi = 0$.
3. *The scalar field potential obeys* $\Psi V'(\Psi) \geq 0$ *everywhere.*

With everything defined, we can now start to do the proof. Let us multiply the KG equation, Equation 1.3, by $\Psi$ and integrate it over all spacetime outside the BH. One finds the following equation,
$$-\int d^4x\sqrt{-g}\left[\partial_\mu\Psi\partial^\mu\Psi + \Psi V'(\Psi)\right] + \int_{\mathcal{H}} d^3\sigma\, n^\mu\Psi\partial_\mu\Psi = 0 , \tag{1.4}$$
where the last term is a boundary term computed at the horizon, $\mathcal{H}$. An additional boundary term calculated at infinity appears on the result above, but it vanishes to ensure asymptotic flatness. If one looks closely at the horizon's boundary term, one will see that this term also vanishes. Such is true because the event horizon is a Killing horizon due to the stationarity and asymptotic flatness of the problem [22]. Thus, the normal vector to the horizon, $n^\mu$, is a linear combination of the Killing vector fields, and since, following Assumption 2 above, the scalar field inherits the spacetime symmetries, $n^\mu\partial_\mu\Psi = 0$. Therefore, the above equation simplifies to,
$$-\int d^4x\sqrt{-g}\left[\partial_\mu\Psi\partial^\mu\Psi + \Psi V'(\Psi)\right] = 0 . \tag{1.5}$$

We shall see now that this result implies that the scalar field must vanish everywhere. For that we evoke firstly Assumption 3, $\Psi V'(\Psi) \geq 0$. This assumption ensures that the second term of the integration, $\Psi V'(\Psi)$, is always non-negative. By evoking Assumption 2 again, one can show that the first term of the integration is non-negative because, due to the inherited symmetries of the scalar field, its gradient is either null or spacelike, thus $\partial_\mu \Psi \partial^\mu \Psi \geq 0$. Therefore, since both terms in the integration are non-negative, Equation 1.5 holds iff $\Psi = 0$. This result shows that a rotating, stationary, axially symmetric, regular on and outside of an event horizon, and asymptotically flat BH cannot support a scalar field.

Bekenstein's theorem was and still is an influential theorem regarding the possibility of having BHs with a scalar field. However, a theorem is as good as its assumptions. If we violate one of the theorem's assumptions, we can find that the theorem's result may or may not still hold. In the case of the theorem above, we shall see that the violation of some of the above assumptions can lead to the breach of the theorem's result, and one will find rotating, stationary, axially symmetric, regular on and outside of an event horizon, and asymptotically flat BHs that can support a scalar field.

### 1.1.1 Violation of Assumption 1

The violation of Assumption 1 implies that the scalar field is now non-minimally coupled with Einstein's gravity. Such non-minimally coupling leads to new equations of motion, different from those presented before, Equation 1.1 and Equation 1.3. In general, the new scalar field equation will have an explicit dependence on the curvature, meaning that the scalar field is strongly correlated with the geometry, instead of only being affected by the spacetime through the metric. Hence, the scalar field is now part of the gravitational interaction instead of being considered only as an energy-matter field. However, such distinction is conformal-frame dependent.

A good starting point to study the violation of Assumption 1 is to consider the Brans-Dicke theory [23]. This theory is a scalar-tensor theory where the scalar field is non-minimally coupled with Einstein's gravity, and its action reads, in the Jordan frame,

$$\mathcal{S}^J_{BD} = \int d^4x \sqrt{-\hat{g}} \left[ \frac{\varphi \hat{R}}{16\pi} - \frac{\omega_0}{\varphi} \hat{\nabla}_\mu \varphi \hat{\nabla}^\mu \varphi + \mathcal{L}_m(\hat{g}_{\mu\nu}, \Psi_m) \right] , \tag{1.6}$$

where $\varphi$ is the scalar field and plays the role of a spacetime varying Newton's constant. $\omega_0$ is the the scalar field coupling constant. The Brans-Dicke metric, $\hat{g}$ is used to obtain the Ricci scalar, $\hat{R}$ and the covariant derivative, $\hat{\nabla}$, and it is minimally coupled to the matter field, $\Psi_m$, described by the matter Lagrangian density, $\mathcal{L}_m$.

Within this theory, Hawking showed that the regular BHs (in electrovacuum) are the same as those in GR, and none of them possesses scalar hair [24]. To illustrate this result, we must write the above action in the Einstein frame. To do so, we must perform the following conformal transformation of the metric and, for clarity, a redefinition of the scalar field,

$$g_{\mu\nu} = \varphi \hat{g}_{\mu\nu} \; , \;\; \Phi = \sqrt{\frac{2\omega_0 + 3}{4}} \ln \varphi \; . \tag{1.7}$$

The new action reads now,

$$\mathcal{S}^{E}_{BD} = \int d^4x\sqrt{-g}\left[\frac{R}{16\pi} - \frac{1}{2}\nabla_\mu\Phi\nabla^\mu\Phi + \mathcal{L}_m\left(\frac{g_{\mu\nu}}{\varphi}, \Psi_m\right)\right] , \tag{1.8}$$

where the Ricci scalar, $R$, and the covariant derivative, $\nabla$, are computed through the new metric $g_{\mu\nu}$. In this frame, Newton's constant is (locally) constant. However, the masses of particles vary with $\varphi^{-1/2}$.

We can compute the scalar field equation from this action, and obtain a similar equation to the Klein-Gordon equation mentioned in the previous section, Equation 1.3, with $V'(\Psi) = 0$. However, the right-hand side of that scalar field equation will not vanish but instead be sourced by the trace of the energy-momentum tensor associated with the matter fields. This way, if the trace of the energy-momentum tensor vanishes, we get Equation 1.3 with $V'(\Psi) = 0$. Therefore, by considering vacuum or only electromagnetic fields such that the trace of the energy-momentum tensor vanishes, we can use the same logic (and the remaining assumptions) used by Bekenstein in his theorem to show that the scalar field must vanish outside of a horizon. Hence, BHs in electrovacuum Brans-Dicke theory do not have scalar hair.

Another example that violates Assumption 1 is the class composed of Horndeski theories [25]. This class represent the most general scalar-tensor theory of gravity in four dimensions that contains only up to second-order derivatives of the scalar field in its action that generates up to second-order field equations to avoid Ostrogradsky instabilities [26]. The action of this theory reads,

$$\begin{aligned}\mathcal{S} = \int d^4x\sqrt{-g}\,\Big\{&G_2(\Phi,X) - G_3(\Phi,X)\Box\Phi + G_4(\Phi,X)R + G_{4,X}\left[(\Box\Phi)^2 - \left(\nabla_\mu\nabla_\nu\Phi\right)^2\right]\\ &+G_5(\Phi,X)G_{\mu\nu}\nabla^\mu\nabla^\nu\Phi - \frac{G_{5,X}}{6}\left[(\Box\Phi)^3 - 3\Box\Phi\left(\nabla_\mu\nabla_\nu\Phi\right)^2 + 2\left(\nabla_\mu\nabla_\nu\Phi\right)^3\right]\Big\} ,\end{aligned} \tag{1.9}$$

where $G_i|_{i=2,3,4,5}$ are generic functions of the scalar field $\Phi$ and $X \equiv -\frac{1}{2}\nabla_\mu\Phi\nabla^\mu\Phi$. $G_{i,X}$ represents the derivative of the functions $G_i$ with respect to $X$, and $G_{\mu\nu}$ is the Einstein tensor. Within this theory, by specifying the several functions $G_i$ in the right way, one can obtain, *e.g.* GR and Brans-Dicke theory.

Within the class of Horndeski theories, an interesting subset of theories are the ones that possess shift-symmetry, *i.e.* the scalar field is invariant under $\Phi \to \Phi + \text{const}$. These can be obtained through the above action by removing the scalar field dependency of the $G_i$ functions [27], [28]. Due to the shift-symmetry, there is a Noether current, $J^\mu$, that it conserved, $D_\mu J^\mu = 0$. Additionally, we can write the scalar field equation through such conservation.

For shift-symmetric Horndeski theories, a no-hair theorem has developed by Hui and Nicolis [29] in 2012, where they show that static and spherically symmetric (but not necessarily asymptotically flat) BHs can not have scalar hair. Their proof reads as follows: first, we start by writing down an ansatz for the static and spherically symmetric spacetime,

$$ds^2 = -f(\rho)dt^2 + f^{-1}(\rho)d\rho^2 + r^2(\rho)d\Omega^2 , \tag{1.10}$$

where $d\Omega^2$ is the metric of the unit 2-sphere. Then, we impose that the scalar field has the same symmetries as the spacetime (similarly to Assumption 2 above); thus, the scalar field shall have only a radial dependency, $\Phi = \Phi(\rho)$. This means that the 4-current only has a radial component $J^\rho$; hence, its norm is given by $J_\mu J^\mu = (J^\rho)^2/f$. If we compute the 4-current's norm at the horizon, where $f \longrightarrow 0$, it diverges, which is an undesired behaviour since any physical quantity must be well-behaved on and outside the horizon. Therefore, to avoid the undesired behaviour, $J^\rho = 0$ on the horizon. The next step is to integrate the scalar field equation, $D_\mu J^\mu = 0$. This computation yields $r(\rho)^2 J^\rho = \text{const}$. Since, on the horizon, $r(\rho)$ is finite (neither infinite nor zero) because it corresponds to the measurement of the area of constant-$\rho$ spheres, then $J^\rho$ must vanish everywhere. Finally, one only has to argue that $J^\rho = 0$ corresponds to $\Phi = \text{const}$ on all spacetime, and, therefore, the metric will satisfy the vacuum Einstein equations (assuming that $G_4(0) = 1$).

From both examples given, we saw that the violation of Assumption 1 did not automatically imply that BHs could have scalar hair within the studied theories. In the first example, Hawking generalised Bekenstein's theorem for the electrovacuum Brans-Dicke theory using a similar method to Bekenstein's. In the second one, Hui and Nicolis provided a theorem saying that static and spherically symmetric BHs can not have hair in shift-symmetric Horndeski theories. However, in this latter example, Sotiriou and Zhou found a loophole in the Hui-Nicolis' no-hair theorem that led them to find hairy BHs in shift-symmetric Horndeski theories [27], [28].

The loophole dwells on the last step mentioned in the previous paragraph, mainly that $J^\rho = 0$ corresponds to $\Phi = \text{const}$ everywhere. The radial component of the 4-current for a static and spherically symmetric BH – *cf.* Equation 1.10 – can be written as,

$$J^\rho = f \partial_\rho \Phi \, F(\partial_\rho \Phi; g, \partial_\rho g, \partial_\rho^2 g) \,, \tag{1.11}$$

where $F$ is an unknown function that asymptotes to a non-zero constant at spatial infinity. This is a minimal requirement for a theory with a standard canonical kinetic term in the weak field regime. Intuitively, if $J^\rho = 0$ everywhere, then $\partial_\rho \Phi = 0$ everywhere as well. But, what if we specify the $G_i$ function on the action such that one or more terms of the $F$ function cancels out $\partial_\rho \Phi$, leading to terms that are $\partial_\rho \Phi$ independency on $J^\rho$? In this case, $\partial_\rho \Phi = 0$ is no longer a solution of $J^\rho = 0$.

Sotiriou and Zhou followed the same idea and found that there is only one way, within shift-symmetric Horndeski theories, to circumvent Hui-Nicolis' no-hair theorem using the above logic: the scalar field $\Phi$ must be linearly coupled with the Gauss-Bonnet (GB) quadratic invariant $R^2_{\text{GB}} \equiv R^2 - 4R_{\mu\nu}R^{\mu\nu} + R_{\mu\nu\alpha\beta}R^{\mu\nu\alpha\beta}$.

To obtain the linear coupling between the scalar field and the GB invariant, one can specify the $G_i$ functions in the following way [28], [30], $G_2 = G_3 = G_4 = 0$ and $G_5 = -4\alpha \ln |X|$, where $\alpha$ is the coupling constant. The action, Equation 1.9, simplifies to

$$\mathcal{S} = \int d^4x \sqrt{-g} \left( \frac{R}{16\pi} - \frac{1}{2} \nabla_\mu \Phi \nabla^\mu \Phi + \alpha \Phi R^2_{\text{GB}} \right) \,. \tag{1.12}$$

After circumventing Hui-Nicolis' theorem, Sotiriou and Zhou obtained static, spherically symmetric solutions with nontrivial scalar hair of the above action, effectively showing that it is possible to acquire hairy solutions within shift-symmetric Horndeski theories. They also found that these BHs have a minimum size which is set by the length scale associated with the coupling constant, $\alpha$. Furthermore, they found that the hairy solutions do not deviate much from the Schwarzschild BH.

The natural step forward is to analyse the possibility of obtaining the rotation generalisation of the Sotiriou-Zhou's solutions, *i.e.* stationary, axially symmetric, regular on and outside of the horizon, asymptotically flat solutions of Equation 1.12. Since these solutions are not static and spherically symmetric, they fall outside the scope of the no-hair theorem of Hui and Nicolis. Therefore there should not be any problem to find rotating solutions. Indeed, it is possible to construct them. Such construction was reported some years after the work done by Sotiriou and Zhou [31]. An analysis of these rotating solutions and a study of their physical properties are presented in chapter 6 of this thesis.

### 1.1.2 Violation of Assumption 2

A different avenue to circumvent Bekenstein's theorem and obtain BHs with scalar hair without needing to go to modified theories of gravity (in other words, still within GR) is to violate Assumption 2. This particular assumption is quite natural and straightforward to impose, given the symmetries of the problem, but it is not essential. The essential part can be understood by looking at the Einstein equations. Suppose a spacetime has a given symmetry. Then, through the Einstein equations, one sees that the energy-momentum tensor must have the same symmetry as the spacetime, but not the matter field itself. This opens a possibility to consider a scalar field that does not have the same symmetries of the BH, but its energy-momentum tensor has. To achieve that, we allow the scalar field to be complex and to possess a harmonic time and axial dependency, $\Psi \propto e^{-i\omega t}e^{im\varphi}$, where $\omega$ is the scalar field frequency and $m \in \mathbb{Z}$ is the azimuthal harmonic index. This form of $\Psi$ allows the energy-momentum tensor to be time- and axial-independent, even though the scalar field is not.

One particular example of this type of solutions, albeit not BHs, are *Boson Stars* (BSs) [32], [33], which were first shown to exist back in the 60s by Kaup [34] and by Ruffini and Bonazzola [35]. They are self-gravitating, solitonic-like exotic stars compose only by a scalar field which has a potential of the form $V(\Psi) = \mu^2\Psi^*\Psi + \ldots$, where $\mu$ represents the mass of the field, the asterisk indicates the complex conjugate, and the ellipsis correspond to higher-order terms on the scalar field (such as the terms describing the self-interactions of the field). They can also be though has a self-gravitating Bose-Einstein condensate.

Since it has been shown that these exotic stars can be obtained, albeit only numerically, one can ask if they have a BH generalisation. In other words, can we have a BH with a surrounding scalar field such that we get a hairy BH with a nontrivial scalar field configuration? The short answer is it depends on the symmetries/assumptions of the spacetime and the scalar field. On the one hand, for the case where we consider that the scalar field only has a harmonic time dependency, $\Psi \propto e^{-i\omega t}$, Peña and Sudarsky showed that static, spherically symmetric, regular on and outside the horizon, and asymptotically flat BHs with matter fields that obey the weak energy condition could not have nontrivial scalar field hair with an arbitrary potential [36].

On the other hand, if we consider a stationary, axially symmetric, regular on and outside the horizon, asymptotically flat BH surrounded by a harmonic time and axial dependent massive complex scalar field, it will not trivialise as in the previous case. Hence, one can construct BHs with nontrivial scalar hair configurations, effectively circumventing Bekentein's theorem by only violating its second assumption. These hair BHs are known as *Kerr black holes with scalar hair* (KBHsSH), and they were first obtained by Herdeiro and Radu in 2014 [37].

## 1.2 Kerr Black Holes with Scalar Hair

KBHsSH, as mentioned above, are stationary, axially symmetric, regular on and outside the horizon, asymptotically flat solutions of the aforementioned action, Equation 1.2, with a massive complex scalar field,

$$\mathcal{S} = \int d^4x\sqrt{-g}\left[\frac{R}{16\pi} - g^{\alpha\beta}\partial_\alpha\Psi^*\partial_\beta\Psi - \mu^2\Psi^*\Psi\right] . \tag{1.13}$$

Varying the above action with respect to the metric and the scalar field yields the equations of motion,

$$R_{\alpha\beta} - \frac{1}{2}g_{\alpha\beta}R = 8\pi T_{\alpha\beta} , \quad \Box\Psi = \mu^2\Psi , \tag{1.14}$$

$$T_{\alpha\beta} = 2\partial_{(\alpha}\Psi^*\partial_{\beta)}\Psi - g_{\alpha\beta}\left(\partial_\gamma\Psi^*\partial^\gamma\Psi + \mu^2\Psi^*\Psi\right) . \tag{1.15}$$

One of the requirement to construct hairy solutions resides on the scalar field's harmonic time and axial dependence. Hence, we write it, in Boyer-Lindquist coordinates $(t, r, \theta, \varphi)$, as,

$$\Psi = \phi(r,\theta)e^{-i\omega t}e^{im\varphi} . \tag{1.16}$$

This way, we will circumvent Bekenstein's theorem by violating its second assumption. One can easily confirm that the energy-momentum tensor is compatible with the symmetries associated with stationarity and axial symmetry, despite the scalar field itself is not.

Let us start by studying the test field analysis. In this analysis, we linearise all equations of motion. In the end, we must solve the massive Klein-Gordon equation (right equation in Equation 1.14) in the background of a vacuum solution of the Einstein equations, $R_{\mu\nu} = 0$. Since we are interested in stationary, axially symmetric, regular on and outside of the horizon, and asymptotically flat solutions, we choose our vacuum solution to be Kerr's. We now solve the Klein-Gordon equation on a background of a Kerr BH. This can be achieved by performing a separation of variables, splitting the unknown scalar function $\phi(r,\theta)$ into two $R_{lm}(r)$ and $S_{lm}(\theta)$,

$$\Psi = R_{lm}(r)S_{lm}(\theta)e^{-i\omega t}e^{im\varphi} , \tag{1.17}$$

where $S_{lm}(\theta)$ are the spheroidal harmonics, and $R_{lm}(r)$ obeys a radial Teukolsky equation [38]. By solving the Teukolsky equation and imposing a purely ingoing boundary condition at the horizon (in a co-rotating frame), the scalar field should acquire a complex angular frequency, $\omega = \omega_R + i\omega_I$, with $\omega_I < 0$ to represent the decaying amplitude of the scalar field as it falls inside the BH's horizon. This is what happens in the case of a Schwarzschild BH. The scalar field always has a negative imaginary angular frequency, which implies that the amplitude of the scalar field decays in time as it falls towards inside the BH. However, for the Kerr case, the same only happens when $\omega_R > m\Omega_H$, where $\Omega_H$ is the horizon angular velocity. When $\omega_R < m\Omega_H$, $\omega_I$ starts to be positive, which increases the amplitude of the scalar field over time by the extraction of rotational energy of the BH. This phenomenon is known as *superradiance* [16], [39]. At the threshold of superradiance, where,

$$\omega = m\Omega_H \ , \tag{1.18}$$

the imaginary part of the angular frequency vanishes, $\omega_I = 0$. Hod showed that *true bound states* with a real angular frequency are indeed possible to obtain in this threshold, and also constructed them analytically first for the extremal Kerr [40], and a few months later, for rapidly rotating Kerr BHs [41]. These true bound states are known as *scalar clouds*, and there are several studies about them on the literature [37], [40]–[43]. They form a discrete set labelled by 3 'quantum' numbers, $(n, m, l)$, where $n \in \mathbb{N}_0$ corresponds to the node number of $R_{lm}$. If one fixes the 'quantum' numbers, one finds a 1-parameter subspace of the 2-dimensional Kerr parameter space, dubbed *existence line* – *cf.* see blue dotted line in Figure 1.1. This line corresponds to Kerr BHs that can support scalar clouds around them.

With the test field analysis study completed, we can move on to the full nonlinear problem. Unfortunately, the construction of exact solutions of the above equations of motion, Equation 1.14, can only be achieved using numerical methods. Herdeiro and Radu used a professional package called FIDISOL/CADSOL [44]–[46]. This solver uses a Newton-Raphson method and a backward finite different method with self-adaptive grid and consistency order – see Appendix A for more details.

We shall now summarise the numerical approach Herdeiro and Radu used to obtain KBHsSH. This approach will be essential throughout this thesis since it will be identical to the one used in all chapters where we construct new solutions. For the cases where the approach is not the same, we shall point out all the differences compared with this approach.

To obtain KBHsSH, Herdeiro and Radu used Equation 1.16 as an ansatz for the scalar field and the following ansatz for the metric,

$$ds^2 = -e^{2F_0} N dt^2 + e^{2F_1}\left(\frac{dr^2}{N} + r^2 d\theta^2\right) + e^{2F_2} r^2 \sin^2\theta \left(d\varphi - W dt\right)^2 \ , \tag{1.19}$$

where $N \equiv 1 - r_H/r$ and $\{F_i, W\}_{i=0,1,2}$ are ansatz functions that depend only on $r$ and $\theta$. Note that this metric is not written in Boyer-Lindquist (BL) coordinates. The relation between this coordinate system and the BL one, for the Kerr case, can be found in Appendix A.

After choosing the ansatz for both the metric functions and the scalar field, one has to define the boundary conditions of all ansatz functions on the proper limits that best suit the problem at hand. Such boundary conditions will be now summarised.

- *Asymptotically boundary conditions*: The asymptotical flatness property of the solutions that we are searching implies that the asymptotic behaviour of the ansatz functions must be,

$$\lim_{r\to\infty} F_i = \lim_{r\to\infty} W = \lim_{r\to\infty} \phi = 0 \,, \tag{1.20}$$

- *Axial boundary conditions*: The axial symmetry property, together with regularity on the symmetry axis, leads to the following boundary condition on the symmetry axis ($\theta = \{0, \pi\}$),

$$\partial_\theta F_i = \partial_\theta W = \partial_\theta \phi = 0 \,. \tag{1.21}$$

  Moreover, the non-existence of conical singularities additionally implies that, on the symmetry axis,

$$F_1 = F_2 \,. \tag{1.22}$$

  We shall focus only on even parity solutions, thus they are symmetric *w.r.t* the equatorial plane, $\theta = \pi/2$. This way, we only have to solve the equations of motion in the range $0 \le \theta \le \pi/2$ and impose, on the equatorial plane, the following boundary conditions,

$$\partial_\theta F_i = \partial_\theta W = \partial_\theta \phi = 0 \,. \tag{1.23}$$

- *Event horizon boundary conditions*: To simplify the study of the boundary conditions at the event horizon, let us introduce the following radial coordinate transformation, $x = \sqrt{r^2 - r_H^2}$. With this coordinate transformation, we can perform a series expansion of the ansatz functions at the horizon, $x = 0$, and find that,

$$F_i = F_i^{(0)}(\theta) + x^2 F_i^{(2)}(\theta) + \mathcal{O}(x^4) \,, \tag{1.24}$$

$$W = \Omega_H + \mathcal{O}(x^2) \,, \tag{1.25}$$

$$\phi = \phi^{(0)}(\theta) + \mathcal{O}(x^2) \,, \tag{1.26}$$

  With these series expansions, we can naturally impose the following boundary conditions, on the horizon,

$$\partial_x F_i = \partial_x \phi = 0 \,, \quad W = \Omega_H \,. \tag{1.27}$$

With the ansatze and all boundary conditions defined, we can proceed to the numerical integration of the equations of motion. To perform such integration, it is useful to rescale the following key quantities,

$$r \to r\mu \,, \;\; \phi \to \phi\sqrt{4\pi} \,, \;\; \omega \to \omega/\mu \,. \tag{1.28}$$

This removes the dependency of the scalar field mass, $\mu$, from the equations of motion. However, all global quantities will now be express in terms of $\mu$.

In this approach, by expanding the equations of motion, we get a set of five coupled, non-linear, elliptic partial differential equations for the ansatz functions, $\mathcal{F}_a = \{F_0, F_1, F_2, W; \phi\}$.

They are composed of the massive Klein-Gordon equation, Equation 1.3, together with the following combination of the Einstein field equations,

$$E_r^r + E_\theta^\theta - E_\varphi^\varphi - E_t^t = 0 \,, \tag{1.29}$$

$$E_r^r + E_\theta^\theta - E_\varphi^\varphi + E_t^t + 2WE_\varphi^t = 0 \,, \tag{1.30}$$

$$E_r^r + E_\theta^\theta + E_\varphi^\varphi - E_t^t - 2WE_\varphi^t = 0 \,, \tag{1.31}$$

$$E_\varphi^t = 0 \,, \tag{1.32}$$

where $E_{\mu\nu} \equiv R_{\mu\nu} - 1/2 g_{\mu\nu} R - 8\pi T_{\mu\nu}$. The remaining two equations, $E_\theta^r = 0$ and $E_r^r - E_\theta^\theta = 0$ are not solved directly, but are, instead, used as constrains equations to monitor the numerical solution. Typically, these constrain equations are satisfied at the level of the overall numerical accuracy. Since these equations are lengthy and convoluted, we shall not explicitly exhibit them here.

The numerical treatment is as follows. First, one restricts the domain of integration to the region outside of the horizon. Using the aforementioned radial coordinate transformation, $x = \sqrt{r^2 - r_H^2}$, one transforms the radial region of integration from $[r_H, \infty)$ to $[0, \infty)$. Then, by introducing a new radial coordinate transformation, $\bar{x} = x/(x+1)$, one maps the semi-infinite region $[0, \infty)$ to the finite region $[0, 1]$. After this, one discretises the equations on a grid in $\bar{x}$ and $\theta$. Most of the results presented here (and in the following chapters) were obtained on an equidistant grid with $251 \times 30$ points. This grid covers the integration region $0 \leq \bar{x} < 1$ and $0 \leq \theta \leq \pi/2$.

The equations of motion were solved subject to the boundary conditions introduced above by using the already mentioned FIDISOL/CADSOL package. Such package provides its users with an error estimate for each unknown function, and, for KBHsSH, the maximal error found for each ansatz function was in the order of $10^{-3}$. Different numerical tests can be performed, such as verifying the Smarr formula – *cf.* see Equation 1.36 below. For this particular test, Herdeiro and Radu found an error estimate of the same order.

By specifying the input parameters of the problem, *i.e.*, the angular frequency of the scalar field, $\omega$; the azimuthal harmonic index, $m$; the radial coordinate of the horizon, $r_H$; and the number of modes, $n$, it is possible to obtain one numerical solution corresponding to a KBHSH. With that solution, one can extract all kind of physical quantities.

Most physical quantities of interest can be obtained through the metric functions at the event horizon or spacial infinity. At the horizon, one computes the Hawking temperature, $T_H$, and horizon area, $A_H$, as [37],

$$T_H = \frac{1}{4\pi r_H} e^{(F_0 - F_1)|_{r_H}} \,, \qquad A_H = 2\pi r_H^2 \int_0^\pi d\theta \sin\theta e^{(F_1 + F_2)|_{r_H}} \,. \tag{1.33}$$

The entropy follows from the Bekenstein-Hawking formula, $S = A_H/4$, and the horizon angular velocity is found evaluating the ansatz function $W$ at the event horizon, $\Omega_H = W|_{r_H}$.

At spatial infinity, on the other hand, the Arnowitt, Deser and Misner [47] (ADM) mass, $M$, and total angular momentum, $J$, are computed from the asymptotic behaviour of the

metric functions:

$$g_{tt} = -e^{2F_0}N + e^{2F_2}W^2r^2\sin^2\theta \;\; \rightarrow \;\; -1 + \frac{2M}{r} + \dots \,, \tag{1.34}$$

$$g_{\phi t} = -e^{2F_2}Wr^2\sin^2\theta \;\; \rightarrow \;\; -\frac{2J}{r}\sin^2\theta + \dots \,. \tag{1.35}$$

The above quantities, together with two new ones, are related by a Smarr-type formula [48],

$$M = 2T_H S + 2\Omega_H (J - J_\Psi) + M_\Psi \,, \tag{1.36}$$

where the new quantities are the scalar field mass, $M_\Psi$, and angular momentum, $J_\Psi$. They are computed through the energy-momentum tensor of the matter existing on the spacetime outside of the horizon,

$$M_\Psi = -2\int_\Sigma dS_\mu \left(T^\mu_\nu \xi^\nu - \frac{1}{2}T\xi^\mu\right) \,, \quad J_\Psi = \int_\Sigma \left(T^\mu_\nu \eta^\nu - \frac{1}{2}T\eta^\mu\right) \,, \tag{1.37}$$

where $\Sigma$ is a spacelike surface, bounded by the 2-sphere at infinity, $S^2_\infty$, and the spatial section of the horizon, $\mathcal{H}$, and $\xi^\mu$ and $\eta^\mu$ are the timelike and rotational Killing vectors, respectively.

The global $U(1)$ transformation $\Psi \rightarrow e^{i\alpha}\Psi$, where $\alpha$ is a constant, leaves the action, Equation 1.13, invariant; thus it is possible to write a scalar-current [37] $j^\mu = -i\left(\Psi^*\partial^\mu\Psi - \Psi\partial^\mu\Psi^*\right)$ which is conserved, $D_\mu j^\mu = 0$. With such conserved 4-current, one finds the conserved Noether charge, $Q$, through the following computation,

$$Q = \int_\Sigma j^t = 4\pi\int_0^\infty dr \int_0^\pi d\theta\, r^2 \sin\theta\, e^{-F_0+2F_1+F_2}\frac{\omega - mW}{N}\phi^2 \,. \tag{1.38}$$

This physical quantity encodes the information about the scalar field and it is not independent. In fact, it is related with the angular momentum of the scalar field, $J_\Psi$, through,

$$J_\Psi = mQ \,. \tag{1.39}$$

Since this relation is always true for all KBHsSH, it is convenient to introduced the dimensionless parameter, $q$, that quantifies how hairy a given BH is,

$$q \equiv \frac{J_\Psi}{J} = \frac{mQ}{J} \,. \tag{1.40}$$

If $q = 0$ the BH has no scalar hair; this is the Kerr BH limit corresponding to the existence line. On the other end of the spectrum, if $q = 1$, all angular momentum is in the scalar hair; in fact, this is no longer a BH but rather an everywhere regular solitonic solution, corresponding to the *boson star* limit. In this case, all angular momentum is quantised in terms of the Noether charge [49]–[51]. In between, when $0 < q < 1$, hairy BHs exist.

So far, since all input parameters of the problem were fixed, only one solution was constructed. However, one can not extract any conclusions regarding the overall behaviour of the physical quantities of these solutions, just from one of them. An analysis of a large set of solutions is required. For that, we only fix some input parameters and vary the others. In the particular case of the work developed by Herdeiro and Radu, they fixed the number of nodes, $n = 0$, which corresponds to fundamental solutions (solutions with a higher number of nodes were constructed in [52], [53]); and the azimuthal harmonic index $m = 1$ (solutions with higher harmonic index were obtained in [54] and they are reported in chapter 3 of this thesis). The input parameters that they varied were the angular frequency of the scalar field, $\omega$, and the radial coordinate of the horizon, $r_H$.

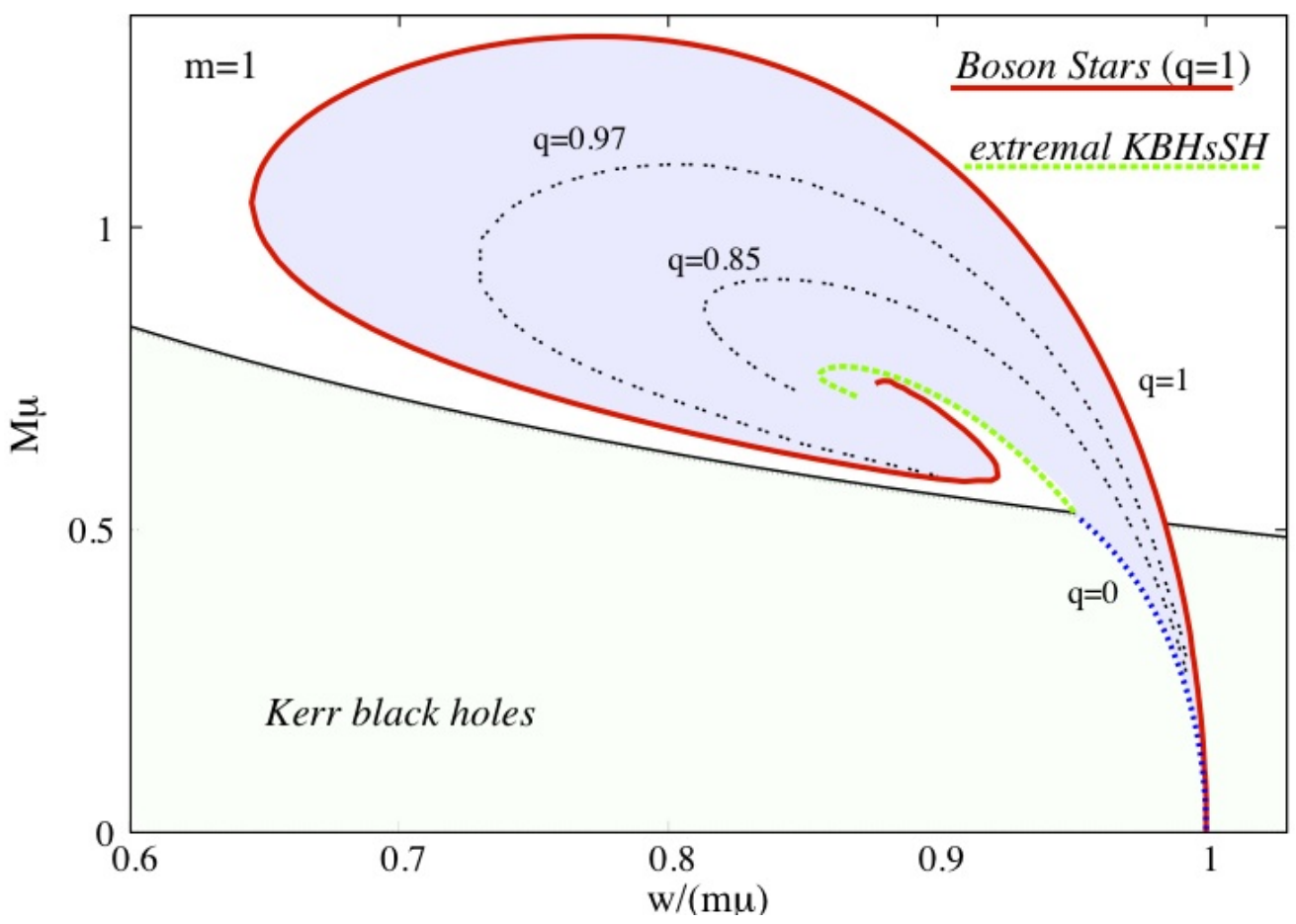

**Figure 1.1:** Domain of existence of KBHsSH in a mass, $M\mu$, *versus*, horizon angular velocity $\Omega_H/\mu = \omega/(m\mu)$ with $m = 1$. Also represented is the domain of existence of Kerr BH in the region below the solid black line, which corresponds to extremal Kerr BHs. Adapted from [16].

One of the most significant advantages of the numerical approach discussed above, and the use of the FIDISOL/CADSOL package, is the possibility to obtain new numerical solutions systematically. Since the numerical package uses a Newton-Raphson method, one can obtain new solutions through a neighbouring first solution via changing the input parameters. This implies that, after constructing one first solution, we can get its neighbour, and then its neighbour's neighbour, and so on. Such a systematic approach lets the user construct a large set of thousand or tens of thousand of numerical solutions through a single first solution. A disadvantage of this approach resides in how to construct the first solution. In some theories, this construction can be tricky. However, having a good understanding of the problem at hand, performing test field analyses, and slow rotation approximations often make it possible to get the first solution.

After constructing a large set of numerical solutions, they will form a region where it is possible to find them. Such a region defines the domain of existence that can be extrapolated into the continuum. For KBHsSH, the domain of existence is shown in Figure 1.1. One can see that the hairy solutions live on a region bounded by three lines,

- The *boson star line* ($q = 1$) – red solid line. In this limit, the horizon of the BH vanishes and we only have a rotating self-gravitating scalar field, *i.e.* a BS. Since the horizon vanishes, the horizon radius and area vanishes as well, $r_H \to 0$ and $A_H \to 0$.
- The *extremal KBHsSH line* – green dashed line. Solutions in this limit possess a vanishing Hawking temperature, $T_H \to 0$. Due to the choice of coordinates for the metric, in this limit, the radial coordinate of the horizon also vanishes, $r_H \to 0$, as in the BS line, but the horizon area does not.

- The *existence line* ($q = 0$) – blue dotted line. These solutions are vacuum Kerr BHs that can support scalar clouds [37], [40]–[43] as discussed before.

This class of hairy BHs is continuously connected to Kerr BHs (existence line) and BSs (BS line), as it is possible to verify through Figure 1.1. Hence, it is possible to predict that the physical properties and phenomenology of KBHsSH will be a mix between the ones already known for Kerr BHs and BSs. It may be expected as well that if we consider a hairy solution that is close to the existence (BS) line, such solution will have similar physical properties and phenomenology to Kerr BHs (BSs).

A variety of physical properties and phenomenology studies for KBHsSH has already been done following the years after their discovery. Mainly, shadows and lensing [55]–[61], X-ray spectrum observables, such as the Iron K-$\alpha$ line [62], horizon geometry [63], [64] – *cf.* chapter 2 –, equatorial timelike circular orbits [65], quasi-periodic oscillations [66], between others. From these studies, it is possible to confirm that KBHsSH can have physical properties and phenomenologies that range from the ones found for BSs to the ones known for Kerr BHs, as predicted before.

## 1.3 Motivation and Outline

We currently live in a golden age of observational gravitational physics. The first detection of gravitational waves in 2015 by the LIGO collaboration [67], and subsequent detections (together with the help of Virgo and KAGRA) [68]–[70], and the first observation of the shadow and lensing of light around the supermassive BH at the centre of the galaxy M87 in 2019 [71]–[76], and subsequent work about the polarisation of light [77], opened the door for the realisation of tests of GR at the strong gravity level.

In particular, it starts to be possible to test a hypothesis that is a direct consequence of the no-hair conjecture – the *Kerr Hypothesis*. This hypothesis states that the Kerr geometry describes all astrophysical BHs in isolation in the Universe. A possible avenue to test this hypothesis is to analyse the juxtaposition between the experimental data obtained by on-going experiments, such as the ones mentioned in the previous paragraph, and the phenomenology of Kerr BHs.

A different avenue is to study the phenomenology of BHs in either GR with additional fields (such as scalar fields, vector fields or others fields), or in modified theories of gravity (such as scalar-tensor theories), and then analyse the juxtaposition of the phenomenology and the experimental data. This type of study can also constrain or rule out certain fields within GR or modified theories of gravity. However, to perform phenomenology studies of such new BHs, one must first construct them and study their physical properties. Here resides the motivation of this thesis. We shall use the numerical approach discussed above to obtain solutions in different theories, mainly GR with scalar fields with various scalar potentials (complex-Einstein-Klein-Gordon) and shift-symmetric Horndeski theory, and analyse their physical properties.

This thesis shall be organised in the following way. In chapter 2, we shall discuss the horizon geometry of Kerr BHs and KBHsSH. In particular, we shall analyse the isometric embedding of the horizon of both classes of BHs in Euclidean 3-space and compute and compare several horizon quantities. This chapter will be based on the work developed in [64].

In chapter 3, we will obtain and study KBHsSH with higher azimuthal harmonic indexes, mainly $m = 2$ and $m = 3$. Here, an overview of the domain of existence and phase space is done, together with the study of some physical quantities and the horizon geometry. The results reported in this chapter are based on [54].

The following two chapters introduce a different potential for the scalar field, namely, a Quantum Chromodynamics (QCD) axion-like potential. With this new potential, it is possible to obtain new stars and BHs. In chapter 4, we construct and analyse bosonic stars composed entirely of an axionic scalar field. An overview of the domain of existence is done together with an analysis of some physical quantities. An investigation of ergoregions, timelike and lightlike circular orbits and energy conditions is also performed. In the following chapter (chapter 5), we study the BH generalisation of the axionic bosonic stars. The domain of existence is shown. The usual physical quantities are computed and examined. The horizon geometry of this class of BHs is analysed together with the ergoregions and timelike and lightlike circular orbits. The former chapter is based on the work done in [78], whereas the latter chapter reports the results presented in [79].

In chapter 6, we provide both spherical and rotating BHs solutions to the shift-symmetric Horndeski theory, where a real scalar field is non-minimally coupled to the GB quadratic curvature invariant, as discussed before. We present the domain of existence where all solutions fall and inspect some physical quantities. We also show studies of the ergoregions and horizon geometry, as well as, the orbital frequency at the innermost stable circular orbit and light-rings. This chapter presents the work developed in [31].

In chapter 7, we study equatorial circular orbits around very generic compact objects. We show that the radial stability of light-rings determines the radial stability of timelike circular orbits around them, and provide several examples where one can see such a result. We also study the efficiency associated with the amount of gravitational energy converted into radiation as a timelike particle falls towards the compact object. The results provided in this chapter are based on [80].

Finally, in chapter 8 we give some final comments and remarks about the work done throughout this thesis.

CHAPTER 2

# Horizon Geometry for Kerr Black Holes with Synchronised Hair

## 2.1 Introduction

For slow rotation, the spatial sections of the Kerr BH [5] event horizon[1] are oblate spheroids. These are closed 2-surfaces, with everywhere positive Gaussian curvature, and with a proper size for the equatorial geodesic circle larger than for meridian ones. This fact is intuitive: rotation typically deforms spherical objects into oblate ones, as it is familiar for the (rotating) Earth. Everywhere positively curved 2-surfaces can be *globally* embedded in Euclidean 3-space, $\mathbb{E}^3$, and the embedding is rigid ($i.e.$ unique up to rigid rotations - see $e.g.$ [81] and references therein). Thus, the horizon geometry of slowly spinning Kerr BHs can be embedded in $\mathbb{E}^3$ and this embedding provides a rigorous and intuitive tool to visualise it – see [81]–[83] and subsection 2.2.1 below.

For sufficiently fast rotation, however, the Kerr horizon geometry becomes more exotic. Smarr [82] first observed that for a dimensionless spin $j > \sqrt{3}/2 \equiv j^{(S)}$ – hereafter the *Smarr point* –, the Gaussian curvature of the horizon becomes negative in a vicinity of the poles. In this regime, an isometric embedding of the Kerr horizon geometry in $\mathbb{E}^3$ is no longer possible. In fact, such embedding fails even *locally* at the negatively curved axi-symmetry fixed points (the poles). The Kerr horizon may still be embedded in other manifolds, such as Euclidean 4-space [83], hyperbolic 3-space [81] or 3-dimensional Minkowski spacetime (see $e.g.$ [84], [85]); these do not provide, however, a simple and intuitive visualisation of the Kerr horizon geometry.

In this chapter, we shall study the horizon geometry of a generalisation of the Kerr solution and, in particular its embedding in $\mathbb{E}^3$. The generalised Kerr BHs we shall be considering are Kerr BHs with synchronised scalar hair [37], [86]. As mentioned on the Introduction of this thesis, chapter 1, this is a family of stationary, non-singular (on and outside) the event horizon, asymptotically flat BH solutions of Einstein-Klein-Gordon theory, that interpolates between Kerr BHs (the "bald" limit) and horizonless, everywhere regular BSs [32] (the solitonic limit). The generic solution possesses an event horizon surrounded by a non-trivial scalar field configuration. It circumvents well known no-hair theorems (see $e.g$ [16], [87]–[89]) due to a synchronisation of the angular velocity of the horizon, $\Omega_H$ with the angular phase velocity of the scalar field (see also [90], [91]). Various other asymptotically flat, four dimensional "hairy" BHs akin to these solutions have been constructed in [92]–[95].

[1]Hereafter we refer to the geometry of the horizon spatial sections as the "horizon geometry", for simplicity.

These hairy BHs possess, at the level of their event horizon, only the same two conserved charges of a Kerr BH: the horizon mass, $M_H$ and the horizon angular momentum, $J_H$, that can be computed as Komar integrals [96]. It has been previously observed that the corresponding dimensionless spin $j_H \equiv J_H/M_H^2$ can exceed unity [63], unlike that of the vacuum Kerr BH. On the other hand, a geometrically meaningful horizon linear velocity $v_H$ ($cf.$ subsection 2.2.1) never exceeds unity ($i.e.$ the velocity of light), for either vacuum Kerr or the hairy BHs, and it only attains unity for extremal vacuum Kerr [63]. This suggests that, by virtue of the coupling between the horizon and the surrounding "hair", these BHs present a different ability to sustain angular momentum; heuristically, they have a different (higher) moment of inertia. Such observation makes these hairy BHs an interesting laboratory to test the relation between rotation, angular momentum and horizon deformability.

Here, we shall compare the embedding of these hairy BHs in $\mathbb{E}^3$ with that of the Kerr BH. For the latter, such embedding is only possible if the horizon dimensionless spin $j_H$ (which for the Kerr case equals the total spacetime dimensionless spin, $j$), the sphericity $\mathfrak{s}$ and the horizon linear velocity $v_H$ are smaller than critical values attained at the Smarr point: $j^{\,(\mathrm{S})}, \mathfrak{s}^{(\mathrm{S})}, v_H^{(\mathrm{S})}$. As we shall see, this family of hairy BHs allows us to disentangle the role of the angular momentum and the linear velocity in terms of being compatible with embeddable solutions. Our results indicate that $j^{\,(\mathrm{S})}$ ($v_H^{(\mathrm{S})}$) provides only a minimum (maximum) value below (above) which the Euclidean embedding is guaranteed (impossible). On the other hand, it is the sphericity $\mathfrak{s}^{\,(\mathrm{S})}$, that remains a faithful diagnosis for the existence of the $\mathbb{E}^3$ embedding, throughout the whole family of solutions, below (above) which the Euclidean embedding is guaranteed (impossible). We shall also observe that sufficiently hairy BHs are always embeddable, even if the *total* dimensionless spin, $j$, is larger than unity.

This chapter is based on the work done in [64] and it is organised as follows. In section 2.2 we review the isometric embedding of Kerr BHs in $\mathbb{E}^3$, in particular presenting the threshold sphericity and horizon linear velocity. The corresponding analysis for hairy BHs is performed in section 2.3, detailing the behaviour of the sphericity, dimensionless spin and horizon linear velocity throughout the parameter space. Concluding remarks are presented in section 2.4.

## 2.2 Kerr Black Holes

### 2.2.1 Isometric embedding

The Kerr BH [5] is the unique 4-dimensional, regular on and outside the event horizon, axisymmetric and asymptotically flat solution of vacuum Einstein's gravity. In BL coordinates [97], the metric reads:

$$ds^2 = -\frac{\Delta}{\Sigma}\left(dt - a\sin^2\theta d\varphi\right)^2 + \Sigma\left(\frac{dr^2}{\Delta} + d\theta^2\right) + \frac{\sin^2\theta}{\Sigma}\left[a dt - \left(r^2 + a^2\right) d\varphi\right]^2 \ , \tag{2.1}$$

where $\Delta \equiv r^2 - 2Mr + a^2$ and $\Sigma \equiv r^2 + a^2\cos^2\theta$, in which, $M$ is the ADM mass and $a \equiv J/M$ is the total angular momentum per unit mass.

The isometric embedding of the Kerr horizon in $\mathbb{E}^3$ has been first considered by Smarr [82]. Let us reconstruct the main result. The 2-metric induced in the spatial sections of the event horizon is obtained as a $t$ = const, $r = r_H$ section of Equation 2.1, where $r_H \equiv M + \sqrt{M^2 - a^2}$ is the largest root of $\Delta = 0$. This 2-metric reads:

$$d\sigma^2 = \Sigma_{r_H}\, d\theta^2 + \frac{\sin^2\theta}{\Sigma_{r_H}}\left(r_H^2 + a^2\right)^2 d\varphi^2 \ , \tag{2.2}$$

where $\Sigma_{r_H} \equiv r_H^2 + a^2\cos^2\theta$. The embedding of any $U(1)$ invariant 2-surface in $\mathbb{E}^3$, with the standard Cartesian metric $d\sigma^2 = dX^2 + dY^2 + dZ^2$, may be attempted by using the following embedding functions,

$$X + iY = f(\theta)e^{i\varphi} \ , \qquad Z = g(\theta) \ , \tag{2.3}$$

so that,

$$d\sigma^2 = \left(f'(\theta)^2 + g'(\theta)^2\right) d\theta^2 + f(\theta)^2 d\varphi^2 \ , \tag{2.4}$$

where the prime indicates the derivative with respect to $\theta$. Comparing with Equation 2.2, we obtain, for the Kerr horizon,

$$f'(\theta)^2 + g'(\theta)^2 = \Sigma_{r_H} \quad , \quad f(\theta) = \frac{\sin\theta\left(r_H^2 + a^2\right)}{\sqrt{r_H^2 + a^2\cos^2\theta}} \ . \tag{2.5}$$

Thus,

$$g'(\theta) = \frac{\sqrt{h(\theta)}}{\left(r_H^2 + a^2\cos^2\theta\right)^{3/2}} \ , \qquad h(\theta) \equiv (r_H^2 + a^2\cos^2\theta)^4 - (r_H^2 + a^2)^4\cos^2\theta \ . \tag{2.6}$$

In order for the Kerr horizon to be embeddable, both embedding functions have to be real. In the case of $g(\theta)$ this requires $h(\theta) \geqslant 0$. Since $h(0) = 0 = h'(0)$ and $h''(0) = 2\left(r_H^2 + a^2\right)^3\left(r_H^2 - 3a^2\right)$, then $h''(0) \leqslant 0$ iff $|a| \geqslant r_H/\sqrt{3}$. In this region of the parameter space the Kerr horizon fails to be embeddable in $\mathbb{E}^3$ (in the neighbourhood of the poles). In the complementary domain

$$|a| \leqslant \frac{\sqrt{3}}{2}M \ , \tag{2.7}$$

the Kerr horizon is embeddable. The threshold of this inequality $|j^{(S)}| = |a|/M = \sqrt{3}/2$, defines a point for the (absolute value of the) dimensionless spin $j \equiv a/M$, which we shall call the *Smarr point*.

The Gaussian curvature for Equation 2.2 is

$$\mathcal{K} = \frac{\mathcal{R}}{2} = \frac{\left(r_H^2 + a^2\right)\left(r_H^2 - 3a^2\cos^2\theta\right)}{(\Sigma_{r_H})^3} \ , \tag{2.8}$$

where $\mathcal{R}$ is the Ricci scalar of the 2-metric, Equation 2.4. Thus, at the poles, the Gaussian curvature has the same sign as $h''(0)$. Therefore, the Euclidean embedding fails precisely when the Gaussian curvature at the poles becomes negative.

### 2.2.2 Sphericity and horizon linear velocity

We now introduce one geometrical and one physical parameter related to the horizon: the sphericity, measuring how much the horizon intrinsic shape deviates from a round sphere, and the horizon linear velocity [63], providing a *linear* velocity measure of how fast the null geodesic generators of the horizon are being dragged with respect to a static observer at spatial infinity.

We define the *sphericity*, $\mathfrak{s}$, of a $U(1)$ invariant 2-surface as

$$\mathfrak{s} \equiv \frac{L_e}{L_p} \,, \tag{2.9}$$

where $L_e$ is the equatorial proper length and $L_p$ is twice the proper length of a meridian ($\varphi$ = const curve for an azimuthal coordinate adapted to the $U(1)$ symmetry). Observe that the $U(1)$ symmetry guarantees all meridians have the same proper size. The Kerr horizon has a $\mathbb{Z}_2$ symmetry and the equator is the set of fixed points of this symmetry. Then, from Equation 2.2,

$$L_e = \int_0^{2\pi} d\varphi \, \frac{r_H^2 + a^2}{r_H} = 2\pi \frac{r_H^2 + a^2}{r_H} \,, \tag{2.10}$$

$$L_p = 2\int_0^{\pi} d\theta \, \sqrt{r_H^2 + a^2\cos^2\theta} = 4\sqrt{r_H^2 + a^2}\, \mathrm{E}\left(\frac{a^2}{r_H^2 + a^2}\right) \,, \tag{2.11}$$

where $\mathrm{E}(k)$ is the complete elliptic integral of second kind: $\mathrm{E}(k) \equiv \int_0^{\pi/2} d\theta \sqrt{1 - k\sin^2\theta}$ [98]. Thus

$$\mathfrak{s}(j) = \frac{\pi}{2}\frac{\sqrt{2 + 2\sqrt{1-j^2}}}{1 + \sqrt{1-j^2}} \mathrm{E}\left(\frac{j^2}{2 + 2\sqrt{1-j^2}}\right)^{-1} \quad \Rightarrow \quad \mathfrak{s}^{\,(\mathrm{S})} \equiv \mathfrak{s}(j^{\,(\mathrm{S})}) = \frac{\pi}{\sqrt{3}}\mathrm{E}\left(\frac{1}{4}\right)^{-1} \approx 1.23601 \,. \tag{2.12}$$

In Figure 2.1 we exhibit the sphericity for the Kerr horizon as a function of the dimensionless angular momentum, $j$. One can see that the sphericity is always $\mathfrak{s} \geqslant 1$, meaning that the length along the poles is always equal or smaller that the length along the equator. Thus, the Kerr event horizon can be a round sphere (for $\mathfrak{s} = 1$, corresponding to the Schwarzschild limit) or an oblate spheroid (for $\mathfrak{s} > 1$). In the figure we highlight the Smarr point $j^{\,(\mathrm{S})}$; the corresponding sphericity, yielding the maximal value of this quantity for which the Kerr horizon is embeddable in $\mathbb{E}^3$, is $\mathfrak{s}^{\,(\mathrm{S})}$.

The *horizon linear velocity* is defined as [63]:

$$v_H \equiv R\Omega_H \,, \tag{2.13}$$

where $\Omega_H$ is the horizon angular velocity and $R \equiv L_e/2\pi$ is the circumference radius of the equator. For the specific case of the Kerr BH [63], [99],

$$v_H(j) = \frac{j}{1 + \sqrt{1-j^2}} \qquad \Rightarrow \qquad v_H^{(\mathrm{S})} \equiv v_H(j^{\,(\mathrm{S})}) = \frac{1}{\sqrt{3}} \approx 0.57735 \,. \tag{2.14}$$

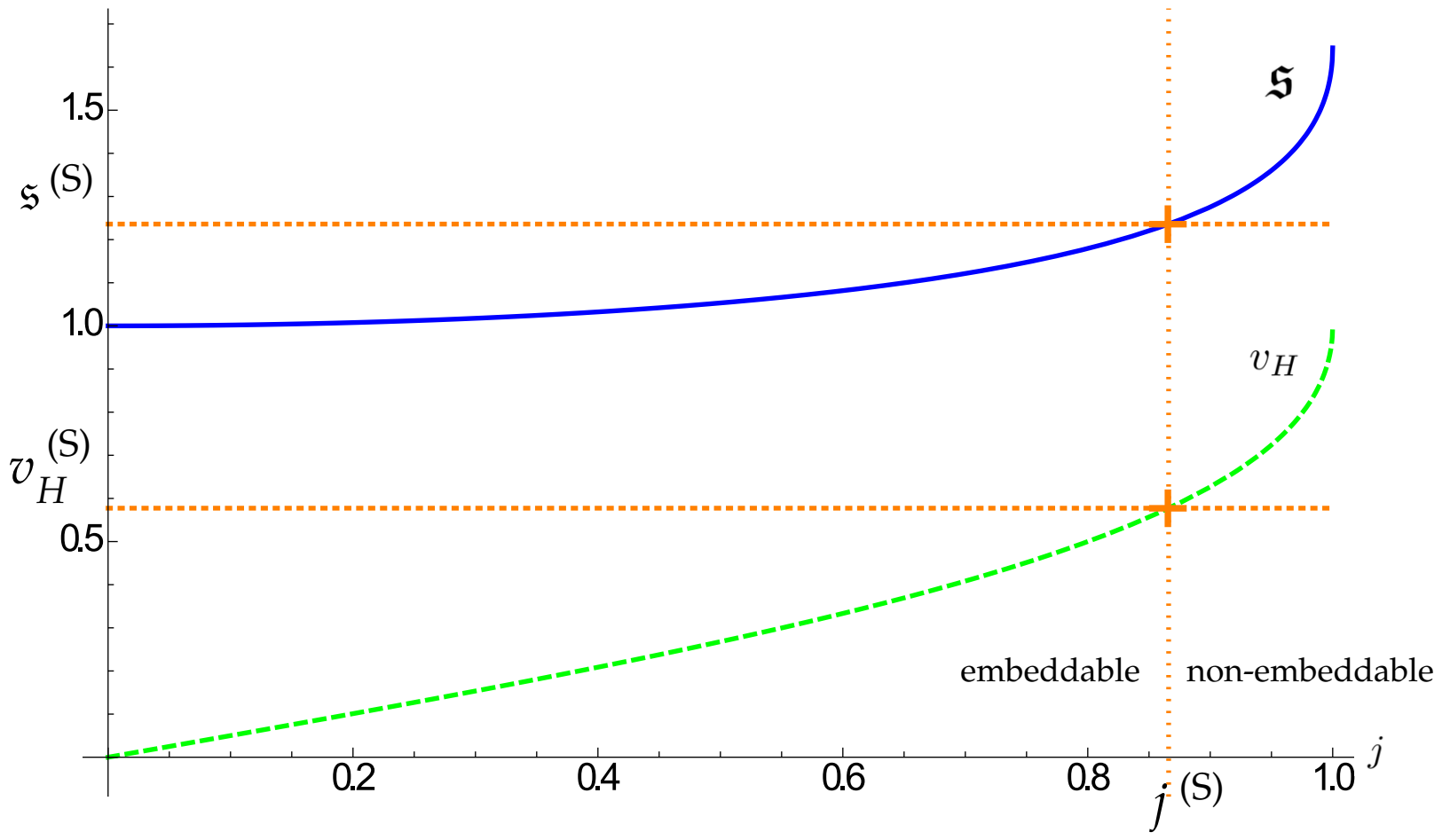

**Figure 2.1:** The sphericity $\mathfrak{s}$ (top solid blue curve) and the horizon linear velocity, $v_H$ (bottom dashed green curve) as a function of the dimensionless angular momentum, $j$, for the Kerr horizon. The vertical line gives the Smarr point dimensionless spin value, $j^{(S)}$.

In Figure 2.1 we also exhibit the horizon linear velocity, $v_H$, as a function of the dimensionless angular momentum, $j$, for the Kerr horizon. This velocity never exceeds the speed of light, $v = 1$, reaching this value in the extremal case. The horizon linear velocity for the Smarr point, $v_H^{(S)}$, is highlighted in the plot.

## 2.3 Kerr BHs with Scalar Hair

KBHsSH [37], as discussed during chapter 1, are 4-dimensional, regular on and outside the event horizon, axisymmetric, asymptotically flat solutions of the (complex-)Einstein-Klein-Gordon action – *cf.* Equation 1.13. These solutions are only known numerically [86] and have been constructed by using the ansatze described in Equation 1.19 for the metric and in Equation 1.16 for the scalar field. The resulting family of solutions depends on three physical quantities: the ADM mass, $M$, the total angular momentum, $J$ and a Noether charge that encodes information about the scalar hair, $Q$, but these last two are not independent – *cf.* Equation 1.39. Of the discrete set of families labelled by the azimuthal harmonic index, $m$, here we shall focus on solutions with $m = 1$.

In Figure 2.2 (left panel) we exhibit a similar plot as shown in Figure 1.1 describing the domain of existence of the solutions in an $M$ *versus* $\Omega_H$ diagram. This domain is bounded by the same three lines as before: the BS line (red solid line); the extreme KBHsSH line (dashed-dotted brown line); and the existence line (dotted blue line) [37], [40]–[43], [100].

The BS line and the extremal KBHsSH line are expected to spiral toward a common (likely singular) point, which however has not yet been reached numerically. It proves convenient to introduce two parameters that quantify the amount of "hair" in the BH:

$$p \equiv \frac{M_\Psi}{M} = 1 - \frac{M_H}{M} , \qquad q \equiv \frac{J_\Psi}{J} = 1 - \frac{J_H}{J} , \tag{2.15}$$

where the right equation is the same as Equation 1.40, and $M_H, J_H$ ($M_\Psi, J_\Psi$) are the horizon (scalar field) mass and angular momentum, that can be computed as Komar integrals on the horizon (volume integrals outside the horizon) - see, *e.g.* [86].

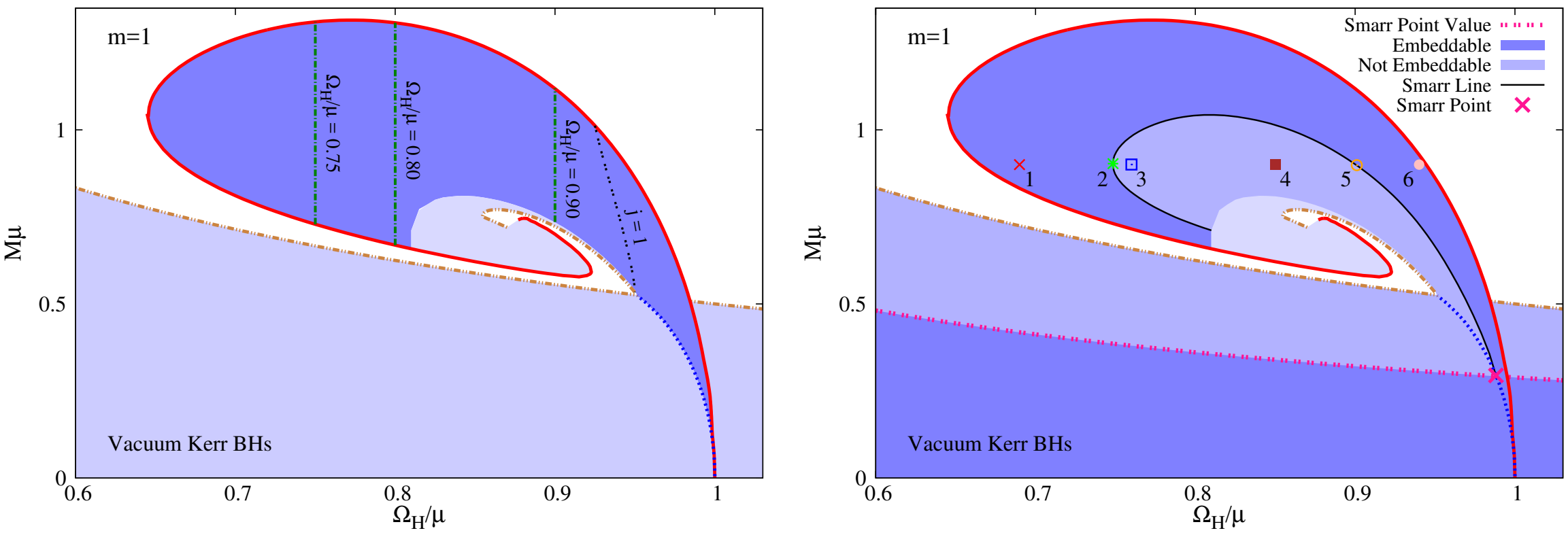

**Figure 2.2:** (Left panel) Domain of existence of KBHsSH solutions with $m = 1$ (dark blue and upper light blue shaded regions) in an $M$ *versus* $\Omega_H$ diagram, both in units of the scalar field mass $\mu$. Three lines of constant $\Omega_H$ solutions are displayed; these sequences will be used below. Also the line of solutions with $j = 1$ is shown, which starts at the extremal vacuum Kerr. The bottom light blue shaded region corresponds to vacuum Kerr BHs which is limited by a dashed-dotted brown line corresponding to extremal Kerr BHs. (Right panel) Same plot showing the embeddable (dark blue) and not embeddable (intermediate blue) regions of KBHsSH and Kerr BHs delimited by the Smarr line (solid black line) and the Smarr point (pink cross – extended as the pink dotted line). The lightest blue region of the domain has not been examined in detail (being a numerically challenging region). Six specific solutions are highlighted, whose embedding diagrams are shown in Figure 2.3. The remaining lines are the same lines as in the left panel.

### 2.3.1 Embedding

To embed the spatial sections of the event horizon of KBHsSH in $\mathbb{E}^3$ we repeat the procedure of subsection 2.2.1. The induced metric of these spatial sections is obtained by setting $t =$ const and $r = r_H$ in metric ansatz:

$$d\sigma^2 = r_H^2 \left[ e^{2F_1(r_H,\theta)} d\theta^2 + e^{2F_2(r_H,\theta)} \sin^2\theta d\varphi^2 \right] \ . \qquad (2.16)$$

Comparison with Equation 2.4 yields:

$$f'(\theta)^2 + g'(\theta)^2 = e^{2F_1(r_H,\theta)} r_H^2 \quad , \qquad f(\theta) = e^{F_2(r_H,\theta)} r_H \sin\theta \ , \qquad (2.17)$$

where the functions $F_1$ and $F_2$ now only depend on $\theta$. It follows that:

$$g'(\theta) = r_H \sqrt{k(\theta)} \ , \qquad k(\theta) \equiv e^{2F_1(r_H,\theta)} - e^{2F_2(r_H,\theta)} \left[ F_2'(r_H,\theta) \sin\theta + \cos\theta \right]^2 \ . \qquad (2.18)$$

As for the Kerr case, the spatial sections of the hairy BHs horizon will only be globally embeddable in $\mathbb{E}^3$ if both embedding functions are real for all $\theta$, thus iff $k(\theta) \geqslant 0$. Observe that $k(0) = e^{2F_1(r_H,0)} - e^{2F_2(r_H,0)}$. In order to avoid conical singularities [86],

$$F_1(r,0) = F_2(r,0) \ . \qquad (2.19)$$

Thus $k(0) = 0$, just as for the Kerr case. Then, $k'(0) = 2F_1'(r_H,0)e^{2F_1(r_H,0)} - 4F_2'(r_H,0)e^{2F_2(r_H,0)}$. For regularity, $F_1'(r_H,0) = 0$ and $F_2'(r_H,0) = 0$ [86]. Thus $k'(0) = 0$, also, as for the Kerr case. Finally, $k''(0) = 2e^{2F_2(r_H,0)}\mathcal{T}$, where $\mathcal{T} \equiv F_1''(r_H,0) - 3F_2''(r_H,0) + 1$, and thus the embedding fails if $\mathcal{T} < 0$.

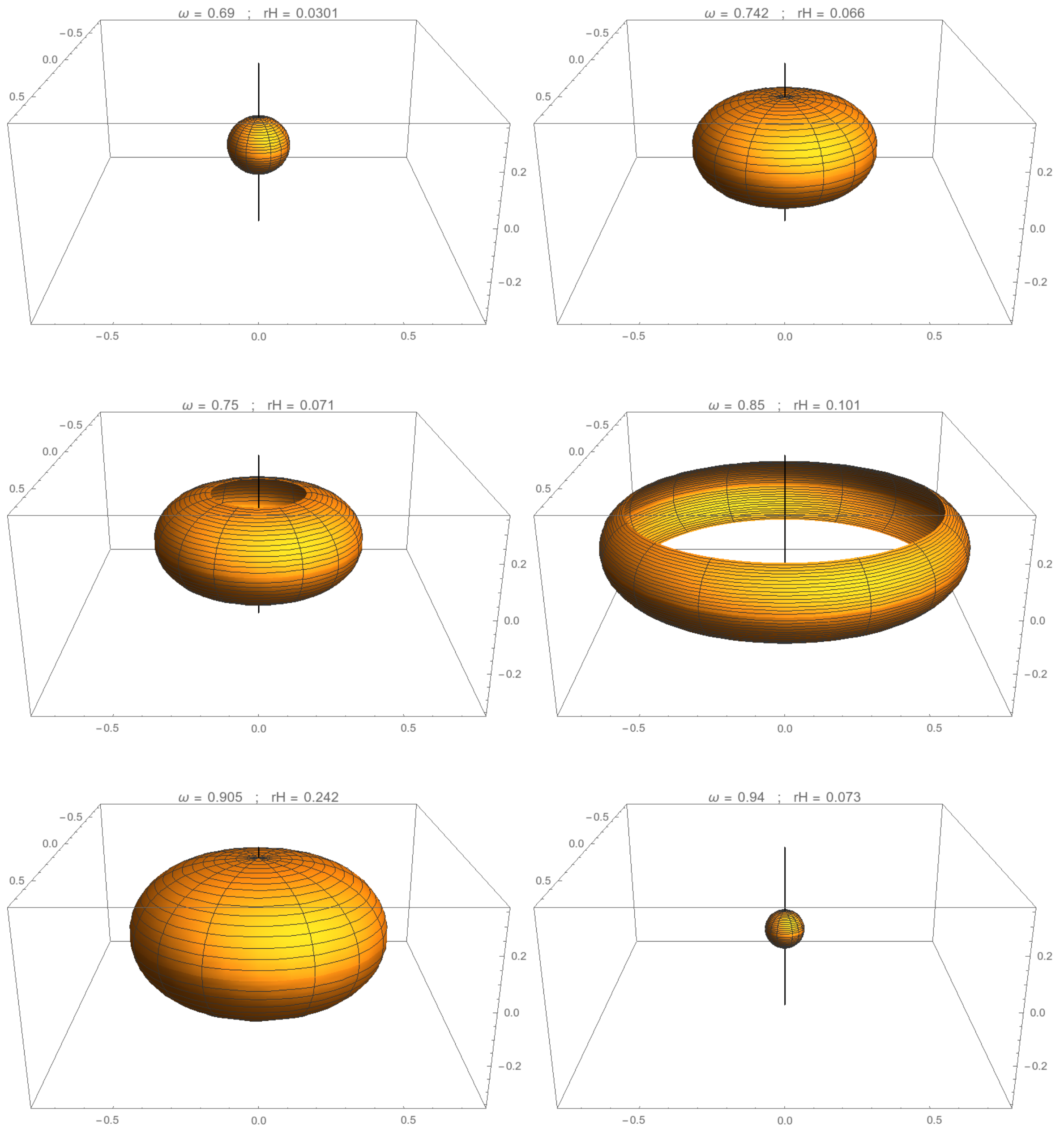

**Figure 2.3:** Embedding diagrams, in $\mathbb{E}^3$, for the horizon of the six highlighted solutions in the right panel of Figure 2.2. First, second and third rows, left to right: solutions 1,2; 3,4; 5,6. Solutions 1-6 have parameters $(j; j_H; \mathfrak{s}; v_H)$, respectively: (0.92; 6.27; 1.04; 0.09); (0.88; 5.53; 1.24; 0.29); (0.88; 4.92; 1.27; 0.33); (0.86; 1.69; 1.46; 0.65); (0.89; 1.16; 1.23; 0.48); (1.08; 1.64; 1.01; 0.08).

Since the analogous condition to $k(\theta) \geqslant 0$ fails for Kerr BHs with $j > j^{(S)}$, and since Kerr BHs with $j \in [0,1]$ occur at one of the boundaries of the domain of existence of KBHsSH, there will be a non-embeddable region in the domain of existence of KBHsSH. This region was obtained by analysing the above condition for the numerical data, and it is exhibited as the intermediate blue shaded region in Figure 2.2 (right panel). It is bounded by a (black solid) line – dubbed *Smarr line* – that terminates at the Smarr point, in the Kerr limit. The Smarr line seems to spiral in a similar fashion to the BS line.

The failure of the embedding of Equation 2.16 in $\mathbb{E}^3$ is again accompanied by the development of a region of negative Gaussian curvature at the poles. To see this, note that the Gaussian curvature for Equation 2.16 is

$$\mathcal{K} = \frac{e^{-2F_1(r_H,\theta)}}{r_H^2}\left\{1 + F_1'(r_H,\theta)\left[F_2'(r_H,\theta) + \cot\theta\right] - F_2'(r_H,\theta)\left[F_2'(r_H,\theta) + 2\cot\theta\right] - F_2''(\theta)\right\} \tag{2.20}$$

At the poles, $\theta = \{0, \pi\}$,

$$\mathcal{K}|_{\text{poles}} = \frac{e^{-2F_1(r_H,0)}}{r_H^2}\mathcal{T}\ . \tag{2.21}$$

Thus we conclude that, just as for the Kerr case, the embedding fails when the Gaussian curvature becomes negative at the poles.

The 3-dimensional embedding diagrams for the sequence of six solutions highlighted in Figure 2.2 are shown in Figure 2.3. These six solutions all have the same $\mu M = 0.9$. Solution 1 is in the embeddable region near the BS line; solution 2 is at the border with the non-embeddable region, wherein solutions 3 and 4 are located. Solution 5 is again at the boundary of the embeddable region and solution 6 is again in the embeddable region close to the BS line. Observe that the horizon of the solutions close to the BS line is quite spherical. The solutions at the threshold of the embeddable region are quite flat near the poles.

Observe that in the representation used in Figure 2.2, a $j$ = const vacuum Kerr solution is actually a line in the lower part of the diagram. Thus, the Smarr point of vacuum Kerr BHs is also represented as a line (dashed pink line in the domain of Kerr BHs), separating the globally embeddable region (lower part - dark blue), from the non-embeddable region (upper part - intermediate blue).

### 2.3.2 Sphericity

For KBHsSH, $L_e$ and $L_p$ can be obtained from the induced metric Equation 2.16, giving:

$$L_e = \int_0^{2\pi} d\varphi\ e^{F_2(r_H,\pi/2)} r_H = 2\pi e^{F_2(r_H,\pi/2)} r_H\ , \qquad L_p = 2r_H \int_0^{\pi} d\theta\ e^{F_1(r_H,\theta)}\ . \tag{2.22}$$

Thus, the sphericity can be written as,

$$\mathfrak{s} \equiv \frac{L_e}{L_p} = \frac{\pi\ e^{F_2(r_H,\pi/2)}}{\int_0^{\pi} d\theta\ e^{F_1(r_H,\theta)}}\ . \tag{2.23}$$

In Figure 2.4 we exhibit the domain of existence of KBHsSH in a $\mathfrak{s}$ *versus* $r_H$ diagram. In such a diagram, all BS solutions and extremal hairy BHs must lie along the $y$ axis, since, as observed in chapter 1, in both these cases $r_H \to 0$. It turns out that all BS solutions fall into a single point, precisely at the origin of this diagram (red cross), corresponding to $r_H = 0$ and $\mathfrak{s} = 1$. Since BSs have no horizon, the sphericity here is defined by continuity, and this result means that the event horizon of a KBHsSH very close to the BS limit will be essentially spherical. This observation resonates with the idea that the heavy scalar environment around the hairy BH horizon endows the horizon with a large momentum of inertia [63], [101]. Dragging such a "heavy" environment results in a slower horizon linear velocity, as will be confirmed in the next subsection, and therefore in a more spherical event horizon, regardless of the $j$ parameter of the spacetime, which can be even larger than unity.

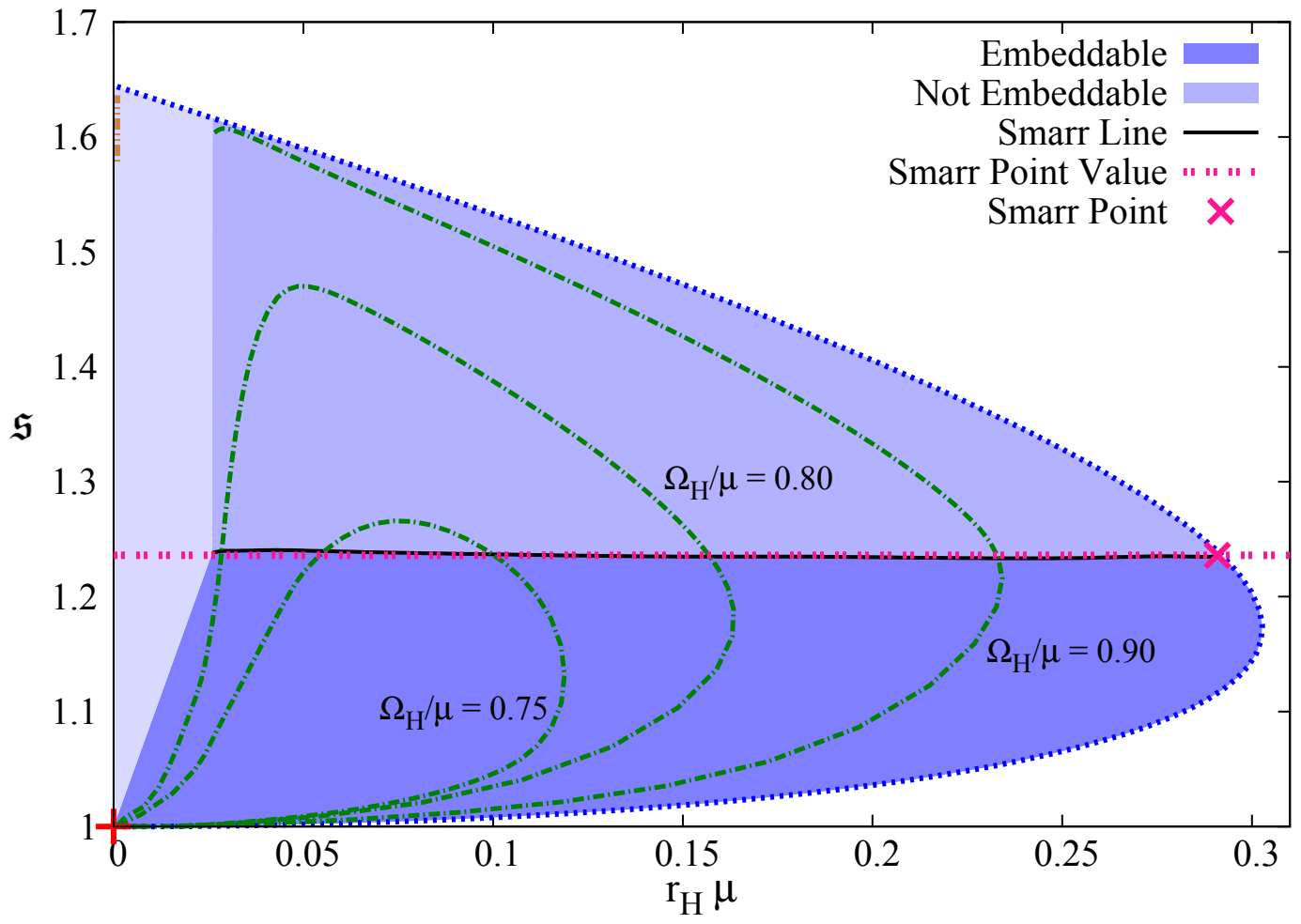

**Figure 2.4:** Domain of existence of KBHsSH solutions in a $\mathfrak{s}$ *versus* $r_H$ diagram. As before, the dark blue region corresponds to embeddable solutions, the intermediate blue region corresponds to non-embeddable solutions and the light blue region to the region we have not analysed. Three lines of constant horizon angular velocity $\Omega_H/\mu = \{0.75; 0.80; 0.90\}$ (dark green dot-dashed lines) are also exhibited. The remaining lines/points are the same as in Figure 2.2.

All the extremal KBHsSH solutions fall onto a line at $r_H = 0$ and $\mathfrak{s} \in [1, 1.64473]$. The largest value in this interval correspond to extremal vacuum Kerr:

$$\mathfrak{s}(j=1)\big|_{\text{Kerr}} = \frac{\pi}{\sqrt{2}} \frac{1}{E(1/2)} \approx 1.64473 \, . \tag{2.24}$$

The region closest to the lowest value in this interval ($\mathfrak{s} = 1$) was not obtained with our numerical solutions, since these solutions correspond to the very central region of the spiral, and hence numerically challenging. The solutions we have actually examined are a subset of these and are represented as a brown dashed-dotted line in Figure 2.4.

Figure 2.4 also shows the Smarr (black solid) line, as well the Smarr point for Kerr BHs (pink cross), which, for reference, is extended for other values of $r_H$ as a pink dotted line. One can see that the Smarr line has the same value of spheroidicity as the Smarr point, for all solutions analysed, within numerical error. The green dot-dash lines in Figure 2.4 correspond to solutions with a constant horizon angular velocity $\Omega_H/\mu = \{0.75; 0.80; 0.90\}$. For each value of $\Omega_H/\mu$ there are two possible values of $r_H$ for the same sphericity. These solutions can be quite different but they are either both embeddable or both non-embeddable, since this property is determined by the value of $\mathfrak{s}$.

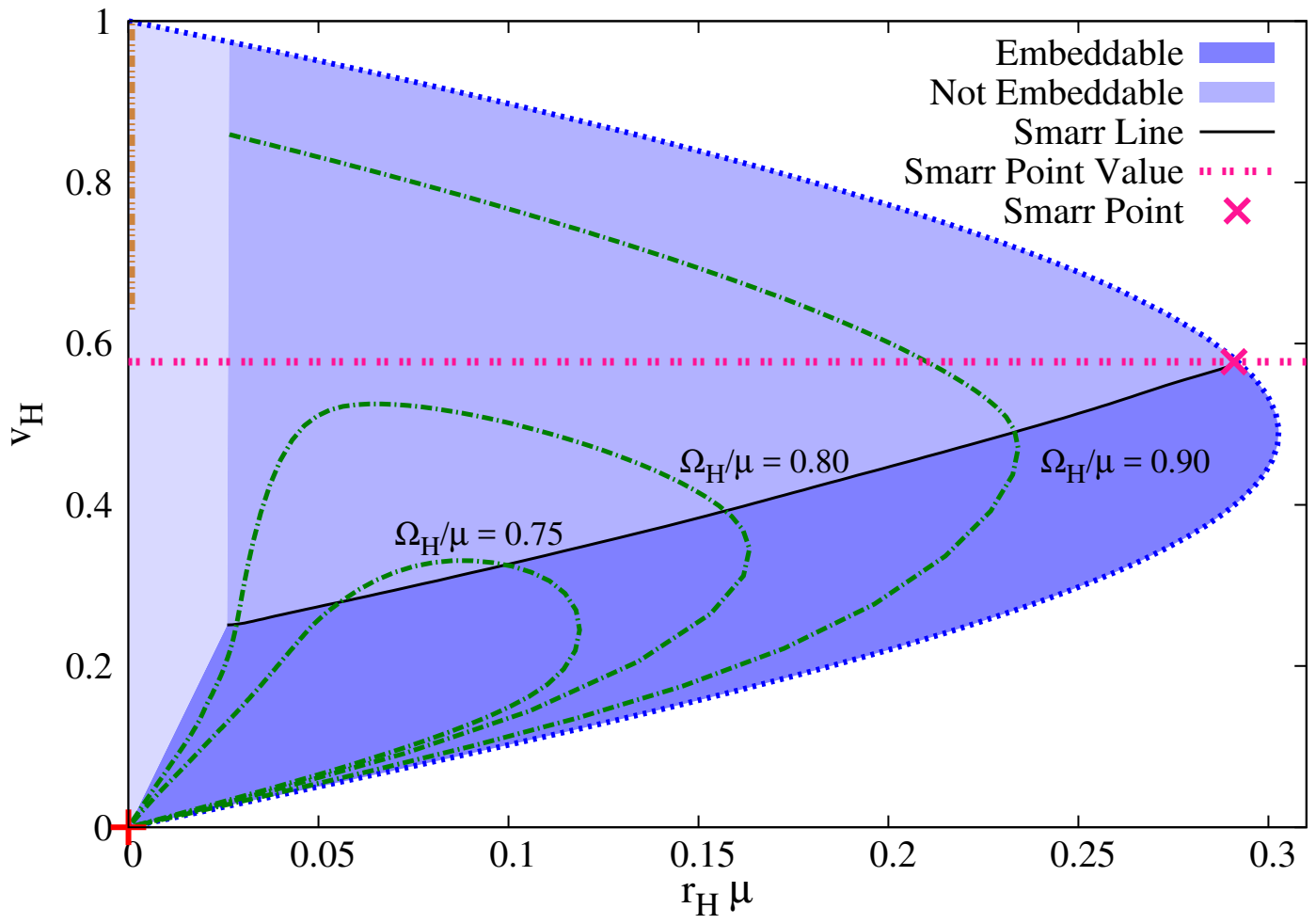

**Figure 2.5:** Domain of existence of KBHsSH solutions in a $v_H$ *versus* $r_H$ diagram. All colour coding is as in Figure 2.4.

### 2.3.3 Horizon linear velocity

Using the same reasoning as in the Kerr case, we obtain

$$v_H = e^{F_2(r_H,\pi/2)} r_H \Omega_H \; . \tag{2.25}$$

In Figure 2.5 we exhibit the domain of existence of KBHsSH solutions in a $v_H$ *versus* $r_H$ diagram. All BSs solutions are, in this diagram, a single point – the origin – represented as a red cross. The remaining points with $r_H = 0$ correspond to the extremal hairy BHs. These extremal BHs span the interval $v_H \in [0,1]$, where $v_H = 1$ corresponds to the Kerr limit and $v_H = 0$ to the conjectured singular solution at the centre of the spiral.[2]

The green dot-dashed lines, as before, represent solutions with constant horizon angular velocities, $\Omega_H/\mu = \{0.75; 0.80; 0.90\}$. Taking, for instance, $\Omega_H/\mu = 0.8$, we see that there is a region of degeneracy where it is possible to have two solutions with the same $v_H$, one being embeddable and the other non-embeddable; $r_H$ or $\mu M$ raise the degeneracy.

Figure 2.5 also shows the Smarr (black solid) line, as well the Smarr point for Kerr BHs (pink cross), again extended for other values of $r_H$ as a pink dotted line. In contrast to the case for the sphericity, the Smarr line only matches the value at the Smarr point in the Kerr limit, where $v_H = 1/\sqrt{3}$ – *cf.* subsection 2.2.2. Here the Smarr line decreases monotonically as $r_H$ decreases. This line is always (for all solutions studied) smaller than the Smarr point value for Kerr BHs. This indicates that $v_H = 1/\sqrt{3}$ is the maximum velocity at which a Kerr BH, with or without scalar hair, can rotate and be globally embeddable in Euclidean 3-space.

[2] Again, our numerical solutions do not extend all the way to $v_H = 0$, for the same reason discussed before.

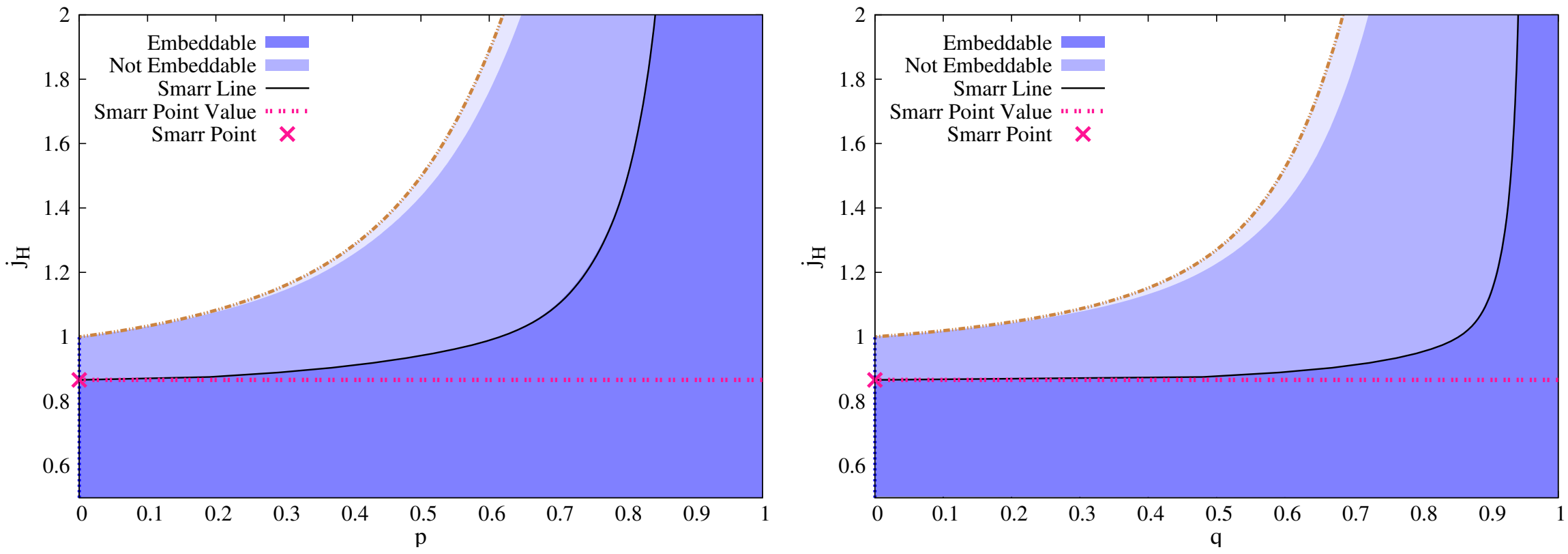

**Figure 2.6:** Horizon dimensionless angular momentum, $j_H$ *versus* the hairiness parameters $p$ (left panel) and $q$ (right panel). The color coding is the same as in previous figures.

### 2.3.4 Horizon angular momentum

Let us also analyse the relation between the angular momentum carried by the BH and the existence of an Euclidean embedding. In the Kerr case, it is irrelevant if one considers this angular momentum to be the horizon one or the asymptotically measured one, since they coincide. But for the case of the hairy BHs this is not the case, as a part of the angular momentum is stored in the scalar field outside the horizon (similarly to what occurs for the energy). Let us consider the horizon dimensionless angular momentum, $j_H$, as this is really the angular momentum stored in the BH.

In Figure 2.6, we present the horizon angular momentum as a function of the hairiness parameters $p$ and $q$. Taking $p = 0 = q$ is the Kerr limit where we see the separation between embeddable and non-embeddable solutions at the Smarr point. The value $p = 1 = q$ is the BS limit where we see all solutions become embeddable. Most importantly, all solutions with $j_H < j^{(S)}$ are embeddable. The Smarr dimensionless spin provides a lower bound below which all hairy BHs can be embedded in Eucliden 3-space.

Concerning the total angular momentum, we would like to point out that in the BS limit it can become larger than unity. This occurs for the BSs between the Newtonian limit (at $\Omega_H/\mu = 1$) and the $j = 1$ point - see Figure 2.2 (left panel). In the vicinity of these solutions, KBHsSH also have $j > 1$ by continuity, and they are always embeddable in $\mathbb{E}^3$, as one may verify from inspection of Figure 2.2 (right panel).

## 2.4 Conclusions

In this chapter we have analysed the horizon geometry of Kerr BHs with scalar hair, a family of solutions that continuously connects to vacuum Kerr BHs. We have been particularly interested in distinguishing solutions which are embeddable in Euclidean 3-space, $\mathbb{E}^3$, as this embedding provides an accurate and intuitive tool to perceive the horizon geometry of these BHs. In the Kerr case, the solutions stop being embeddable at the Smarr point, when the dimensionless spin $j$, the sphericity $\mathfrak{s}$ and the horizon linear velocity $v_H$ are smaller than critical values attained at the Smarr point: $j^{(S)}, \mathfrak{s}^{(S)}, v_H^{(S)}$. Our analysis of the hairy BHs shows that:

- all hairy solutions with a *horizon* dimensionless spin $j_H \leqslant j^{(S)}$ are embeddable in $\mathbb{E}^3$; but there are also hairy BHs with $j_H > j^{(S)}$ that are embeddable;
- $\mathfrak{s}^{(S)}$ remains the threshold sphericity for the embeddable solutions all over the explored domain of existence of hairy BHs. Thus hairy BHs with $\mathfrak{s} < \mathfrak{s}^{(S)}$ ($\mathfrak{s} > \mathfrak{s}^{(S)}$) are (are not) embeddable;
- there are hairy solutions with $v_H < v_H^{(S)}$ which are not globally embeddable in $\mathbb{E}^3$ and all hairy BHs with $v_H > v_H^{(S)}$ are not embeddable.

The analysis of this family of hairy solutions suggestively disentangles the role of these three parameters in the embedding properties of the solutions. For the general family of hairy BHs, $j_H < j_H^{(S)}$ is a sufficient, but not necessary, condition for being embeddable; $v < v_H^{(S)}$ is a necessary, but not sufficient, condition for being embeddable and $\mathfrak{s} < \mathfrak{s}^{(S)}$ is a necessary and sufficient condition for being embeddable in $\mathbb{E}^3$. It would be interesting to test the generality of these results in other families of BH solutions. In this respect, we have verified the threshold value of the sphericity $\mathfrak{s}^{(S)}$ still holds for the Kerr-Newman [6] and Kerr-Sen BH solutions [102].

Another possible avenue for further work would be to investigate the horizon geometry within the framework of the "analytic effective model" for hairy BHs recently proposed in [103]. In this model, the horizon quantities (such as horizon area, Hawking temperature and horizon angular velocity) of the hairy BH are well approximated by those of a Kerr BH, but with the replacements $(M, J) \to (M_H, J_H)$. As discussed in [103], [104], this model works well in the neighbourhood of the existence line wherein the hairy BHs reduce to the vacuum Kerr solution. Our preliminary results suggest that the "analytic effective model" holds also at the level of the horizon geometry. That is, in a region close to the existence line, the horizon geometry Equation 2.16 is well approximated by the following expressions

$$g_{\theta\theta} = r_H^2 e^{2F_1(r_H,\theta)} = 2M_H^2 \left(1 + \sqrt{1 - j_H^2} - \frac{j_H^2}{2}\sin^2\theta\right), \tag{2.26}$$

$$g_{\varphi\varphi} = r_H^2 e^{2F_2(r_H,\theta)} \sin^2\theta = \frac{4M_H^2\left(1 + \sqrt{1 - j_H^2}\right)\sin^2\theta}{1 + \sqrt{1 - j_H^2} + (1 - \sqrt{1 - j_H^2})\cos^2\theta}.$$

This approximation holds within the same errors as those reported in [103], [104] for various physical quantities.

CHAPTER 3

# Kerr Black Holes with Synchronised Scalar Hair and Higher Azimuthal Harmonic Index

## 3.1 Introduction

Kerr BHs with synchronised scalar hair [37], also known as KBHsSH, as seen in chapter 1, are a counterexample to the no-hair conjecture [12] – see [16], [88], [89] for reviews – occurring in a simple and physically sound model: (complex-)Einstein-Klein-Gordon theory. Many related solutions, relying on a similar synchronisation mechanism, have been found in the last few years, in different setups and approximations. An incomplete list of references, including also various studies of physical properties, is [40]–[43], [53], [55]–[60], [62], [64], [66], [86], [90]–[95], [100], [104]–[142].

These hairy BH solutions have a relation with the physical phenomenon of superradiance [39], from which they can form dynamically from the Kerr solution [104], [125], [126] - see also [133], [135] for a discussion on the metastability of these solutions against superradiance. They also reduce to Kerr BHs and BSs [32], [33], in appropriate limits. BSs are a sort of gravitating soliton interpreted as a Bose-Einstein condensate of an ultra-light scalar field, that could be a dark matter candidate [143], [144]. Moreover, the existence of the hairy BH solutions does not rely on particular choices of scalar field potentials that violate energy conditions, unlike other examples of asymptotically flat BHs with scalar hair, see $e.g.$ [145], [146]. Thus, besides the issue of the no-hair conjecture in BH physics, these hairy BHs contain different angles of interest.

Kerr BHs with synchronised hair comprise a family that, besides the continuous parameters mass, angular momentum and Noether charge, is labelled by two discrete numbers: the azimuthal harmonic index of the scalar field $m \in \mathbb{Z}^+$ and its node number $n$. Most of the studies of the solutions have focused on the fundamental solutions, $n = 0$, with the smallest value of $m = 1$. Excited solutions ($n \neq 0$) have also been constructed [53]. Solutions with $m > 1$, on the other hand, have only been considered in the solitonic (BS) limit [86], [147], with the exception of the non-minimal model studied in [92]. The purpose of this chapter is to construct solutions with $m > 1$ in the minimal, simplest model, and to study some of the basic physical properties of these new solutions.

One motivation to study the higher $m$ solutions is that the superradiant instability of a given $m$ solution could drive it to migrate to an $m + 1$ solution, in an asymptotic cascading process leading to $m \to \infty$ [90]. This process is, likely, non-conservative, ejecting some energy and, especially, angular momentum towards infinity; but for particular solutions with a given $m$, if a neighbouring solution (in terms of global quantities) exists for $m + 1$, the process could be approximately conservative. In fact, this approximate conservativeness has been observed in the transition from the Kerr BH (which corresponds to $m = 0$) to the $m = 1$ hairy solution in [104], [125], [126]. For this approximately conservative migration to be possible, the higher $m$ neighbouring solution would have to be entropically favoured. As we shall see herein, this is always the case: comparing solutions with consecutive values of $m$ with the same global quantities, the higher $m$ solution has a larger horizon area.

Another motivation for studying this higher $m$ solutions is to assess the universality of some physical properties. For instance, it was observed in [106] that, when scanning the domain of existence, these BHs exhibit a more diverse structure of ergo-regions than the standard one of the Kerr BH. The existence of these ergo-regions is at the origin of the superradiant instability. So, a natural question is if a similar structure is present for higher $m$. We shall see here that this is the case. Moreover, the horizon geometry of these hairy BHs was studied in the previous chapter and in [64], where it was found that the key property for deciding whether the horizon is embeddable in Euclidean 3-space is the horizon sphericity. Again, we shall see that this is also the case for the higher $m$ solutions. Both these analyses provide evidence that the properties observed for $m = 1$ solutions are universal throughout the whole discrete family labelled by $m$.

This chapter is based on the work developed in [54] and it is organised as follows. The model is presented in section 3.2. The construction of the domain of existence of the $m = 2, 3$ solutions is presented in section 3.3, where they are compared with the $m = 1$ case. In section 3.4 the phase space is discussed and the entropy comparison shown. In section 3.5 other physical properties, in particular, the ergo regions and horizon geometry, are discussed. section 3.6 wraps up the chapter with a discussion.

## 3.2 The Model

KBHsSH are solutions of the (complex-)Einstein-Klein-Gordon action, Equation 1.13, and they represent a Kerr BH in equilibrium with a massive scalar field configuration. As mentioned in the previous chapters, they were obtained numerically by solving the Einstein-Klein-Gordon field equations, Equation 1.14 with the help of Equation 1.16 and Equation 1.19 as ansatze for the scalar field and metric, respectively.

As discussed in the chapter 1, the existence of these solutions relies on the so-called *synchronisation condition* – *cf.* Equation 1.18. This condition can be interpreted as a synchronisation between the horizon angular velocity of the BH, $\Omega_H$, and the phase angular velocity of the scalar field, $\omega/m$, hence justifying its name.

Our goal here is to study fundamental solutions, $n = 0$, with larger azimuthal harmonic index, namely $m = 2, 3$, as all previous studies for the model Equation 1.14 have focused on $m = 1$ solutions. As such, the numerical framework will be precisely the same as the one discussed in section 1.2, with the only difference being the use of $m = 2, 3$ instead of $m = 1$.

## 3.3 Domain of Existence

Fixing $n = 0$, the domain of existence spanned by the hairy BHs is a 2-dimensional space. In our framework to construct the solutions, this domain is scanned by varying the angular frequency of the scalar field, $\omega$, and the radial coordinate of the event horizon, $r_H$. Such 2-dimensional region can, however, be exhibited in several more physically meaningful ways, as $r_H$ is not physically meaningful *per se*. In Figure 3.1, following previous literature, the domain of existence is shown in an ADM mass *vs.* horizon angular velocity (Figure 3.1a) and in an ADM mass *vs.* scalar field angular frequency (Figure 3.1b) plots. Both panels exhibit the domain of existence of the hairy BHs with $m = 1$, $m = 2$ and $m = 3$. For the $m = 3$ case, only a part of the domain of existence is shown, corresponding to a region of interest for the entropic comparison. The left panel shows, moreover, the region where vacuum Kerr BHs exist – below the black solid line in Figure 3.1a.

The domain of existence of the hairy BHs is shown in Figure 3.1 as the shaded blue regions, corresponding to the extrapolation to continuum of isolated numerical points. It is bounded by the same three curves that we already saw for the KBHsH with $m = 1$ – *cf.* Figure 1.1 and Figure 2.2: the BS line (red solid line) corresponding to the solitonic limit; the extremal line (green dashed line) corresponding to extremal hairy BHs; and the existence line (blue dotted line) corresponding to the subset of vacuum Kerr BHs that can support scalar clouds.

Firstly, consider the right panel (Figure 3.1b). As $m$ increases, the domain of existence broadens up in its frequency range, allowing hairy BHs with lower angular frequencies and larger ADM masses. Each $m$ family can overlap with the previous $m - 1$ family, where is possible to have hairy BHs with the same angular frequency and ADM mass but with different $m$. Observe, however, that the regions of overlap for $m = 1, 2$ and $m = 2, 3$ solutions are distinct. Thus, three consecutive $m$ families do not overlap.

In Figure 3.1a, on the other hand, one observes that there is no region of overlapping $m = 1, 2$ solutions. Two hairy BHs with different $m$, and $m = 1, 2$, can have the same ADM mass, but not the same horizon angular velocity. The same can not be said for $m = 2, 3$ solutions: there is a region of overlap. Nonetheless, by cross-checking information from Figure 3.1b and Figure 3.1a one can establish that no two hairy BHs with the same ADM mass, angular frequency, $\omega$, and horizon angular velocity, $\Omega_H$ exist, in the $m = 2, 3$ overlap. This overlap in Figure 3.1b occurs for large angular frequencies, which correspond to solutions close to $\Omega_H \sim 0.5$; in Figure 3.1a, for $m = 2, 3$, on the other hand, one can see that the overlapping solutions occur only around $\Omega_H \sim 0.3$.

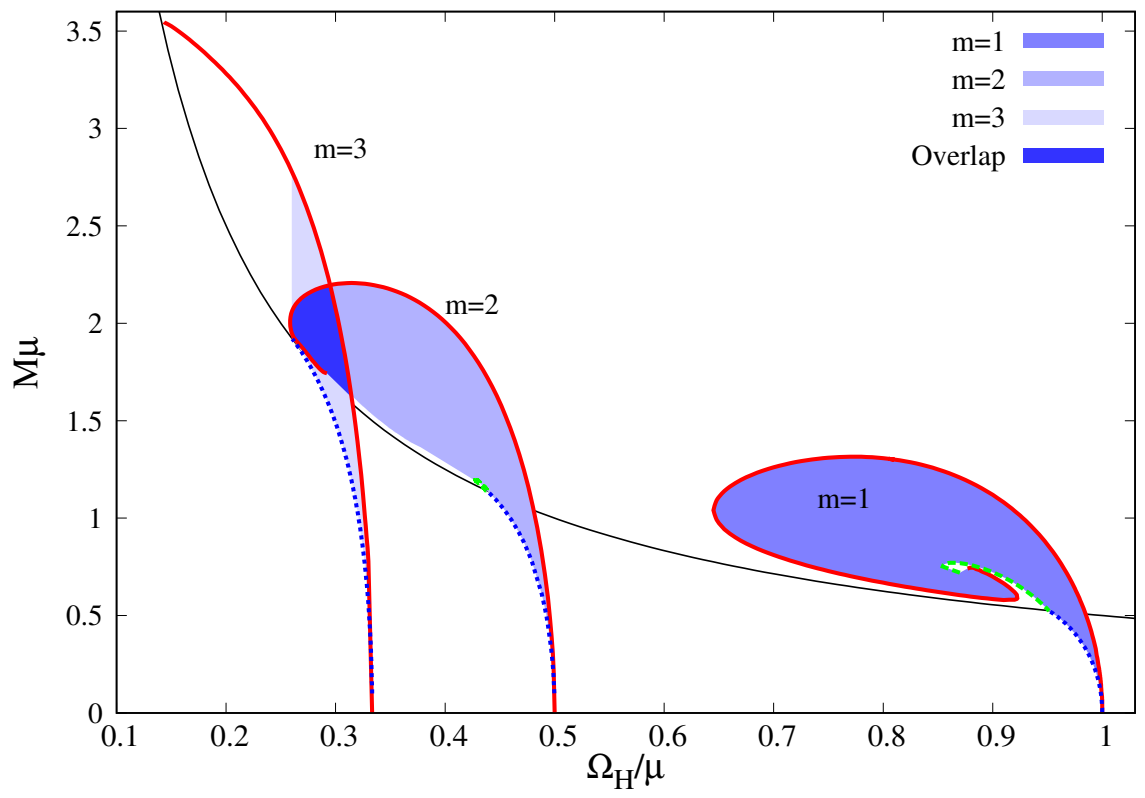

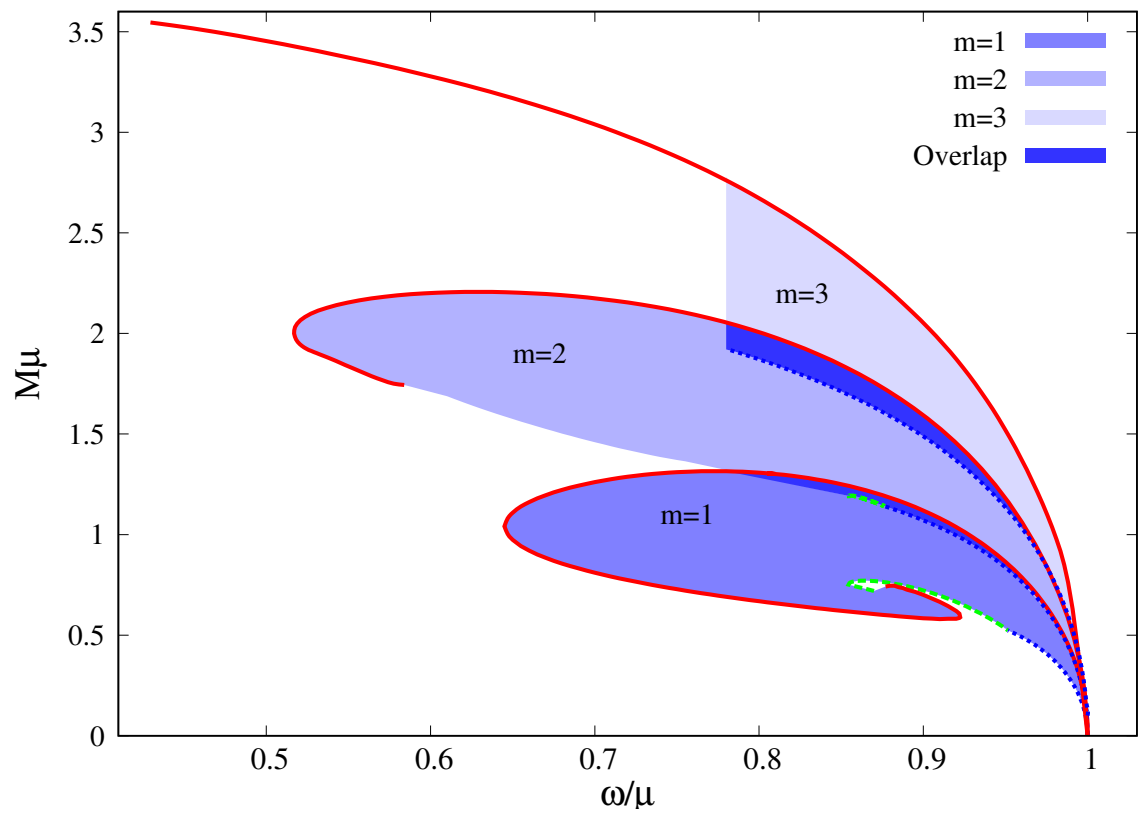

**Figure 3.1:** Domain of existence in an ADM mass *vs.*: **(a)** Event horizon angular velocity, $\Omega_H$; **(b)** Scalar field angular frequency, $\omega$. The red line represents the BS line, the blue dotted line is the existence line, and the green dashed line is the line of extremal hairy BHs. Solutions exist in the domain (shaded blue regions) bounded by these three lines. The black solid line (left panel) describes extremal Kerr BHs: the Kerr family of BHs exists on and below that line. The colour scheme is kept in the subsequent figures.

In Figure 3.1 $m = 2$ ($m = 3$) solutions have an horizon angular velocity which is half (one third) of the allowed angular frequency – cf. Figure 3.1b. For $m = 1$, the scalar field angular frequency is equal to the horizon angular velocity, and the domain of existence of hairy BHs with $m = 1$ is exactly the same in both plots.

## 3.4 Phase space

Let us now analyse the domain of existence in the total $(M, J)$ space, *i.e.* phase space. This is represented in Figure 3.2a and Figure 3.3a. Figure 3.1 already made manifest that solutions with higher $m$ are allowed to be more massive; this is confirmed in Figure 3.2a and Figure 3.3a. The latter, moreover, shows that higher $m$ solutions can have larger angular momentum, thus broadening the domain of existence. Furthermore, a region of overlapping solutions is again manifest: there are hairy BHs with different $m$ but with the same $(M, J)$. A natural question is then, which amongst these degenerate solutions, in terms of global quantities, is entropically preferred.

In Figure 3.2b the reduced horizon area, $a_H \equiv A_H/16\pi M^2$ is shown as a function of the reduced spin, $j \equiv J/M^2$, for hairy BHs belonging to the $m = 1, 2$ families (orange lines represent $m = 1$; black lines represent $m = 2$), with two illustrative values for the ADM mass, $M\mu = 0.3$ (dashed lines) and $M\mu = 0.5$ (solid lines). Observe that the existence line (dashed blue line) is common to both families. These lines follow the Kerr relation,

$$a_H^{\rm Kerr} = \frac{1}{2}\left(1 + \sqrt{1 - j_{\rm Kerr}^2}\right) \ . \tag{3.1}$$

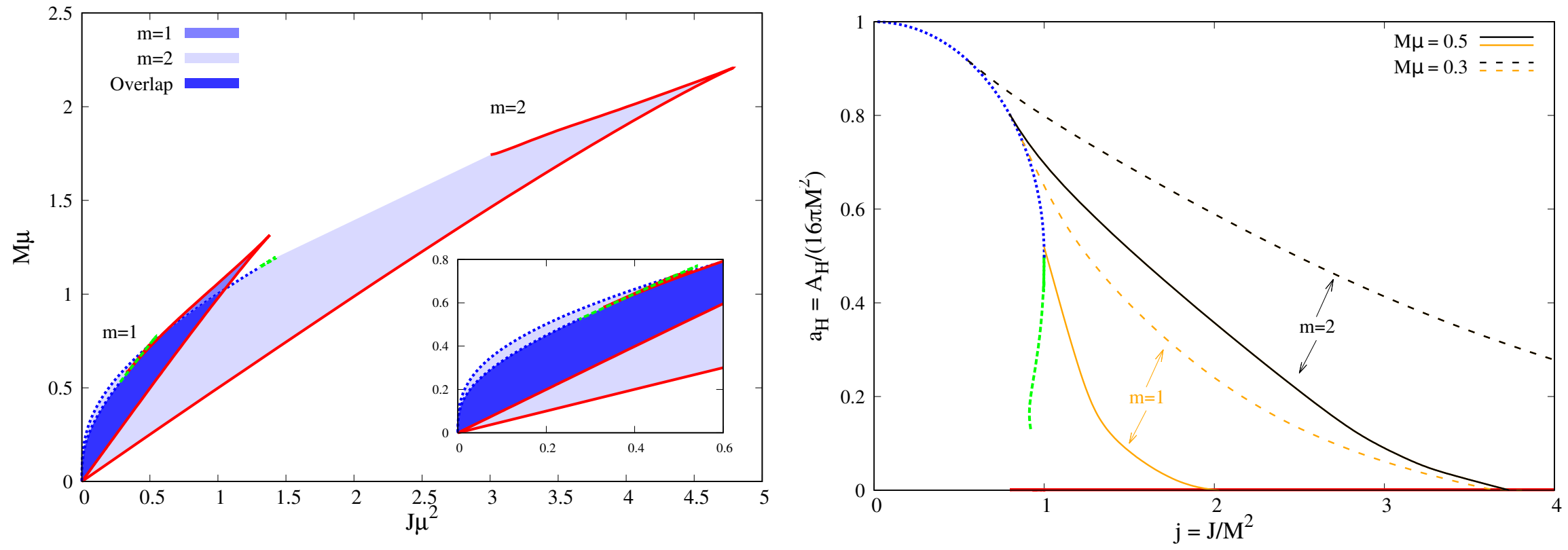

**(a)** **(b)**

**Figure 3.2:** **(a)** ADM mass *vs*. total angular momentum for the $m = 1$ and $m = 2$ families; **(b)** Reduced horizon area, $a_H$, *vs*. reduced spin, $j$. The (orange, for $m = 1$ and black, for $m = 2$) curves correspond to solutions with constant ADM mass: dashed (solid) lines correspond to $M\mu = 0.3$ ($M\mu = 0.5$).

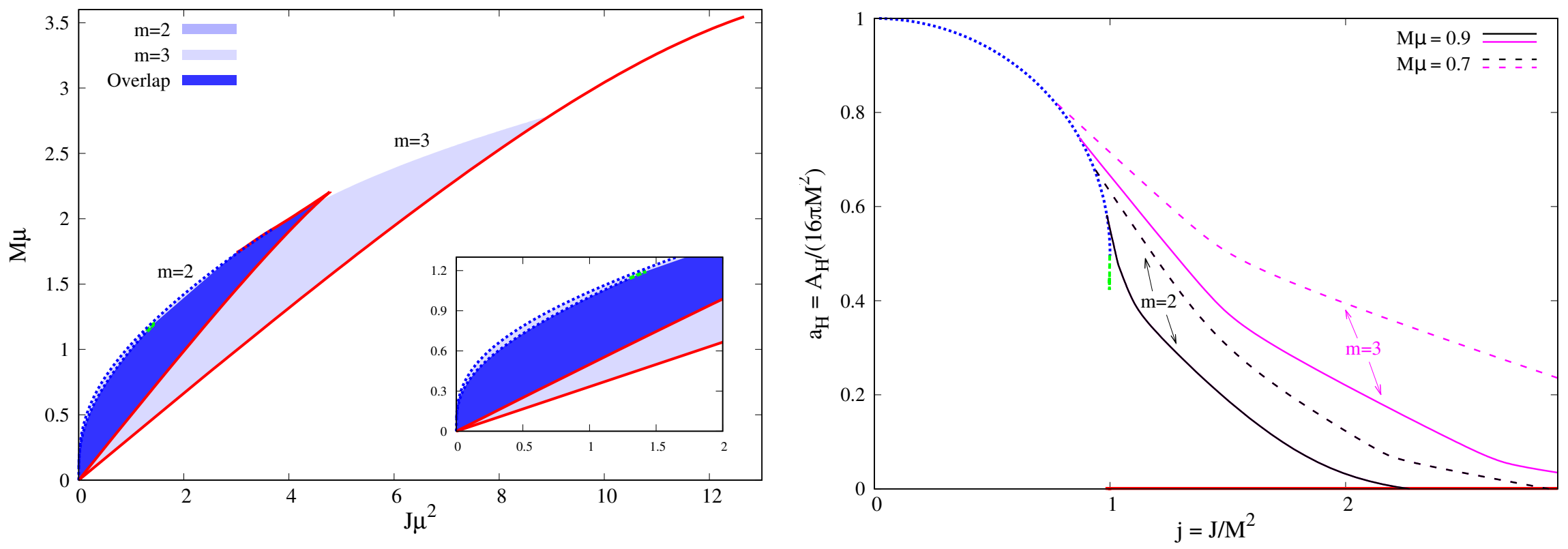

**(a)** **(b)**

**Figure 3.3:** **(a)** ADM mass *vs*. total angular momentum for the $m = 2$ and $m = 3$ families; **(b)** Reduced horizon area, $a_H$, *vs*. reduced spin, $j$. The (black, for $m = 2$ and pink, for $m = 3$) curves correspond to solutions with constant ADM mass: dashed (solid) lines correspond to $M\mu = 0.7$ ($M\mu = 0.9$).

The extremal BH line (dashed green lines) of both $m$ families, on the hand, overlap only at the point wherein they touch the existence line. Beyond this point, both lines are close but do not overlap and most of the green line seen in Figure 3.2b corresponds to the $m = 1$ solutions. The figure also exhibits two illustrative pairs of lines corresponding to sequences of hairy BHs with the same $M$ (solid black and orange lines for $M\mu = 0.5$, or dashed black and orange line for $M\mu = 0.3$), demonstrating that the solutions with $m = 2$ will always have a larger horizon area and hence a larger entropy, when both solutions have the same $j$. A similar analysis is performed in Figure 3.3b, for $m = 2, 3$ solutions with similar conclusions. We remark that in this case only the $m = 2$ extremal line is shown, as this line was not computed in the $m = 3$ case.

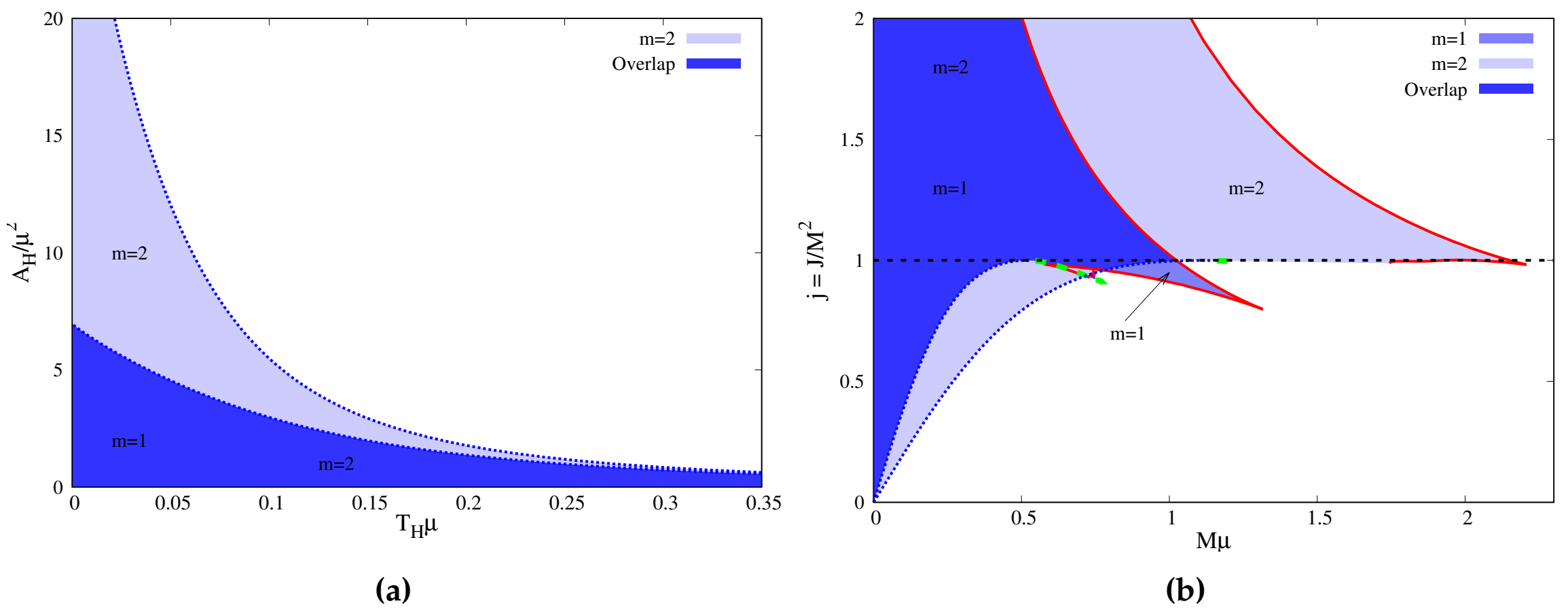

**Figure 3.4: (a)** Horizon area, $A_H$ *vs*. the Hawking temperature, $T_H$; **(b)** Reduced spin, $j = J/M^2$ *vs*. the ADM mass. The black dotted line corresponds to the Kerr bound where $j = 1$. Hairy solutions with both $m = 1$ and $m = 2$, can violate this bound.

## 3.5 Other physical properties

Let us now briefly consider other salient properties of the hairy BHs with $m > 1$.

### 3.5.1 Temperature distribution and Kerr bound violation

In Figure 3.4a we exhibit the horizon area, $A_H$ of $m = 1, 2$ hairy BHs *vs*. their Hawking temperature, $T_H$. Fixing $T_H$, there are always hairy BHs with $m = 2$ with larger horizon area and hence entropically preferred. Likewise, fixing $A_H$, there are always $m = 2$ solution with a larger Hawking temperature than $m = 1$ solutions.

In Figure 3.4b, the reduced spin, $j = J/M^2$, is exhibited in terms of the ADM mass of the hairy BHs. This confirms a result already manifest in Figure 3.2b. For Kerr BHs there is a limit to the reduced spin they can carry; if a Kerr BH rotates too fast, no event horizon is possible. This is the *Kerr bound*, $j \leqslant 1$. Figure 3.2b, Figure 3.3b and Figure 3.4b, confirm that the existence line (vacuum Kerr BHs that can support scalar clouds) only extends to $j = 1$, obeying the Kerr bound, but hairy BHs of both $m$ families, can violate the Kerr bound. In fact, for constant $M$, larger $m$ solutions have stronger violations of the bound.

### 3.5.2 Ergoregions

Kerr BHs are well known to possess an ergoregion [148], wherein the asymptotically timelike Killing vector field becomes spacelike outside the event horizon. In such region, the BH has to perform work on any causally moving object [149], which by energy conservation means the BH transfers some of its rotational energy to such an object. The existence of an ergoregion is at the source of the Penrose process, superradiant scattering and superradiant instabilities; the latter trigger the migration of the Kerr BH and hairy BH solutions towards higher $m$ in (complex-)Einstein-Klein-Gordon models.

The typology of ergoregions in the $m = 1$ hairy solutions is richer than in Kerr [106]. In the former case, BHs can have two different types of ergoregions: an ergo-sphere – the same as Kerr BHs; or an ergo-Saturn. The latter is the superposition of the standard BH ergo-sphere and an ergo-torus known to be present in some fast rotating BSs [150].

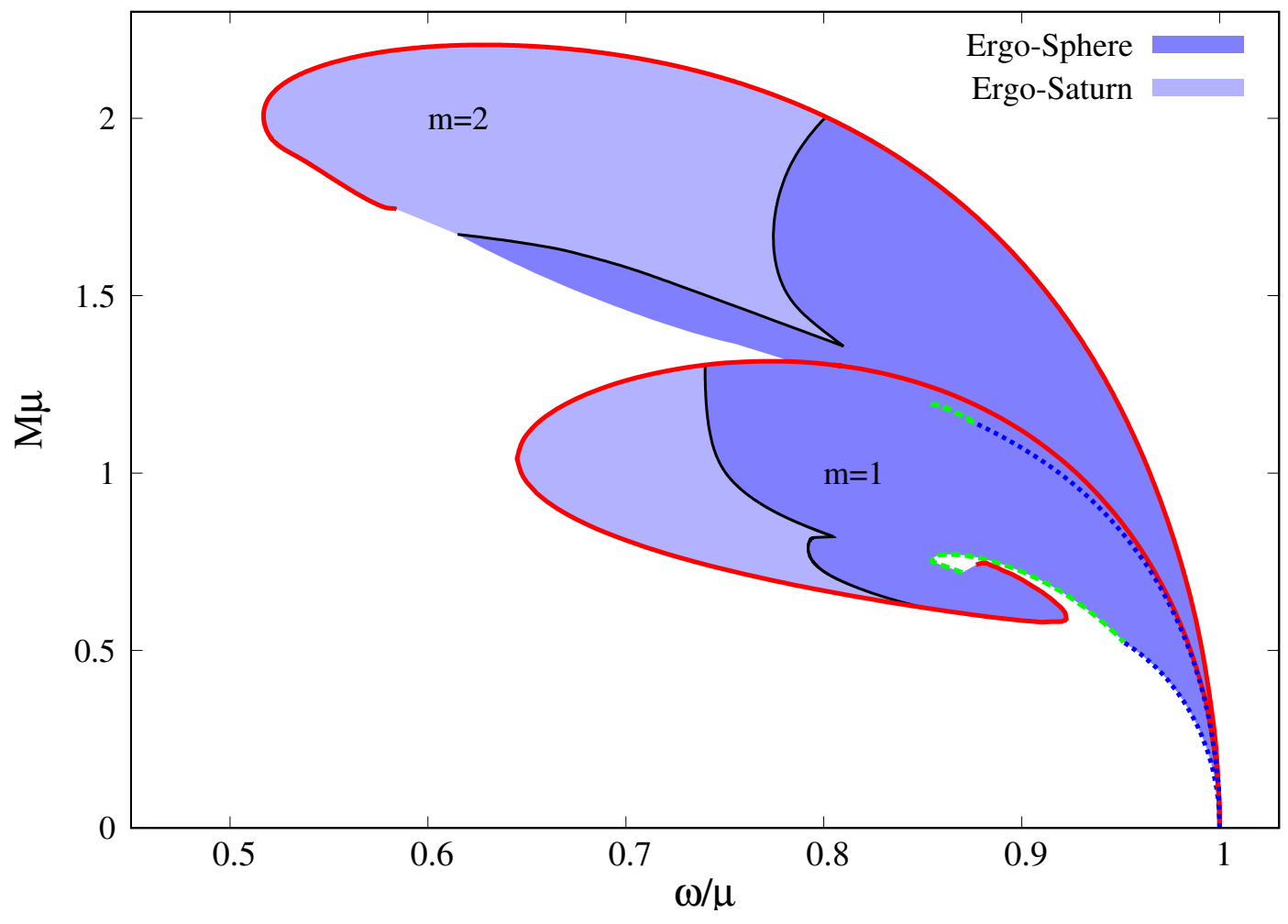

**Figure 3.5:** Ergo-regions typologies. Hairy BHs develop an ergo-sphere in the dark blue shaded region and an ergo-Saturn in the light blue shaded region.

In Figure 3.5 we show how the typology of ergoregions is distributed in the domain of existence of hairy BHs with $m = 1, 2$. The distribution is qualitatively similar in both cases. Ergo-spheres exist in the hairy BHs that connect to BSs without ergoregions and also in the vicinity of the Kerr limit. Ergo-Saturns, on the other hand, only exist in the parts of the domain of existence of lower frequency, in the neighbourhood of the BS solutions that possess an ergo-torus. The transition from solutions that possess only an ergo-sphere to the ones with the composite structure of an ergo-Saturn is similar to that found in the $m = 1$ case and which is detailed in Fig. 3 in [106]. Similar ergo-Saturns were recently reported in a different model of BHs with synchronised hair [151].

### 3.5.3 Horizon isometric embedding

As a final physical aspect let us consider the horizon geometry of the higher $m$ hairy BHs. In subsection 2.3.1 of the last chapter we analysed the horizon geometry of KBHsSH with $m = 1$ [64]. We found that, in order to totally embed the horizon in an Euclidean 3-space, $\mathbb{E}^3$, the following function must be always non-negative,

$$k(\theta) = e^{2F_1(r_H,\theta)} - e^{2F_2(r_H,\theta)} \left[F_2'(r_H,\theta)\sin\theta + \cos\theta\right]^2 , \tag{3.2}$$

where the prime denotes the angular derivative. This can be assured iff the second derivative of the $k(\theta)$ function evaluated at the poles is non-negative, *i.e.*, $k''(0) \geq 0$. Solving this inequality yields,

$$F_1''(r_H, 0) - 3F_2''(r_H, 0) + 1 \geqslant 0 . \tag{3.3}$$

On the other hand, the Gaussian curvature of the horizon at the poles is given by,

$$\mathcal{K}|_{\theta=\{0,\pi\}} = \frac{e^{-2F_1(r_H,0)}}{r_H^2} \left[F_1''(r_H, 0) - 3F_2''(r_H, 0) + 1\right] . \tag{3.4}$$

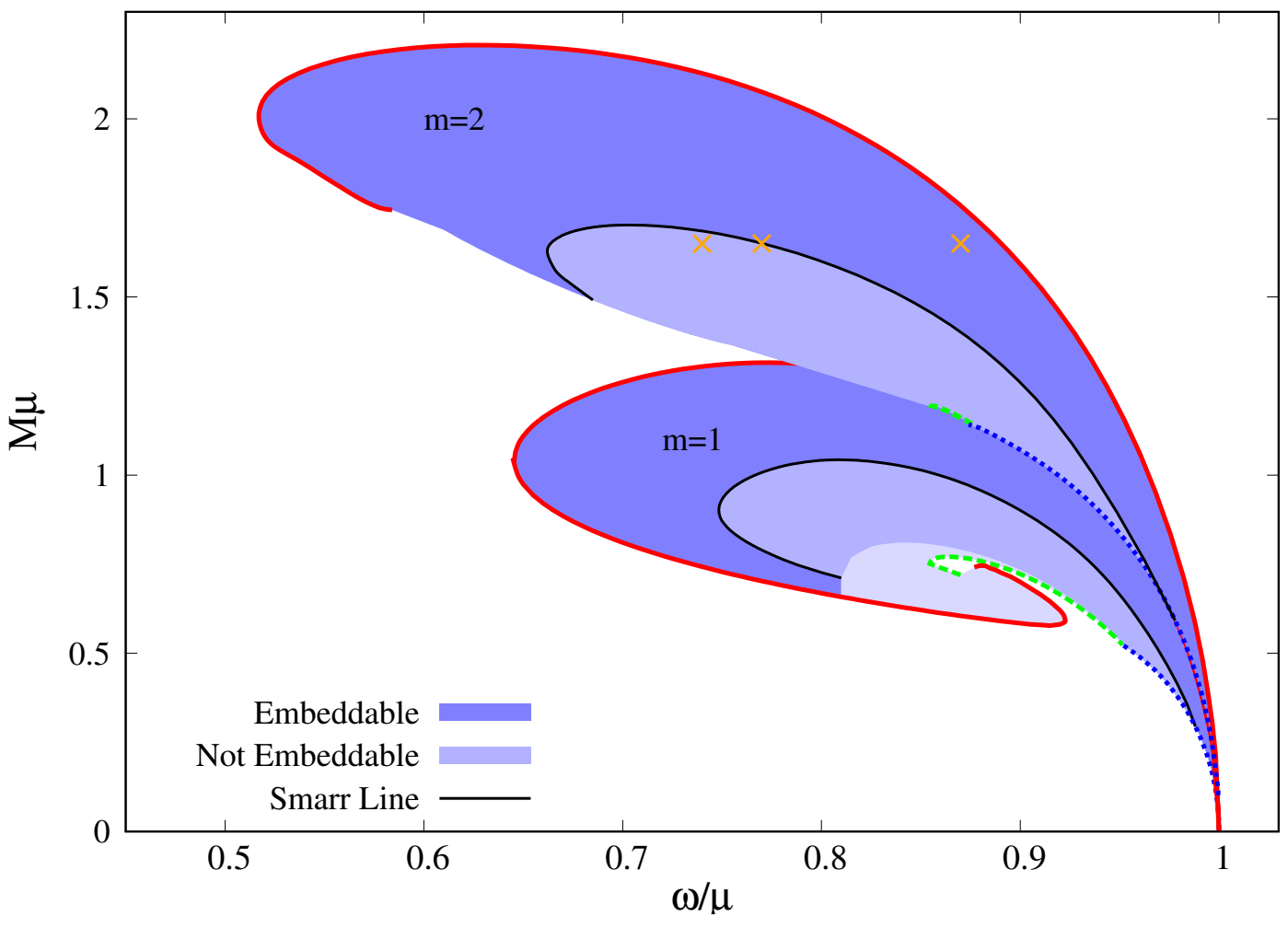

**Figure 3.6:** Smarr line in the domain of existence, for both $m = 1, 2$ families. The dark (medium) blue regions correspond to embeddable (non-embeddable) solutions. The light blue region for $m = 1$ solutions corresponds to solutions that were not analysed due to numerical accuracy. Three illustrative solutions are highlighted (with crosses) all with the same ADM mass, $M\mu = 1.65$.

Thus a necessary and sufficient condition for a global embedding in $\mathbb{E}^3$ to exist is that the curvature at the poles is non-negative. This conclusion was first obtained in the Kerr-Newman case by Smarr [82]. Thus, the threshold of embeddability occurs when the Gaussian curvature vanishes at the poles. The sequence of hairy solutions that occur at this threshold composed the Smarr line [64].

In Figure 3.6 we present the Smarr (black solid) line in the domain of existence of both $m = 1, 2$ solutions. The Smarr line divides the domain of existence into the embeddable region (medium blue) and non-embeddable region (dark blue). This division is qualitatively similar for both the $m = 1, 2$ families. Both Smarr lines start at the existence line, in the exact point where the Kerr BH is no longer embeddable, and both have an inspiral behaviour, attaining first a maximum value of the ADM mass, then a minimum value of the angular frequency of the scalar field and backbends into the opposite direction.

A visualisation of the isometric embedding of the horizon in $\mathbb{E}^3$ is shown in Figure 3.7 for the three hairy BH solutions with the same ADM mass, $M\mu = 1.65$, highlighted in Figure 3.6. The first solution is within the non-embeddable region, so the embedding misses the region close to the poles. The second solution is on the Smarr line, thus this solution will have a zero Gaussian curvature at the poles, therefore such region will appear flat. The third and final solution is within the embedding region, so it will be possible to draw completely the horizon. Concerning the latter, we remark that, as this solution is close to the BS line, where the solutions have a vanishing horizon area, $A_H \to 0$, its horizon is smaller than that of the previous two solutions.

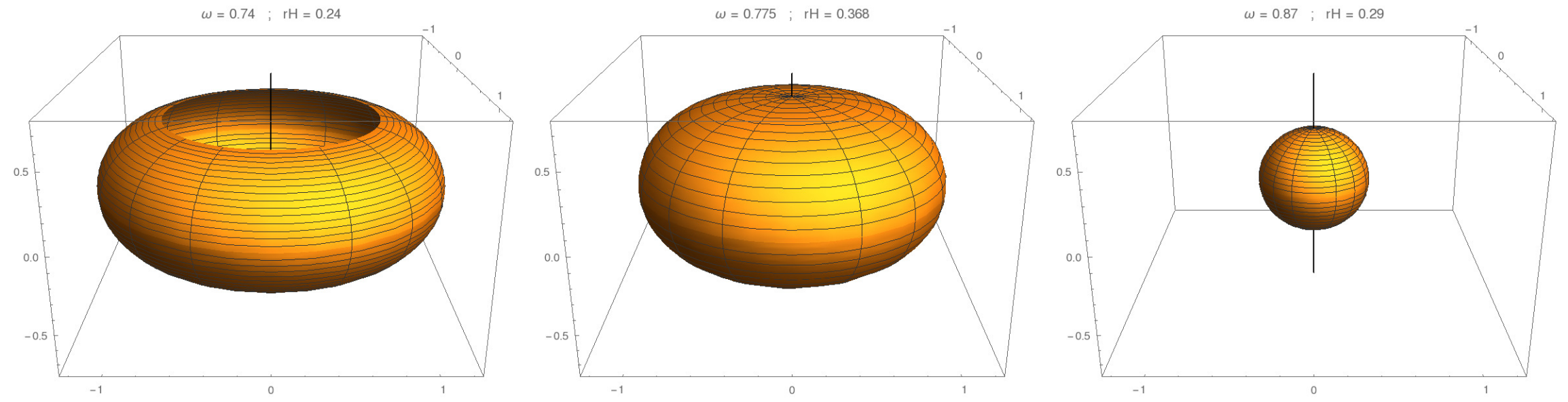

**Figure 3.7:** Isometric embedding in $\mathbb{E}^3$ of the three highlighted solution in Figure 3.6. Left panel: Non-embeddable solution; Middle panel: Solution on the Smarr Line; Right panel: Embeddable solution.

### 3.5.4 Horizon sphericity and linear velocity

To conclude the horizon analysis, following the same analysis done in [64] and at the end of the previous chapter, we consider the sphericity, $\mathfrak{s}$, and the horizon linear velocity, $v_H$, in order to assess what is the key property to determine the global embeddability of the horizon. As seen previously, the sphericity measures the deformation of a $U(1)$ invariant compact and simply connected 2-surface when compared to a round sphere, and is defined as the ratio between the proper lengths of the horizon measured around the equator and the poles – *cf.* Equation 2.9. For the hairy BHs it amounts to,

$$\mathfrak{s} = \frac{\pi \, e^{F_2(r_H,\pi/2)}}{\int_0^\pi d\theta e^{F_1(r_H,\theta)}} \; . \tag{3.5}$$

The horizon linear velocity [63] measures how fast the null geodesics generators of the horizon rotate relatively to a static observer at spatial infinity and is defined as the multiplication between the perimetral radius of the circumference at the equator and the horizon angular velocity – *cf.* Equation 2.13. For a hairy BH,

$$v_H = e^{F_2(r_H,\pi/2)} r_H \Omega_H \; . \tag{3.6}$$

Both quantities are exhibited in Figure 3.8 as a function of the radial coordinate of the horizon, $r_H$. In this representation all hairy solutions are enclosed by the existence line and the vertical line $r_H = 0$, which correspond to both the extremal line – green dashed line – and the BS line – red cross. The Smarr line is also plotted in both figures, as well as its Kerr limit (the Smarr point). The value at the Smarr point is then extrapolated as a benchmark – Smarr point value (dashed pink) line. For the sphericity, the Smarr point has a value of $\mathfrak{s}^{(S)} \approx 1.23601$ and for the horizon linear velocity, the Smarr point has a value of $v_H^{(S)} \approx 0.57735$, as we discussed in the previous chapter.

Consider first Figure 3.8a. We see that the Smarr line has the same value of sphericity as the Smarr point, within numerical accuracy. Therefore, the sphericity is a faithful diagnosis for embeddability also for $m = 2$ solutions: if $\mathfrak{s}$ is lower or equal than $\mathfrak{s}^{(S)}$ than the hairy BH will be embeddable; otherwise, it will be non-embeddable. The same was seen for hairy BHs with $m = 1$.

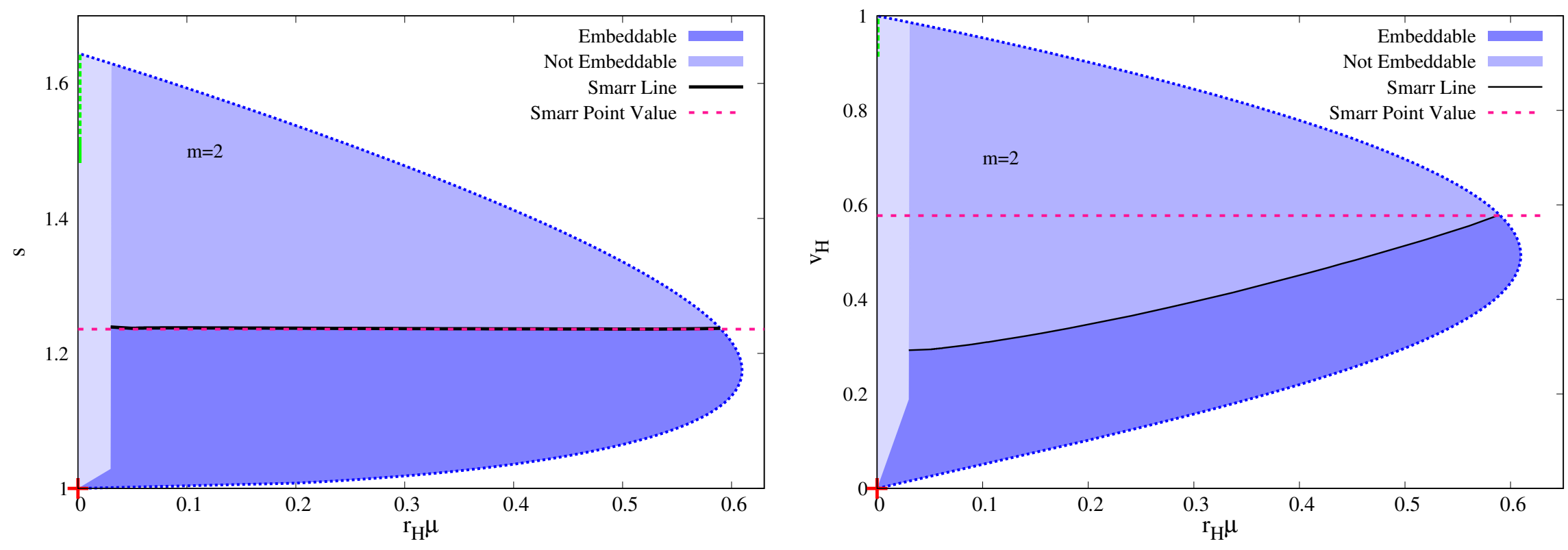

**(a)** **(b)**

**Figure 3.8:** **(a)** Sphericity, $\mathfrak{s}$, and **(b)** horizon linear velocity, $v_H$, $vs.$ the radial coordinate of the event horizon, $r_H$. As before, the dark blue region corresponds to embeddable solutions, and the medium blue represent non-embeddable solutions. The new region of light blue corresponds to solutions that were not analysed due to numerical accuracy.

Now consider Figure 3.8b. None of the hairy solutions exceeds $v_H = 1$, $i.e$ the speed of light. The limit of $v_H = 1$ is only attained by the extremal vacuum Kerr BH. Concerning the Smarr line, unlike what we saw in Figure 3.8a, the Smarr line only matches the Smarr point value at in the Kerr limit. The remaining Smarr line solutions will always have a lower $v_H$ than the one obtained at the Smarr point, $v_H^{(S)}$. Thus, the value of horizon linear velocity of the Smarr point – pink dashed line on Figure 3.8b – is an upper bound, above which all hairy solution with $m = 2$ are non-embeddable. Below that bound, both embeddable and non-embeddable solutions exist. The same results were found for $m = 1$.

## 3.6 Discussion

In this chapter, we have constructed and analysed Kerr BHs with synchronised hair and higher azimuthal harmonic index $m$. Specifically, solutions with $m = 2, 3$ were constructed and contrasted with the $m = 1$ solutions.

There are two results from the analysis that should be emphasised. Firstly, consecutive $m$ families can have degenerate solutions in terms of the global quantities $(M, J)$. When this occurs, the higher $m$ solutions are entropically favoured. This supports the possibility that migrations between such families, triggered by the superradiant instability, could be approximately conservative. This possibility, however, is by no means guaranteed to occur dynamically, as significant gravitational radiation and scalar ejection could take place in this migration. Secondly, there is a high degree of universality in all physical properties that have been unveiled for $m = 1$ solutions, that our analysis shows extend *mutatis mutandis* for the higher $m$ solutions. These properties include, in particular, the typology of ergo-regions and the event horizon geometry. There is no reason to expect new qualitative features concerning these physical properties would emerge for even higher $m$ values.

Similar solutions will also exist in other models of BHs with synchronised hair, for instance including self-interactions [93], or with a Proca field [94]. The results herein indicate no significant differences are to be expected with respect to the $m = 1$ case in these models. It would, nonetheless, be interesting to study some phenomenological properties of this higher $m$ solutions, such as BH shadows [55], [57], $X$-ray spectrum [62], accretion disk morphology [140] or star trajectories [66], since these higher $m$ solutions could play a role in the dynamical evolution of the BH/fundamental field system, in case such fundamental, ultra-light, scalar or vector fields exist in Nature.

# CHAPTER 4

# Rotating Axion Boson Stars

## 4.1 Introduction

Bosonic stars have attracted much interest in strong gravity research. Various reasons support this interest. Their original construction was motivated by the search of *geons*, suggested by Wheeler [152]. In modern terminology, this is what one may call self-gravitating solitons. Electrovacuum does not seem to allow solitons (see [153] for an overview), but they exist in Einstein-Klein-Gordon theory. The original spherical scalar BSs [34], [35], obtained for a massive complex scalar field minimally coupled to Einstein's General Relativity, allowed many generalisations. These include the addition of several types of scalar self-interactions, *e.g.* [51], [92], [93], [150], [154], the inclusion of rotation [49], [50], see also *e.g.* [93], [147], [155], the construction of vector analogues (Proca stars) [156]–[158], including other asymptotics, *e.g.* [159], [160], multi-field configurations [161], alongside many others - see the reviews [32], [33].

Bosonic stars are, moreover, dynamically interesting solutions, in the sense some of the models are perturbatively stable [156], [162]–[164] and allow all sorts of dynamical studies, in particular of their dynamical formation [165], [166] and of their evolution in binaries, producing gravitational waveforms of interest for ongoing LIGO/Virgo searches, *e.g.* [129], [167]–[171]. Moreover, since bosonic stars can achieve a compactness comparable to that of BHs, they provide a case study example of BH mimickers, *e.g.* [172]–[177]. Finally, in the last few years there has been considerable interest in a possible role of BSs in the dark matter problem, in particular in connection to ultralight bosonic dark matter candidates [143], [144].

Amongst the theoretically suggested ultralight dark matter particles, axion like particles (ALPs) are alongside the best motivated ones. The QCD axion [178], [179] is a pseudo Nambu-Goldstone boson suggested by the Peccei-Quinn mechanism [180] proposed to solve the strong CP problem in QCD [181]–[183]. More recently, the axion started to be used as a prototype for weakly-interacting ultralight bosons beyond Standard Model [184]–[187] and considered as a plausible dark matter candidate. In particular, ALPs are ubiquitous in string compactifications, wherein they can be seen both as particles beyond Standard Model and signature of extra dimensions [188], [189].

Recently, spherical BSs with self-interactions determined by an axion-type potential were discussed in [190] and dubbed *axion boson stars*.[1] The goal of this chapter is to construct and study the basic physical properties of *rotating axion boson stars* (RABSs), the spinning generalisation of the solutions in [190]. On the one hand, astrophysical objects in the Universe generically have angular momentum; thus, it is important to consider RABSs to assess the physical plausibility of axion bosonic stars. On the other hand, numerical evolutions have recently revealed an instability of the simplest model of spinning BSs [193]. This instability could be mitigated by self-interactions, thus adding another motivation to construct RABS.

This chapter is based on the work presented in [78] and it is organised as follows: In section 4.2, we present the full model, the equations of motion, the ansatz used to solve them, and the QCD axion potential. In section 4.3, we provide the numerical framework in which we base our results, discuss the boundary conditions, the quantities of interest and the numerical approach. In section 4.4, we present the numerical results, describing the space of RABSs solutions and analysing some physical properties. We conclude the chapter in section 4.5 with some remarks.

## 4.2 The model

We will work within the (complex-)Einstein-Klein-Gordon theory, which describes a massive complex scalar field, $\Psi$, minimally coupled to Einstein's gravity.

$$\mathcal{S} = \int d^4x\sqrt{-g}\left[\frac{R}{16\pi} - g^{\mu\nu}\partial_\mu\Psi^*\partial_\nu\Psi - V(|\Psi|^2)\right] , \tag{4.1}$$

where $V$ is the scalar self-interaction potential. The equations of motion resulting from the variation of the action with respect to the metric and scalar field, are,

$$E_{\mu\nu} \equiv R_{\mu\nu} - \tfrac{1}{2}g_{\mu\nu}R - 8\pi\, T_{\mu\nu} = 0 , \tag{4.2}$$

$$\Box\Psi - \frac{\partial V}{\partial|\Psi|^2}\Psi = 0 , \tag{4.3}$$

where

$$T_{\mu\nu} = 2\partial_{(\mu}\Psi^*\partial_{\nu)}\Psi - g_{\mu\nu}\left(\partial^\alpha\Psi^*\partial_\alpha\Psi + V\right) , \tag{4.4}$$

is the energy-momentum tensor associated with the scalar field.

Similarly to the hairy BHs studied in the previous chapter, a conserved Noether charge, $Q$, is possible to be computed due to the $U(1)$ symmetry of the scalar field. This quantity can be interpreted as the number of scalar particles in a given solution, albeit this relation only becomes rigorous upon field quantisation. A simple computation, moreover, shows that for solitonic solutions (without an event horizon[2]) $Q$ is related with the total angular momentum

[1] Axionic stars with a real scalar field (oscillatons) had been previously considered, *e.g.* [191], [192].

[2] For solutions with an event horizon, as the ones studied in the previous chapters, we already know that $Q$ is related with only the angular momentum associated to the scalar field, $J^\Psi = mQ$ – *cf.* Equation 1.39

as [49], [50]

$$J = mQ \ . \tag{4.5}$$

This is a generic relation for rotating BSs, already observed in other models with a self-interactions potential - see, *e.g.* [150].

We are interested in horizonless, everywhere regular, axisymmetric and asymptotically flat solutions of the above field equations. Due to the horizonless and everywhere regular conditions, we can not use the same ansatz for the metric as before since such ansatz was optimised for solutions with an event horizon. Thus, we choose a new ansatz for the metric obtained through the line element in Equation 1.19 by imposing $r_H = 0$, which yields $N = 1$. We further do the following replacement $W \longrightarrow W/r$ as we find it preferable to work with this new $W$ function when dealing with horizonless objects due to their boundary conditions at the origin. Therefore, the new metric reads,

$$ds^2 = -e^{2F_0}dt^2 + e^{2F_1}\left(dr^2 + r^2 d\theta^2\right) + e^{2F_2} r^2 \sin^2\theta \left(d\varphi - \frac{W}{r}dt\right)^2 , \tag{4.6}$$

where, again, $\{F_i, W\}_{i=0,1,2}$ are ansatz functions that only depend on $(r, \theta)$ coordinates. The ansatz for the scalar field is the same as in Equation 1.16.

To obtain solutions one has to specify the scalar field potential $V$. Following [190], we shall use the QCD axion potential - see [194] for a discussion -, added to a constant term, in order to construct asymptotically flat RABSs. The potential to be considered is

$$V(\phi) = \frac{2\mu_a^2 f_a^2}{B}\left[1 - \sqrt{1 - 4B\sin^2\left(\frac{\phi}{2f_a}\right)}\right] , \tag{4.7}$$

where $B$ is a constant, defined in terms of the up and down quarks' masses, $m_u, m_d$, as $B = \frac{z}{(1+z)^2} \approx 0.22$, where $z \equiv m_u/m_d \approx 0.48$. The potential has two free parameters, $\mu_a$ and $f_a$. To see their meaning, we expand the potential around $\phi = 0$, yielding

$$V(\phi) = \mu_a^2 \phi^2 - \left(\frac{3B-1}{12}\right)\frac{\mu_a^2}{f_a^2}\phi^4 + \ldots \tag{4.8}$$

where one can identify the ALP mass and quartic self-interaction coupling, respectively,

$$m_a = \mu_a \ , \qquad \lambda_a = -\left(\frac{3B-1}{12}\right)\frac{\mu_a^2}{f_a^2} \ . \tag{4.9}$$

Thus, $\mu_a$ defines the mass of the ALP and $f_a$ defines the strength of the self-interactions. We shall refer to $\mu_a$ and $f_a$ as the ALP mass and decay constant, respectively. We remark the above expansion is only valid if

$$f_a \gg \phi \ . \tag{4.10}$$

In fact, to leading order, only the mass term remains. Thus, we can anticipate that the properties of the solutions will match those of the standard spinning mini-BSs [49], [50] when $f_a$ is large.

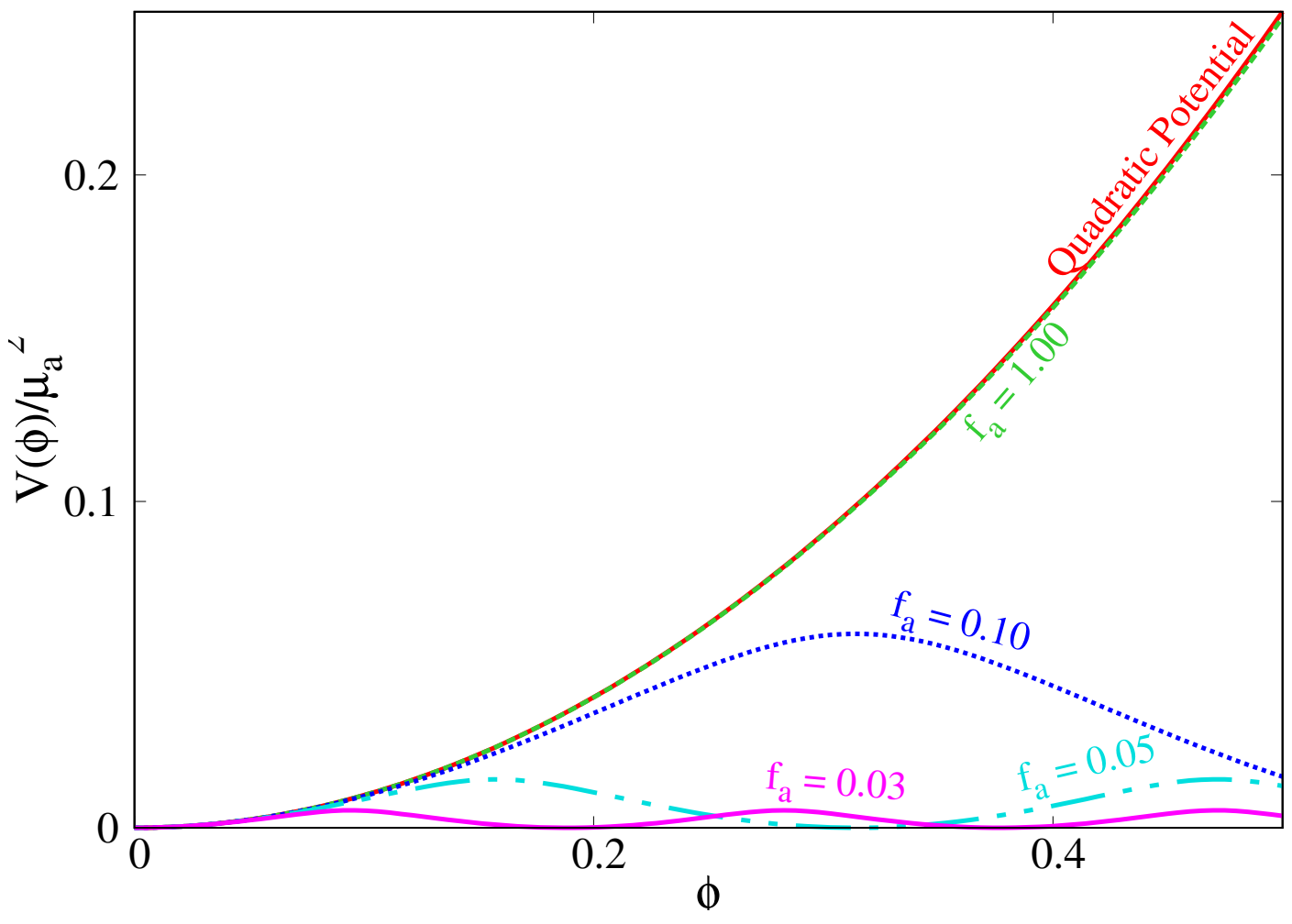

**Figure 4.1:** Axion potential Equation 4.7 for several values of the decay constant $f_a$ together with the quadratic potential.

In Figure 4.1 we show the axion potential Equation 4.7 for several values of $f_a$, together with the quadratic potential - first term in Equation 4.8. It can be seen that for $f_a = 1.0$ the quadratic potential is a good approximation for the range of values of $\phi$ displayed, which will be in the range obtained in the RABSs solutions below, *cf.* top panel in Figure 4.5 below.

We remark that the potential (Equation 4.7) satisfies the conditions in Ref. [195] allowing for the existence of solitonic solutions even in the absence of the gravity term in the action (Equation 4.1). Our results indicate that indeed, such configurations exist in a flat spacetime background, sharing all the basic properties of the (non-gravitating) Q-balls with a sextic potential [51]. A discussion of these solutions will be reported elsewhere.

## 4.3 Framework

### 4.3.1 Boundary conditions

With the metric and scalar field ansatz defined, we now need to specify the boundary conditions for the ansatz functions that best suit the problem at hand. As in the previous chapters, these solutions are asymptotically flat and axially symmetric. Therefore, their ansatz functions will have the same behaviour on those boundaries like the ones obtained previously. However, there is a critical difference. Since these solutions are horizonless, we impose boundary conditions at the origin instead of at the event horizon. Moreover, regularity implies that we must impose the following boundary conditions at the origin, $r = 0$,

$$\partial_r F_i = W = \phi = 0 \,. \tag{4.11}$$

### 4.3.2 Extracting physical quantities

For RABSs, two of the key quantities of interest are encoded in the decay of the metric functions towards spatial infinity. Such quantities are the ADM mass $M$ and angular momentum $J$ which are read off from the asymptotic behaviour of the following metric functions, in a similar way as described in the chapter 1,

$$g_{tt} = -e^{2F_0} + e^{2F_2} W^2 \sin^2\theta \quad \to \quad -1 + \frac{2M}{r} + \ldots \ , \tag{4.12}$$

$$g_{\varphi t} = -e^{2F_2} W r \sin^2\theta \quad \to \quad -\frac{2J}{r}\sin^2\theta + \ldots \ . \tag{4.13}$$

Alternatively, both quantities can be computed through their Komar integrals [196] as in Equation 1.37. However, the spacelike surface $\Sigma$ is now only bounded by a 2-sphere at infinity, $S^2_\infty$, since there is no event horizon. A good test for the numerical results amounts to checking the match between the mass and angular momentum obtained from the asymptotic behaviour and from the Komar integrals.

Another quantity of interest is the compactness of the RABSs. Since the scalar field decays exponentially, these stars do not have a well defined surface, wherein a discontinuity in the matter distribution occurs. Nonetheless, a standard definition of "surface" in this context is as follows [86]. Firstly, we compute the perimeteral radius $R_{99}$, which contains 99% of the total mass of the BS, $M_{99}$. The perimetral radius is a geometrically significant radial coordinate $R$, such that a circumference along the equatorial plane has perimeter $2\pi R$, and it is related to the radial coordinate $r$ in the ansatz metric by $R = e^{F2} r$. Secondly, we define the inverse compactness as the ratio between $R_{99}$ and the Schwarzschild radius associated with 99% of the RABS's mass, $R_{Schw} = 2M_{99}$,

$$\text{Compactness}^{-1} = \frac{R_{99}}{2M_{99}} \ . \tag{4.14}$$

We expect stars to be less compact than BHs and thus this quantity to be larger than unity for RABSs.

### 4.3.3 Numerical approach

The numerical approach used to obtain RABS is quite similar to the one done to obtain KBHsSH (both with $m = 1$ or with higher azimuthal harmonic index discussed in the previous chapters). After rescaling some key quantities by the ALP mass, $\mu_a$, in a similar way as in Equation 1.28, together with the rescale of $f_a \to f_a\sqrt{4\pi}$, one has to solve a set of differential equations consisted by the Klein-Gordon equation, Equation 4.3, together with the same combination of the Einstein equations described in Equation 1.29 through Equation 1.32.

This set is discretised on a 2D grid for the radial and angular part. Regarding the radial part, only the coordinate transformation $x \equiv r/(r+1)$ is performed to map the semi-infinite region $[0, \infty)$ to the finite region $[0, 1]$. For the majority of the results in this chapter, we used an equidistant grid with 401×40 points. The grid covers the integration region $0 \leqslant x \leqslant 1$ and $0 \leqslant \theta \leqslant \pi/2$.

All differential equations were solved using the same professional package used in the previous chapters. For the RABSs solutions, the maximal numerical error is estimated to be of the order of $10^{-3}$. As mentioned in the previous subsection, a further test of the numerical accuracy of the solutions can be obtained by comparing the ADM mass and angular momentum with the corresponding Komar integral. This comparison yields an error estimate of the same order as the one given by the solver.

In our scheme, there are three input parameters: **i**) the decay constant $f_a$ in the potential, Equation 4.7; **ii**) the angular frequency of the scalar field $\omega$ and **iii**) the azimuthal harmonic index $m$. The number of nodes $n$ of the scalar field, as well as all other quantities of interest mentioned before, are computed from the numerical solution. For simplicity, we have restricted our study to fundamental configurations, *i.e.* with a nodeless scalar field, $n = 0$ and with $m = 1$.

## 4.4 Numerical Results

We obtained around sixty thousand solutions, considering a set of illustrative values of the decay constant $f_a$. For each choice of $f_a$, $\omega$ spans an interval of values. For all solutions, the metric functions, together with their first and second derivatives with respect to both $r$ and $\theta$, have smooth profiles. This leads to finite curvature invariants on the full domain of integration. The profile of the metric functions, together with the Ricci and Kretschmann scalars, $R$ and $K \equiv R_{\alpha\beta\mu\nu}R^{\alpha\beta\mu\nu}$, as well as the components $T_t^t$ and $T_\varphi^t$ of energy-momentum tensor, of a typical solution are exhibited in Figure 4.2 and Figure 4.3. In particular, observe the toroidal shape of the scalar field (Figure 4.2, bottom left panel). Thus, similarly to mini-BSs, RABSs are mass tori in General Relativity. There is a clear imprint of this toroidal distribution in the curvature and energy-momentum tensor, as can be appreciated from Figure 4.3.

### 4.4.1 The domain of existence

In scanning the space of solutions of RABSs, after fixing the number of nodes $n = 0$ and the azimuthal harmonic index $m = 1$, such as to look for fundamental states, we have to vary the axion decay constant $f_a$. For each $f_a$, the domain of existence is obtained by varying the angular frequency of the scalar field $\omega$. When the decay constant is large, $f_a \to \infty$, the model reduces to that of a massive, free complex scalar field – *cf.* Equation 4.8 – and the solutions match the usual spinning mini-BSs. The opposite limit, wherein $f_a \to 0$, is trickier and solutions cannot be obtained for arbitrarily low $f_a$. For the following discussion, we have chosen a sample of illustrative values for the decay constant $f_a$ and have analysed the resulting solutions. The sample of $f_a$ values is

$$f_a = \{1.00, 0.10, 0.05, 0.03\} \,. \tag{4.15}$$

In Figure 4.4 the two global charges, mass and angular momentum, are exhibited *vs.* the angular frequency, for the chosen sample of $f_a$ values. As a reference, all panels include as well the corresponding plot for mini-BSs. Outstanding features include the following. Firstly, for the largest value of $f_a$ considered, $f_a = 1.00$, the domain of existence of the RABSs is indistinguishable from that of mini-BSs. This is consistent with the potential profile seen in Figure 4.1 and the range of scalar field values shown in the illustrative behaviour in Figure 4.2, bottom left panel; one can see that $f_a = 1.00$ is large enough so that Equation 4.10 holds. This is further confirmed by the analysis in Figure 4.5 (top panels), where the maximum value of the scalar field, $\phi_{\max}$ is shown against $\omega$. Following each line starting from $\omega = 1$, $\phi_{\max}$ increases monotonically, but only up to around $\phi_{\max} \sim 0.5$. This holds for all numerically accurate solutions obtained. Moreover, $\phi_{\max}$ is never much larger, in fact always comparable, to that in the mini-BS case. This confirms why $f_a \gtrsim 1.00$ RABSs are essentially equivalent to mini-BSs.

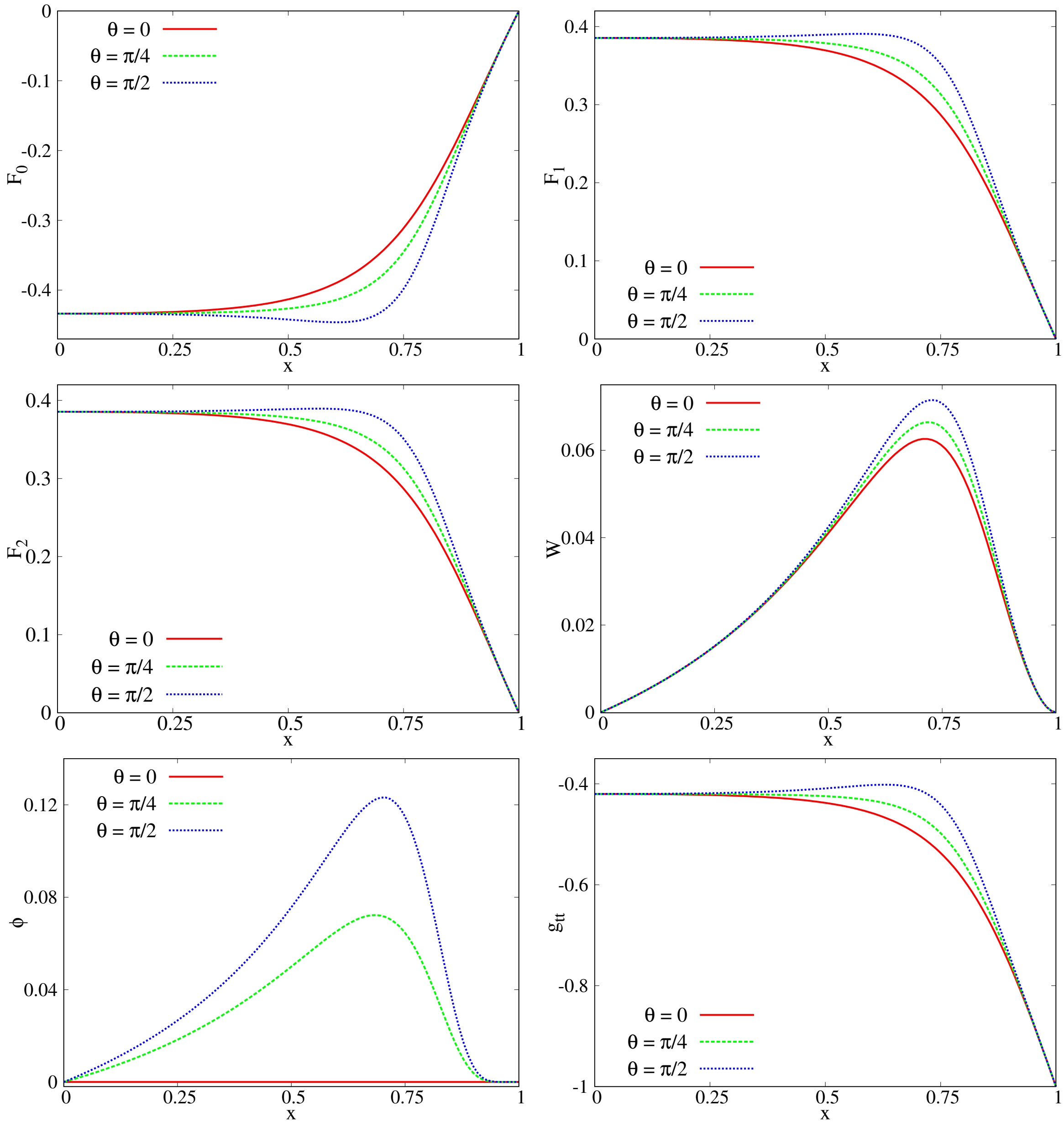

**Figure 4.2:** Profile functions of a typical solution with $f_a = 1.00$ and $\omega = 0.80$, $vs.$ $x = r/(1+r)$, which compactifies the exterior region, for three different polar angles $\theta$.

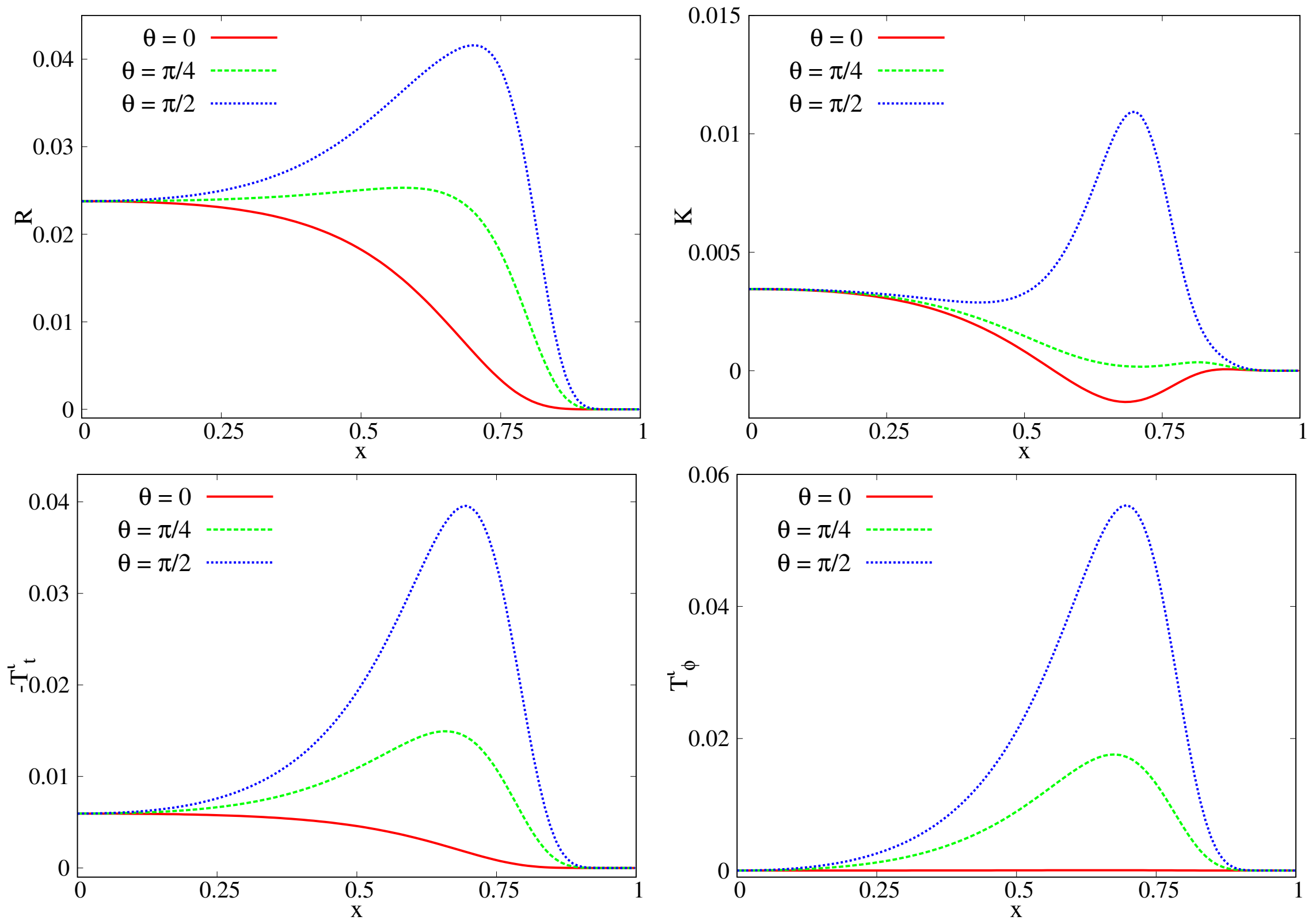

**Figure 4.3:** The Ricci $R$ and Kretschmann $K$ scalars and the components $T_t^t$ and $T_\varphi^t$ of the energy-momentum tensor *vs*. $x \equiv r/(1+r)$, of a typical solution with $f_a = 1.00$ and $\omega = 0.80$ for three different polar angles $\theta$.

The second feature we would like to emphasise about Figure 4.4 is that for all value of $f_a$, solutions exist within an interval of angular frequencies, which depends on $f_a$. As $f_a$ decreases, the lower end of this interval decreases; then, RABSs exist with lower $\omega$ than in the mini-BS case. This feature is instructive. In the case of quartic self-interactions, the frequency range of spinning BSs decreases, with increasing strength of the self-interactions – see Figure 1 in [93] – since the frequency at the lower end *increases*. Such trend for the angular frequency range is never observed in the panels of Figure 4.4. This implies that the quartic approximation Equation 4.8 to the potential Equation 4.7 is never a good approximation for RABSs.[3] In other words, when the quartic term becomes relevant, higher order terms become relevant as well for these solutions, in particular impacting decisively on the frequency range of the space of solutions.

A third observation concerning Figure 4.4, is that for small enough values of $f_a$, the line of solutions deviates from the typical spiral, as one can see for the cases of $f_a = 0.05$ and $f_a = 0.03$. The more intricate space of solutions impacts, in particular, on the global maximum of the mass. For mini-BSs, the ADM mass increases from the Newtonian limit (as $\omega \to 1$) until the global maximum is attained. But for sufficiently small $f_a$, this is only a local maximum; as the frequency continues to decrease, the mass has a non-monotonic behaviour, and the global maximum of the mass is attained at the lower end of the frequency interval. Similar trends occur for the angular momentum. This sort of non-spiralling curves, with the maximal mass appearing at the lower end of the frequency interval, can be seen for BSs with a $Q$-ball (sextic) potential, wherein the quadratic, quartic and sextic terms alternate in sign [51], [150].

[3]We remark that the sign of the quartic term in [93] and in Equation 4.8 is the opposite.

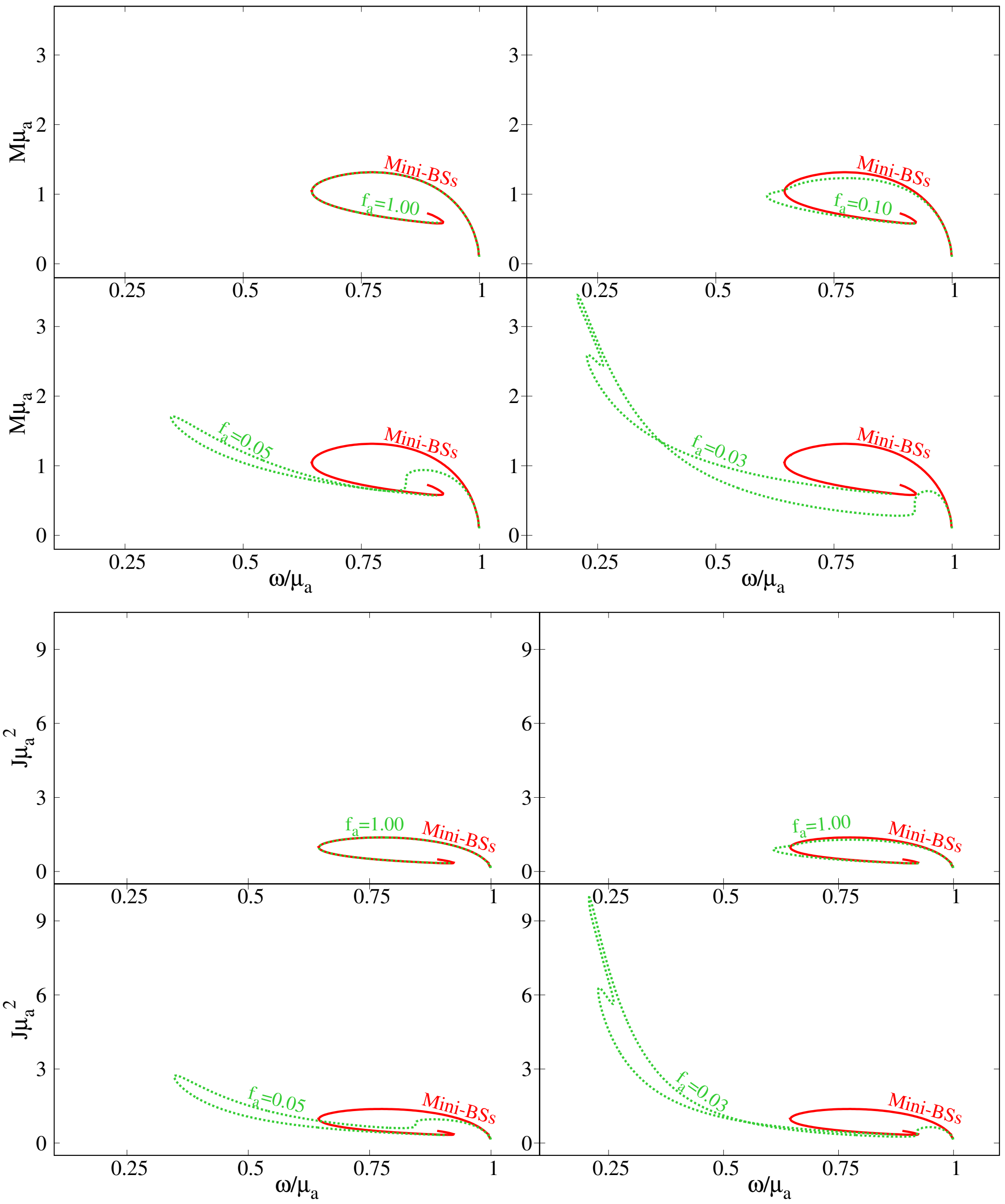

**Figure 4.4:** ADM mass $M$ (top four panels) and total angular momentum $J$ (bottom four panels) *vs*. the angular frequency of the scalar field $\omega$.

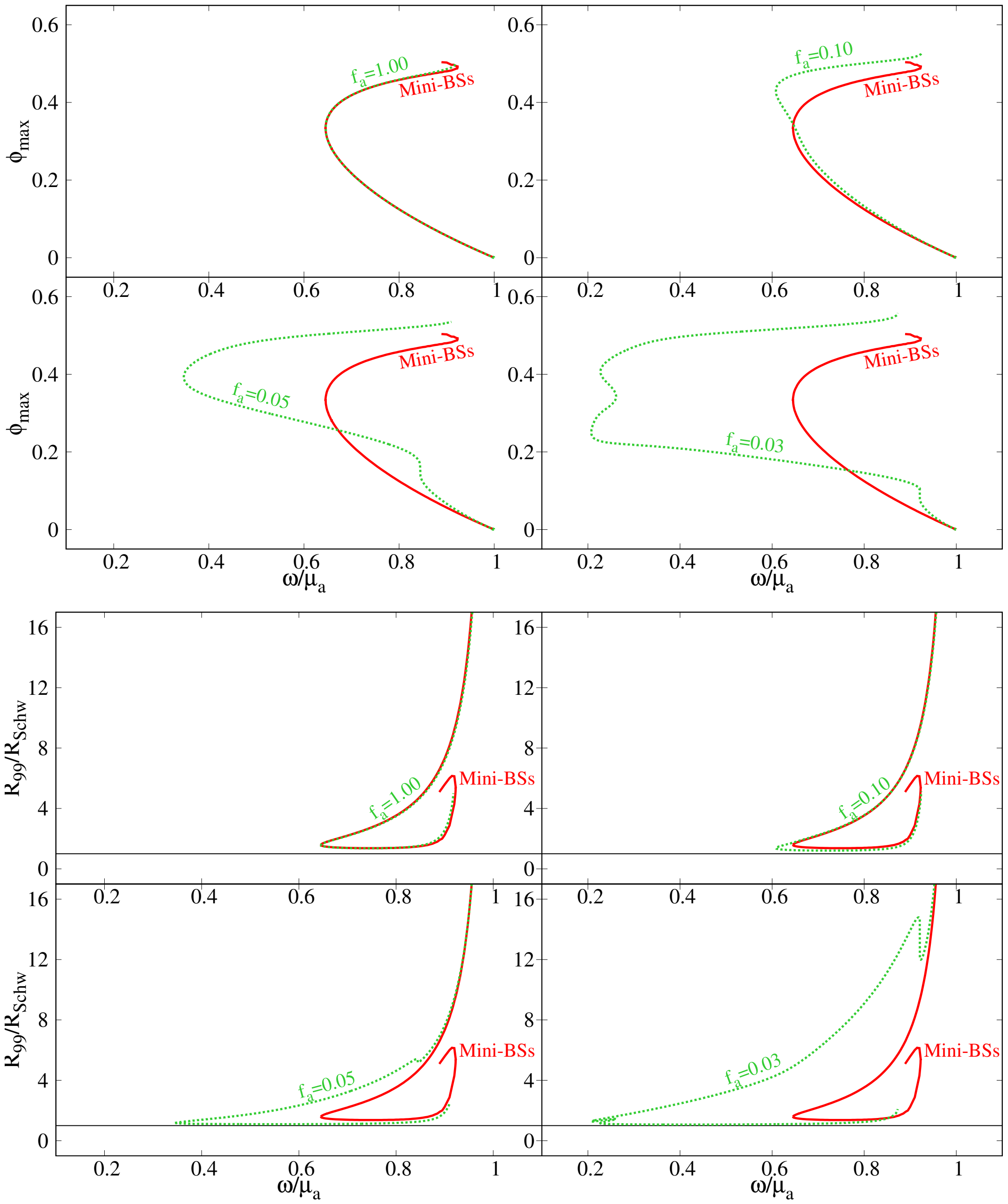

**Figure 4.5:** Maximum value of the scalar field (top four panels) and inverse compactness (bottom four panels) as a function of the angular frequency $\omega$. In the latter, the black horizontal lines correspond to the BH limit.

Finally, a curious and intriguing feature in Figure 4.4 is that despite the $f_a$ dependence of the curves, for all $f_a$, the space of solutions always appears to converge to similar values, albeit not exactly the same, of $M, J, \omega$ as one moves along the curve towards the strong gravity region (the "centre" of the spiral).

The bottom panels in Figure 4.5 show the inverse of the compactness, $R_{99}/R_{\rm Schw}$ *vs.* the angular frequency. All solutions, regardless of the decay constant $f_a$, have an inverse compactness larger than unity, meaning that they are always less compact than a BH, as expected. Moreover, decreasing $f_a$, more compact RABSs can be obtained, reaching closer to the BH limit, $R_{99}/R_{\rm Schw} = 1$. Finally, we remark that, unlike $\phi_{\rm max}$, the compactness is not a monotonic function along the solutions curve.

Figure 4.6 analyses some further features of the total angular momentum, $J$. In the top panels, $J$ is shown against ADM mass. This phase space confirms that $M, J$ are positively correlated: they either both increase or both decreasing, as one moves along the solution space, regardless of $f_a$. The bottom panels exhibit how the dimensionless spin $j \equiv J/M^2$ changes with the angular frequency. In the Newtonian limit, RABSs, as most macroscopic objects, can violate the Kerr bound, $j \leqslant 1$. As one moves along the solution space away from this limit, the solutions become more compact and violations of the Kerr bound may cease. As $f_a$ decreases, the Kerr bound abiding solutions seem to become less pronounced in solution space.

### 4.4.2 Other properties

#### *4.4.2.1 Ergoregions*

The existence of an ergoregion is an important property of the Kerr spacetime, with remarkable physical consequences, namely the possibility of energy extraction, as first pointed out by Penrose [149], [197]. For horizonless ultracompact objects, like spinning BSs, the presence of an ergoregion means, generically, an instability [198]. Thus, it is relevant to analyse the occurrence of an ergoregion for RABSs. Previous analysis of ergoregions in models of spinning BSs can be found in, *e.g.* [52], [106], [150], [151], [199].

In asymptotically flat spacetimes, ergoregions are defined as the spacetime domain in which the norm of $\xi = \partial_t$ becomes positive. An ergoregion is bounded by the surface $\xi^2 = 0$, or in terms of the metric functions,

$$g_{tt} = -e^{2F_0} + W^2 e^{2F_2} \sin^2\theta = 0 \, . \tag{4.16}$$

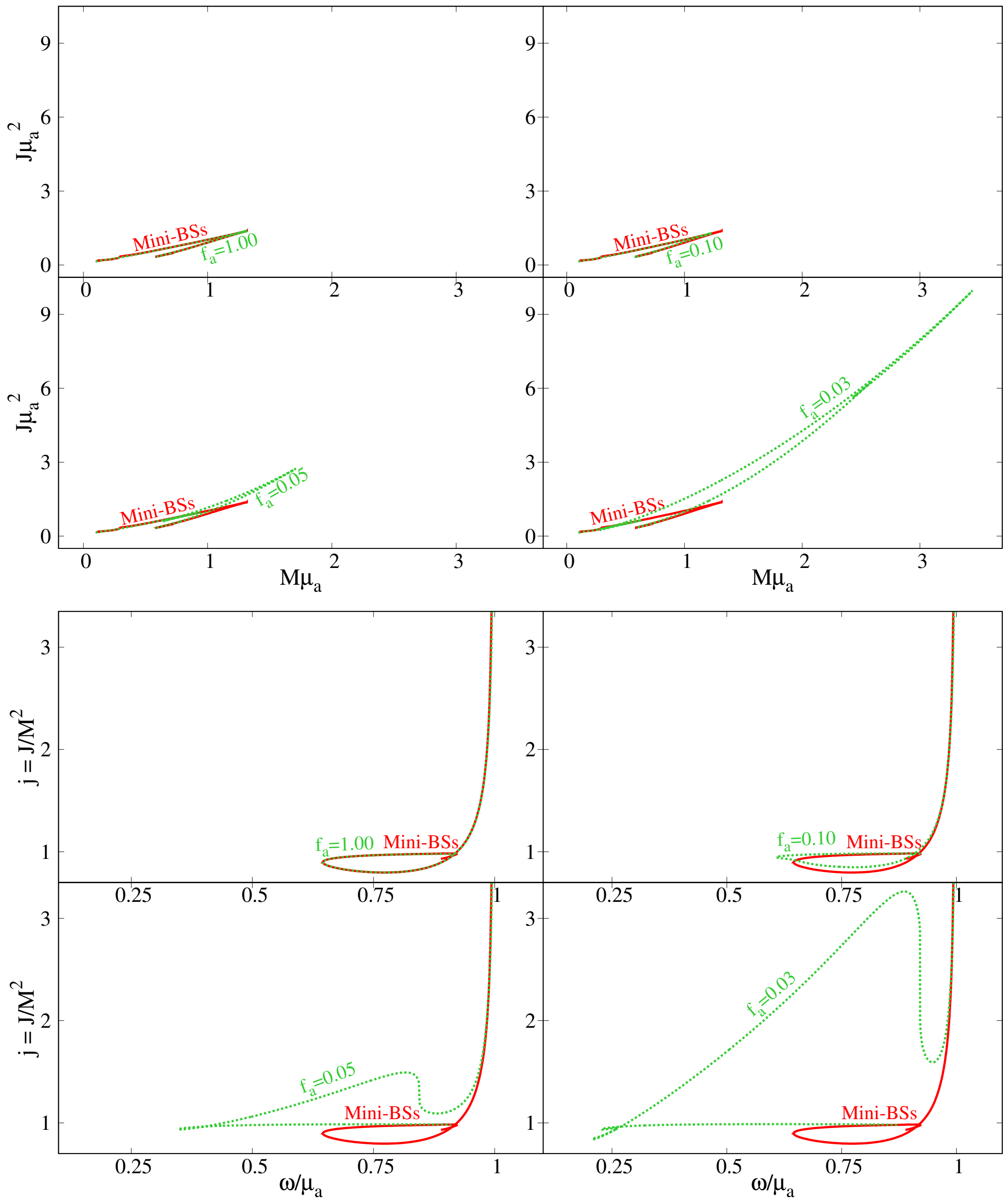

**Figure 4.6:** (Top panels) Total angular momentum *vs*. ADM mass. (Bottom panels) Dimensionless spin $j \equiv J/M^2$ *vs*. the angular frequency $\omega$.

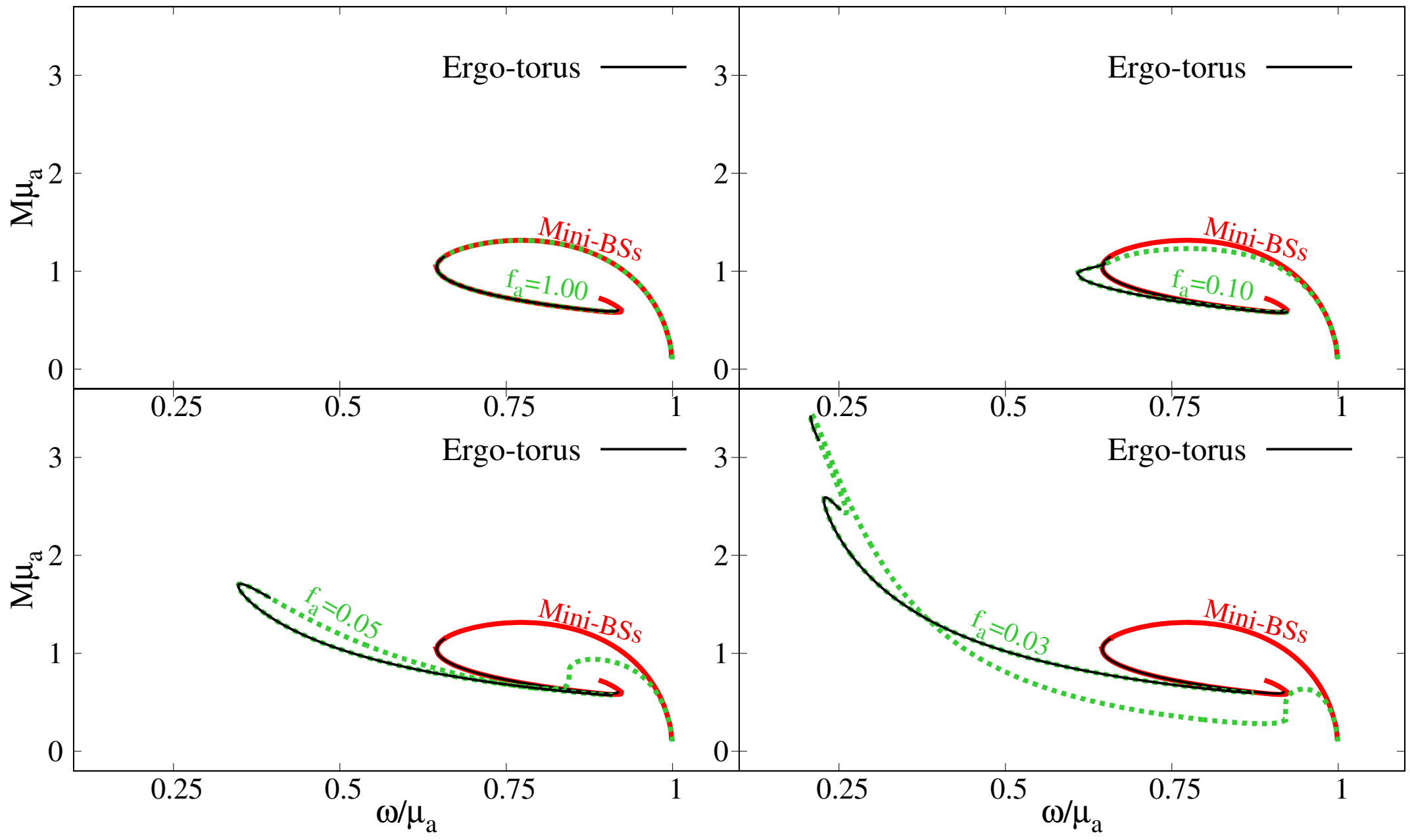

**Figure 4.7:** Occurrence of an ergo-torus in RABSs.

Ergoregions for spinning BSs have a toroidal topology, dubbed ergo-torus [106]. They are not present in the Newtonian limit, as $\omega \to 1$. For the case of spinning mini-BSs, the first solution showing an ergo-region occurs at a lower frequency than the maximal mass solution - Figure 4.7.

For the case of RABSs, we have found that the behaviour is analogous to that of mini-BSs. For large values of $f_a$, the solutions start to develop an ergo-torus around the same value of $\omega$ as in the case of mini-BSs. Following the solutions line towards the strong gravity regime, they have always an ergo-torus. Decreasing $f_a$, the first solutions that exhibit an ergo-torus occur for smaller values of $\omega$. This behaviour is shown in Figure 4.7, wherein the part of the solution line that contains ergo-regions is shown as a black solid line. A qualitative novelty is that for sufficiently small values of $f_a$, it is possible to have two disconnected regions where the solutions have ergo-torus. This behaviour is seen for the case with $f_a = 0.03$ in Figure 4.7, and is distinct from the remaining lines. As a quantitative reference, in Table 4.1 we provide the data of the first solution exhibiting an ergo-region, when moving along the solution line starting from the Newtonian limit.

### *4.4.2.2 Light rings and innermost stable circular orbits*

Besides ergoregions, other two strong gravity features found for geodesic motion around Kerr BHs are the existence of an innermost stable circular orbit (ISCO) for timelike geodesics, and light rings (LRs), for null geodesics [200]. Again, these features can be found for some, but not all, BSs. For instance, spherical mini-BSs do not have an ISCO; but spinning mini-BSs can have, see *e.g.* [174]. On the other hand, BSs, both static and spinning, only have LRs for sufficiently strong gravity configurations [55] - see also [201]. Such ultracompact BSs are perturbatively unstable [202]. Moreover, when LRs are present for BSs, they always come in pairs [203].[4] Also, the interplay between LRs and the ergoregion can lead to interesting lensing features [58].

| $f_a$ | Ergo-region | | LR | |
|---|---|---|---|---|
| | $\omega/\mu_a$ | $M\mu_a$ | $\omega/\mu_a$ | $M\mu_a$ |
| 1.00 | 0.658 | 1.154 | 0.747 | 1.308 |
| 0.10 | 0.651 | 1.078 | 0.747 | 1.224 |
| 0.05 | 0.396 | 1.563 | 0.646 | 0.843 |
| 0.03 | 0.207 | 3.428 | 0.336 | 1.712 |

**Table 4.1:** Data of the first RABS solution exhibiting an ergo-region or a LR, for several decay constants $f_a$.

RABSs with large $f_a$ are identical to mini-BSs; thus they must have an ISCOs and LRs. We shall now analyse how these features change when varying $f_a$. We will follow the standard method to obtain such orbits.

The effective Lagrangian for equatorial, $\theta = \pi/2$, geodesic motion on the geometry Equation 4.6 is

$$2\mathcal{L} = e^{2F_1}\dot{r}^2 + e^{2F_2}r^2\left(\dot{\varphi} - \frac{W}{r}\dot{t}\right)^2 - e^{2F_0}\dot{t}^2 \equiv \epsilon \ , \tag{4.17}$$

where all ansatz functions depend only on the radial coordinate $r$, the dot denotes derivative *w.r.t.* the proper time and $\epsilon = \{-1, 0\}$ for timelike and lightlike particles, respectively. Due to the isometries, we can write $\dot{t}$ and $\dot{\varphi}$ in terms of the energy $E$ and angular momentum $L$ of the test particle,

$$E = \left(e^{2F_0} - e^{2F_2}W^2\right)\dot{t} + e^{2F_2}rW\dot{\varphi} \ , \tag{4.18}$$

$$L = e^{2F_2}r^2\left(\dot{\varphi} - \frac{W}{r}\dot{t}\right) \ . \tag{4.19}$$

The effective Lagrangian, Equation 4.17, then yields an equation for $\dot{r}$, which defines a potential $V(r)$,

$$\dot{r}^2 = V(r) \equiv e^{-2F_1}\left[\epsilon - e^{-2F_2}\frac{L^2}{r^2} + e^{-2F_0}\left(E - L\frac{W}{r}\right)^2\right] \ . \tag{4.20}$$

For circular orbits (COs), both the potential and its first derivative must be zero, $V(r) = 0$ and $V(r)' = 0$. For lightlike particles, the former equation gives two algebraic equations, defining two possible impact parameters for the particle, $b_+ \equiv L_+/E_+$ and $b_- \equiv L_-/E_-$, corresponding to prograde and retrograde orbits, respectively; the latter equation, together with the impact parameters, yields the two radial coordinates of the prograde and retrograde LRs. Whenever it is possible to obtain a real solution for the radial coordinate, the RABS possesses LRs.

For all RABSs obtained in this chapter, no solution possesses a prograde LR. Retrograde LRs, on the other hand, exist for sufficiently strong gravity configurations. This is shown in Figure 4.8. For the largest value of $f_a$, LRs start to occur for roughly the same $\omega$ as for mini-BSs [55]. Moving along the solution line, after LRs first occur, they are always present along the solutions line (shown as the black solid line). Decreasing $f_a$, varies the frequency of the solution for which LRs first occur. As a quantitative reference, in Table 4.1 we provide the data of the first solution exhibiting a LR, when moving along the solution line starting from the Newtonian limit.

[4]The pair refers to one saddle point and one minimum of the effective potential [203]. Typically they are both prograde or retrograde. There may, or may not exist another pair, in the opposite sense of rotation.

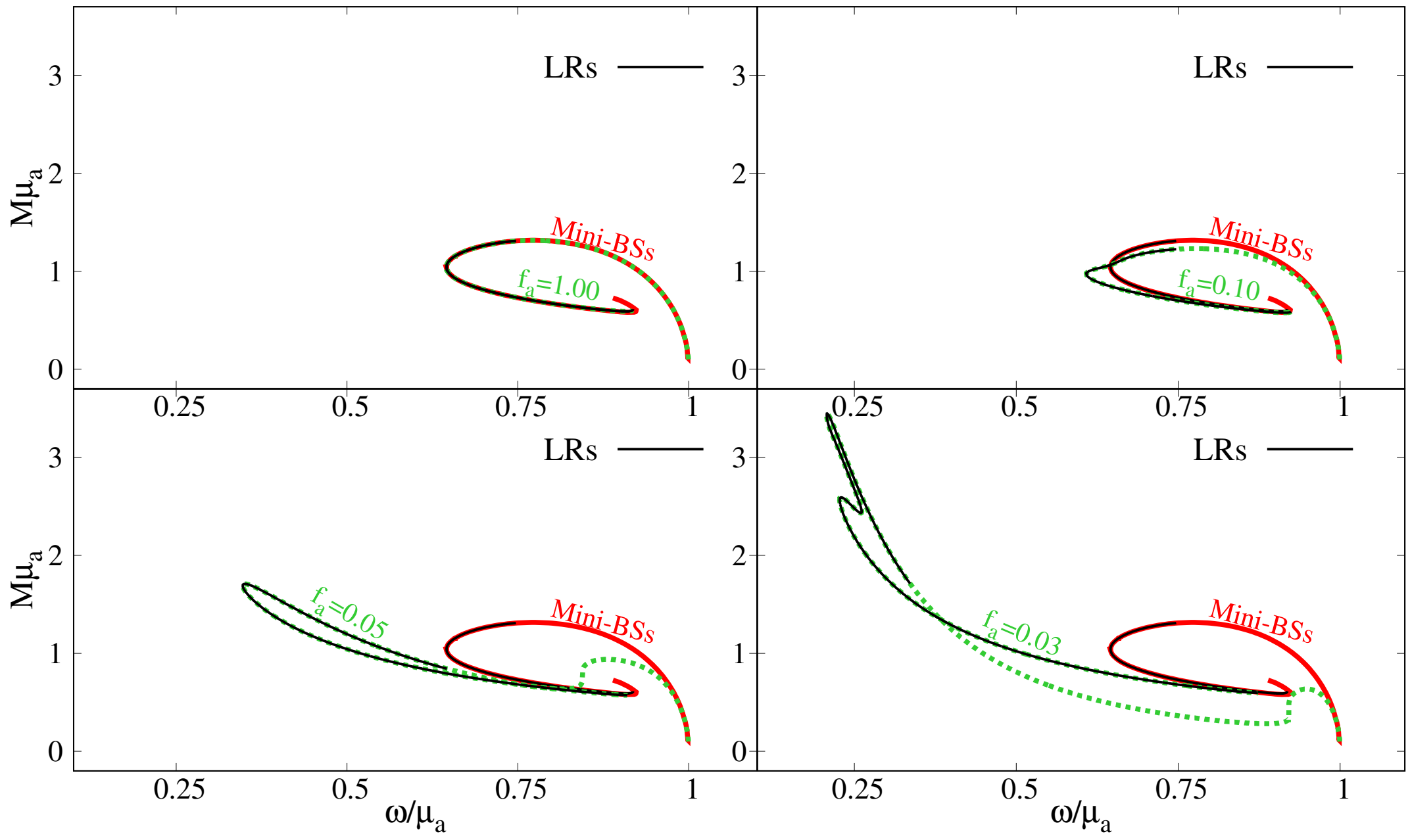

**Figure 4.8:** Occurrence of LRs in RABSs.

Now we turn to timelike orbits. In this case, the vanishing of the potential $V(r)$ and of its first derivative yields two algebraic equations for the energy and angular momentum of the particle, $\{E_+, L_+\}$ and $\{E_-, L_-\}$ corresponding to prograde and retrograde orbits, respectively. The stability of such orbits is determined by the sign of the second derivative of the potential $V(r)$. A negative (positive) sign corresponds to stable (unstable) COs. These are denoted as SCOs and UCOs, respectively.

By studying the stability of the equatorial timelike COs, we found a quite different structure than the one which is commonly found for BHs, and in particular for the Kerr solution. This different structure is exhibited in Figure 4.9, for mini-BSs and RABSs with $f_a = \{0.10, 0.05, 0.03\}$. The mini-BS case is essentially identical to the RABS case with $f_a = 1.00$. But since no such plot has been exhibited in the literature for the paradigmatic case of spinning mini-BSs we emphasise it here.

In Figure 4.9 we have chosen to parameterise the RABSs along the line of solutions, for any $f_a$ by the maximum value of the scalar field, since this is a monotonic quantity along this line, *cf.* Figure 4.5 (top panels): $\phi_{max}$ increases monotonically from the dilute/Newtonian regime ($\omega \sim 1$ and $\phi_{max} \sim 0$) until the strong gravity regime. Therefore, any horizontal line corresponds to a unique RABS solution.

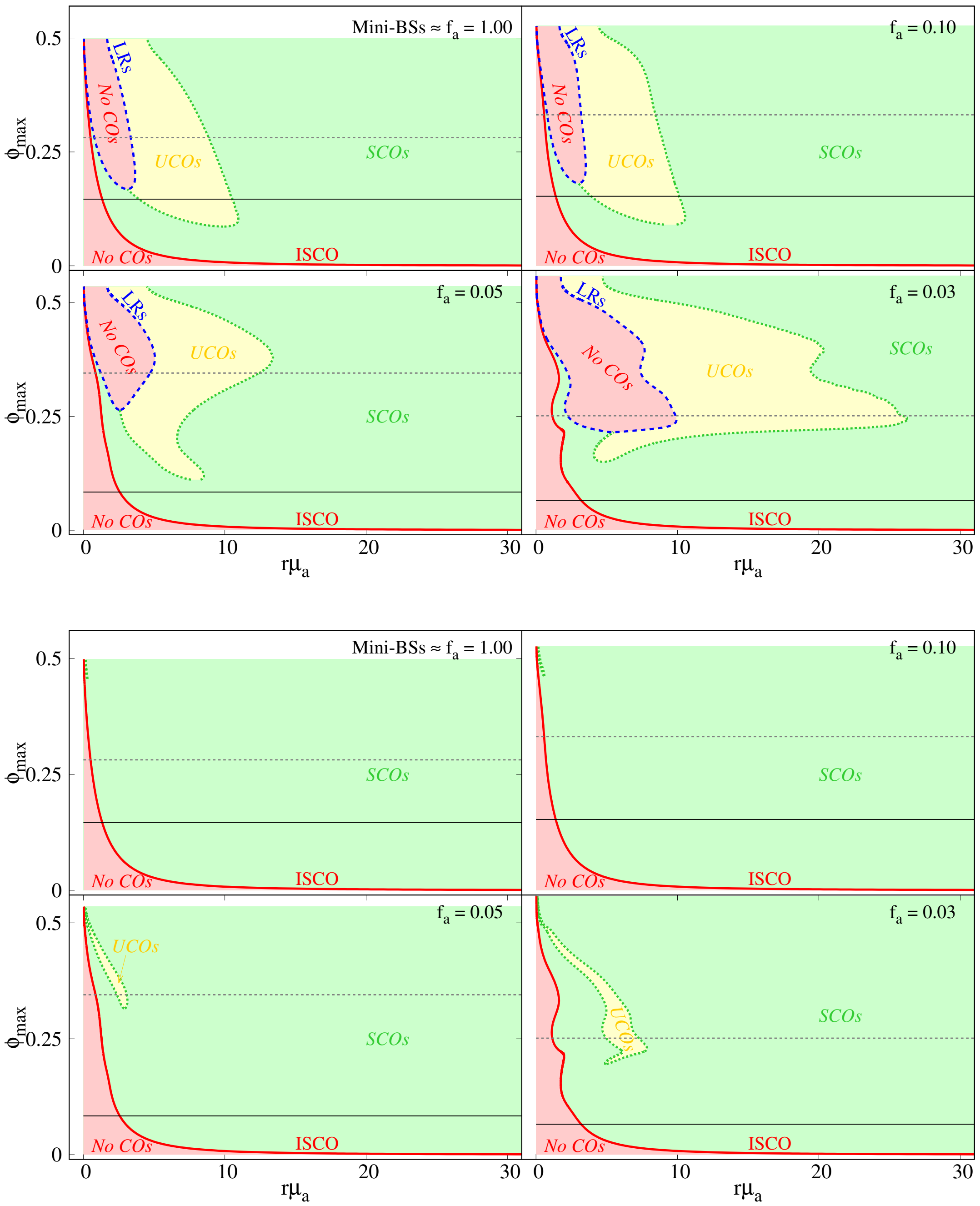

**Figure 4.9:** Structure of retrograde (top panels) and prograde (bottom panels) COs for the case of mini-BSs and RABSs with $f_a = \{0.10, 0.05, 0.03\}$. The light (yellow) green region corresponds to timelike SCOs (UCOs). The light red region corresponds to no circular orbits (No COs). The ISCO is represented as a solid red line and the LRs are represented as a dashed blue line. The horizontal black line corresponds to the solution wherein, moving along the line of solutions from the Newtonian limit, the first local maximum of the mass is attained. The horizontal grey dotted lines correspond to the first solution which develops an ergo-torus.

Figure 4.9 exhibits the ISCO, as a red solid line; the LRs, as a blue dashed line; the region of timelike SCOs is filled by a light green colour; the region of timelike UCOs is filled by a light yellow colour; and the region where no timelike COs exist (No COs) is filled by a light red colour. The black solid horizontal line provides a reference in the space of solutions, denoting the solution where the first local maximum of the ADM mass occurs. The grey dotted horizontal line corresponds to the first solution to develop an ergo-torus.

Let us focus first on the retrograde orbits – Figure 4.9 (top panels). Dilute solutions ($\phi_{max} \sim 0$) only have two different regions: for sufficient small radii it is impossible to have COs since the energy and angular momentum of the particle become imaginary (No COs). But after some radius, it possible to have COs, and they are stable (SCOs). The two regions are separated by the ISCO. For less dilute solutions ($\phi_{max} \sim 0.12$, for mini-BSs), a new region with UCOs emerges. This region occurs between a minimum and maximum radius and it is not connected with the ISCO: there are still SCOs between the ISCO and the region with UCOs. This is a Kerr-unlike feature and may be associated with the toroidal structure of spinning BSs. Moving further into the strong gravity region LRs emerge, as a pair, delimiting a new region with no COs between them. This new region is connected to UCOs for large radii and to SCOs for small radii. Moving further into the strong gravity regime, at larger values of $\phi_{max}$, the two regions with no COs appear to converge, but there is always a small region of SCOs for all solutions we obtained.

All features described in the previous paragraph apply to both to mini-BSs and RABSs with different $f_a$. The main difference resides in the size of the different regions. Decreasing $f_a$, the regions of no COs and UCOs get broader and their shapes change, with an increased complexity associated with the several back-bendings that appear for solutions with smaller $f_a$ – see *e.g.* Figure 4.4. Another difference is related to the solution with the first local maximum of the mass (horizontal black line in Figure 4.9). For sufficiently small values this solution has no UCOs.

The case of prograde orbits – Figure 4.9 (bottom panels) – is much simpler. The key features are the following. Firstly, as mentioned above, prograde LRs do not exist for RABSs. Secondly, albeit hardly noticeable for mini-BSs and RABSs with $f_a = 0.10$, there is always a region of UCOs for the solutions in the strong gravity regime. This region of UCOs starts to get broader and more complex when moving into the strong gravity regime, analogously to the retrograde case. Thirdly, all solutions below the horizontal black lines have the same structure, meaning that they only have a region of no COs and another of SCOs.

#### 4.4.2.3 *Energy conditions*

Possible violations of energy conditions are informative about exotic properties of the matter-energy content necessary to support some spacetime configuration. Let us briefly analyse the weak, dominant and strong energy condition, WEC, DEC and SEC, respectively, for the axion-like potential model and RABS solutions considered in this chapter.

The WEC is defined as the requirement that $T_{\mu\nu}X^{\mu}X^{\nu} \geqslant 0$, for any timelike vector field $X^{\mu}$. It means the energy density measured by any timelike observer must non-negative. The DEC amounts to the WEC plus the requirement that for every future directed causal vector field $X^\mu$, the vector $T_{\mu\nu}X^\nu$ is causal and future directed. This extra requirement means that the energy flux is causal (timelike or null). The SEC is defined as $\left(T_{\mu\nu} - \frac{1}{2}g_{\mu\nu}T\right) X^\mu X^\nu \geqslant 0$, and it means matter gravitates towards matter. Let us test these conditions for RABSs. Previous considerations on energy conditions for spinning BSs can be found, *e.g.* in [204].

Consider a generic unit timelike vector, $X^\mu X_\mu = -1$. Then, for our model:

$$T_{\mu\nu}X^\mu X^\nu = 2X^\mu \partial_\mu \Psi^* X^\nu \partial_\nu \Psi + \partial^\alpha \Psi^* \partial_\alpha \Psi + V \ . \tag{4.21}$$

Since the three terms on the right-hand side of the above equation are non-negative, their sum will be non-negative. Therefore the WEC is never violated. To analyse the DEC, we compute the norm of the vector $T_{\mu\nu}X^\nu$, which is,

$$||T_{\mu\nu}X^\nu||^2 = 2\partial_\mu \Psi^* \partial^\mu \Psi^* (X^\nu \partial_\nu \Psi)^2 - 2X^\mu \partial_\mu \Psi^* X^\nu \partial_\nu \Psi \left(\partial^\alpha \Psi^* \partial_\alpha \Psi + 2V\right) - \left(\partial^\alpha \Psi^* \partial_\alpha \Psi + V\right)^2 . \tag{4.22}$$

This expression is not manifestly sign-definite. But specialising it for the ansatz used in this chapter, we obtain,

$$||T_{\mu\nu}X^\nu||^2 = -\frac{4}{r^2}e^{-2(F_0+F_1)}\left(\omega - m\frac{W}{r}\right)^2 \left[(\partial_\theta \phi)^2 + r^2\left(e^{2F_1}V + (\partial_r \phi)^2\right)\right] - \left(\partial^\alpha \Psi^* \partial_\alpha \Psi + V\right)^2 \ . \tag{4.23}$$

The norm of $T_{\mu\nu}X^\nu$ is now manifestly non-positive; thus the energy flux is timelike or null and consequently the DEC is obeyed. To analyse the SEC we consider

$$\left(T_{\mu\nu} - \frac{1}{2}g_{\mu\nu}T\right) X^\mu X^\nu = 2X^\mu \partial_\mu \Psi^* X^\nu \partial_\nu \Psi - V. \tag{4.24}$$

Since the potential $V$ is always non-negative, the SEC is violated if the potential is large enough. It is well known that the SEC can be violated by a free massive scalar field - see *e.g.* [205]. Here we show that such violation is also possible for the axion-like complex scalar field.

As an example to study the energy conditions, consider an observer whose 4-velocity is orthogonal to $t$ = const hypersurfaces. Such observers are commonly called Zero Angular Momentum Observers (ZAMO). Their 4-velocity is defined as $u_\mu = (u_t, 0, 0, 0)$. For a ZAMO, the energy conditions are,[5]

$$\text{WEC:} \quad -T^t_t - T^t_\varphi \frac{g^{\varphi t}}{g^{tt}} \geqslant 0 \ . \tag{4.25}$$

$$\text{SEC:} \quad -\left(T^t_t - \frac{1}{2}T\right) - T^t_\varphi \frac{g^{\varphi t}}{g^{tt}} \geqslant 0 \ . \tag{4.26}$$

We found that, for a ZAMO, the WEC and DEC is never violated, as shown in general in the previous paragraph; but the SEC is violated for some solutions with small values of $f_a$. This is illustrated in Figure 4.10. Therein, a RABS solution, with $f_a = 0.030$ and $\omega = 0.250$ belonging to the first branch (top branch) is chosen, and one can see it has some space-time regions wherein the SEC is violated. This illustration raises two questions: *i*) for which values of $f_a$ is it possible to obtain solutions that violate the SEC? ($ii$) For each line of constant $f_a$ that has solutions that violate the SEC, where do these solutions occur and how do they depend on $f_a$?

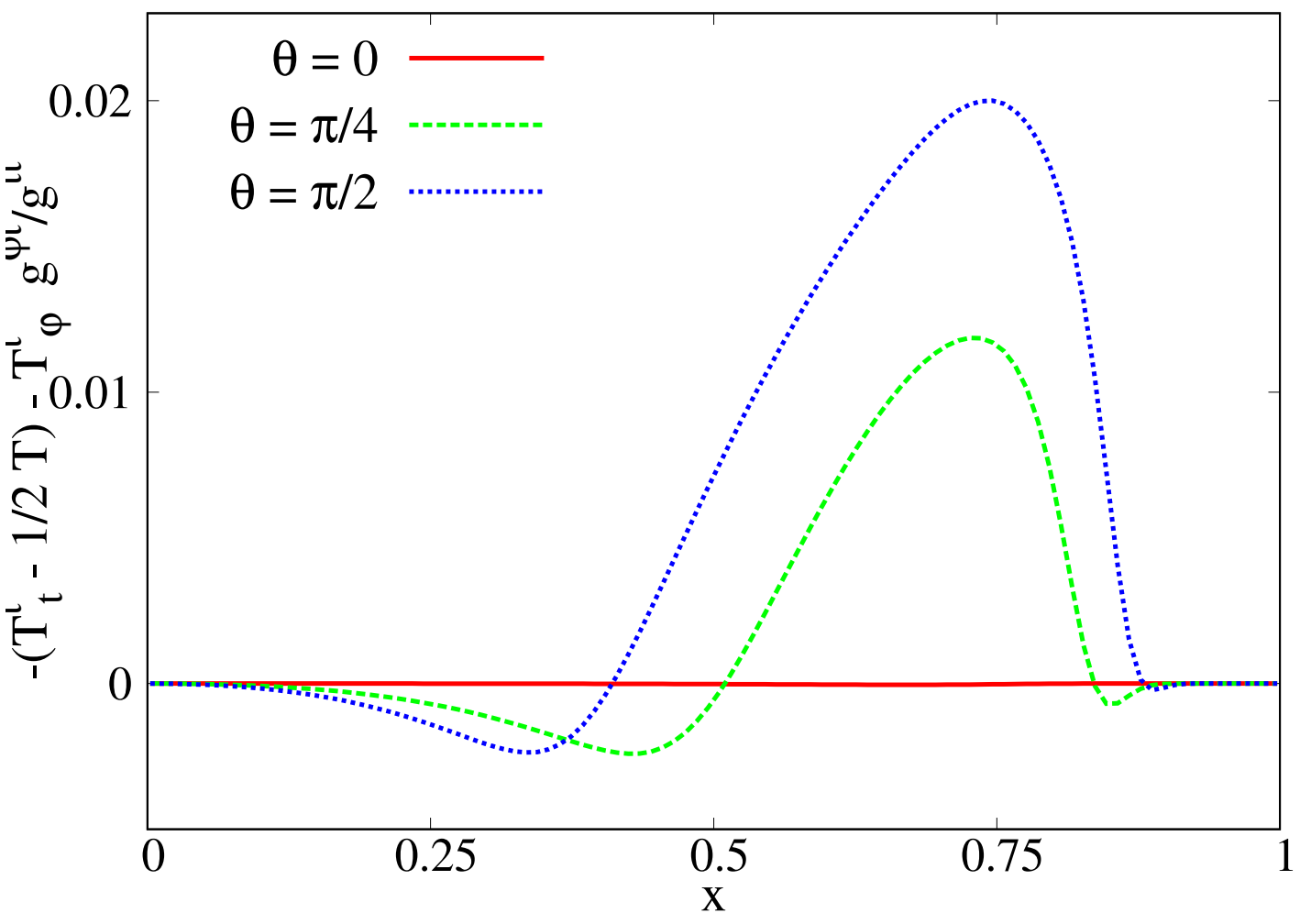

**Figure 4.10:** SEC as a function of the compact radial coordinate $x = r/(1+r)$, for three different polar angles $\theta$, of a ABS solution with $f_a = 0.030$ and $\omega = 0.250$ on the first branch.

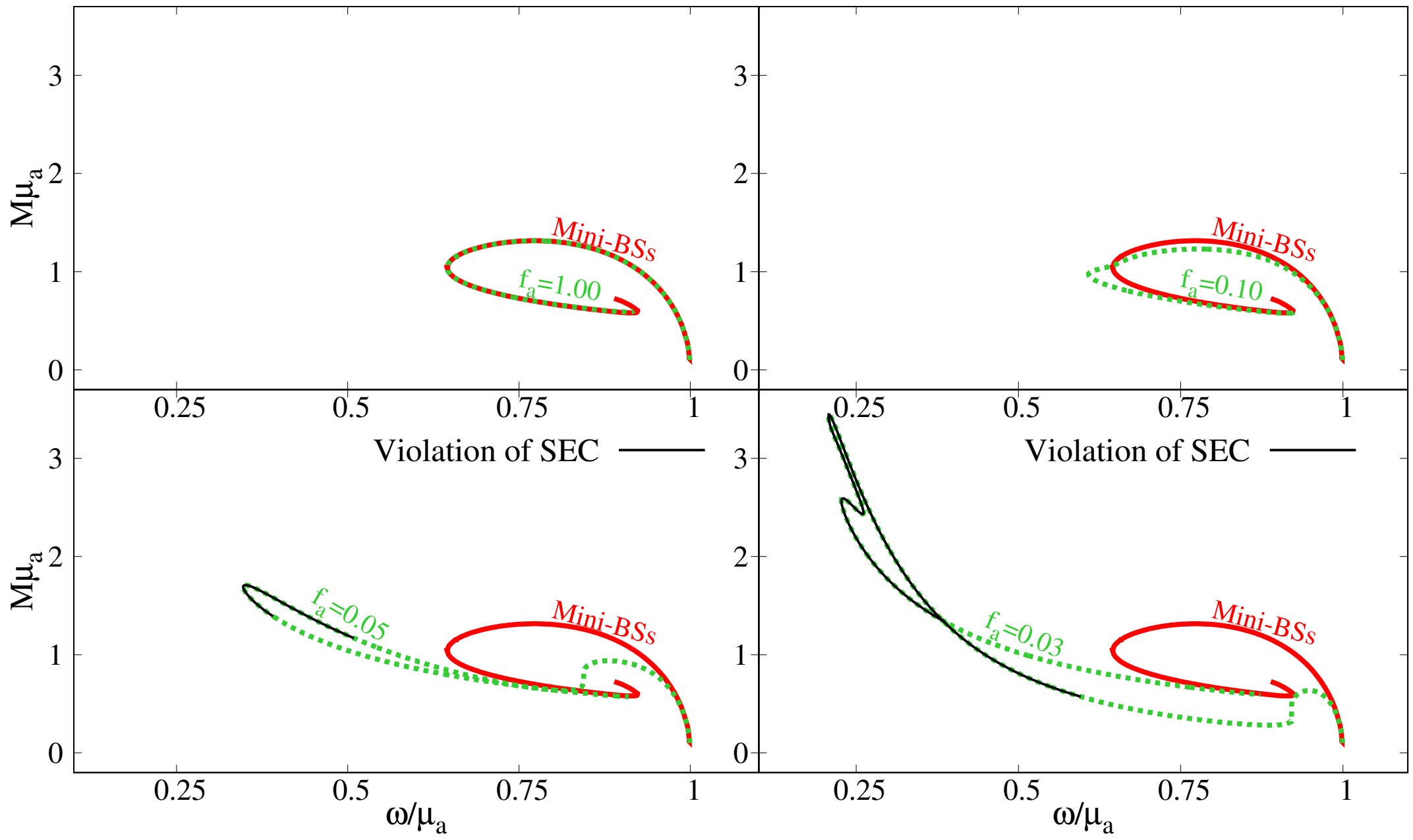

**Figure 4.11:** SEC violations in RABSs.

As for the first question, we found that solutions that violate the SEC start to appear for $f_a \lesssim 0.057$, on a small region of solutions close to the end of the first branch (close to the first back bending). The second question is addressed in Figure 4.11. One can observe that for the line with constant $f_a = 0.050$, there are solutions in the second branch that violate the SEC. By decreasing further $f_a$, the region of solutions that violate the SEC gets broader, as shown in the bottom right panel ($f_a = 0.030$). Further decreasing $f_a$, one may extrapolate that, for small enough $f_a$, a large portion of the solutions with that $f_a$ may violate the SEC. We emphasise this analysis was performed assuming a physical ZAMO observer.

## 4.5 Conclusions and remarks

In this chapter we have constructed RABSs, which are the spinning generalisation of static axion BSs recently found in [190]. These are stationary, axially symmetric, asymptotically flat and everywhere regular solutions of the (complex-)Einstein-Klein-Gordon theory with a QCD axion potential defined in Equation 4.7. The resulting solutions are described by two parameters: the angular frequency of the scalar field $\omega$ and the decay constant of the QCD potential $f_a$. For large $f_a$, the solutions reduce to the standard mini-BSs. In practice, $f_a = 1.00$ is already large.

By comparing RABS with the limiting case of mini-BSs, we observed that the solutions with the smaller values of the decay constant can possess lower values of the angular frequency of the scalar field, be more massive and have larger values for angular momentum and scalar field, leading to more compactness configurations. At the level of the phenomenology, both RABSs can develop LRs if they are sufficiently in the strong gravity part of the space of solutions. The structure of timelike, equatorial COs is also similar for RABSs and mini-BSs; but for low values of $f_a$ the structure starts to be more complex. An important difference arises at the level of the strong energy condition. For a ZAMO, mini-BSs never violate the strong energy condition, but for RABS they can violate that energy condition if their decay constant is low enough.

As a follow up study, it would be interesting to examine the stability of these solutions using fully non-linear numerical relativity simulations, following [193], to assess the impact of varying $f_a$ on the development of the non-axisymmetric instabilities observed for spinning mini-BSs. Finally, another interesting study would be to consider a small value of the decay constant, *e.g.* $f_a = 0.05$, and add a BH horizon in the middle of the RABS, under a synchronisation condition, as in [37]. This would lead to BHs with axionic hair that can possess different phenomenology properties compared to the ones already studied before. Such study shall be present in the next chapter.

[5] We do not present the inequality associated to the DEC which is too complicated to display here.

CHAPTER 5

# Kerr Black Holes with Synchronised Axionic Hair

## 5.1 Introduction

Testing the Kerr hypothesis is a central goal of current strong gravity research [206], [207]. This is the hypothesis that astrophysical BHs, when near equilibrium, are well described by the Kerr metric [5]. This working assumption is intimately connected with the no-hair conjecture [12], which states that the dynamically formed equilibrium BHs have no other macroscopic degrees of freedom beyond those associated with Gauss laws (and hence gauge symmetries) – see *e.g.* [16], [88], [89], [115] for reviews.

Over the last few years it has been realized that the Kerr hypothesis can be challenged *even* within GR and *even* with physically simple and reasonable energy-matter contents, due to the discovery that free, massive complex scalar or vector fields can endow the Kerr BH with synchronised bosonic hair [37], [86], [94], [208], see, *e.g.* [52]–[54], [93], [95], [139], [151], [199], [209] for generalizations. Moreover, if these fields are sufficiently ultra-light, this hair could occur in the mass range of astrophysical BH candidates, spanning the interval from a few solar masses, $\sim M_\odot$ (stellar mass BHs), to $\sim 10^{10}\ M_\odot$ (supermassive BHs).

The existence of such hairy BHs circumvents various no-scalar (and no-Proca) hair theorems - see *e.g.* [14], [16], [87]. Notwithstanding, the key test to the no-hair conjecture is a dynamical one. Are these hairy BHs dynamically robust? That is, can they form dynamically and be sufficiently stable? The current understanding is that: 1) there are dynamical formation channels, namely: *a*) the superradiant instability [39] of the Kerr solution [104], [125] and *b*) mergers of bosonic stars [210]; 2) these hairy BHs may, themselves, still be afflicted by superradiant instabilities [106], [133] but these can be very long-lived, with a time scale that can exceed a Hubble time [135]. The evidence so far, therefore, indicates BHs with synchronised ultralight bosonic hair are an interesting challenge to the Kerr hypothesis even in GR. This has led to different phenomenological studies, including the study of their shadows [55], [57], [61] and their $X$-ray phenomenology, namely the iron K$\alpha$ line [62], [211] and quasi-periodic oscillations [66].

An important outstanding issue is the fundamental physics nature of the putative ultralight bosonic field that could endow BHs with hair. Possible embeddings in high energy physics have been suggested, such as the string axiverse [189], which proposes a landscape of ultralight ALPs emerging from string compactifications. If these particles have sufficiently small couplings with standard model particles, they are dark matter[1], only observable via their gravitational effects, one of which would be endowing spinning BHs with synchronised hair. In the last chapter, we introduced the QCD axion, in which these ALPs are inspired. It is a pseudo Nambu-Goldstone boson suggested by the Peccei-Quinn mechanism [180] to solve the strong CP problem in QCD [181], [182] and it possesses a self-interaction potential [178], [179], characterised by two parameters: the mass of the scalar field $m_a$ and the decay constant $f_a$ – *cf.* Equation 4.7. The QCD-axion has inspired the study of other weakly-interacting ultralight bosons beyond Standard Model [184]–[187], which are regarded as plausible dark matter candidates. Such axion-like particles occur, for instance, in string compactifications [188], [189]. Thus, it would be interesting to assess the gravitational effects of such a potential for the ALPs that could endow spinning BHs with synchronised hair. This is the goal of the present chapter[2].

As seen in the first three chapters of this thesis, Kerr BHs with synchronised bosonic hair have a solitonic limit, obtained when taking the BH horizon to zero size, wherein they reduce to spinning boson stars. In the case of the original model with a free, complex scalar field, the solitonic limit yields the so-called, spinning mini BSs [32], [49], [50]. In the previous chapter, axion BSs have been constructed, wherein the complex scalar field is under the action of a QCD axion-like potential [78], [190], with the two aforementioned parameters.[3] In the limit when $f_a \to \infty$, the potential reduces to a simple mass term, and the axion BSs reduce to mini BSs. Thus, in the same way that the hairy BHs in [37] are the BH generalisation of mini BSs, we shall construct here the BH generalisations of the rotating axion BSs in [78]. Moreover, we shall study the basic physical properties and phenomenology of these BHs, hereafter dubbed *Kerr BHs with synchronised axionic hair* (KBHsAH), or simply "axionic BHs".

This chapter is based on the work developed in [79] and it is organised as follows. In section 5.2 we introduce the model. In section 5.3 we show the numerical results, presenting the domain of existence of the numerical solutions together with an analysis of some of their physical properties and phenomenology. In section 5.4 we present conclusions and some final remarks.

[1] "Dark matter" here is understand as any particle that does not interact with particles from the Standard Model.

[2] See also [212], [213] for other potential impacts of axionic self-interactions which may be used to constrain such self-interactions.

[3] Axionic stars with real fields have been considered, *e.g.* in [191], [192].

## 5.2 The model

The theory for which we shall obtain hairy BH solutions is, again, the (complex-)Einstein-Klein-Gordon model, given by Equation 4.1, where the scalar self-interacting potential, $V(\psi)$, follows the same QCD axion-like potential [194] discussed in the last chapter, Equation 4.7. As noted in the previous section, for sufficiently large decay constant $f_a$, we expect the BHs solutions to become very similar to the original Kerr BHs with synchronised scalar hair [37][4]. The equations of motion resulting from the above action are the same as the ones used to obtained rotating axionic BSs, *i.e.* Equation 4.2 and Equation 4.3, with a energy-momentum tensor given by Equation 4.4.

As BH generalisations of the work done on RABS, we are interested in stationary, regular on and outside the event horizon, axi-symmetric and asymptotically flat solutions. Therefore, these solutions can be obtained by using the same ansatze for the metric and the scalar field as for the hairy solutions in the first three chapters of this thesis, *i.e.* Equation 1.19 for the line element, and Equation 1.16 for the scalar field. Likewise the solutions obtained in the chapter are possible due to the synchronisation between the angular frequency of the scalar field and the horizon angular velocity, described by Equation 1.18.

The numerical framework shall be the same as the one discussed in section 1.2 for KBHsSH. The only difference resides in the introduction of the ALP mass, $\mu_a$, and the decay constant, $f_a$, rescaled in the same way we did for RABS. Furthermore, the decay constant is also used as an input parameter that can be set freely. Hence, there are four input parameters: **i)** the decay constant $f_a$ in the potential, Equation 4.7; **ii)** the angular frequency of the scalar field $\omega$; **iii)** the azimuthal harmonic index $m$; and **iv)** the radial coordinate of the event horizon $r_H$. For simplicity, we have restricted our study to the fundamental configurations, *i.e.* with a nodeless scalar field, $n = 0$ and with $m = 1$. Also, from the results presented in the previous chapter and in [78], where we studied axion BSs with 4 different values for the decay constant, $f_a = \{1.0, 0.1, 0.05, 0.03\}$, and showed that only when the decay constant is small enough ($f_a = \{0.05, 0.03\}$), the axion potential starts to affect greatly the solutions, we shall illustrate the effect[5] of the axion potential on the hairy BHs by performing a thorough study of the solutions with the specific decay constant $f_a = 0.05$.

Finally, let us mention that, as remarked in the previous chapter and in [78], the potential (Equation 4.7) allows for the existence of solutions even in the absence of the gravity term in the action (Equation 4.1). The simplest case corresponds to (non-gravitating) Q-ball-like solitons in a flat spacetime background. As expected, these solutions possess generalizations on a Kerr BH background. These bound states are in synchronous rotation with the BH horizon, *i.e.* they still obey the condition (Equation 1.18) and share most of the properties of the non-linear Q-clouds in [214]. A discussion of these aspects will be reported elsewhere.

[4]Since in the model herein considers a complex scalar field, there are two light scalar degrees of freedom, interacting with each other. This is different from the prototypical ALP, which represents the light phase degree of freedom of a complex scalar, while the magnitude is a heavy degree of freedom which can be integrated out.

[5]We have confirmed the existence of KBHsAH for various $f_a$ ranging from 0.02 to 10.

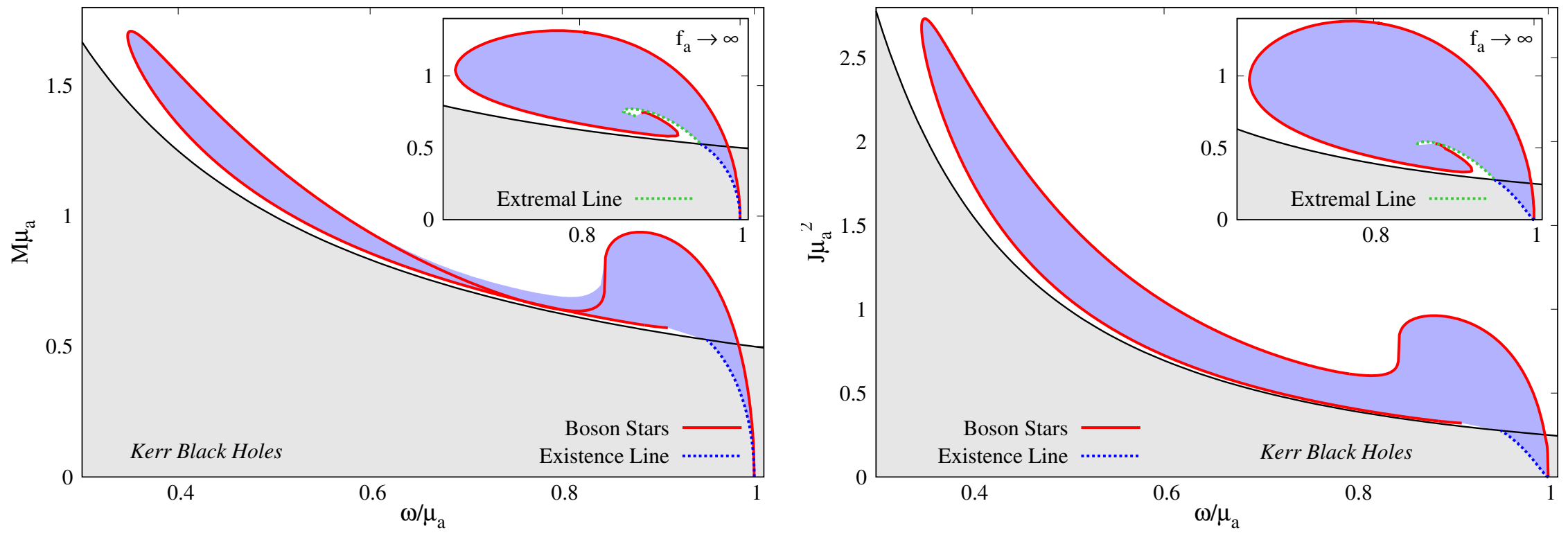

**Figure 5.1:** Domain of existence of KBHsAH for $f_a = 0.05$ in the $M\mu_a$ *vs.* $\omega/\mu_a$ plane (left panel) and in the $J\mu_a^2$ *vs.* $\omega/\mu_a$ plane (right panel). In both panels the insets correspond to the analogous domain of existence for the original KBHsSH (no scalar self-interactions), corresponding to $f_a \to \infty$. For the latter family of solutions we also present the extremal line, composed of BHs with vanishing Hawking temperature.

## 5.3 Numerical Results

### 5.3.1 The domain of existence

At the end of the previous section, we fixed two of the four input parameters of the problem $(f_a, m)$. Thus, the full domain of existence is obtained by varying the remaining two input parameters: the angular frequency of the scalar field, $\omega$; and the radial coordinate of the event horizon, $r_H$. Since it is impossible to obtain all possible BH solutions, we obtained a very large number of them ($\sim$ 30000) and we have extrapolated this large discrete set of solutions into the continuum, which defines the region where one can find the BH solutions with axionic hair.

Such a region can be expressed and plotted in various ways. In the left panel of Figure 5.1 (main panel), we show it in an ADM mass $M\mu_a$ *vs.* angular frequency $\omega/\mu_a$ plane. We can observe that *most* (but not all) of the numerical solutions region is bounded by two specific lines[6],

- The *axion BS line* - corresponding to the solitonic limit, in which both the event horizon radius and area vanish, $r_H = 0$ and $A_H = 0$, and the solutions have no BH horizon; therefore $q = 1$ – see Equation 1.40. Such line is represented in both panels of Figure 5.1 as a red solid line.
- The *existence line* - corresponding to specific subset of vacuum Kerr BHs which can support stationary scalar clouds (with an infinitesimally small $\phi$), first discussed by Hod [40], [41], thus having $q = 0$. These solutions are obtained by linearising the theory, and since, on that regime, the axion self-interacting potential reduces to the mass potential – *cf.* Equation 4.8 – the existence line will be the same as the one obtained for the family of KBHsSH. Such line is represented in both panels of Figure 5.1 as a blue dotted line.

[6] In fact, there is a third line corresponding to extremal hairy BHs, which possess vanishing Hawking temperature. Here we have only studied the neighbouring solutions of these extremal BHs, but not the latter *per se*, which would require a different metric ansatz.

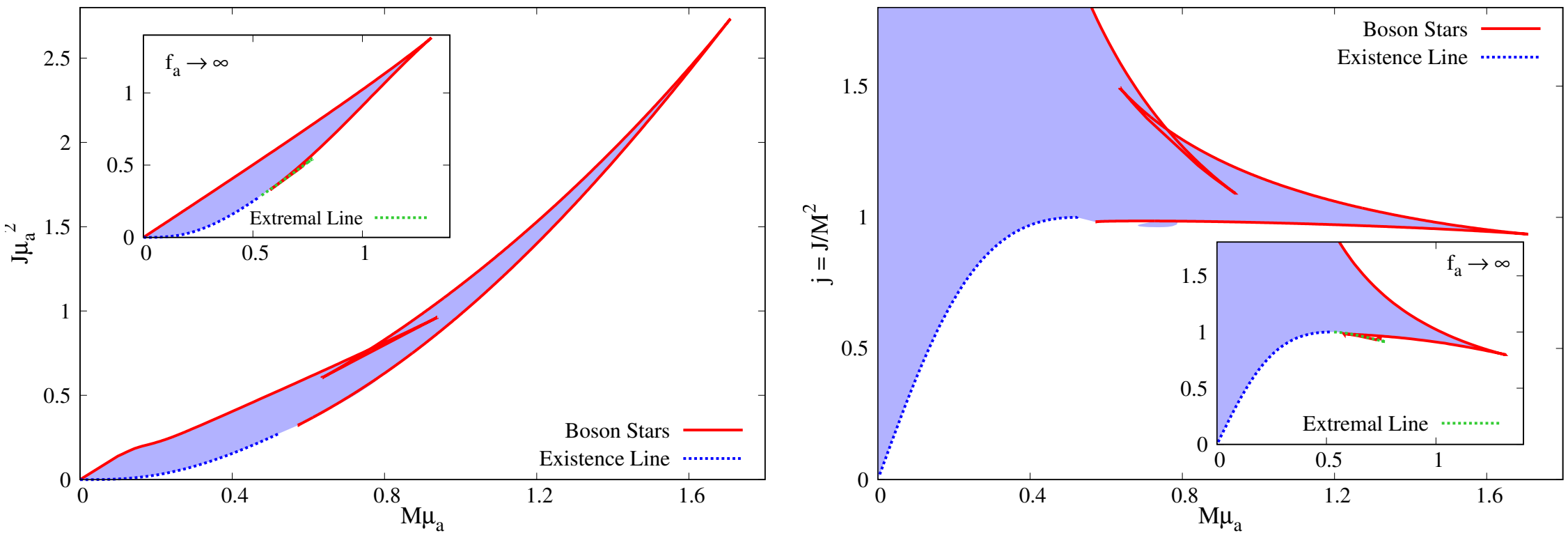

**Figure 5.2:** ADM angular momentum, $J$, (left panel) and dimensionless spin, $j = J/M^2$, (right panel) as a function of the ADM mass $M$. As in Figure 5.1 the insets show the free scalar field limit $f_a \to \infty$, where the extremal line is also shown.

Figure 5.1 (left panel) exhibits a novel property of this class of BHs: the solutions' region is no longer totally bounded by the BS line in this particular representation. For large decay constant, $f_a$, we recover the family of Kerr BHs with synchronised scalar hair [37] (inset), for which the solutions' region *is* totally bounded by the BS line (together with an existence line – scalar clouds –, and extremal line – zero temperature BHs). The consequence of this observation is that for certain frequencies, the ADM mass is not maximised by a boson star, but rather by a hairy BH.

By changing the representation, however, and plotting the domain of existence in the ADM angular momentum $J\mu_a^2$ *vs.* angular frequency $\omega/\mu_a$ plane – right panel of Figure 5.1 –, we see that, for the angular momentum, the BS line bounds all axionic BHs with decay constant $f_a = 0.05$. Thus for all frequencies, the angular momentum is maximised by a boson star.

Another distinctive feature of the domain of solutions of the axionic BHs is the existence of a local maximum for the mass and angular momentum at $\omega \sim 0.9$, which is not the global maximum. For solutions with smaller angular frequency, it is possible to have BHs with more mass and angular momentum than the ones near the local maximum; in fact, these quantities are maximised for the solutions with the smallest possible value of angular frequency. Such is not the case in the absence of the axionic potential (inset of both panels in Figure 5.1).

In Figure 5.2 both the ADM angular momentum, $J$ and the dimensionless spin, $j$, defined as $j \equiv J/M^2$, are exhibited *vs*. the ADM mass, $M$, on the left and right panels, respectively. In the former, we can see that the axionic BHs can have a higher mass and angular momentum than their $f_a \to \infty$ counterparts (inset). We can also see a zigzag of the BS line. This behaviour is explained by the sudden drop on the ADM mass and angular momentum around $\omega/\mu_a \approx 0.84$ – *cf.* Figure 5.1. In the right panel, we see a considerable violation of the Kerr bound, $j \leqslant 1$ in part of solution space. This already occurred for the $f_a \to \infty$ limit

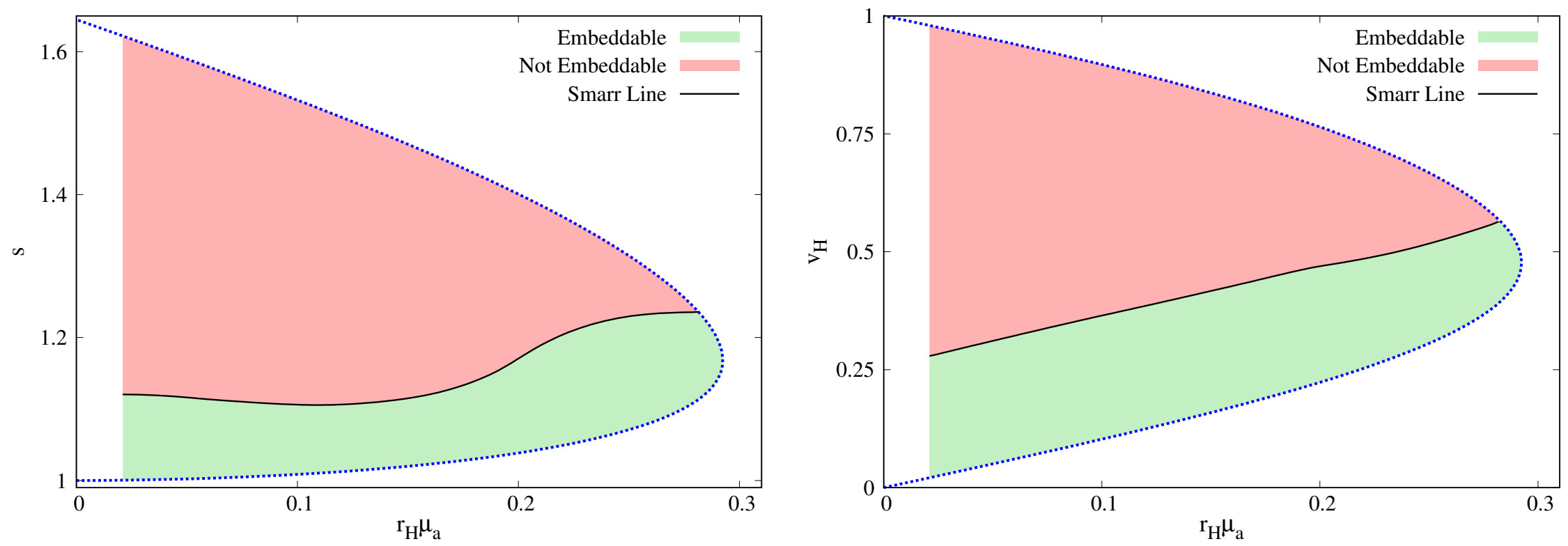

**Figure 5.3:** The sphericity, $\mathfrak{s}$ (left panel), and the horizon linear velocity, $v_H$ (right panel), as a function of the event horizon radial coordinate, $r_H\mu_a$. The black solid line corresponds to the Smarr line, which, in contrast with the $f_a \to \infty$ limit, is no longer constant for the sphericity. Below (above) it, we have (do not have) embeddable BH horizons in Euclidean 3-space. The white region in the far left region of each panel corresponds to solutions outside our scan (challenging due to the small $r_H$).

(inset). Again, it is possible to visualise the zigzag behaviour of the BS line.

Let us now study the horizon geometry of the axionic BHs. Their event horizon has a spherical topology but a spheroidal geometry, similarly to Kerr BHs. This can be seen by studying the spatial cross-section of the horizon. Due to the rotation of the solutions, the horizon is squashed at the poles. To show this, we compute the horizon circumference along the equator, $L_e$, and along the poles, $L_p$, given by Equation 2.22. Then, we define the sphericity as the ratio of both circumferences above [64] – *cf.* Equation 2.23 and Equation 3.5. For values in which the sphericity is greater (lower) than 1, the horizon will be squashed (elongated) at the poles, leading to an oblate (prolate) spheroid. From the left panel in Figure 5.3 we see that all solutions have a sphericity larger than the unity; thus all solutions have an oblate horizon, as expected.

Another physical quantity of interest associated with the horizon is its linear velocity $v_H$ [54], [63], [64]. Such quantity measures how fast the null geodesics generators of the horizon rotate relatively to a static observer at spatial infinity. Its definition is quite simple, only taking into account the perimetral radius of the circumference located at the equator, $R_e \equiv L_e/2\pi$, and the horizon angular velocity, $\Omega_H$ – *cf.* Equation 2.25 and Equation 3.6. The horizon linear velocity is presented in the right panel of Figure 5.3. The central feature in this plot is the fact that all solutions have horizon linear velocity smaller than the unity, which, in the units we are using, corresponds to the speed of light. Therefore, null geodesics generators of the horizon never rotate relatively to the asymptotic observer at superluminal speeds, even though some solutions strongly violate the Kerr bound $j \leqslant 1$.

A final insight about the horizon geometry of the axionic BHs is obtained from investigating whether an isometric embedding of the spatial sections of the horizon is possible in Euclidean 3-space $\mathbb{E}^3$. As shown in chapter 2, for a Kerr BH, such embedding is possible iff its dimensionless spin obeys $j \leqslant j^{(S)}$ [82], where $j^{(S)} \equiv \sqrt{3}/2$ was dubbed the Smarr point [64].

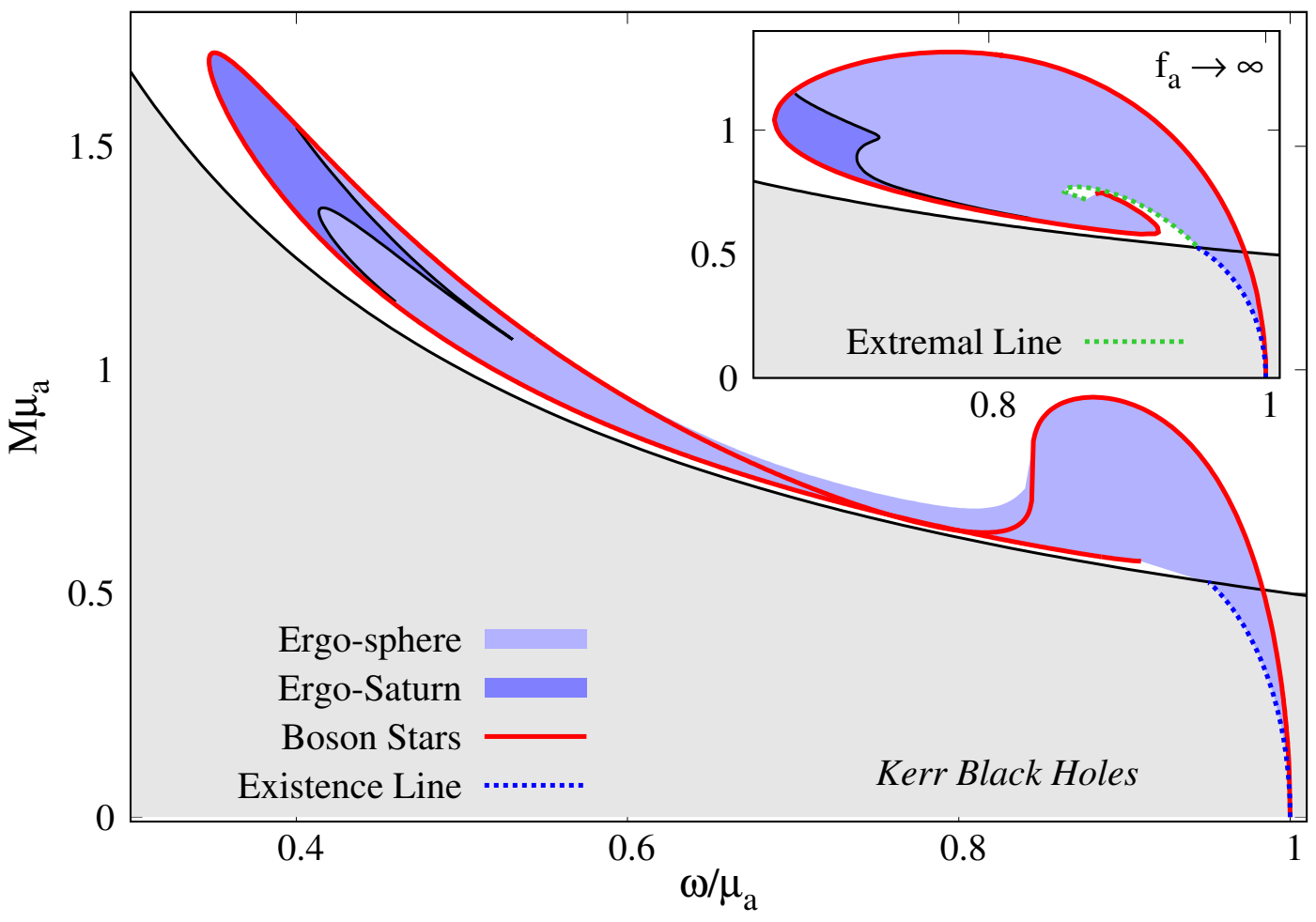

**Figure 5.4:** Ergo-regions. BHs with axionic hair have an ergo-sphere in the light blue region and an ergo-Saturn in the dark blue region. The inset shows the free scalar field case ($f_a \to \infty$), for comparison.

For $j > j^{(S)}$ the Gaussian curvature of the horizon becomes negative in the vicinity of the poles [82], which, prevents the embedding (due to occurring at a fixed point of the axi-symmetry). As expected, this feature also occurs for the axionic BHs. Due to the existence of scalar hair around the BH, we have an extra degree of freedom, which converts the Smarr point into a Smarr line. Such line is represented in both panels of Figure 5.3 as a solid black line. One observes that, for both the sphericity and the horizon linear velocity, the Smarr line is not constant. This contrasts with the behaviour for $f_a \to \infty$; in that case, the sphericity of the Smarr line was constant and equal to the value of the Smarr point in Kerr [64] – *cf.* chapter 2. Thus, the axion potential destroys the constancy of the sphericity along the Smarr line.

### 5.3.2 Other properties

#### *5.3.2.1 Ergo-regions*

An ergo-region is a part of a spacetime, outside the event horizon, wherein the norm of the asymptotically timelike Killing vector $\xi = \partial_t$ becomes positive and thus the vector becomes spacelike. Ergoregions are associated with the possibility of energy extraction from a spinning BH, via the Penrose process [149], [197], or superradiant scattering [39]. In the context of BHs with synchronised hair, the superradiant instability of vacuum Kerr BHs in the presence of ultralight scalar fields is one of the possible channels of formation of these hairy BHs [104], [125]. Thus, it is of relevance to analyse ergo-regions for the axionic BHs.

Kerr BHs possess an ergo-region whose boundary has a spherical topology and touches the BH horizon at the poles – such surface is called an ergo-sphere. For the axionic BHs in the $f_a \to \infty$ limit, the ergo-regions can be more complicated [106]. As seen in subsection 3.5.2 of chapter 3, some solutions have a Kerr-like ergo-region; but others have a more elaborate ergo-region topology, with two disjoint parts, one Kerr-like and another of toroidal topology. The latter were dubbed ergo-Saturns. The toroidal ergo-region is inherited from the mini

BS environment around the horizon, since these stars develop such ergo-regions, when sufficiently compact. Such ergo-torii also occur for spinning axion BSs [78] – *cf.* Figure 4.7. Thus, we expect some axionic BHs to develop an ergo-Saturn. This is confirmed in Figure 5.4. We have found that the ergo-region of axionic BHs follows a qualitatively similar distribution to that of their $f_a \to \infty$ limit: there are solutions which possess an ergo-sphere and others that develop an ergo-Saturn. The latter occur on the far left of the domain of existence in Figure 5.4, where the most massive BHs exist, having the lowest possible values of the angular frequency of the scalar field.

#### 5.3.2.2 *Light rings and timelike innermost stable circular orbits*

A phenomenological aspect of importance is the structure of circular orbits (COs) of both massless and massive particles around a BH. In particular, the (timelike) ISCO and the (null) LRs are of special relevance. The former is associated with a cut-off frequency of the emitted synchrotron radiation generated from accelerated charges in accretion disks; the latter is related to the real part of the frequency of BH quasi-normal modes [215], as well as to the BH shadow [60].

Using a similar approach as the one described for the axionic BSs in the previous chapter, we can use the line element associated with these BHs, Equation 1.19, to compute the effective Lagrangian for equatorial geodesic motion,

$$2\mathcal{L} = \frac{e^{2F_1}}{N}\dot{r}^2 + e^{2F_2}r^2\left(\dot{\varphi} - W\dot{t}\right)^2 - e^{2F_0}N\dot{t} = \epsilon \ , \tag{5.1}$$

and the energy and angular momentum of a test particle,

$$E = \left(e^{2F_0}N - e^{2F_2}r^2W^2\right)\dot{t} + e^{2F_2}r^2W\dot{\varphi} \ , \tag{5.2}$$

$$L = e^{2F_2}r^2\left(\dot{\varphi} - W\dot{t}\right) \ . \tag{5.3}$$

With that we can obtain an equation for $\dot{r}$, which defines a potential $V(r)$,

$$\dot{r}^2 = V(r) \equiv e^{-2F_1}N\left[\epsilon + \frac{e^{2F_0}}{N}\left(E - LW\right)^2 - e^{2F_2}\frac{L^2}{r^2}\right] \ . \tag{5.4}$$

In order to obtain COs, both the potential and its derivative must be zero, *i.e.* $V(r) = V'(r) = 0$. Depending on whether we are considering massless ($\epsilon = 0$) or massive ($\epsilon = -1$) particles, these two equations will yield different results.

For massless particles, the first equation, $V(r) = 0$, will give two algebraic equation for the impact parameter of the particle, $b_+ \equiv L_+/E_+$ and $b_- \equiv L_-/E_-$, corresponding to prograde and retrograde orbits, respectively. The second equation, $V'(r) = 0$, together with the impact parameters, will give the radial coordinate of the prograde and retrograde LRs. Whenever it is possible to obtain a real solution for the radial coordinate, the BH possesses LRs.

In Figure 5.5, we show the distribution of hairy BHs with different number of LRs in the angular frequency, $\omega/\mu_a$ *vs.* angular momentum, $J\mu^2$ plane. The left panel shows the retrograde case, whereas the right panel shows the prograde case. In both cases, it is always possible to have at least one LR, as for the Kerr BH. In the retrograde case, however, if the surrounding scalar field is compact enough, an extra pair of LRs emerge, leading to a hairy BH with 3 LRs. This is the case of a large set of hairy BHs with low values of $\omega/\mu_a$. In fact, we see the same behaviour for the free scalar field case (inset plot); even the solid black line separating the two regions has a qualitatively similar shape. The radial coordinate of the several LRs is shown in Figure 5.6 and Figure 5.7 as a blue dashed line.

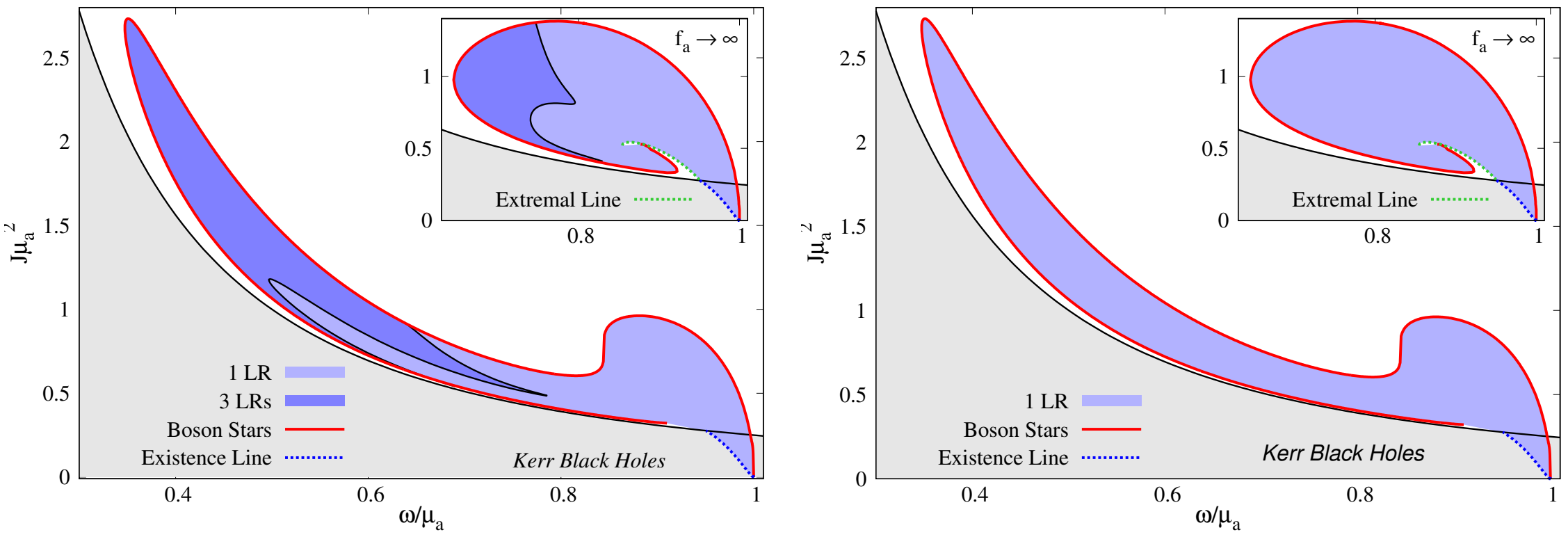

**Figure 5.5:** Number of LRs. The left (right) panel shows the retrograde (prograde) case. Hairy BHs can have 1 LR, as for the case of Kerr, in the light blue region, or 3 LRs, in the dark blue region. The inset follows the same description but for $f_a \to \infty$.

For massive particles, $V(r) = V'(r) = 0$ will yield two algebraic equations for the energy and angular momentum of the particle, $\{E_+, L_+\}$ and $\{E_-, L_-\}$ corresponding to prograde and retrograde orbits, respectively. Their stability can be verified by analysing the sign of the second derivative of the potential $V(r)$. Given a BH solution it will have the same three distinct regions concerning (timelike) COs as its BS counterpart: a region of No COs (region shaded in light red, in Figure 5.6 and Figure 5.7); a region of UCOs (region shaded in light yellow, in Figure 5.6 and Figure 5.7); and a region of SCOs (region shaded in light green, in Figure 5.6 and Figure 5.7).

The ISCO, as the name entails, is the innermost stable circular orbit, *i.e.*, the stable CO with the smallest radial coordinate. At this orbit, the second derivative of the potential vanishes, as it corresponds to the transition between the SCOs region and the UCOs region. The ISCO is represented as a solid red line in Figure 5.6 and Figure 5.7.

The first observation from Figure 5.6 is that, by increasing $q$, it is possible to obtain solutions with larger $\phi_{\max}$. This is reasonable since by increasing $q$ the BHs become hairier. Let us now analyse each individual plot. For the case where $q = 0.5$ (right bottom panel), the structure of the retrograde COs for all BHs is identical to the one for Kerr. There is only one LR and one ISCO, which separates the No COs region from the UCOs region, and separates the UCOs region from the SCOs region, respectively. At large enough $x$ we only have SCOs, but as we approach the horizon, we reach the ISCO and consequently enter the UCOs region. If we continue towards the horizon, we eventually cross the LR and enter the No COs region. The radial coordinate of the LR and ISCO increases monotonically with the increase of the maximal value of the scalar field.

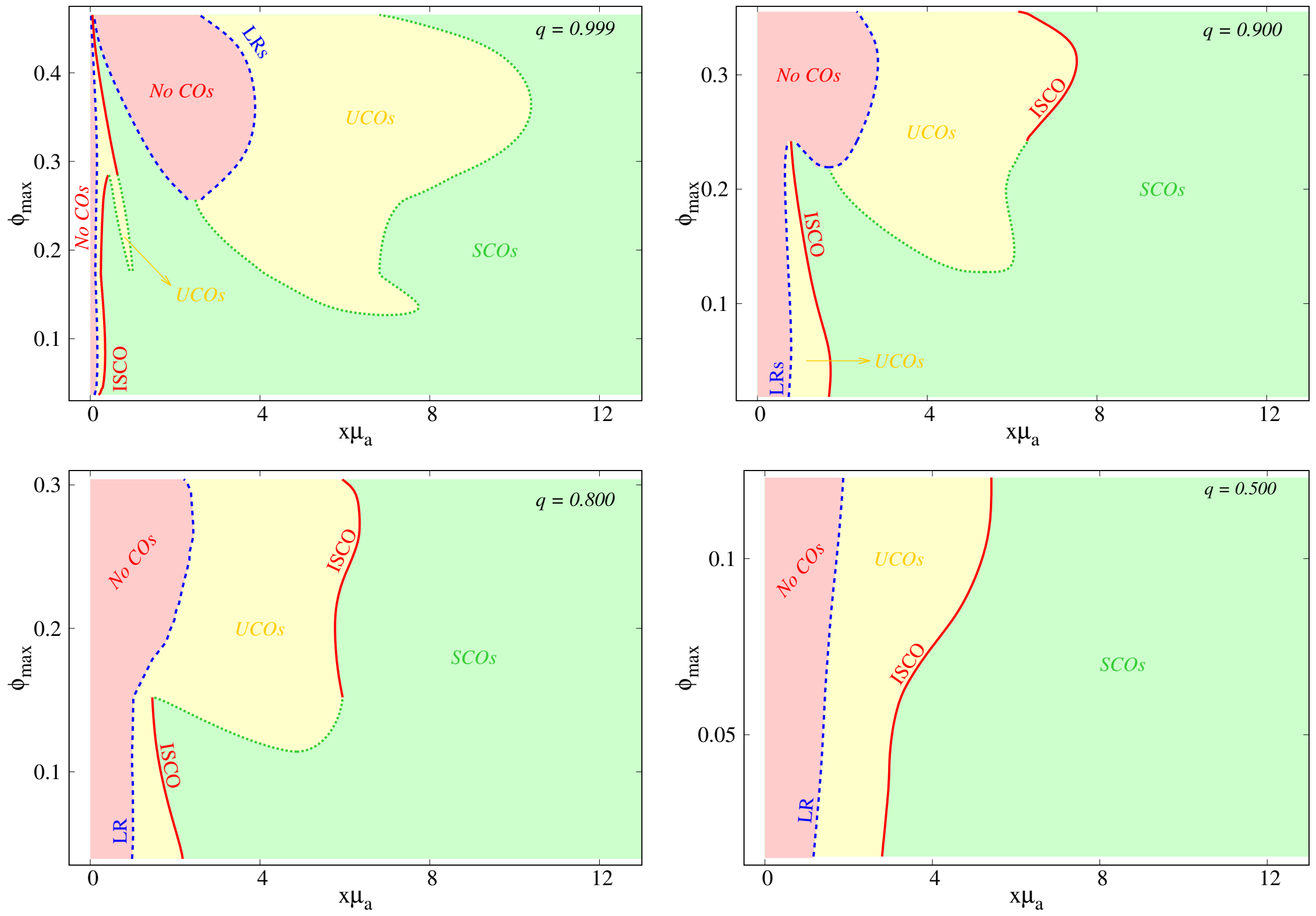

**Figure 5.6:** Structure of retrograde COs for four sets of axionic BHs with constant $q = \{0.5, 0.8, 0.9, 0.999\}$. The maximal value of the scalar field $\phi_{\rm max}$ varies with $q$; thus the vertical scale changes for the four plots.

For the case where $q = 0.8$ (left bottom panel) most of the structure is similar, but a new feature emerges. Starting at the regime where there are BHs with a dilute scalar field ($\phi_{\rm max} < 0.11$), their structure is the same as for Kerr, but for BHs with $\phi_{\rm max} > 0.11$ and $\phi_{\rm max} < 0.15$ an extra region of UCOs appears besides the one already present between the LR and the ISCO, and disconnected from the latter. The new region appears as a single point around $x\mu_a \approx 4.8$ and increases in size as $\phi_{\rm max}$ increases until its inner boundary merges with the ISCO. For $\phi_{\rm max} > 0.15$, the structure is again the same as that of Kerr, but now the radial coordinate of the ISCO is significantly larger than for BHs with a more dilute scalar field. In fact, there is a discontinuity when we study the evolution of the radial coordinate of the ISCO as it changes with the maximum value of the scalar field, as one can see in the left bottom panel of Figure 5.6.

In the $q = 0.9$ case (right top panel), we have a structure similar as for the $q = 0.8$ when we consider BHs with $\phi_{\rm max} < 0.21$, but, for the remaining BHs, the scalar field environment is compact enough to develop a pair of extra LRs. Such pairs give rise to a new region of No COs, disconnected from the already existing one between the event horizon and innermost LR. It starts as a single point for a BH with $\phi_{\rm max} \approx 0.21$; then it grows in size until its inner boundary (one of the LRs) merges with the innermost LR, connecting both regions of No COs. At the same time, the ISCO also merges with the LRs, meaning that such BH has a degenerate point in which two LRs and the ISCO converge. For BHs with a larger value of $\phi_{max}$, a Kerr-like structure again emerges, but now both the LR and the ISCO appear at a larger radial coordinate than for small $\phi_{max}$.

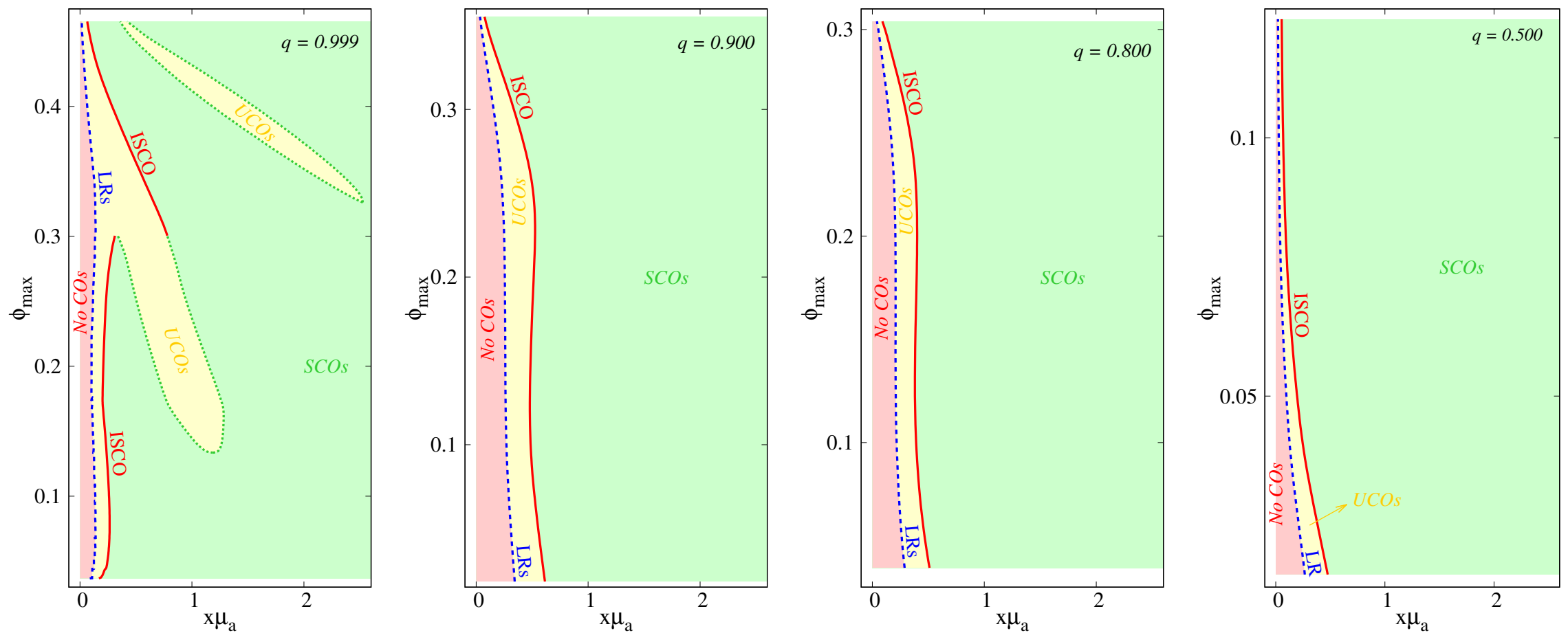

**Figure 5.7:** Structure of prograde COs for four sets of axionic BHs with constant $q = \{0.5, 0.8, 0.9, 0.999\}$. Note that the maximal value of the scalar field $\phi_{max}$ varies with $q$, thus the vertical scale changes for the four plots.

Lastly, in the $q = 0.999$ case (left top panel), the most complex structure is observed. This case inherits the several features discussed in the previous cases, such as the existence of two different and disconnected UCOs and No COs regions, but also presents a new one. For BHs with $\phi_{max} > 0.18$, a third region of UCOs between the two already existing ones. This third region starts as a single point and increases slowly in size as $\phi_{max}$ increases until its inner boundary connects with the ISCO, and the innermost region of UCOs merges with this new region. Although it is not visible in the left top panel of Figure 5.6, one can draw the conclusion that there is a BH with larger $\phi_{max}$ for which the innermost two LRs together with the ISCO converge to the same single point, akin to what happened in the $q = 0.9$ case.

Now we turn to the prograde COs. In Figure 5.7 we followed the same idea that we used to study the retrograde COs and plotted four sets of BHs solutions with constant $q = \{0.5, 0.8, 0.9, 0.999\}$ in the $\phi_{max}$ *vs.* $x$ plane. Figure 5.7 manifests that the structure of prograde COs is much simpler than the retrograde one; only the $q = 0.999$ case presents significant differences from the other ones; the structure of the latter is the same as for a Kerr BH. Moreover, the radial coordinate of the prograde LR and ISCO is always smaller than the one of the retrograde LR and ISCO, as it is for the Kerr BH [200].

Let us comment on the qualitatively different $q = 0.999$ case (first panel of Figure 5.7). For BHs with a dilute scalar field, a Kerr-like structure is observed; but above $\phi_{max} \approx 0.14$ a second region of UCOs emerges. This region exists until we reach a BH with $\phi_{max} \approx 0.30$, where both regions of UCOs merge together. For BHs with slighter larger $\phi_{max}$ a second region of UCOs again emerges, but now this region occurs at large radii, quickly decreasing to smaller radii as $\phi_{max}$ increases. Although not shown in this panel, we can predict that this new region of UCOs will converge, as the previous one, with the already existing and closer to the event horizon UCOs region.

## 5.4 Conclusions and Remarks

In this chapter, we have constructed and analysed BHs with synchronised axionic hair, which are BH generalisations of the RABS discussed in the previous chapter and in [78] - see also [190]. These are stationary, axi-symmetric, asymptotically flat and regular on and outside the event horizon solutions of the Einstein-Klein-Gordon equations of motion with a QCD axion-like potential, Equation 4.7. This family of axionic BHs is described by three parameters: the radial coordinate of the event horizon, $r_H$, the angular frequency of the scalar field, $\omega$ and the decay constant of the QCD potential, $f_a$. Here we have thoroughly scanned the space of solutions with decay constant $f_a = 0.05$, since, from the results found for the RABS in chapter 4 and in [78], this yields a case with considerable impact of the axion potential, and hence considerable differences from the free scalar field case, obtained as the $f_a \to \infty$ limit. The latter yields the original example of Kerr BHs with synchronised scalar hair [37]. Even larger deviations from this original example may occur for even smaller $f_a$, but then the numerics to obtain such solutions becomes more challenging.

When comparing the axionic BHs with their free scalar field counterparts [37], there are both differences and similarities. Some key differences are: (i) axionic BHs can have more mass and angular momentum and lower values of the angular frequency of the scalar field; (ii) the existence of a local maximum for the mass and angular momentum, that does not exists for the $f_a \to \infty$ case; (iii) the presence of a small region of frequencies where the mass of axionic BHs is no longer bounded by the axion BSs. In such region, we have a degeneracy of solutions with the same angular frequency and ADM mass, but such degeneracy is easily lifted by specifying any other physical quantity; (iv) the variation of the sphericity along the Smarr line, which, in the free scalar field case was constant and equal to the sphericity of the Smarr point for Kerr BHs, but varies for axionic BHs.

Concerning the similarities, we may emphasise: (i) the clear violation of the Kerr bound, $j \leqslant 1$; (ii) the sphericity and horizon linear velocity are bounded by the same existence line. Thus, both families can only have the same values of sphericity and horizon linear velocity as the Kerr ones, which implies that all BHs have a horizon which is an oblate spheroid and the rotation of its null generators (relatively to a static observer at spatial infinity) never exceeds the speed of light; (iii) the topology of the ergo-regions is either an ergo-sphere or an ergo-Saturn. In both families, the ergo-Saturn only appears for the solutions with the lower values of angular frequency.

In this chapter, we also presented a study of the structure of COs. Retrograde COs show a more complex structure than the prograde COs counterpart. Concerning the former, we saw that solutions with low or moderate amounts of hair (*e.g.* $q = 0.5$) will have a structure similar to the Kerr one. For very hairy solutions ($q \geqslant 0.8$), a new region of UCOs can emerge, as well as a new region of No COs when the axionic hair outside the event horizon is compact enough to yield an extra pair of LRs. In the extreme case, where most of the BH solution is composed of axionic hair ($q = 0.999$), it possible to have a third region of UCOs, leading to an intercalation of SCOs and UCOs regions.

Similarly to the previous chapter, let us briefly comment on energy conditions. For RABS, it was shown that the WEC and the DEC always hold, whereas the SEC could be violated, and in fact, it is violated for a ZAMO [78]. For the BH generalisation we can prove, following [78], that the WEC and DEC are never violated, but the SEC can be violated, for instance, for a ZAMO.

Let us also make some brief remarks on the dynamical properties of these solutions, although an in-depth analysis is beyond the scope of this chapter. In the $f_a \to \infty$ these hairy BHs reduce to the solutions in [37]. These are known to be afflicted from superradiant instabilities but these can be very long lived [106], [133], [135]. On the other hand such BHs can form from the superradiant instability of Kerr [104], [125] or, potentially, even from mergers of bosonic stars [210]. On the other hand, in the limit of vanishing horizon, the solutions described herein reduced to the axionic spinning bosonic stars [78]. Recently it has been argued that the axion-like potential has a stabilizing effect for these stars [216], [217], in some regions of the parameter space, as the mini spinning bosonic stars are afflicted by a non-axisymmetric instability [193]. As such, it may be that the axion-like potential also helps the dynamical robustness of the hairy BHs discussed here. In the future, we hope to address this open question.

Due to the groundbreaking results presented by the LIGO-Virgo collaboration detecting several gravitational waves events, starting with [67], as well as the results by the EHT collaboration on the first image of a shadow of the M87 supermassive BH [71], two possible and interesting directions to follow up on this chapter are: (1) to study the possible gravitational waves generated by the collision of two axionic hairy BHs, both in the head-on scenario, as well as, in the more realistic scenario of an in-spiral binary system. Recently [218] such collisions were made for Proca stars [156] obtaining a suggestive agreement with the particular gravitational wave event GW190521 [219]; (2) and to study the shadow of axionic hairy BHs, to analyse the impact of the QCD axion-like potential in the BH shadows. This would be a generalisation of the analysis in [55].

CHAPTER 6

# Spinning Black Holes in Shift-Symmetric Horndeski Theory

## 6.1 Introduction

Scalar-tensor theories of gravity have attracted much attention since the pioneering example of Brans-Dicke theory [23]. The physical relevance of such models could be tested, in particular, in strong gravity systems, namely BHs. On the one hand, as it turns out, the BH solutions in Brans-Dicke theory, as well as in a large class of models where the scalar field is non-minimally coupled to the Ricci scalar, are the same as in GR [24], [220], as demonstrated in the introduction of this thesis. On the other hand, BHs in extended scalar-tensor models, namely those with higher curvature corrections are, generically, different from those of GR [16].

Within the class of scalar-tensor theories that possess higher curvature corrections, those including a real scalar field, $\phi$, with a canonical kinetic term, non-minimally coupled to the GB quadratic curvature invariant,

$$R^2_{\rm GB} \equiv R_{\alpha\beta\mu\nu}R^{\alpha\beta\mu\nu} - 4R_{\mu\nu}R^{\mu\nu} + R^2 \ , \tag{6.1}$$

have attracted considerable interest. This is the class of *Einstein-scalar-GB* (EsGB) models described by the action

$$\mathcal{S} = \int d^4x\sqrt{-g}\left[\frac{R}{16\pi} - \frac{1}{2}\partial_\mu\phi\partial^\mu\phi + \alpha f(\phi)R^2_{\rm GB}\right] , \tag{6.2}$$

where $\alpha$ is a dimensionful coupling constant and $f(\phi)$ is a dimensionless coupling function. In these models, the GB term becomes dynamical in four spacetime dimensions, and the equations of motion remain second order, which is typically not the case when higher curvature corrections are included in the action. Moreover, the GB term as a higher order correction is suggested from string theory [221].

The status of BHs in the family of models in Equation 6.2 depends on the properties of $f(\phi)$; its choice determines if $\phi = 0$ is a consistent truncation of the equations of motion. There are two generic cases. Following the classification in [222] for a cousin model, we call models where $\phi = 0$ is *not* a consistent truncation of the equations of motion *class I or dilatonic-type.* In this class of EsGB models $\phi = 0$ does *not* solve the field equations. Thus the Schwarzschild/Kerr BH is not a solution. In terms of the coupling function, this class of models obeys (from the scalar field equation – Equation 6.12 – below)

$$f_{,\phi}(0) \equiv \frac{df(\phi)}{d\phi}\bigg|_{\phi=0} \neq 0 \,. \tag{6.3}$$

A representative example of coupling for this class is the standard dilatonic coupling, $f(\phi) = e^{\gamma\phi}$, which emerges in Kaluza-Klein theory, string theory and supergravity. In this case $\phi$ is often referred to as the *dilaton* field. BHs in the Einstein-dilaton-GB model were constructed in [84], [223], [224], where they were shown to have a qualitatively novel feature: a minimal BH size, determined by the coupling constant $\alpha$. Some of these BHs are perturbatively stable [225] and aspects of their phenomenology has been considered in *e.g.* [226]–[228].

Models where $\phi = 0$ is a consistent truncation are called *class II or scalarised-type.* In this case $\phi \equiv 0$ solves the field equations and thus Schwarzschild and Kerr BHs are solutions of the full model. This demands that

$$f_{,\phi}(0) \equiv \frac{df(\phi)}{d\phi}\bigg|_{\phi=0} = 0 \,. \tag{6.4}$$

This condition holds, for instance, if one requires the model to be $\mathbb{Z}_2$-invariant under $\phi \longrightarrow -\phi$. The Schwarzschild/Kerr BH solution is not, in general, unique. These EsGB models may contain a second set of BH solutions, with a nontrivial scalar field profile – *the scalarised BHs.* Such second set of BH solutions may, or may not, continuously connect with GR BHs. Models within this class have been recenly under scrutiny in relation to BH spontaneous scalarisation - see *e.g.* [229]–[233]. Two reference examples of coupling functions in this case are $f_1(\phi) = \gamma\phi^2$ and $f_2(\phi) = e^{\gamma\phi^2}$ . Although $f_1$ is the linearisation of $f_2$ (the constant term is irrelevant here) these two models have qualitatively different properties. Namely, the spherical scalarised BHs with the former coupling function are unstable against perturbations; but the ones with the latter coupling function can be stable [234].

In this chapter we are interested in a model of class I, the linear coupling or *shift symmetric* model. The coupling function is

$$f(\phi) = \phi \,, \tag{6.5}$$

which implies the existence of a shift symmetry: the equations of motion are invariant under the transformation

$$\phi \longrightarrow \phi + \phi_0 \,, \tag{6.6}$$

with $\phi_0$ an arbitrary constant. This invariance results from the fact that in four spacetime dimensions the GB term alone is a total divergence. BHs in the model presented in Equation 6.2 with Equation 6.6 have been first discussed by Sotiriou and Zhou (SZ) [27], [28]. This model falls within the Horndeski class [25], [30], for which a no-scalar-hair theorem had been established [29]. However, as we discussed in subsection 1.1.1, the SZ solution circumvents this theorem [235]. The SZ solution has a minimal size, such as the BHs in Einstein-dilaton-GB. In fact, the model Equation 6.2 with Equation 6.6 can be seen as a linearisation of the Einstein-dilaton-GB model, and thus one expects similar properties for the BH solutions of both models. However, as pointed out above, models with a certain coupling function and its linearisation may have different properties. It has also been argued that the SZ could emerge dynamically in a gravitational collapse scenario [236].

The goal of this chapter is to construct and study the basic physical properties of the spinning generalisation of the SZ solution. Astrophysical BHs have angular momentum. Thus, considering spinning BHs is fundamental to assess the physical plausibility of any BH model. This is, however, technically more challenging than for spherical BHs, in particular in the presence of higher curvature corrections, such as the GB invariant, as described below.

This chapter is based on the work done in [31] and it is organised as follows. In section 6.2 we briefly discuss the equations of motion and some relevant properties of the model. In section 6.3 we provide a short review of the spherical SZ solutions, as a warm up for the spinning case. In section 6.4 we introduce the framework for the construction of spinning BHs, discussing the ansatz, boundary conditions, the physical quantities of interest and the numerical procedure. In section 6.5 we describe the spinning BH solutions, its domain of existence, and the behaviour of different physical quantities. In section 6.6 we present conclusions and remarks. One appendix with two sections – Appendix B – gives some technical details on the construction of perturbative and extremal solutions.

## 6.2 The model

We consider a general EsGB model with the action Equation 6.2. Observe that the coupling constant has physical dimension $[\alpha] \sim [L]^2$, where $L$ represents "length". Varying the action with respect to the metric tensor $g_{\mu\nu}$, we obtain the Einstein field equations

$$E_{\mu\nu} \equiv R_{\mu\nu} - \frac{1}{2} g_{\mu\nu} R - 8\pi T_{\mu\nu} = 0 \, . \tag{6.7}$$

The *effective* energy-momentum tensor has two distinct components,

$$T_{\mu\nu} = T^{(s)}_{\mu\nu} - 2\alpha T^{(GB)}_{\mu\nu} \, . \tag{6.8}$$

The first one is due to the scalar kinetic term in Equation 6.2

$$T^{(s)}_{\mu\nu} = \partial_\mu \phi \partial_\nu \phi - \frac{1}{2} g_{\mu\nu} \partial_\alpha \phi \partial^\alpha \phi \; ; \tag{6.9}$$

the second one is due to the scalar-GB term in Equation 6.2, and reads

$$T^{(GB)}_{\mu\nu} = P_{\mu\gamma\nu\alpha} \nabla^\alpha \nabla^\gamma f(\phi) \, , \tag{6.10}$$

where we have defined

$$\begin{aligned} P_{\alpha\beta\mu\nu} &\equiv -\frac{1}{4}\varepsilon_{\alpha\beta\rho\sigma}R^{\rho\sigma\gamma\delta}\varepsilon_{\mu\nu\gamma\delta} \\ &= R_{\alpha\beta\mu\nu} + g_{\alpha\nu}R_{\beta\mu} - g_{\alpha\mu}R_{\beta\nu} + g_{\beta\mu}R_{\alpha\nu} - g_{\beta\nu}R_{\alpha\mu} + \frac{1}{2}\left(g_{\alpha\mu}g_{\beta\nu} - g_{\alpha\nu}g_{\beta\mu}\right)R \ . \end{aligned} \tag{6.11}$$

Here, $\varepsilon_{\alpha\beta\rho\sigma}$ is the Levi-Civita tensor. The equation for the scalar field is

$$\Box\phi + \alpha\frac{df(\phi)}{d\phi}R^2_{\rm GB} = 0 \ . \tag{6.12}$$

As pointed out in the introduction of this chapter, the GB term is a total divergence:

$$R^2_{\rm GB} = \nabla_\mu P^\mu \ , \tag{6.13}$$

where the vector $P^\mu$ takes a particularly simple form [237] for a spacetime possessing a Killing vector $\partial/\partial t$

$$P^\mu = 4P_\nu{}^{\alpha\mu t}\Gamma^\nu_{t\alpha} \ . \tag{6.14}$$

Thus the transformation in Equation 6.6 does not change the equations of the model. Moreover, Equation 6.13 implies that the equation for the scalar field, Equation 6.12, can be written as

$$\nabla_\mu J^\mu = 0 \ , \qquad \text{with } J^\mu = \partial^\mu\phi + \alpha P^\mu \ . \tag{6.15}$$

As we shall see, a consequence of this relation is that the scalar "charge" (as read off from the asymptotically leading monopolar mode) is just the Hawking temperature of BH [238].

In this chapter we shall be interested in stationary, axially symmetric solutions of the above action. They possess two asymptotically measured global charges: the mass $M$ and the angular momentum $J$. There is also a scalar charge $Q_s$, but it is not an independent quantity. Differently from the solutions studied so far on this thesis, this scalar charge is not a Noether charge and it is not related with the angular momentum of the scalar field. It depends on the BH mass and angular momentum through the Hawking temperature of the BH – *cf.* Equation 6.36 below. Thus the scalar hair is of secondary type [16]. Also, note that the shift symmetry, Equation 6.6, is broken by imposing $\phi(\infty) = 0$. Horizon quantities of physical interest, on the other hand, include the Hawking temperature $T_H$, the horizon area $A_H$ and the entropy $S$, whose concrete expressions are given below.

Since the equations of the model are invariant under the transformation

$$r \to \lambda r \ , \qquad \alpha \to \lambda\alpha \ , \tag{6.16}$$

where $\lambda > 0$ is an arbitrary constant, the most meaningful physical quantities must be invariant under Equation 6.16. Considering how the various global quantities transform under this scaling ($e.g.$ $M \to \lambda M$, $J \to \lambda^2 J$, $etc.$) we normalise the various quantities $w.r.t.$ the mass of the solutions. In this way, we define the *reduced* angular momentum $j$, horizon area $a_H$, entropy $s$ and Hawking temperature $t_H$ as

$$j \equiv \frac{J}{M^2} \ , \qquad a_H \equiv \frac{A_H}{16\pi M^2} \ , \qquad s \equiv \frac{S}{4\pi M^2} \ , \qquad t_H \equiv 8\pi T_H M \ . \tag{6.17}$$

Alternatively, one can define dimensionless reduced variables $w.r.t.$ the coupling constant $\alpha$ (we recall that $[\alpha] \sim [L]^2$).

## 6.3 Spherically symmetric black holes

Before discussing the case of spinning BHs, it is of interest to review the construction and basic properties of the static, spherically symmetric BHs, the SZ solutions [27], [28]. As we shall see, they contain valuable information, and share some key properties with their rotating counterparts, being easier to study since they are found by solving a set of ordinary differential equations. Moreover, a perturbative *exact* solution is available in the static case, which is discussed in subsection B.1.1 of Appendix B.

### 6.3.1 The equations and boundary conditions

The spherical BHs of Equation 6.2 with Equation 6.6 can be found using Schwarzschild-like coordinates, with a metric ansatz containing two unknown functions,

$$ds^2 = -N(r)\sigma^2(r)dt^2 + \frac{dr^2}{N(r)} + r^2 d\Omega_2^2\,, \qquad \text{with}\ \ N(r) \equiv 1 - \frac{2m(r)}{r}\,, \tag{6.18}$$

where $r$ and $t$ are the radial and time coordinate, respectively, $d\Omega_2^2$ is the metric on the unit round $S^2$ and $m(r)$ is the Misner-Sharp mass [239], which obeys $m(r) \to M$ as $r \to \infty$. The scalar field $\phi$ is a function of $r$ only. The Schwarzschild BH corresponds to $\phi = 0$, $m(r) = r_H/2 = \text{const}$, $\sigma(r) = 1$. One can easily verify that for $\alpha \neq 0$ this is not a solution of the model.

The advantage of this metric gauge choice is the simple form of the Einstein equations, Equation 6.7, which yield the generic relations

$$m' = -\frac{r^2}{4}T_t^t\,, \qquad \frac{\sigma'}{\sigma} = \frac{r}{4N}(T_r^r - T_t^t)\,. \tag{6.19}$$

For the considered EsGB model, the diagonal components of the effective energy-momentum tensor contain second derivatives of the metric functions $N$ and $\sigma$. However, one can find a suitable combination of the field equations such that the functions $m$ and $\sigma$ still solve first order equations. These equations are

$$\left[1 + 2\alpha(1-3N)\frac{\phi'}{r}\right]m' - \left\{\frac{N}{8}r^2\phi'^2 + \alpha(1-N)\left[(1-3N)\frac{\phi'}{r} + 2N\phi''\right]\right\} = 0\,, \tag{6.20}$$

$$\frac{\sigma'}{\sigma}\left[1 + 2\alpha(1-3N)\frac{\phi'}{r}\right] - \frac{1}{4r}\left[r^2\phi'^2 + 8\alpha(1-N)\phi''\right] = 0\,. \tag{6.21}$$

The Einstein equations contain also a second order equation which provides a constraint, being a linear combination of Equation 6.20 and Equation 6.21 together with their first derivatives.

The scalar field $\phi$ is a solution of a 2nd order equation in terms of $N$ and $\phi'$ only

$$\phi''\Bigg[1+\frac{2\alpha}{r}(1-7N)\phi'-\frac{24\alpha^2}{r^4}\left[2(1-N)^2+r^2N(1-3N)\phi'^2\right]+\frac{8\alpha^3N\phi'}{r^5}\Big[24(1-N)^2 + r^2\{1+3N(2-5N)\}\phi'^2\Big]\Bigg]+\frac{1}{r}\Bigg[\left(1+\frac{1}{N}\right)\phi'+\frac{2\alpha}{r^3N}\left[6(1-N)^2+r^2(1-N-12N^2)\phi'^2 -\frac{1}{8}r^4N^2\phi'^4\right]-\frac{8\alpha^2\phi'}{r^4}\left[6(1+N^2)-r^2\phi'^2(1+21N^2)-N\left(12-10r^2\phi'^2+\frac{1}{8}r^4\phi'^4\right)\right] +\frac{8\alpha^3}{r^3}(1-3N)^2(1-5N)\phi'^4\Bigg]=0\,. \tag{6.22}$$

This approach leads to a good accuracy of the numerical results, and can easily be generalized for an arbitrary coupling function $f(\phi)$.

The approximate form of the solutions valid for large-$r$ reads

$$N(r)=1-\frac{2M}{r}+\frac{Q_s^2}{4r^2}+\dots\,,\qquad \sigma(r)=1-\frac{Q_s^2}{8r^2}+\dots\,,\qquad \phi(r)=-\frac{Q_s}{r}-\frac{Q_sM}{r^2}+\dots\,, \tag{6.23}$$

in terms of mass $M$ and a scalar "charge" $Q_s$. Close to the event horizon, located at $r=r_H$, the solutions possess an approximate expression as a power series in $r-r_H$, with

$$\begin{aligned} N(r) &= N_1(r-r_H)+\dots\,,\qquad \sigma(r)=\sigma_H+\sigma_1(r-r_H)+\dots\,,\\ \phi(r) &= \phi_H+\phi_1(r-r_H)+\phi_2(r-r_H)^2+\dots\,, \end{aligned} \tag{6.24}$$

where

$$N_1=\frac{1}{2\alpha\phi_1+r_H}\,,\qquad \sigma_1=\frac{(16\alpha\phi_2+\phi_1^2r_H^2)\sigma_H}{4(2\alpha\phi_1+r_H)}\,, \tag{6.25}$$

while $\phi_2$ is a complicated function of $\phi_1$, $r_H$ and $\alpha$. The Hawking temperature, horizon area and entropy of the solutions, as computed from the formalism in the next section, are given by

$$T_H=\frac{N_1\sigma_H}{4\pi}\,,\qquad A_H=4\pi r_H^2\,,\qquad S=\pi r_H^2+4\pi\alpha\phi_H\,. \tag{6.26}$$

The field equations imply that the first derivative of the scalar field, $\phi_1$, is a solution of the quadratic equation

$$\phi_1^2+\frac{r_H}{2\alpha}\phi_1+\frac{6}{r_H^2}=0\,, \tag{6.27}$$

which implies the following condition for the existence of a real root

$$\frac{\alpha}{r_H^2}<\frac{1}{4\sqrt{6}}\simeq 0.10206\,. \tag{6.28}$$

This requirement translates into the following coordinate independent condition between the horizon size and the coupling constant $\alpha$

$$A_H>16\pi\sqrt{6}\alpha\,. \tag{6.29}$$

We remark that $A_H=4\pi r_H^2$ for the metric ansatz employed here. Thus, for a theory with a given value of the input parameter $\alpha>0$, the BHs are not smoothly connected with the Minkowski vacuum. There is minimal horizon size and a mass gap [27], [28], just as for BHs in the Einstein-dilaton-GB model [84], [223], [224].

### 6.3.2 The solutions

The parameter space of solutions can be scanned by starting with the Schwarzschild BH ($\alpha = 0$) and increasing the value of $\alpha$ for fixed $r_H$. When appropriately scaled, they form a line, starting from the smooth GR limit and ending at a *critical* solution where the condition Equation 6.29 is violated, and where the maximal value of the ratio $\alpha/M^2$ (around 0.32534) is achived. Once the critical configuration is reached, the solutions cease to exist in the parameter space. Physically this means that the EsGB BHs have a minimal size and mass, for given $\alpha$. A possible interpretation is that the GB term provides a repulsive contribution, becoming overwhelming for sufficiently small BHs, thus preventing the existence of an event horizon. The full set of static solutions will be shown below in Figure 6.3 (the blue dotted line with $j = 0$) as a function of the dimensionless parameter $\alpha/M^2$.

As discussed in subsection B.1.1 of Appendix B, a simple perturbative solution can be found as a power series in the parameter

$$\beta \equiv \frac{\alpha}{r_H^2} = \frac{4\pi\alpha}{A_H} \, . \tag{6.30}$$

The results in Appendix subsection B.1.1 imply the following expressions

$$a_H = \frac{A_H}{16\pi M^2} = 1 - \frac{98}{5}\beta^2 + \frac{146378}{1925}\beta^4 - \frac{42468831605804}{13266878625}\beta^6 + \dots \, , \tag{6.31}$$

$$t_H = 8\pi T_H M = 1 + \frac{146}{15}\beta^2 + \frac{1410898}{17325}\beta^4 + \frac{72356439488}{57432375}\beta^6 + \dots \, , \tag{6.32}$$

$$s = \frac{S}{4\pi M^2} = 1 + \frac{146}{15}\beta^2 - \frac{13451026}{51975}\beta^4 + \frac{25584053312}{57432375}\beta^6 + \dots \, ,$$

$$q = \frac{Q_s}{M} = 8\beta - \frac{1184}{15}\beta^3 - \frac{4614784}{17325}\beta^5 + \dots \, ,$$

$$\phi(r_H) = \frac{22}{3}\beta + \frac{40516}{675}\beta^3 - \frac{7057522938136377682}{119373478599375}\beta^7 + \dots \, .$$

Interestingly, all corrections to the reduced temperature $t_H$ are positive. That is, for the same mass, the shift symmetric Hordenski BH is "hotter". For the other quantities, no clear generic pattern emerges.

We have found that the perturbative solution provides a very good approximation to the numerical results. This follows from the smallness of the parameter $\beta$. In fact, the condition in Equation 6.28 implies $\beta_{\max} \simeq 0.102062$. As such, the contribution of the higher order terms in $\beta$ quickly becomes irrelevant.

## 6.4 Spinning black holes: the framework

### 6.4.1 Ansatz and boundary conditions

To obtain stationary and axi-symmetric BH spacetimes, possessing two commuting Killing vector fields, $\xi$ and $\eta$, we use a coordinate system adapted to these symmetries. Then $\xi = \partial_t$, $\eta = \partial_\varphi$, and we consider a metric ansatz which has been employed in the past for the study of KBHsSH [37] and the study of several BHs solutions already mentioned throughout this

thesis – Equation 1.19. Furthermore, the boundary conditions of the ansatz functions of the line element and the scalar field (despite being real instead of complex) will be the same as those discussed in section 1.2.

### 6.4.2 Quantities of interest and a Smarr relation

Most of the quantities of interest, such as the total mass and angular momentum, the horizon temperature and area, can be obtained by using the same methods as the ones described in section 1.2. The main difference to the previous cases resides on the Smarr law. The rotating solutions in this scalar-tensor theory obey,

$$M + 2\Omega_H J + M_s = 2T_H S \ , \tag{6.33}$$

where $S$ is the entropy as computed from Wald's formula [240],

$$S = S_E + S_{sGB} \ , \qquad S_E = \frac{A_H}{4} \ , \qquad S_{sGB} = \frac{\alpha}{2}\int_{\mathcal{H}} d^2x\sqrt{h}\phi\mathcal{R} \ , \tag{6.34}$$

and $\mathcal{R}$ is the Ricci scalar of the induced horizon metric $h$. In the Smarr-type law, $M_s$ is a contribution of the scalar field,

$$M_s = \frac{1}{2}\int_{\Sigma} d^3x\sqrt{-g}\partial_\mu\phi\partial^\mu\phi \ , \tag{6.35}$$

which can also be expressed as an integral of $\phi R^2_{\rm GB}$ term.

Also, by integrating Equation 6.15 over an hypersurface bounded by the event horizon and the sphere at infinity one can prove the following interesting relation,

$$Q_s = 16\pi\alpha T_H \ . \tag{6.36}$$

This proportionality between the scalar charge and the Hawking temperature is a unique feature of the shift symmetric EsGB model – a discussion about this relation is present in [238].

The EsGB BHs satisfy also the first law of thermodynamics,

$$dM = T_H dS + \Omega_H dJ \ . \tag{6.37}$$

### 6.4.3 The numerical approach

In our approach, the field equations reduce to a set of five coupled non-linear elliptic partial differential equations for the functions $\mathcal{F}_a = (F_0, F_1, F_2, W; \phi)$, which are found by plugging the ansatz, Equation 1.19, together with $\phi = \phi(r, \theta)$ into the field equations, Equation 6.7 and Equation 6.12. They consist of the Klein-Gordon equation, Equation 6.12, together with suitable combinations of the Einstein equations, Equation 6.7, which is now different from the combination used on the previous chapters,

$$E_r^r + E_\theta^\theta = 0 \ , \tag{6.38}$$

$$E_\varphi^\varphi = 0 \ , \tag{6.39}$$

$$E_t^t = 0 \ , \tag{6.40}$$

$$E_\varphi^t = 0 \ . \tag{6.41}$$

The explicit form of the equations solved in practice is too complicated to display here; each equation containing around 250 independent terms. Also, in a similar way as it was done for the solutions in previous chapters, the remaining equations $E_\theta^r = 0$ and $E_r^r - E_\theta^\theta = 0$ are not solved directly. They yielding two constraints which are monitored in numerics. Typically they are satisfied at the level of the overall numerical accuracy. We remark that one can verify that the remaining equations vanish identically, $E_r^\varphi = E_r^t = E_\theta^\varphi = E_\theta^t = 0$, the circularity condition being satisfied. We emphasise this is an exact result, for our framework, not a numerical approximation. As such, the employed ansatz is consistent, a fact which is not *a priori* guaranteed (see [241] for a discussion in an Einstein-scalar field model which leads to a non-circular metric form).

Our numerical treatment shall be an analogous treatment to the one presented in chapter 1, chapter 3 and chapter 5. The only difference is the grid size. Here, most of the results were found for an equidistant grid with $300 \times 40$ points.

The equations for $\mathcal{F}_a$ have been solved subject to the boundary conditions mentioned. All numerical calculations are performed by using the professional package already mentioned throughout this thesis. For the solutions in this chapter, the maximal numerical error for the functions is estimated to be on the order of $10^{-3}$. The Smarr relation (Equation 6.33) provides a further test of the numerical accuracy, leading to error estimates of the same order.

In our numerical scheme, there are three input parameters: **i**) the event horizon radius $r_H$; **ii**) the event horizon angular velocity $\Omega_H$ in the metric ansatz (Equation 1.19) and **iii**) the coupling constant $\alpha$ in the action (Equation 6.2).

The results reported in this chapter are obtained from around twenty thousand solution points. For all these BHs we have monitored the Ricci and the Kretschmann scalars, and, at the level of the numerical accuracy, we have not observed any sign of a singular behaviour on and outside the horizon (see, however, the discussion below on the limiting solutions).

## 6.5 Spinning black holes: numerical results

### 6.5.1 General properties and limiting behaviour

In an approach based on the Newton-Raphson method – *cf.* Appendix A – a good initial guess for the profile of the various functions is an essential condition for a successful implementation. The spinning solutions in this chapter can be constructed by using two different routes. In the first approach, one uses the profile of a Kerr BH with given $r_H, \Omega_H$ as an initial guess for EsGB solutions[1] with a small value of the ratio $\alpha/r_H^2$. The iterations converge and, repeating the procedure, one obtains in this way solutions with large $\alpha$. In the second approach, one starts instead with spherically symmetric solutions of EsGB, either obtained numerically or from the perturbative expansion. These can also be studied within the ansatz, Equation 1.19, with $W = 0$, $F_i$ being functions of $r$ only and with $F_1 = F_2$. Then, starting with an EsGB spherical BH with a given $r_H$ and $\alpha \neq 0$, rotation is introduced by introducing and slowly increasing $\Omega_H$.

[1] We mention that, similar to the static limit, the scalar field equation, Equation 6.12, possesses a nontrivial solution in a fixed Kerr background, which inherits most of the basic properties of the backreacting generalization. In particular, the scalar charge-Hawking temperature relation, Equation 6.36, holds also in this case, while the scalar field appears to diverge as the extremal Kerr limit is approached.

For all solutions we have found, the metric functions $\mathcal{F}_a$, together with their first and second derivatives with respect to both $r$ and $\theta$ have smooth profiles. This leads to finite curvature invariants on the full domain of integration, in particular at the event horizon. The shape of the metric functions $F_0, F_1, F_2$ and $W$ is similar to those in the $\alpha = 0$ case. The maximal deviation from the Einstein gravity profiles (with the same input parameters $r_H, \Omega_H$) is near the horizon. At the same time, the scalar field may possess a complicated angular dependence, in particular for fast spinning configurations.

The profile functions of a typical solution are exhibited in Figure 6.1. The insets show the same curves for Kerr with the same $r_H$, $\Omega_H$, for comparison. The Ricci and the Kretschmann scalars, $R$ and $K \equiv R_{\alpha\beta\mu\nu}R^{\alpha\beta\mu\nu}$, together with the components $T_t^t$ and $T_\varphi^t$ of the *effective* energy-momentum tensor are shown in Figure 6.2. In these plots, the corresponding functions are shown in terms of the (inverse) radial variable $r$ for three different values of the angular coordinate $\theta$. One observes, for instance, that $g_{tt}$ becomes positive along the equator, near the horizon, thus manifesting the existence of an ergo-region (see next subsection). One also notices that both $R$ and $K$ stay finite everywhere, in particular at the horizon. From the components of the effective energy-momentum tensor one observes, in particular, that $-T_t^t < 0$ for a region in the vicinity of the symmetry axis, manifesting a breakdown of the weak energy condition for the effective energy-momentum tensor.

Returning to the construction of the solutions, we have noticed the existence of a critical set of input parameters for which the numerical process fails to converge. Neither a singular behaviour nor a deterioration of the numerical accuracy in the vicinity of this set was observed. An explanation for this behaviour, similar to that justifying the critical configurations found in the static case, is based on the analysis of the field equations in the vicinity of the event horizon. After some algebra, one finds that the second order term $\phi_2(\theta)$ in the expansion of the scalar field $\phi(x,\theta) = \phi_0(\theta) + \phi_2(\theta)x^2 + \ldots$ is a solution of a quadratic equation,

$$a\phi_2^2 + b\phi_2 + c = 0\,, \tag{6.42}$$

where the coefficients $a, b, c$ depend on the values of $F_i$, $W$ and their derivatives at the horizon. Then, a real solution to the above equation exists only if $\Delta = b^2 - 4ac > 0$. In practice, we have monitored this discriminant and observed that the numerical process fails to converge[2] when $\Delta$ takes small values close to zero at $\theta = 0, \pi$. As in the spherically symmetric case, we have found no evidence for the emergence of a secondary branch of solutions in the vicinity of the critical solutions.

A different limiting behaviour is found when varying the value of the horizon velocity $\Omega_H$ for fixed $(r_H, \alpha)$. As for the vacuum Kerr family, following this method one finds two branches of solutions, which join for a maximal value of $\Omega_H$. The first branch emerges from the corresponding static configuration. The second branch, on the other hand, ends, as for $\alpha = 0$, at *extremal configurations*. These have vanishing Hawking temperature and nonvanishing global charges, horizon area and entropy. We must emphasise, however, that only near extremal solutions, as opposed to exactly extremal BHs, can be constructed within the framework proposed in this chapter. As such, the results for the extremal solutions reported here result from extrapolating the data found in the near-extremal case. Moreover, unlike the extremal vacuum Kerr BH which yields a perfectly regular geometry [242], the extremal EsGB solutions appear to not be regular, with the Ricci scalar tending to diverge at the poles of the horizon. A partial understanding of this behaviour is given in section B.2 of Appendix B, based on a perturbative construction of the near-horizon configurations.

[2] The values of $a, b, c$ becomes very large as the value of the reduced temperature decreases, which complicates their accurate extraction and the evaluation of $\Delta$ in the vicinity of the extremal set.

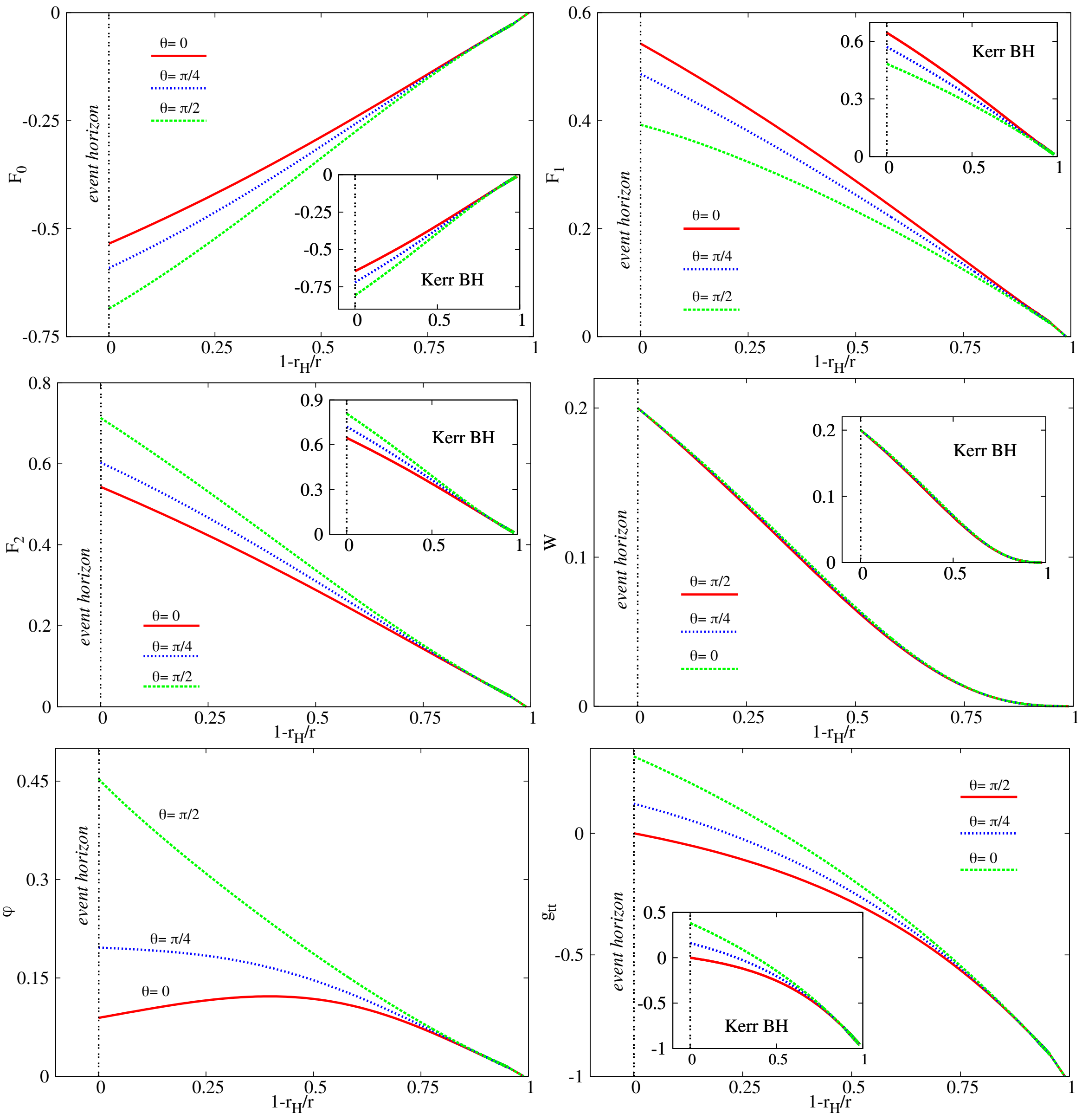

**Figure 6.1:** Profile functions of a typical solution with $r_H = 1.38$, $\Omega_H = 0.2$, $\alpha = 0.4$, *vs*. $1 - r_H/r$, which compactifies the exterior region, for three different polar angles $\theta$. The insets show the corresponding functions for a Kerr BH with the same $r_H, \Omega_H$. The behaviour is qualitatively similar for both cases, with small quantitative differences.

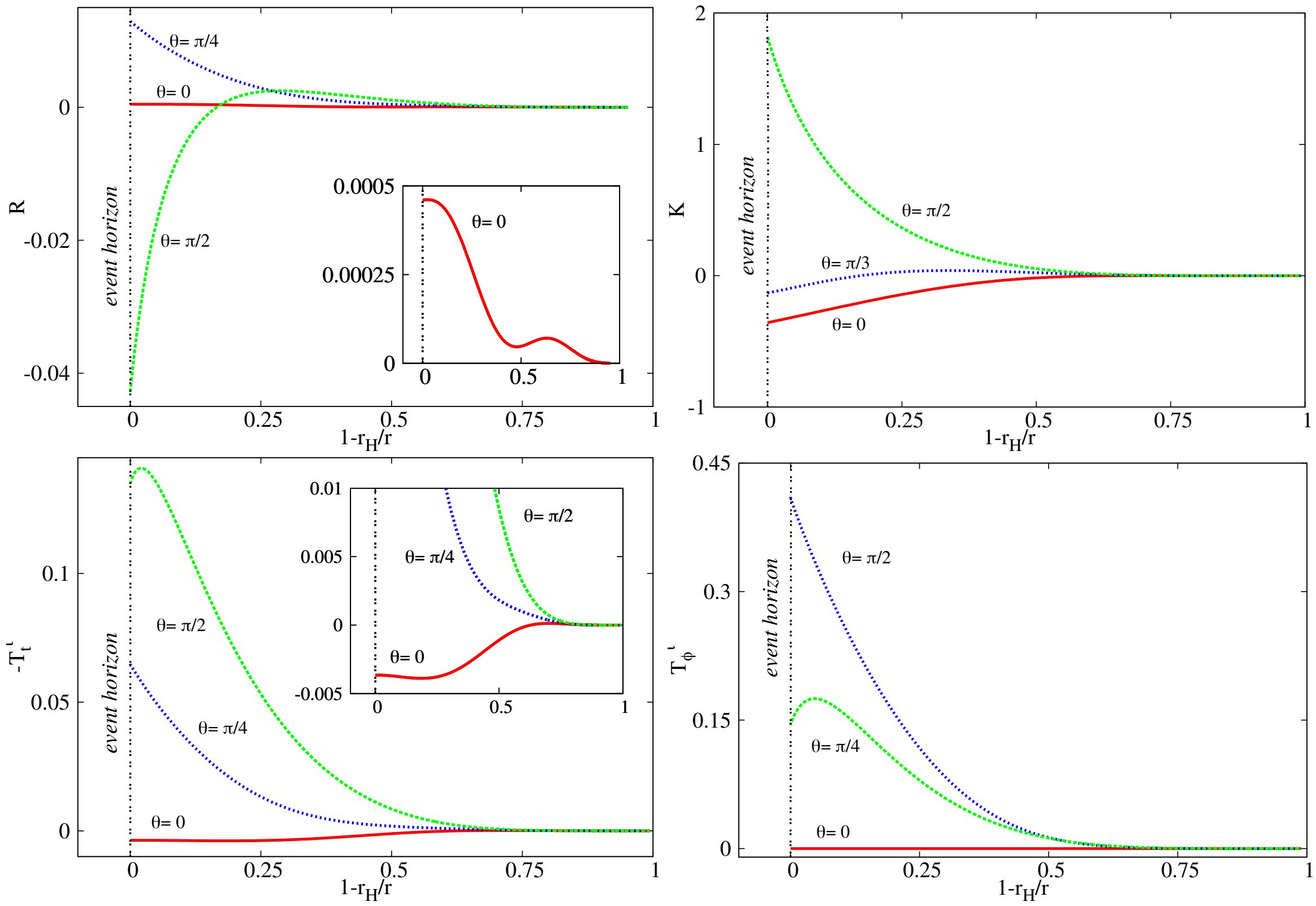

**Figure 6.2:** The Ricci $R$ and Kretschmann $K$ scalars and the components $T_t^t$ and $T_\varphi^t$ of the *effective* energy-momentum tensor, $vs.$ $1 - r_H/r$, for three different polar angles $\theta$ and the same solution as in Figure 6.1. The inset of the bottom left panel shows the existence of a region of negative energy densities around the axis. The inset of the top left panel shows a zoom of the $\theta = 0$ curve.

### 6.5.2 The domain of existence

Let us now address the domain of existence of the EsGB solutions. There are two fundamental scales, the coupling constant $\alpha$, and the BH mass of the solutions $M$. In what follows we display various quantities of interest as a function of the dimensionless coupling constant $\alpha/M^2$. This parameter measures the impact of non-GR features, due to the GB contribution. The analysis is also performed in terms of the dimensionless angular momentum $j = J/M^2$. This parameter measures the impact of non-staticity. The link between these two quantities is provided by the Figure 6.3, where we plot the domain of existence (shaded blue region) in a $j$ *vs*. $\alpha/M^2$ plot. Therein, all data points which were found numerically are also explicitly shown. The blue shaded region is the extrapolation of these points into the continuum. The figure shows that the domain of existence is delimited by:

- the set of static BHs ($j = 0$, blue dotted line);
- the set of extremal BHs (black dotted line);
- the set of critical solutions (green line);
- the set of GR solutions – the Kerr/Schwarzschild BHs ($\alpha/M^2 = 0$, red line).

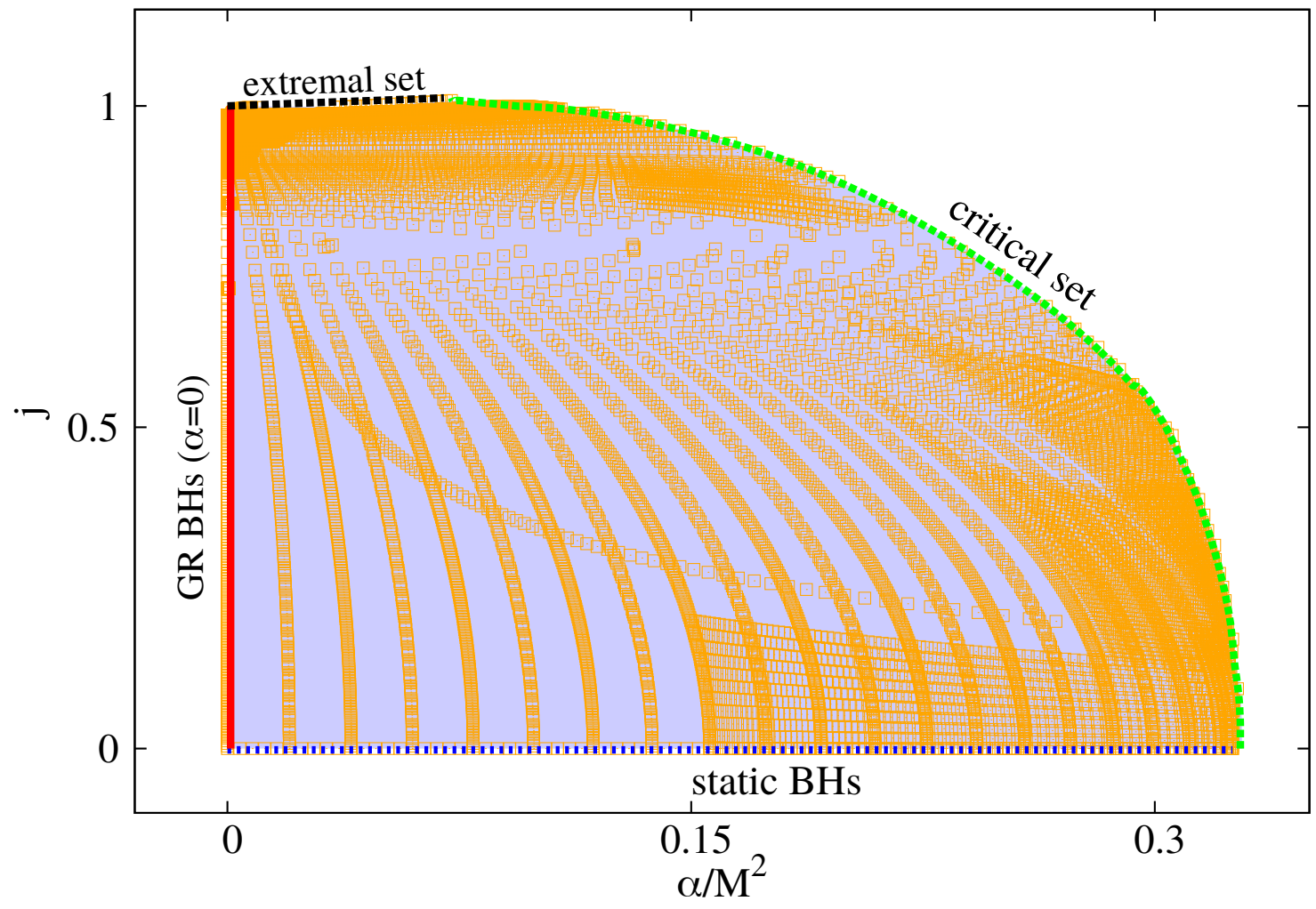

**Figure 6.3:** Domain of existence of EsGB spinning BHs in a $j$ *vs*. $\alpha/M^2$ diagram. Here and in Figure 6.4, all quantities are normalised *w.r.t.* the mass of the solutions. The domain is obtained by extrapolating into the continuum over twenty thousand numerical points. Each such point corresponds to an individual BH solution, and is represented in this plot as a small orange circle.

Two comments on Figure 6.3. First, the Kerr bound $j \leqslant 1$ is violated for spinning EsGB BHs in a small region of the domain of existence close the extremal set. However, this violation is rather small, with $j^{(max)} \sim 1.013$ for all (accurate enough) solutions studied so far. Second, along $j$ fixed lines, the critical solution is attained at a smaller $\alpha/M^2$ as $j$ is increased. A possible interpretation is that both the GB contribution and the spin are repulsive effects. Thus, in the presence of rotation, BHs cease to exist for a smaller GB contribution.

In Figure 6.4 (left panels) the reduced horizon area $a_H \sim A_{\rm H}/M^2$, entropy $s \sim S/M^2$ and temperature $t_H \sim T_{\rm H}M$ of all solutions are shown as functions of the dimensionless coupling constant $\alpha/M^2$. A complementary picture is found when exhibiting the same data as a function of the reduced angular momentum $j$ - Figure 6.4 (right panels).

Let us comment on some features resulting from Figure 6.4. For fixed $j$, the BH area decreases as $\alpha/M^2$ increases; but the corresponding reduced BH entropy *increases*. This provides a clear example how BH entropy deviates from the Hawking-Bekenstein formula in this modified gravity: when the GB contribution becomes larger, the BH becomes smaller but it carries more entropy (for fixed $j$). On the other hand, fixing the EsGB dimensionless coupling constant $\alpha/M^2$, both the reduced area and the reduced entropy decrease as $j$ increases. Thus, for any fixed EsGB model, spin reduces the size and the entropy of BHs. The BH temperature, on the other hand, increases with $\alpha/M^2$ for fixed $j$ and decreases with $j$ for fixed $\alpha/M^2$.

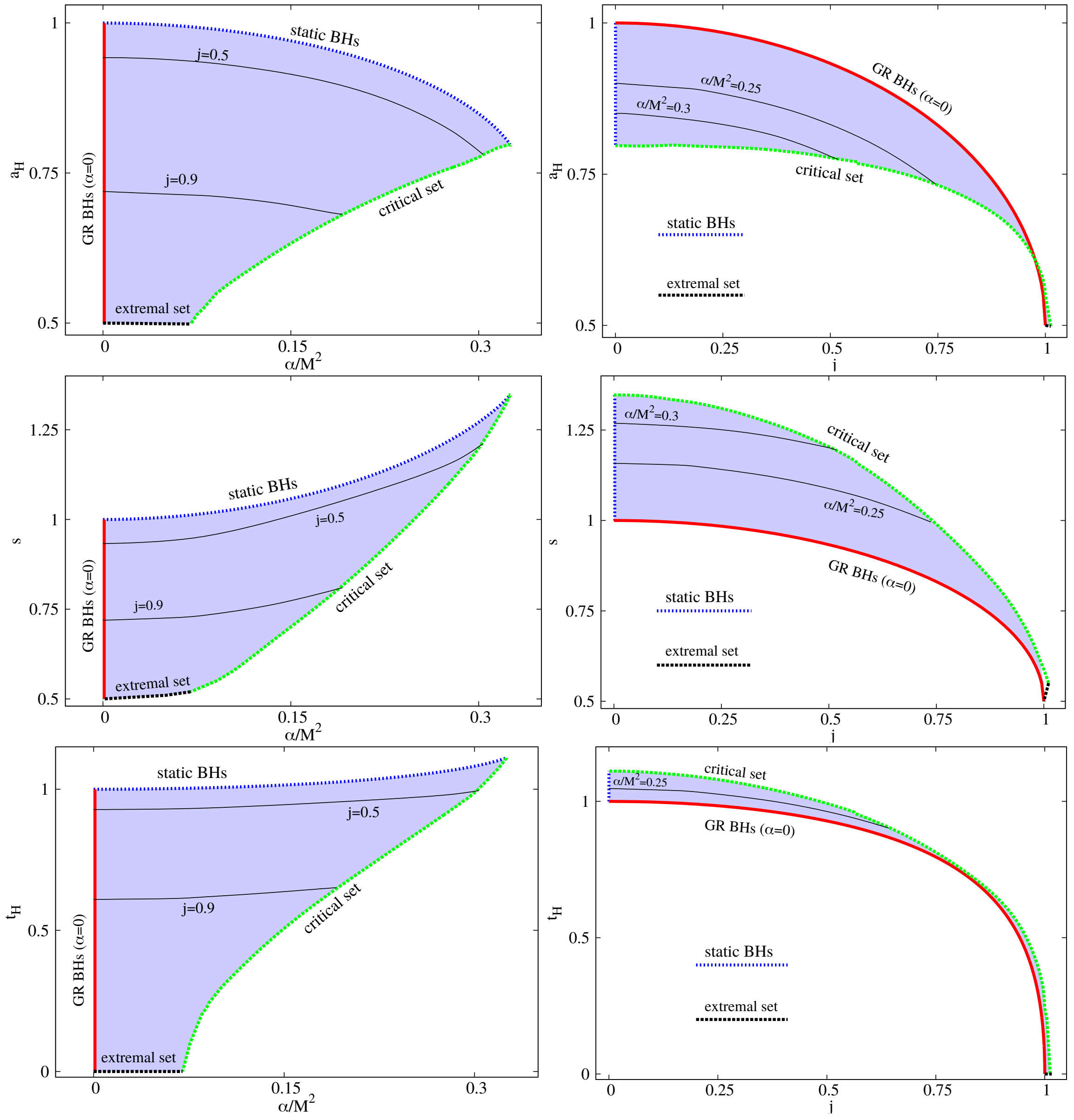

**Figure 6.4:** Domain of existence of spinning EsGB BHs in a reduced horizon area (top panels), entropy (middle panels) and Hawking temperature (bottom panels) *vs*. the dimensioness coupling $\alpha/M^2$ (left panels) or angular momentum $j$ (right panels).

### 6.5.3 Other properties

#### *6.5.3.1 Ergoregion and horizon properties*

All spinning EsGB BHs have an ergoregion, defined as the domain in which the norm of $\xi = \partial_t$ becomes positive outside the horizon. This region is bounded by the event horizon and by the surface where,

$$g_{tt} = -e^{2F_0}N + W^2 e^{2F_2} r^2 \sin^2\theta = 0 \; . \tag{6.43}$$

For the Kerr BH, this surface has a spherical topology and touches the horizon at the poles. As discussed in [106] and in several chapters of this thesis, the ergoregion can be more complicated for other models, notably for BHs with synchronised scalar hair, with the possible existence of an additional $S^1 \times S^1$ ergo-surface (ergo-torus) – see also [52]. We have found that this is not the case for EsGB BHs, where all solutions are Kerr-like in the sense they possess a single topologically $S^2$ ergosurface.

Let us now consider the horizon geometry. Similarly to the GR Kerr solution and to all solutions studied in this thesis, EsGB BHs have an event horizon of spherical topology. Geometrically, however, the horizon is a squashed, rather than round, sphere. This is shown by computing the horizon circumference along the equator, $L_e$, and along the poles, $L_p$, with Equation 2.22, and by using them to calculate the sphericity, already defined in Equation 2.23 and Equation 3.5. In Figure 6.5 (left panel) the sphericity is shown as a function of the dimensionless coupling constant $\alpha/M^2$. An interesting feature there is that $\mathfrak{s}$ can exceed the maximal GR value for a set of EsGB solutions close to extremality. Roughly, the EsGB can become more oblate than Kerr. Also, as expected, the squashing of the horizon produced by the rotation is such that $\mathfrak{s}$ is always larger than unity. That is, the solutions are always deformes towards oblatness, rather than prolatness.

Another physical quantity of interest is the horizon linear velocity $v_H$ [54], [63], [64]. As already discussed several times, $v_H$ measures how fast the null geodesics generators of the horizon rotate relatively to a static observer at spatial infinity. It is obtained through the product between the perimetral radius of the circumference located at the equator, $R_e \equiv L_e/2\pi$, and the horizon angular velocity $\Omega_H$. Such definition is provided in Equation 2.25 and Equation 3.6. As seen in Figure 6.5 (right panel), all studied EGBs solutions have $v_H < 1$, just like for Kerr, and despite the (small) violations of the Kerr bound. Thus, the null geodesics generators of the horizon rotate relatively to the asymptotic observer at subluminal speeds.

Further insight into the horizon geometry is obtained by considering the isometric embedding of the spatial sections of the horizon in an Euclidean 3-space $\mathbb{E}^3$. A well-known feature of the Kerr horizon geometry is that for a dimensionless spin $j > \sqrt{3}/2 \equiv j^{(S)}$ (known as Smarr point) the Gaussian curvature of the horizon becomes negative in a vicinity of the poles [82]. In this regime, an isometric embedding of the Kerr horizon geometry in $\mathbb{E}^3$ is no longer possible. As expected, this feature also occurs for the solutions in this chapter, even though the position of the Smarr point now depends on the value of the dimensionless coupling constant $\alpha/M^2$. Following the discussions bestowed in previous chapters and [54], [64], the collection of Smarr points as $\alpha/M^2$ is varied is the Smarr line. Figure 6.5 displays also the position of the Smarr line as a function of $\alpha/M^2$. One observes that, as for the Kerr limit, an isometric embedding of the horizon geometry in $\mathbb{E}^3$ is possible only up to a maximal value of $\mathfrak{s}$ and $v_H$. Also, notice that both the sphericity $\mathfrak{s}$ and $v_H$ are not constant along the Smarr line and slighly larger values of both these quantities are allowed for embeddable BHs when $\alpha/M^2$ is increased.

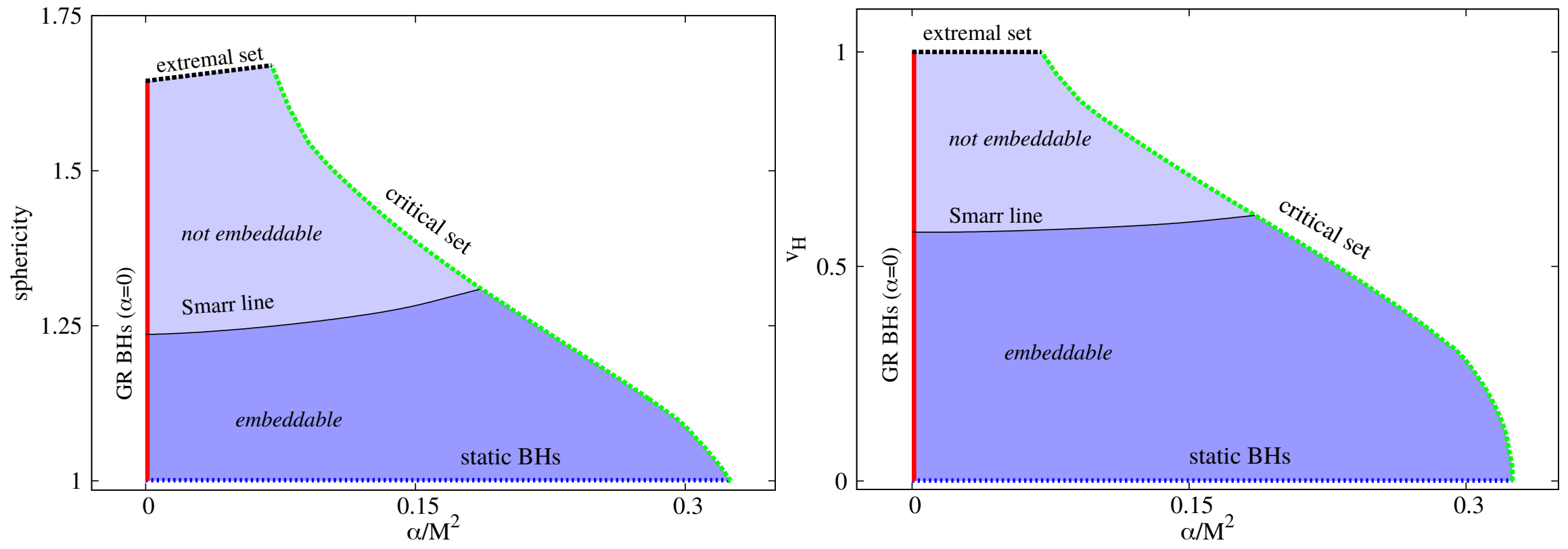

**Figure 6.5:** The sphericity $\mathfrak{s}$ (left panel), and the horizon linear velocity $v_H$ (right panel) $vs.$ $\alpha/M^2$ for the full set of EsGB BHs.

#### 6.5.3.2 Orbital Frequency at the ISCO and Light Rings

A phenomenologically relevant aspect of any BH concerns the angular frequency at both the ISCO and the LR. As mentioned during a similar discussion for Kerr BHs with axionic hair, the former is associated to a cut-off frequency of the emitted synchrotron radiation generated from accelerated charges in accretion disks, whereas, the latter is related to the real part of the frequency of BH quasi-normal modes [215]. The LRs are also key in determining the BH shadow [60].

Following the standard method, one finds that the angular frequency of a test particle with energy, $E$, and angular momentum, $L$, on the equatorial plane, $\theta = \pi/2$, is,

$$\omega = \frac{\dot{\varphi}}{\dot{t}} = W - \frac{e^{2(F_0-F_2)}L}{r^2(L\,W - E)}N\,. \tag{6.44}$$

The radial coordinate, $r$, of such particle obeys the same equation as the one discussed in Equation 5.4.

In the case of massive test particles, circular orbits require that both the potential $V(r)$ and its derivative vanish, $V(r) = V'(r) = 0$. This yields two algebraic equations for $E$ and $L$, which can be solved analytically. These have two distinct pairs of solutions, $(E_+, L_+)$ and $(E_-, L_-)$, corresponding, respectively, to prograde and retrograde orbits. It is then possible to assess the stability of the circular orbits by computing the second derivative of the potential. The ISCO will correspond to the orbit in which the test particle has energy and angular momentum that solves $V(r) = V'(r) = 0$ and the radial coordinate that solves $V''(r) = 0$. Having obtained the energy, angular momentum and radial coordinate of the ISCO, the corresponding angular frequency is computed using Equation 6.44.

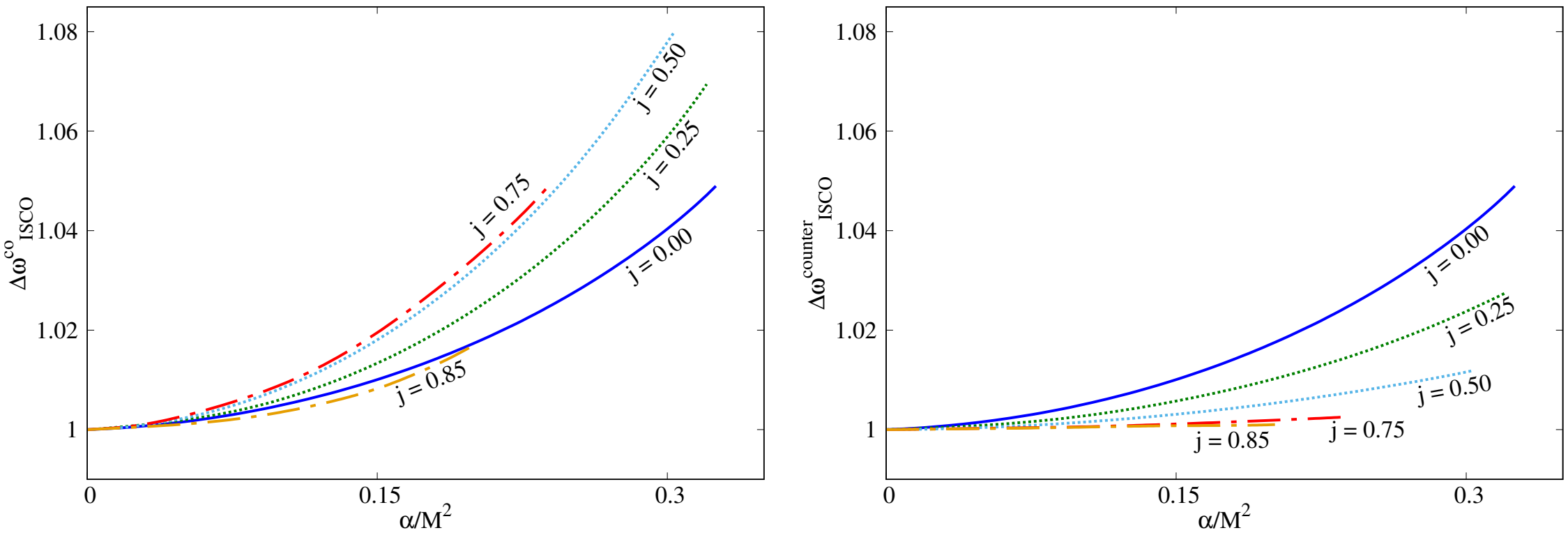

**Figure 6.6:** Ratio between the angular frequency at the ISCO between EsGB BHs and Kerr BHs for prograde orbits (left panel) and retrograde orbits (right panel).

In Figure 6.6, we present the ratio between the angular frequency at the ISCO between EsGB BHs and Kerr BHs, for both prograde, $\Delta\omega^{\rm pro}_{\rm ISCO}$ and retrograde orbits, $\Delta\omega^{\rm retro}_{\rm ISCO}$, fixing $j$, as a function of the reduced coupling constant, $\alpha/M^2$:

$$\Delta\omega^{\rm pro}_{\rm ISCO}(j, \alpha/M^2) = \frac{\omega^{\rm pro}_{\rm ISCO}(j, \alpha/M^2)}{\omega^{\rm pro}_{\rm ISCO}(j, \alpha/M^2 = 0)} \,, \qquad \Delta\omega^{\rm retro}_{\rm ISCO}(j, \alpha/M^2) = \frac{\omega^{\rm retro}_{\rm ISCO}(j, \alpha/M^2)}{\omega^{\rm retro}_{\rm ISCO}(j, \alpha/M^2 = 0)} \,. \tag{6.45}$$

Several illustrative values of $j$ are exhibited.

For both the prograde and retrograde cases, by definition, the ratio converges to unity in the Kerr limit. For all fixed $j$ and for both pro and retrograde orbits, the ratio diverges away, monotonically, from unity as $\alpha/M^2$ increases. How the ratio goes away from unity depends, however, on $j$ and on the direction of the orbital motion.

For $j = 0$ the distinction between prograde and retrograde orbits is meaningless. The ratio grows away from unity as $\alpha/M^2$ increases – solid blue line in Figure 6.6. Naively, this is related to the fact that the static BH size decreases with increasing $\alpha/M^2$, making the ISCO also decrease and hence its frequency increase. Introducing $j$ raises the degeneracy between prograde and retrograde orbits. For prograde (retrograde) orbits and small $j \neq 0$, the ratio is always larger (smaller) than that for the static BHs ($j = 0$) – dotted lines in Figure 6.6 (left and right panels). One may interpret these behaviours as a consequence of frame dragging, which enhances (damps) motion along prograde (retrograde) orbits. In the retrograde case this trend remains for large $j$ – dashed lines in Figure 6.6 (right panel). In the prograde case, however, an unexpected behaviour emerges. For sufficiently large $j$ the ratio stops being enhanced with respect to the static case, and eventually becomes *suppressed* with respect to it – dashed lines in Figure 6.6 (left panel).

A possible explanation for this unexpected behaviour is found by studying the angular velocity of the horizon, $\Omega_H$. This quantity is a better measure of dragging effects than the spacetime angular momentum. Indeed, the fact that a BH has a large $j$ does not imply that it has a large horizon angular velocity.[3] Let us then consider the reduced horizon angular velocity, $\omega_H \equiv \Omega_H M$, and its difference beween EsGB and Kerr BHs with the same $j$, defined as:

$$\delta\omega_H(j, \alpha/M^2) \equiv \omega_H(j, \alpha/M^2) - \omega_H(j, \alpha/M^2 = 0) \, . \tag{6.46}$$

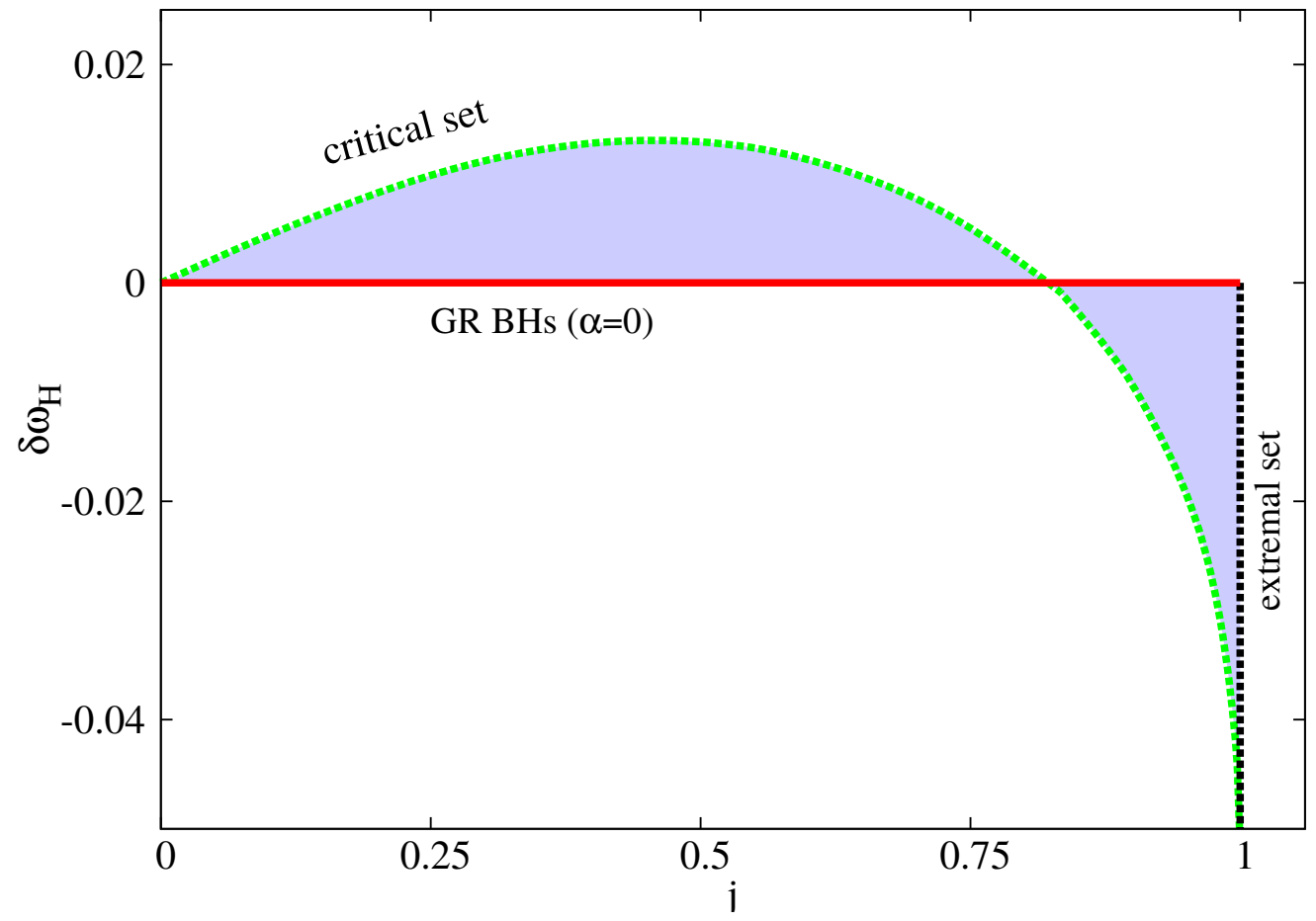

**Figure 6.7:** Reduced horizon angular velocity difference between EsGB BHs and Kerr BHs in a $\delta\omega_H$ *vs.* $j$ plot. For small $j$ the difference is positive, meaning that EsGB BHs spin faster. But for large $j$ the difference is negative, meaning that EsGB BHs spin slower.

This quantity is plotted against the reduced angular momentum $j$ in Figure 6.7. One observes that, for small enough fixed $j$, the EsGB BHs have larger $\omega_H$ than Kerr ones. This support the idea that dragging effects are stronger and should enhance the angular frequency at the ISCO. However, after a given spin $j$, the EsGB BHs have smaller $\omega_H$ than Kerr BHs. That is, albeit having a larger spacetime angular momentum, large $j$ EsGB BHs spin more slowly, and thus source weaker frame dragging, than Kerr BHs. Qualitatively, at least, this provides an explanation for the behaviour observed in Figure 6.6 (left panel).

Quantitatively, for prograde orbits, the maximal deviation from Kerr is $\Delta\omega^{\text{pro}}_{\text{ISCO}} \sim 8\%$ and occurs for $j \sim 0.5$ and the maximal value of $\alpha/M^2$. For retrograde orbits, on the other hand, the ratio is maximised, for any $\alpha/M^2$, by the static case.

In the case of massless particles, a similar analysis can be done. Now, solving $V(r) = 0$, we obtain an algebraic equation for the impact parameter, $b_p = L/E$, which yields two distinct solutions $b_p^+$ and $b_p^-$ corresponding to prograde and retrograde orbits, respectively. Using this result, and solving $V'(r) = 0$, yields the radial coordinate of the LR. Having computed the impact parameter and the radial coordinate of the LRs, one can again compute their angular frequency, using Equation 6.44.

Figure 6.8 shows the ratio between the angular frequency at the LR of EsGB BHs and Kerr BHs, for both prograde and retrograde orbits defined in an analogous way to Equation 6.45, with different values of spin, $j$, as a function of the reduced coupling constant, $\alpha/M^2$. The overall behaviour is very similar to the one discussed above for the ISCO frequency. The main difference for the LR case is that the maximal deviation for both types of orbits is smaller than the corresponding orbits at the ISCO.

[3] The relation between the two quantities should be determined by a moment of inertia. See [101] for an attempt to introduce this notion in BH physics.

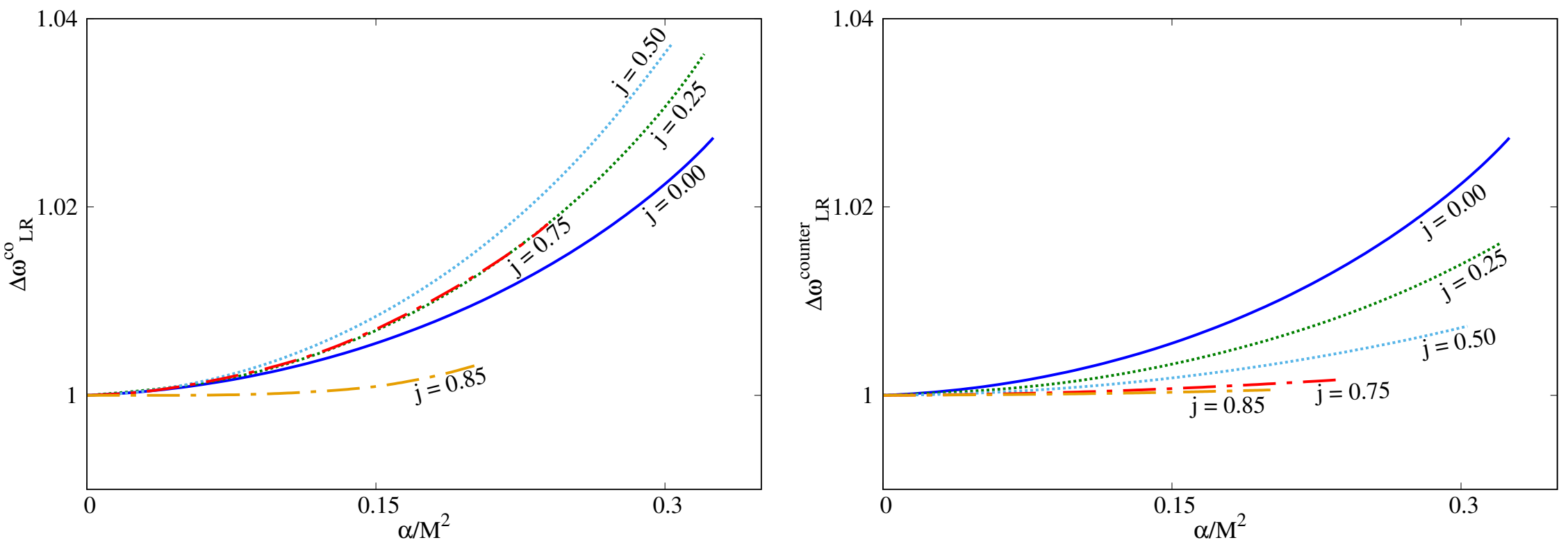

**Figure 6.8:** Ratio between the angular frequency at the LR between EsGB BHs and Kerr BHs, for prograde orbits (left panel) and retrograde orbits (right panel).

## 6.6 Conclusions and remarks

In this chapter we have constructed the spinning generalisations of the static BHs in the shift symmetric Hordenski model. This is a family of asymptotically flat, stationary, axially symmetric BHs, that are non-singular on and outside an event horizon. The domain of existence of these solutions is naturally described by two dimensionless parameters: the dimensionless coupling constant of the model, $\alpha/M^2$, and the dimensioness spin of the BHs, $j = J/M^2$. Then, the domain of existence is bounded by four special limiting behaviours: the GR limit (when $\alpha = 0$), the static limit (when $j = 0$), the extremal limit, when the surface gravity of the solutions vanishes, and a critical set of solutions for which a horizon ceases to exist. This last boundary has an important implication. For non-zero $\alpha$ it means there is minimum mass (and hence) size for BHs. Thus there is a mass gap with respect to the Minkowski vacuum, which is also a solution of the theory.

This non-GR property also occurs for the Einstein-dilaton-GB model discussed $e.g.$ in [84], [224]. Other properties of the BHs we have constructed and analysed in this chapter also parallel the solutions found in the Einstein-dilaton-GB model. This similarity of properties was antecipated by the observation made in the introduction: the linearisation of the action of Einstein-dilaton-GB model

$$S = \int d^4x\sqrt{-g}\left[R - \frac{1}{2}\partial_\mu\phi\partial^\mu\phi + \alpha e^\phi R^2_{\rm GB}\right] , \tag{6.47}$$

reduces to (Equation 6.2) in the limit of small $\phi$, $i.e.$ $e^\phi \simeq 1 + \phi$, by virtue of Equation 6.13. Since the scalar field takes rather small values for typical Einstein-dilaton-GB BHs, the shift symmetric EsGB BHs with the same input parameters provide a reasonable approximation - see, for instance, the bottom left panel of Figure 6.1 for the scalar field magnitude of a typical solution. Thus, the domain of existence of the Einstein-dilaton-GB and EsGB BHs are indeed quite similar, as confirmed by the results in this chapter.

Yet, there are both qualitative and quantitative differences between the two models. An intriguing property of the model we have focused on, that does not occur for the Einstein-dilaton-GB model, is the scalar charge-temperature relation (Equation 6.36). In fact, also the Smarr law is different in both models. Quantitatively, the correspondence between the two models holds only for small enough values of $\alpha/M^2$ and $j$. For example, the critical value of the ratio $\alpha/M^2$ is 0.3253 for the spherically symmetric solutions in this chapter (being fixed by an algebraic condition between the horizon size and the coupling constant $\alpha$, (Equation 6.29)) and 0.1728 for Einstein-dilaton-GB BHs (in which case the generalization of (Equation 6.29) includes, as well, a dependence on the value of the scalar field at the horizon, see $e.g.$ Ref.[223]). Moreover, a specific feature of the Einstein-dilaton-GB model is the occurrance, near the critical configuration of a small secondary branch of BH solutions [243]–[245]. Along this branch, the mass increases with decreasing horizon radius. This secondary branch appears to be absent in the EsGB case.

Finally, let us remark that the way the SZ solutions circumvent the no-scalar-hair theorem also applies to the model herein [29]. This occurs by violating the assumption that the current associated to the shift-symmetry should be finite at the horizon. For the static SZ BHs this current diverges on the horizon. This, however, does not induce any physical pathologies. We have checked that this current (squared) diverges at the horizon also in the spinning BHs reported in this chapter.

CHAPTER 7

# Equatorial Timelike Circular Orbits around Generic Ultracompact Objects

## 7.1 Introduction

The discovery that *quasars* are powerful extragalactic radio sources [246] raised the intriguing question of how their luminosities are produced. Eventually, supermassive BHs emerged as the widely acknowledged engines for such extreme energy outputs [247], due to their deep gravitational potential wells. This was a turning point in the history of BHs, which slowly started to be considered as realistic physical objects by the wider astrophysics community.

The paradigmatic GR BH, described by the Kerr metric [5], has an equatorial ISCO, below which material particles trapped in the BH's potential well are expected to plunge into the horizon.[1] Thus, computing the energy per unit mass of a particle at the ISCO, $E_{\rm ISCO}$, gives an estimate of the rest mass to radiation energy conversion by the BH. The rationale is that a particle in, say, an equatorial thin accretion disk, moves towards smaller and smaller stable timelike circular orbits (TCOs) losing angular momentum (due to turbulence in the disk) and converting its energy into radiation (by heating up), starting off at a large radius until it reaches the ISCO, from which it plunges into the BH. Thus, the *efficiency*

$$\epsilon \equiv 1 - E_{\rm ISCO} \; , \tag{7.1}$$

provides an estimate of the energy conversion into radiation by particles spiralling down the BH's potential well.[2] The first mention of efficiency was done by Shakura and Syunyaev [248]. For a Kerr BH, the efficiency increases monotonically as one increases its spin. For the non-spinning case (Schwarzschild), $\epsilon \sim 5.7\%$, whereas for the extremal Kerr case it reaches $\epsilon \sim 42\%$ [249], [250]. Such dramatic rest mass to radiation energy conversion, well above that observed in typical nuclear reactions (which is smaller than 1%), explains why BHs could source powerful luminosities like those observed in quasars.

[1]This in contrast with the analogue Keplerian problem, wherein stable circular orbits are admissible at any radius.

[2]This simple estimate ignores the energy conversion during the plunge.

The estimates just quoted for $\epsilon$ rely on the structure of stable TCOs around Kerr BHs. How does this structure change for more generic BHs or even for horizonless compact objects? This is a timely question, in view of the ongoing BH astrophysics precision era, triggered by gravitational wave detections [67], [68], horizon scale electromagnetic observations [71], [75]–[77] amongst other observational developments that impact on our knowledge of strong gravity systems. Within the goal of testing the *Kerr hypothesis*, *i.e.* that (near equilibrium) astrophysical BHs are well described by the Kerr metric, it becomes instructive to consider more general models of BHs, motivated by beyond General Relativity gravitational theories or beyond the standard model of particle physics matter models, as well as consider their phenomenology, in particular concerning $\epsilon$.

In this chapter, we shall investigate the structure of TCOs around a generic class of equilibrium BHs or even horizonless, but sufficiently compact (*i.e. ultra*-compact), objects that could imitate BHs in some observables. For this purpose, we start off from two recent theorems on the existence and structure of *light rings* (LRs), *i.e.*, null circular orbits around compact objects. Firstly, it was shown by using a topological argument that for a stationary, axisymmetric, asymptotically flat, 4-dimensional horizonless compact object that can be smoothly deformed into flat spacetime, LRs always come in pairs, one being stable and the other unstable [203]. Secondly, an adaptation of the same sort of topological argument, established that stationary, axisymmetric, asymptotically flat, 4-dimensional BHs always have, at least, one (unstable) LR [251]. With this starting point, in the first part of this chapter, we shall show that for any given ultra-compact object, *i.e.* any object with LRs, possessing also a (north-south) $\mathbb{Z}_2$ symmetry, the structure of the LRs determines, to a large extent (but not fully), the structure of the TCOs on its equatorial plane (section 7.2 and section 7.3). This allows us to establish a simple picture, identifying a small set of building blocks, whose combinations compose the structure of the equatorial TCOs around a generic equilibrium ultra-compact object - *cf.* Figure 7.1-Figure 7.3 below. We note that we shall study the structure of TCO by looking into the *radial* stability of the orbits. A more complete study of the full stability shall be left for a subsequent work.

The study of the full stability of TCOs has been already considered in the literature for the case of static (spherical or axisymmetric) spacetimes. In particular, Vieira *et. al.* [252] (see also the references therein) have shown that, for such Ricci-flat spacetimes, the sum of the radial and angular epicyclic frequencies squared on the equatorial plane, measured by an observer at infinity, vanishes at the LRs. This implies that, for either an unstable or a stable LR, in its adjacent region supporting TCOs one of the epicyclic frequencies squared must be negative, leading always to an unstable region of TCOs. Dropping the Ricci flatness assumption, Vieira *et. al.* [252] were still able to show that the aforementioned sum of the radial and angular epicyclic frequencies squared is always positive if the Strong Energy Condition is obeyed.

Then, in section 7.4, we illustrate these generic structures of TCOs by considering a sample of models of alternative BHs (to the Kerr solution) and horizonless ultra-compact objects, also computing their efficiency. These generalised models unveil an ambiguity related to the proper definition of the efficiency parameter for objects with a more complex structure of TCOs, that we shall discuss. Amongst the explicit examples considered, there are several family of bosonic scalar and vector stars [50], [78], [95], [156]–[158], [190], [253], as well as two different families of "hairy" BHs [28], [31], [79]. In section 7.5 we present a closing discussion about our results. All results shown in this chapter are based on the ones reported in [80].

## 7.2 Circular geodesics on the equatorial plane

We assume a stationary, axi-symmetric, asymptotically flat, 1+3 dimensional spacetime, $(\mathcal{M}, g)$, describing an ultra-compact object. $(\mathcal{M}, g)$ may, or may not, have an event horizon. Our description is theory agnostic: we do not assume $(\mathcal{M}, g)$ solves any particular model.

Let the two Killing vectors associated to stationarity and axi-symmetry be, respectively, $\{\eta_1, \eta_2\}$. Then, a theorem by Carter guarantees that (due to asymptotic flatness), $[\eta_1, \eta_2] = 0$, $i.e.$ the Killing vector fields commute [254]. Consequently, a coordinate system adapted to *both* Killing vectors can be chosen: $(t, r, \theta, \varphi)$, such that $\eta_1 = \partial_t, \eta_2 = \partial_\varphi$. In addition, we assume that the metric: $(i)$ admits a north-south $\mathbb{Z}_2$ symmetry; and $(ii)$ is circular. For asymptotically flat spacetimes, circularity implies that the geometry possesses a 2-space orthogonal to $\{\eta_1, \eta_2\}$ $(c.f.$ theorem 7.1.1 in [255]). Thus $g$ admits the discrete symmetry $(t, \varphi) \to (-t, -\varphi)$.

The spherical-like coordinates $(r, \theta)$ in the orthogonal 2-space are assumed to be orthogonal (which amounts to a gauge choice). If an event horizon exists, another gauge choice guarantees the horizon is located at a constant (positive) radial coordinate $r = r_H$; thus the exterior region is $r_H < r < \infty$. If not, the radial coordinate spans $\mathbb{R}_0^+$ $(0 \leqslant r < \infty)$. Under such choices, $g_{r\theta} = 0$, $g_{rr} > 0$ and $g_{\theta\theta} > 0$ (outside the possible horizon). The $(r, \theta)$ coordinates match the standard spherical coordinates asymptotically $(r \to \infty)$; thus, $\theta \in [0, \pi]$, $\varphi \in [0, 2\pi[$ and $t \in ]-\infty, +\infty[$. The rotation axis, $i.e.$ the set of fixed points of $\eta_2$, is $\theta = \{0, \pi\}$; the equatorial plane, $i.e.$ the set of fixed points of the $\mathbb{Z}_2$ symmetry, is $\theta = \pi/2$. Outside the possible horizon, causality requires $g_{\varphi\varphi} \geqslant 0$. Thus, our generic metric, which has signature $(-, +, +, +)$, is[3]

$$ds^2 = g_{tt}dt^2 + 2g_{t\varphi}dtd\varphi + g_{\varphi\varphi}d\varphi^2 + g_{rr}dr^2 + g_{\theta\theta}d\theta^2 \ . \tag{7.2}$$

Observe that outside a possible horizon, wherein the coordinate system is valid,

$$B(r, \theta) \equiv g_{t\varphi}^2 - g_{tt}g_{\varphi\varphi} > 0 \ , \tag{7.3}$$

which follows from the condition $\det(-g) > 0$ together with a positive signature for the $(r, \theta)$-sector of the metric. Another combination of interest, as it will become clear below, is,

$$C(r, \theta) \equiv (g'_{t\varphi})^2 - g'_{tt}g'_{\varphi\varphi} \ , \tag{7.4}$$

[3]In the following we shall consider that the radial coordinate is a faithful measure of the distance to the central object. That is, that as r increases then, say, the circunferential radius of equatorial orbits of $\eta_2$ also increases.

where the prime denotes the derivative *w.r.t* the radial coordinate. As we shall see, there are ultra-compact objects for which this quantity becomes negative in the domain of outer communication. This impacts in the structure of circular geodesics.

Test particle motion in the generic geometry (Equation 7.2) is ruled by the effective Lagrangian (dots denote derivatives with respect to an affine parameter, which is proper time in the timelike case),

$$2\mathcal{L} = g_{\mu\nu}\dot{x}^\mu\dot{x}^\nu = \xi \ , \tag{7.5}$$

where $\xi = -1, 0, +1$ for timelike, null and spacelike geodesics, respectively. The equatorial plane is a totally geodesic submanifold, wherein the effective Lagrangian simplifies to:

$$2\mathcal{L} = g_{tt}(r, \theta = \pi/2)\dot{t}^2 + 2g_{t\varphi}(r, \theta = \pi/2)\dot{t}\dot{\varphi} + g_{rr}(r, \theta = \pi/2)\dot{r}^2 + g_{\varphi\varphi}(r, \theta = \pi/2)\dot{\varphi}^2 = \xi \ . \tag{7.6}$$

Dropping (for notation ease) the explicit radial dependence of the metric functions, and introducing the two integrals of motion associated to the Killing vectors, the energy, $E$, and the angular momentum, $L$,

$$-E \equiv g_{t\mu}\dot{x}^\mu = g_{tt}\dot{t} + g_{t\varphi}\dot{\varphi} \ , \quad L \equiv g_{\varphi\mu}\dot{x}^\mu = g_{t\varphi}\dot{t} + g_{\varphi\varphi}\dot{\varphi} \ , \tag{7.7}$$

the Lagrangian can be recast as

$$2\mathcal{L} = -\frac{A(r, E, L)}{B(r)} + g_{rr}\dot{r}^2 = \xi \ , \tag{7.8}$$

where

$$A(r, E, L) \equiv g_{\varphi\varphi}E^2 + 2g_{t\varphi}EL + g_{tt}L^2 \ , \tag{7.9}$$

and $B(r)$ is the function in Equation 7.3 restricted to $\theta = \pi/2$. This suggests introducing an effective potential $V_\xi(r)$ as,

$$V_\xi(r) \equiv g_{rr}\dot{r}^2 = \xi + \frac{A(r, E, L)}{B(r)} \ . \tag{7.10}$$

Then, a particle follows a circular orbit at $r = r^{\rm cir}$ iff the following two conditions are simultaneously obeyed throughout the orbit:

$$V_\xi(r^{\rm cir}) = 0 \quad \Leftrightarrow \quad A(r^{\rm cir}, E, L) = -\xi B(r^{\rm cir}) \ , \tag{7.11}$$

and

$$V'_\xi(r^{\rm cir}) = 0 \quad \Leftrightarrow \quad A'(r^{\rm cir}, E, L) = -\xi B'(r^{\rm cir}) \ , \tag{7.12}$$

where prime denotes radial derivative and we have used Equation 7.11 to obtain the last equation in Equation 7.12. Moreover, the *radial*[4] stability of such a circular orbit is determined by the sign of $V''_\xi(r^{\rm cir})$, which reads, upon using Equation 7.11 and Equation 7.12:

$$V''_\xi(r^{\rm cir}) = \frac{A''(r^{\rm cir}, E, L) + \xi B''(r^{\rm cir})}{B(r^{\rm cir})} \ . \tag{7.13}$$

[4]Henceforth, all mentions to stability shall be understood as radial stability.

Then,

$$V''_\xi(r^{\rm cir}) > 0 \;\Leftrightarrow\; \text{unstable}\,; \qquad V''_\xi(r^{\rm cir}) < 0 \;\Leftrightarrow\; \text{stable}\,. \tag{7.14}$$

As such the transition between stable and unstable circular orbits will be determined by $V''_\xi = 0$. In the Kerr family, for TCOs and for each rotation direction, there is only one radius for which $V''_{-1} = 0$. For a generic ultra-compact object there may be more solutions. Thus we define the location of the:

- Marginally Stable Circular Orbit (MSCO)

$$V''_{-1}(r^{\rm MSCO}) = 0 \quad \wedge \quad V'''_{-1}(r^{\rm MSCO}) < 0\,. \tag{7.15}$$

The MSCO should be understood as the stable circular orbit with the smallest radius that is continuously connected to spatial infinity by a set of stable TCOs. For objects that only have one solution satisfying the condition in Equation 7.15 (such as the Kerr case), MSCO corresponds to the well-known ISCO. However, for more generic (non-Kerr) objects, a more intricate structure of TCOs may be present, with other regions of stable TCOs that are *not* continuously connected to spatial infinity by a set of stable TCOs. In such cases, we can define the ISCO, which will be different from the MSCO.

To motivate the definition of the latter, we observe that along circular geodesics, the angular velocity (as measured by an observer at infinity) is

$$\Omega = \frac{d\varphi}{dt} = \frac{\dot\varphi}{\dot t} = -\frac{E g_{t\varphi} + L g_{tt}}{E g_{\varphi\varphi} + L g_{t\varphi}}\,. \tag{7.16}$$

If $\Omega$ is real, circular orbit are possible (timelike, null or spacelike). Then, in a stationary, but not static, spacetime one distinguishes between prograde/retrograde orbits, which are co-rotating/counter-rotating with the spacetime. The angular velocity, energy and angular momentum of the former [latter] are denoted as $(\Omega_+, E_+, L_+)$ $[(\Omega_-, E_-, L_-)]$ and depend on $r^{\rm cir}$. If, however,

$$C(r) < 0\,, \tag{7.17}$$

where, $C(r)$ is the function in Equation 7.4 restricted to $\theta = \pi/2$, then $\Omega$ is not real and no equatorial circular geodesics exist (of any causal character) – *cf.* Equation 7.22 and Equation C.10 on Appendix C below. This possibility, if it occurs, typically arises close to the centre of the ultra-compact object as shown in the examples below. Then, an ISCO could emerge which is different from the MSCO defined above. We thus define the location of the:

- ISCO as the smallest $r$ for which

  $$C(r^{\rm ISCO}) = 0\,, \tag{7.18}$$

  in case this corresponds to a stable TCO *and* occurs in the domain of outer communication; or else, the smallest $r$ for which

  $$V''_{-1}(r^{\rm ISCO}) = 0 \quad \wedge \quad V'''_{-1}(r^{\rm ISCO}) < 0\,, \tag{7.19}$$

  in case there is more than one radial solution of Equation 7.19. The ISCO is determined by either Equation 7.18 or Equation 7.19, whatever is smaller.

Below we shall give examples wherein an ISCO arises from Equation 7.18 and other examples where it arises from Equation 7.19 (and ISCO $\neq$ MSCO).

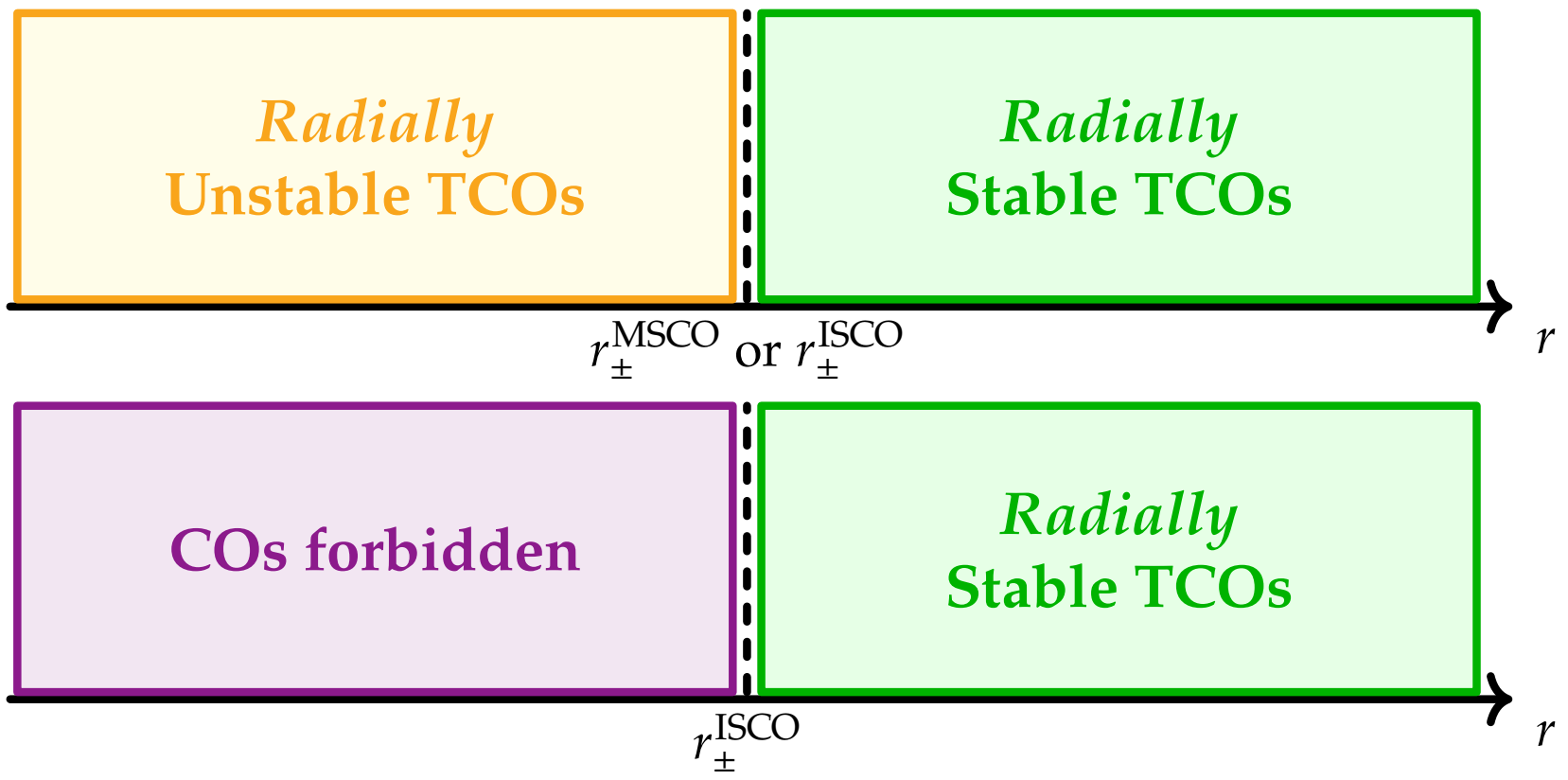

**Figure 7.1:** Structure of the equatorial TCOs around the MSCO and ISCO. (Top panel) The MSCO is determined by the largest radius at the threshold of the stability condition, *cf*. Equation 7.15. In principle this equation can have several radial solutions, so that the ISCO could also be determined by the smallest radius of the same condition, *cf*. Equation 7.19. As we shall see in an example below, the latter could be at the origin, so that the yellow region of unstable TCOs is absent. (Bottom panel) The ISCO can, alternatively, be determined by threshold condition for the absence of any circular orbits, *cf*. Equation 7.18.

### 7.2.1 TCOs

For timelike particles, $\xi = -1$, the condition in Equation 7.11 together with Equation 7.16, determine the energy and angular momentum for circular orbits in terms of the angular velocity as

$$E_\pm = -\left.\frac{g_{tt} + g_{t\varphi}\Omega_\pm}{\sqrt{\beta_\pm}}\right|_{r^{\mathrm{cir}}}, \qquad L_\pm = \left.\frac{g_{t\varphi} + g_{\varphi\varphi}\Omega_\pm}{\sqrt{\beta_\pm}}\right|_{r^{\mathrm{cir}}}, \tag{7.20}$$

where we have defined

$$\beta_\pm \equiv \left(-g_{tt} - 2g_{t\varphi}\Omega_\pm - g_{\varphi\varphi}\Omega_\pm^2\right)\Big|_{r^{\mathrm{cir}}} = -A(r^{\mathrm{cir}}, \Omega_\pm, \Omega_\pm)\,. \tag{7.21}$$

Then, the remaining condition (Equation 7.12) yields $\Omega_\pm$ in terms of the derivatives of the metric functions at $r^{\mathrm{cir}}$

$$\Omega_\pm = \left[\frac{-g'_{t\varphi} \pm \sqrt{C(r)}}{g'_{\varphi\varphi}}\right]_{r^{\mathrm{cir}}}. \tag{7.22}$$

This confirms that the angular velocity ceases to be real when $C(r) < 0$.

Asymptotically, TCOs are essentially Keplerian, and thus stable. Then, as already anticipated in the previous subsection, two important orbits amongst the TCOs emerge: the MSCO and the ISCO. For Kerr-like objects, both orbits are one and the same, and they lie at the threshold of the stability condition (Equation 7.15), $V''_{-1}(r^{\mathrm{MSCO}}) = 0$, *and* that it is continuously connected by stable TCOs to spatial infinity. Thus, it is determined by

$$\left(g''_{\varphi\varphi}E_\pm^2 + 2g''_{t\varphi}E_\pm L_\pm + g''_{tt}L_\pm^2\right)\Big|_{r^{\mathrm{MSCO}}} = \left(g_{t\varphi}^2 - g_{tt}g_{\varphi\varphi}\right)''\Big|_{r^{\mathrm{MSCO}}}. \tag{7.23}$$

In generic ultra-compact objects, however, and as illustrated in the examples below, there may be further disconnected regions with stable TCOs closer to the centre of the compact object. In particular, rather than ending at a transition to a region of unstable TCOs, they can end at a limiting orbit below which *no* circular geodesics are possible (with $C(r) < 0$). Thus, the ISCO can occur at the threshold of this region, which is given by Equation 7.18:

$$C(r^{\rm ISCO}) = [(g'_{t\varphi})^2 - g'_{tt}g'_{\varphi\varphi}]_{r^{\rm ISCO}} = 0\ , \quad \text{and} \quad V''_{-1}(r^{\rm ISCO} + |\delta r|) < 0\ , \tag{7.24}$$

where $|\delta r| \ll 1$. An illustration of the structure of TCOs around the MSCO and ISCO is shown in Figure 7.1. We remark that below we will see ultra-compact objects where both MSCO and ISCO are present.

### 7.2.2 LRs

For lightlike particles, $\xi = 0$, circular orbits are LRs. The condition in Equation 7.11 is a quadratic equation for the *inverse impact parameter*,

$$\sigma_\pm \equiv \frac{E_\pm}{L_\pm}, \tag{7.25}$$

*i.e.* it reads

$$A(r^{\rm LR}, \sigma_\pm, \sigma_\pm) = \left[g_{\varphi\varphi}\sigma_\pm^2 + 2g_{t\varphi}\sigma_\pm + g_{tt}\right]_{\rm LR} = 0\ , \tag{7.26}$$

with the solutions (for prograde, $\sigma_+$, and retrograde, $\sigma_-$, LRs),

$$\sigma_\pm = \left[\frac{-g_{t\varphi} \pm \sqrt{g_{t\varphi}^2 - g_{tt}g_{\varphi\varphi}}}{g_{\varphi\varphi}}\right]_{\rm LR}\ . \tag{7.27}$$

The second condition (Equation 7.12), on the other hand, yields

$$\left[g'_{\varphi\varphi}\sigma_\pm^2 + 2g'_{t\varphi}\sigma_\pm + g'_{tt}\right]_{\rm LR} = 0\ . \tag{7.28}$$

This determines LR's radial coordinate and cannot be solved if Equation 7.17 holds. The stability of the LRs is evaluated by checking the sign of $V''_0(r^{\rm LR})$ given by Equation 7.13; explicitly

$$V''_0(r^{\rm LR}) = L_\pm^2\left[\frac{g''_{\varphi\varphi}\sigma_\pm^2 + 2g''_{t\varphi}\sigma_\pm + g''_{tt}}{g_{t\varphi}^2 - g_{tt}g_{\varphi\varphi}}\right]_{\rm LR}\ . \tag{7.29}$$

The sign is determined by the numerator; if it is positive (negative) the motion is unstable (stable).

## 7.3 TCOs in the vicinity of LRs

We now assume the existence of a LR (which, from the last section, requires $C(r_{\rm LR}) \geqslant 0$).[5] Then, we wish to determine if TCOs exist in its immediate neighbourhood and whether they are stable or unstable.

[5] The case $C(r_{\rm LR}) = 0$ is rather special; although it may be realized in the examples below it corresponds to a zero measure set in the space of solutions. We will further comment on it below, but for now, we shall assume the generic case $C(r_{\rm LR}) > 0$.

### 7.3.1 Allowed region

First, we connect the description of timelike and null orbits. The connection amounts to observe that LRs are determined by

$$\beta_\pm\big|_{\rm LR} = 0\,, \qquad \text{and noting that} \quad \Omega_\pm\big|_{\rm LR} = \sigma_\pm\,. \tag{7.30}$$

Indeed, from Equation 7.21, the condition $\beta_\pm = 0$ becomes equivalent to Equation 7.26 and Equation 7.28 is solved by virtue of Equation 7.22.

The function $\beta_\pm$ will guide us in the connection between LRs and TCOs.[6] From the continuity of $\beta_\pm$ – see Appendix C for more details – one expects that (generically) in the neighbourhood of the LR ($r$ immediately above or below $r_{\rm LR}$) $\beta_\pm$ may become negative. In that case the energy and angular momentum (Equation 7.20) of a timelike particle along such a putative circular orbit become imaginary: such region will *not* contain TCOs (rather it will have spacelike circular orbits).

We will show now that, for either rotation sense, there is always one side in the immediate vicinity of a LR, wherein TCOs are forbidden, whose relative location with respect to the LR depends solely on the stability of the latter.

Assume a LR exists[7] at $r = r_\pm^{\rm LR}$ such that $\beta_\pm(r_\pm^{\rm LR}) = 0$. The first order Taylor expansion of $\beta_\pm$ around the LR reads

$$\beta_\pm(r) = \beta'_\pm(r_\pm^{\rm LR})\delta r + \mathcal{O}(\delta r^2)\,, \tag{7.31}$$

where $\delta r \equiv r - r_\pm^{\rm LR}$. Thus, the sign of $\beta_\pm$ in the vicinity of the LR is determined by $\delta r$ and

$$\beta'_\pm\big|_{\rm LR} = -2\left[\Omega'_\pm\left(g_{t\varphi} + \Omega_\pm g_{\varphi\varphi}\right)\right]_{\rm LR}\,, \tag{7.32}$$

where we have made use of Equation 7.28 (or equivalently Equation 7.22). Explicitly computing $\Omega'_\pm$ (from the quadratic equation leading to Equation 7.22),

$$\Omega'_\pm\big|_{\rm LR} = -\frac{1}{2}\left[\frac{g''_{tt} + 2g''_{t\varphi}\Omega_\pm + g''_{\varphi\varphi}\Omega_\pm^2}{g'_{t\varphi} + \Omega_\pm g'_{\varphi\varphi}}\right]_{\rm LR} \overset{Equation\ 7.29}{=} -\frac{1}{2}\frac{V''_0(r_\pm^{\rm LR})}{L_\pm^2}\left[\frac{g_{t\varphi}^2 - g_{tt}g_{\varphi\varphi}}{g'_{t\varphi} + \Omega_\pm g'_{\varphi\varphi}}\right]_{\rm LR}\,. \tag{7.33}$$

Thus, we can rewrite Equation 7.32 as

$$\beta'_\pm(r_\pm^{\rm LR}) = \frac{V''_0(r_\pm^{\rm LR})}{L_\pm^2}\left[\frac{g_{t\varphi} + \Omega_\pm g_{\varphi\varphi}}{g'_{t\varphi} + \Omega_\pm g'_{\varphi\varphi}}\left(g_{t\varphi}^2 - g_{tt}g_{\varphi\varphi}\right)\right]_{\rm LR}\,. \tag{7.34}$$

Using Equation 7.26 and Equation 7.28 to simplify this result, the first order Taylor expansion of $\beta_\pm$ can be finally written as,

$$\beta_\pm(r) = \frac{V''_0(r_\pm^{\rm LR})}{L_\pm^2}\left[\frac{\left(g_{t\varphi}^2 - g_{tt}g_{\varphi\varphi}\right)^3}{(g'_{t\varphi})^2 - g'_{tt}g'_{\varphi\varphi}}\right]_{\rm LR}^{1/2}\delta r + \mathcal{O}(\delta r^2)\,. \tag{7.35}$$

Thus, the sign of $\beta_\pm$ is determined by the signs of $V''_0(r_\pm^{\rm LR})$ ( stability of the LR) and $\delta r$ (upper or lower neighbourhood of the LR). It follows that in the vicinity of:

[6] This function can be regarded as proportional to the mass squared of the particle along the corresponding circular orbit; thus it is positive, zero and negative, for TCOs, null circular orbits and spacelike circular orbits.

[7] One can consider either a prograde or retrograde LR or both.

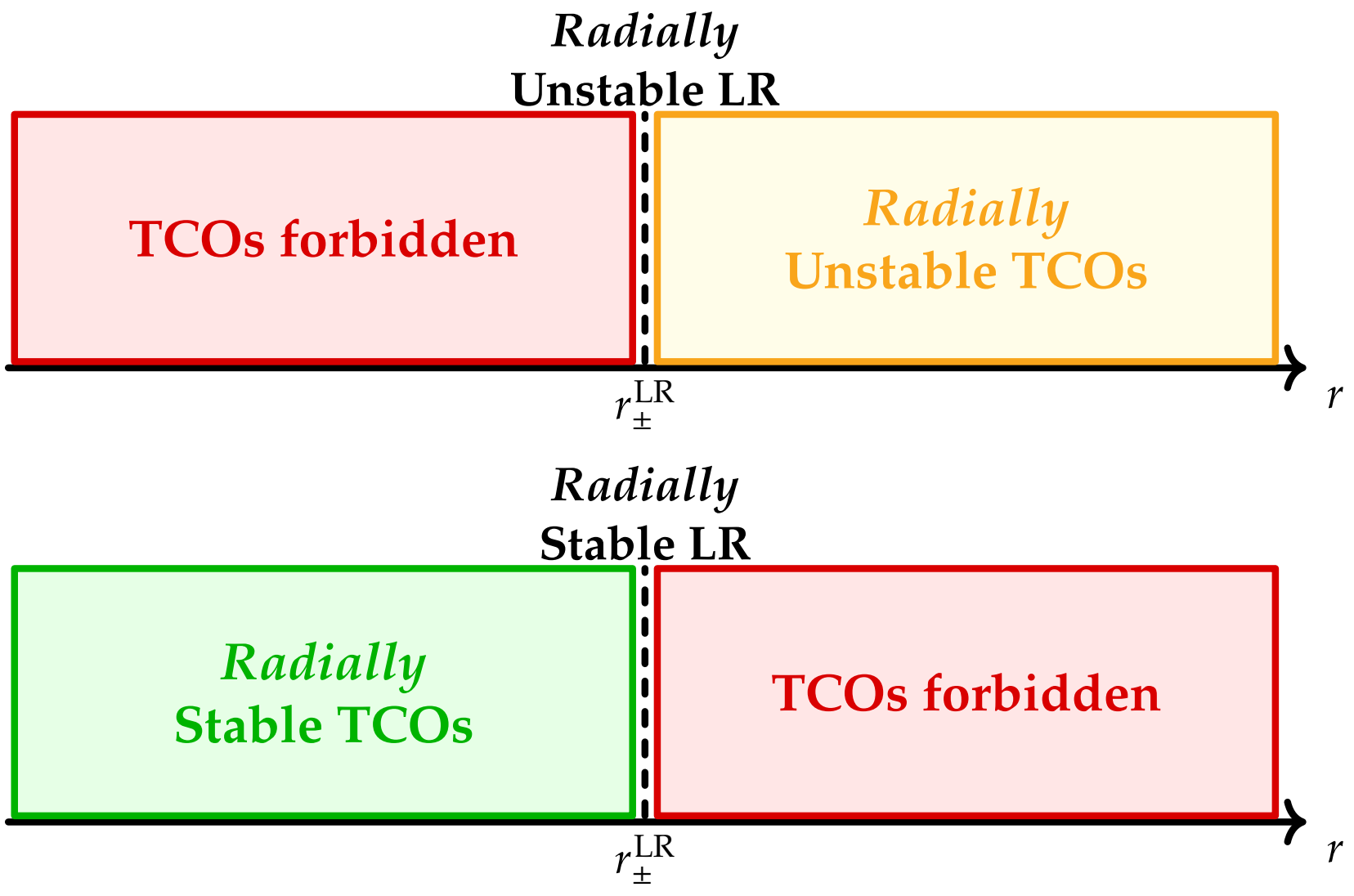

**Figure 7.2:** Structure of the equatorial TCOs in the vicinity an unstable (top panel) and stable (bottom panel) LR.

- An unstable LR ($V_0''(r_\pm^{\mathrm{LR}}) > 0$), $\beta_\pm(r) < 0$ in the region below the LR, *i.e.*, $r < r_\pm^{\mathrm{LR}} \Leftrightarrow \delta r < 0$, wherein no TCOs are thus possible. No obstruction exists for TCOs on the other side.
- A stable LR ($V_0''(r_\pm^{\mathrm{LR}}) < 0$), a symmetric reasoning holds. Thus, no TCOs are possible in the region above the LR, *i.e.*, $r > r_\pm^{\mathrm{LR}} \Leftrightarrow \delta r > 0$.

### 7.3.2 Stability

It is possible to extend further this analysis and determine the stability of the TCOs that occur in the neighbourhood of the LR. We will now show that the region above (below) an unstable (stable) LR always harbours unstable (stable) circular orbits.

To consider the stability of TCOs we examine $V_{-1}''(r)$. Using the definitions of energy and angular momentum, Equation 7.20, we can write,

$$V_{-1}''(r) = \frac{g_{tt}''(g_{t\varphi} + \Omega_\pm g_{\varphi\varphi})^2 - 2g_{t\varphi}''(g_{tt} + \Omega_\pm g_{t\varphi})(g_{t\varphi} + \Omega_\pm g_{\varphi\varphi}) + g_{\varphi\varphi}''(g_{tt} + \Omega_\pm g_{t\varphi})^2}{\beta_\pm(g_{t\varphi}^2 - g_{tt}g_{\varphi\varphi})} - \frac{(g_{t\varphi}^2 - g_{tt}g_{\varphi\varphi})''}{g_{t\varphi}^2 - g_{tt}g_{\varphi\varphi}} \,. \quad (7.36)$$

$V_{-1}''(r_\pm^{\mathrm{LR}})$ diverges, since $\beta_\pm(r_\pm^{\mathrm{LR}}) \to 0$ features in the denominator of the first term. Thus we need to understand with which sign it diverges (and we can ignore the finite second term).

Approaching the LR from the side wherein TCOs are allowed ($\beta_\pm > 0$), the denominator of the first term in Equation 7.36, is positive (recall Equation 7.3). Hence, $\beta_\pm(g_{t\varphi}^2 - g_{tt}g_{\varphi\varphi}) > 0$ and the sign of the term is dictated by the numerator.

Considering now the numerator of the first term in Equation 7.36. Using similar manipulations as before and using Equation 7.29, the numerator can be written, at $r = r_\pm^{\mathrm{LR}}$, as,

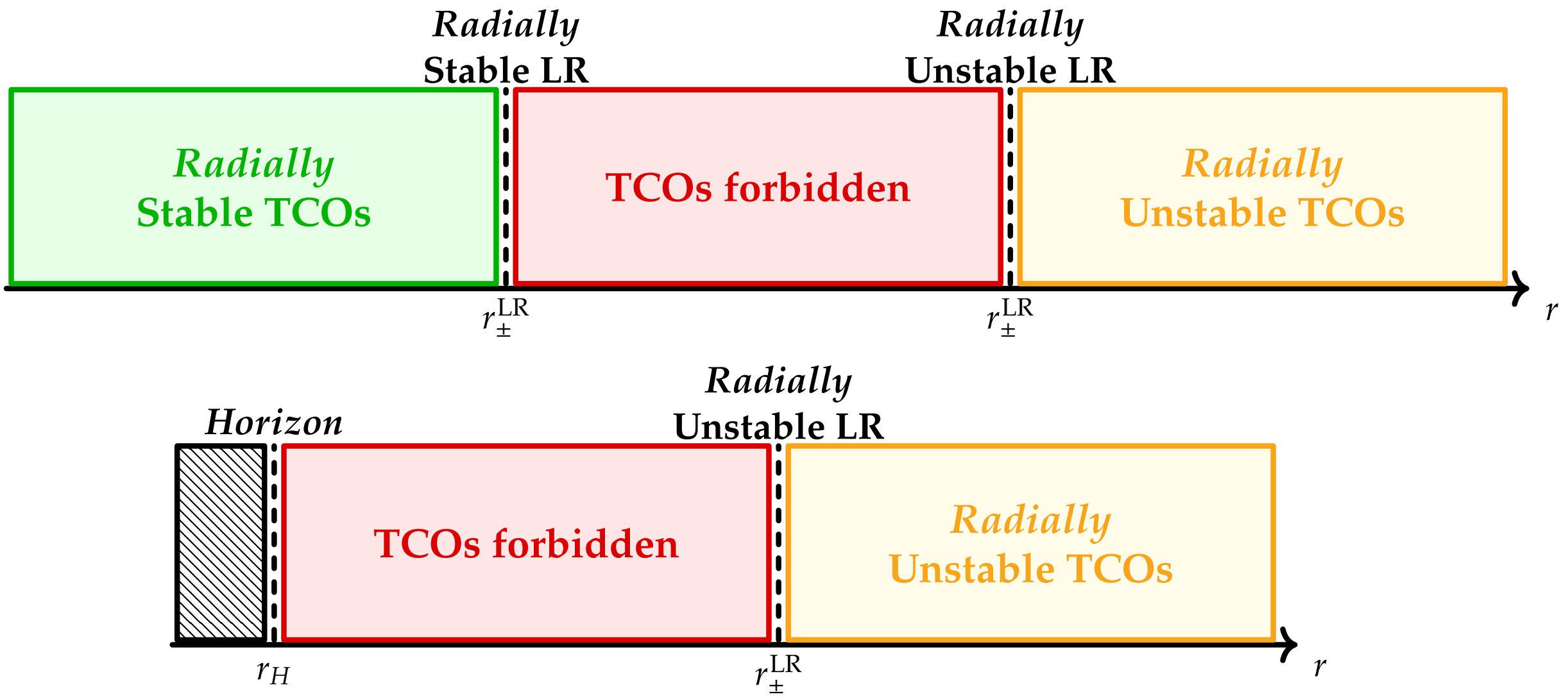

**Figure 7.3:** Structure of the equatorial TCOs for a stationary, axisymmetric, asymptotically flat and $\mathbb{Z}_2$ symmetric: (top panel) horizonless ultra-compact object around its pair of LRs; (bottom panel) BH around its unstable LR. These illustrations are universal, regardless of the direction of rotation of the LR and of the timelike particle.

$$g_{tt}''(g_{t\varphi}+\Omega_\pm g_{\varphi\varphi})^2-2g_{t\varphi}''(g_{tt}+\Omega_\pm g_{t\varphi})(g_{t\varphi}+\Omega_\pm g_{\varphi\varphi})+g_{\varphi\varphi}''(g_{tt}+\Omega_\pm g_{t\varphi})^2 = V_0''(r_\pm^{\rm LR})\frac{(g_{t\varphi}^2-g_{tt}g_{\varphi\varphi})^2}{L_\pm^2}\,. \tag{7.37}$$

Thus, the sign of the numerator is dictated by $V_0''(r_\pm^{\rm LR})$. We concluded that, when approaching the LR from the allowed region: $V_{-1}''(r_\pm^{\rm LR}) \to +\infty$ if the LR is unstable ($V_0''(r_\pm^{\rm LR}) > 0$), and $V_{-1}''(r_\pm^{\rm LR}) \to -\infty$ if the LR is stable ($V_0''(r_\pm^{\rm LR}) < 0$). In short:

- Near an unstable LR, $V_0''(r_\pm^{\rm LR}) > 0$, the allowed region for TCOs harbours unstable orbits - Figure 7.2 (top panel).
- Near a stable LR, $V_0''(r_\pm^{\rm LR}) < 0$, the allowed region for TCOs harbours stable orbits - Figure 7.2 (bottom panel).

### 7.3.3 Generality

The analysis above has two interesting corollaries.

First, it was shown in [203] that for asymptotically flat stationary and axisymmetric horizonless ultracompact objects, that can be smoothly deformed into flat spacetime, LRs come in pairs with one stable and one unstable LR. The proof presented above shows that, for such objects with a $\mathbb{Z}_2$ symmetry, the region between the LRs has no TCOs. Otherwise, there would be a subregion between the LRs wherein $\beta_\pm > 0$, which would imply, by continuity, two points with $\beta_\pm = 0$, *i.e.* another pair of LRs – Figure 7.3 (top panel).

A second corollary applies to BHs. It has been shown that a stationary, axisymmetric and asymptotically flat black hole always has (at least) one unstable LR in each sense of rotation [251]. This statement, together with our proof, implies that the region between the event horizon and the unstable LR is always a region without TCOs. This follows from a similar argument to that presented before for horizonless objects. As in the previous paragraph, this can be easily established by contradiction, and relying on the continuity of $\beta_{\pm}$ – Figure 7.3 (bottom panel).

## 7.4 Illustrations and efficiency

As discussed in the introduction of this chapter, the efficiency of a given compact object can be understood as the amount of gravitational energy which is converted into radiation as a timelike particle falls down from infinity. If one assumes that all radiation escapes towards infinity, then the efficiency is computed as the difference between the energy per unit mass measured at infinity and at the ISCO, as given by Equation 7.1. This definition of efficiency is only an approximation; the real efficiency should take into account how much of the converted radiation effectively reaches infinity and how much falls back into the BH. However, this simple estimate provides an intuition about the magnitude of the process. Moreover, it provides a simple estimate to compare different models to the Kerr BH, which has a maximal efficiency of 42%, probing if alternative models of compact objects could produce even larger energy conversions.

To compute the efficiency, $cf$. Equation 7.1, we need to compute the energy of the TCO at the ISCO. For Kerr BHs, the location of the ISCO is unambiguous: there is only one solution of Equation 7.15 (for each rotation direction) and no solution of Equation 7.18, thus the ISCO is the same as the MSCO. However, for more generic models, there can be several disconnected regions with stable TCOs - see $e.g$. Figure 4.9, Figure 5.6, Figure 5.7 and [65], [78], [79]. For those objects, the ISCO is no longer the same as the MSCO.

The rationale that the efficiency is related to the energy conversion by particles moving along a continuous sequence of stable TCOs, from large distances until the last stable TCO, suggests the efficiency should be computed on that last stable TCO that is continuously connected by a sequence of stable TCOs to infinity. This corresponds to MSCO and the corresponding efficiency is denoted by $\epsilon_{\rm MSCO}$. In parallel, we shall also consider an alternative efficiency computed at the ISCO, denoting the corresponding efficiency by $\epsilon_{\rm ISCO}$. As we shall see

$$\epsilon_{\rm ISCO} \geqslant \epsilon_{\rm MSCO} \, . \tag{7.38}$$

In the following we will analyse the efficiency of several stationary, axisymmetric and asymptotically flat spinning horizonless compact objects (that we shall generically refer to as "stars") as well as of spinning BHs. The examples include: mini-boson stars [50], [253], gauged boson stars [95], axionic boson stars [78], [190] (obtained and discussed in chapter 4), Proca stars [156]–[158], Kerr BHs with synchronised axionic hair [79] (obtained and discussed in chapter 5) and BHs of the shift-symmetric Horndeski theory (Einstein-scalar-Gauss-Bonnet BHs) [28], [31] (obtained and discussed in chapter 6).

For completeness, let us briefly comment on how the considered configurations have been found. For both stars and BHs, the same methodology has been used to obtain the solutions. Such methodology is the same that was used on the previous chapters were several different solutions were obtained. In particular, for each case, one starts with the action of the theory and obtained first the equations of motion. That is, after defining an appropriate ansatz, we have computed the set of partial differential equations for the metric (and matter fields), together with the corresponding boundary conditions. Unfortunately, in all cases, no analytical solutions are known to exist. Therefore, for all models, the solutions were found by employing a professional numerical solver [44]–[46], which uses a Newton-Raphson method. A detailed review of these aspects can be found in the papers where the solutions were initially reported (or their corresponding chapter of this thesis). Once the (numerical) components of the metric are known, one can do various physical and phenomenological studies.

With the knowledge learned from the previous Sections, we have considered the stability of TCOs for all solutions in all mentioned examples, and investigated in which regions of the spacetime it was possible to have stable TCOs ($V''_{-1} < 0$), unstable TCOs ($V''_{-1} > 0$), no TCOs ($\beta_\pm < 0$), or no circular orbits at all ($\Omega_\pm \in \mathbb{C}$). For solutions that possess LRs, we also computed their radii, by solving Equation 7.28 together with Equation 7.27, as well as their stability, Equation 7.29. With all regions defined, we have analysed their boundaries, mainly the boundaries between regions of stable and unstable TCOs, as well as, between regions without any circular orbits and stable TCOs. At each boundary of interest, we have computed the energy of such circular orbit together with the efficiency.

### 7.4.1 Stars

All star solutions discussed herein are only known numerically (no analytic form is known, although in some cases perturbative expansions are possible, *e.g.* [31]). The solutions are computed specifying an ansatz for the metric and remaining fields. For the problem at hand, however, we only need the metric. Thus, we shall only specify the ansatz metric, which is the same for all stars considered and is it the same used to obtain rotating axionic boson stars, Equation 4.6. The correspondence with the ansatz Equation 7.2 is:

$$g_{tt} = -e^{2F_0} + e^{2F_2} W^2 \sin^2\theta \ , \quad g_{rr} = e^{2F_1} \ , \ g_{\theta\theta} = e^{2F_1} r^2 \ , \tag{7.39}$$

$$g_{\varphi\varphi} = e^{2F_2} r^2 \sin^2\theta \ , \quad g_{t\varphi} = -e^{2F_2} r W \sin^2\theta \ . \tag{7.40}$$

For each family of star solutions we shall present four plots. On the one hand, the top (bottom) two plots exhibit the results for prograde (retrograde) orbits. On the other hand, the left plots illustrate the structure of TCOs and LRs *vs.* the radial coordinate $r$ (which is normalise for each family), in the space of solution. For that, the specific solution is labelled by the maximal value of the scalar field $\phi_{\rm max}$ (except for the Proca stars). In this way, each horizontal line corresponds exactly to one star solution. Then, for each plot there are four different coloured regions: in violet, we have a region in which no (timelike, null or spacelike) circular orbits exist (labelled *No COs*); in red, we have a region in which no TCOs exist (labelled *No TCOs*); In yellow, the region of unstable TCOs (labelled *UTCOs*); in green, the region of stable TCOs (labelled *STCOs*). The plots also exhibit the MSCO, ISCO, LRs and a solid black horizontal line representing the first solution for which MSCO ≠ ISCO. In all cases, the results were found by extrapolation into the continuum the data corresponding to a large number (from a few hundreds to thousands) of individual points.

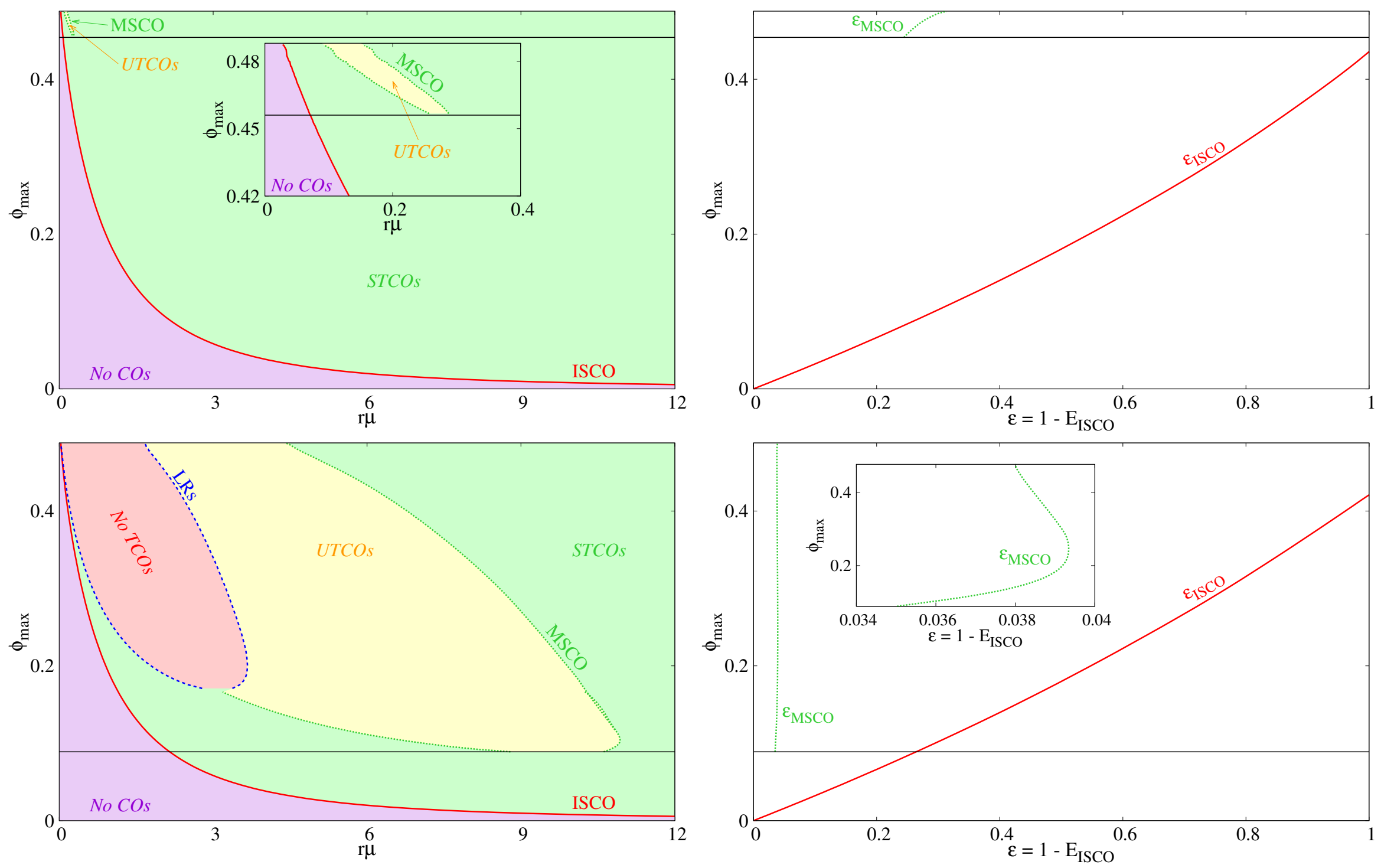

**Figure 7.4:** Structure of TCOs and LRs (*left column*) and efficiency (*right column*) for mini-boson stars with $m = 1$. prograde orbits are presented in the *top row*; retrograde orbits are presented in the *bottom row*.

For all stars studied in this chapter, the structure of TCOs close to LRs follows exactly the patterns deduced in the previous section. In particular, they only possess a pair of retrograde LRs, in which the LR with the largest (smallest) radii is always unstable (stable). Then, the region above (below) an unstable (stable) LR is a region of unstable (stable) TCOs, and the region between the pair of LRs is a region without TCOs.

Finally, the right plots exhibit the efficiencies $\epsilon_{\rm MSCO}$ and $\epsilon_{\rm ISCO}$ *vs*. $\phi_{\rm max}$: $\epsilon_{\rm ISCO}$ is given by the solid red line, whereas $\epsilon_{\rm MSCO}$ is given by dashed green line. For convenience, we keep also the same black solid line as in the left plots.

#### 7.4.1.1 Mini-Boson Stars

Mini-boson stars are regular everywhere solutions of the (complex-)Einstein-Klein-Gordon theory, where a massive free scalar field $\Psi$ is minimally coupled to Einstein's gravity. The action is given by Equation 1.13. These solutions can be consider as a macroscopic version of a Bose-Einstein condensate and were initially developed (in spherical symmetry) by Kaup [34] and Ruffini and Bonazzala [35] - see also, *e.g.* [32], [157]. Later, due to the efforts of Schunck and Mielke [253] and Yoshida and Eriguchi [50], spinning generalisation of the previous static solutions were found - see also, *e.g.* [155].

Here we will consider three families of mini-boson stars with two different values of the azimuthal harmonic index, $m = \{1, 2\}$, which appears in the scalar field ansatz, Equation 1.16.

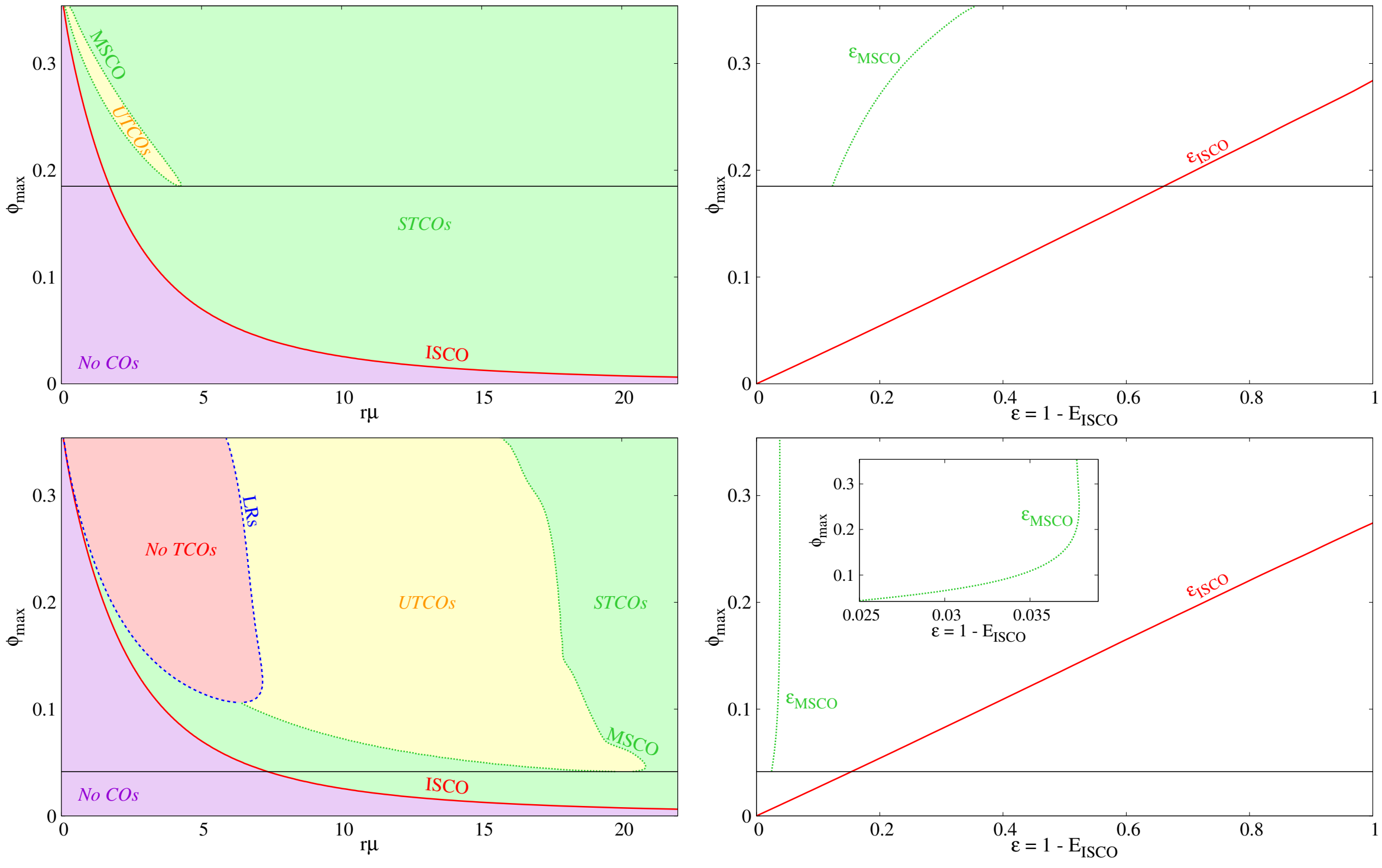

**Figure 7.5:** Same as Figure 7.4 but for mini-boson stars with $m = 2$.

We focus on the fundamental states only (even parity, nodeless BSs).

For the $m = 1$ solutions, the study of the efficiency was already done in [65]. These authors, however, focused on solutions for which MSCO = ISCO. Here we will also consider solutions for which MSCO ≠ ISCO and compute $\epsilon_{\rm MSCO} \neq \epsilon_{\rm ISCO}$.

The right panels in Figure 7.4 exhibit the efficiency for prograde (top) and retrograde (bottom) TCOs for mini-boson stars with $m = 1$. In both cases, $\epsilon_{\rm ISCO}$ (red solid line) increases monotonically with $\phi_{\rm max}$ reaching unity. In fact, for very compact solutions, the gravitational potential can be deep enough to yield efficiencies greater than one; truncating these plots (and the upcoming ones) at $\epsilon = 1$ is, however, enough to show that larger efficiencies than the ones found for Kerr are attained. Such behaviour is explained by the increasingly smaller radii for the ISCO, which, in turn, leads to progressively smaller energies for TCOs therein. This is consistent with the results in [65] (in the region analysed therein).

Now consider $\epsilon_{\rm MSCO}$. In the case of prograde orbits, MSCO ≠ ISCO only in the strong gravity regime, wherein the solutions start to develop a small region of unstable TCOs. Then, $\epsilon_{\rm MSCO}$ ranges from ∼ 25% up to ∼ 30%. For retrograde orbits, MSCO ≠ ISCO for lower values of $\phi_{\rm max}$ than in the prograde case. Thus, for solution which are not very compact (fairly small value of $\phi_{\rm max}$), one can compute $\epsilon_{\rm MSCO}$ which is about ∼ 4% and stabilises around this value even for more compact solutions.

In Figure 7.5 the same analysis as in Figure 7.4 is repeated for mini-boson stars with $m = 2$. The overall structure of TCOs and LRs as well as the efficiencies are very similar as for the $m = 1$ case. The most notorious difference is that the region of unstable TCOs, both for prograde and retrograde orbits, occurs for smaller value of $\phi_{\rm max}$. For prograde orbits (top), MSCO ≠ ISCO above $\phi_{\text{max}} \sim 0.18$, wherein, $\epsilon_{\text{ISCO}} \sim 66\%$ and $\epsilon_{\text{MSCO}} \sim 12\%$. For more compact solutions, $\epsilon_{\text{MSCO}}$ increases monotonically until $\sim 35\%$, whereas $\epsilon_{\text{ISCO}}$ reaches unity. For retrograde orbits (bottom), MSCO ≠ ISCO for lower values of $\phi_{\text{max}}$. For the first solution with $\epsilon_{\text{MSCO}} \neq \epsilon_{\text{ISCO}}$, $\epsilon_{\text{ISCO}} \sim 15\%$ and $\epsilon_{\text{MSCO}} \sim 2.5\%$. Increasing the compactness, these values increase to around $\epsilon_{\text{MSCO}} \sim 3.8\%$ and $\epsilon_{\text{ISCO}} \sim 100\%$.

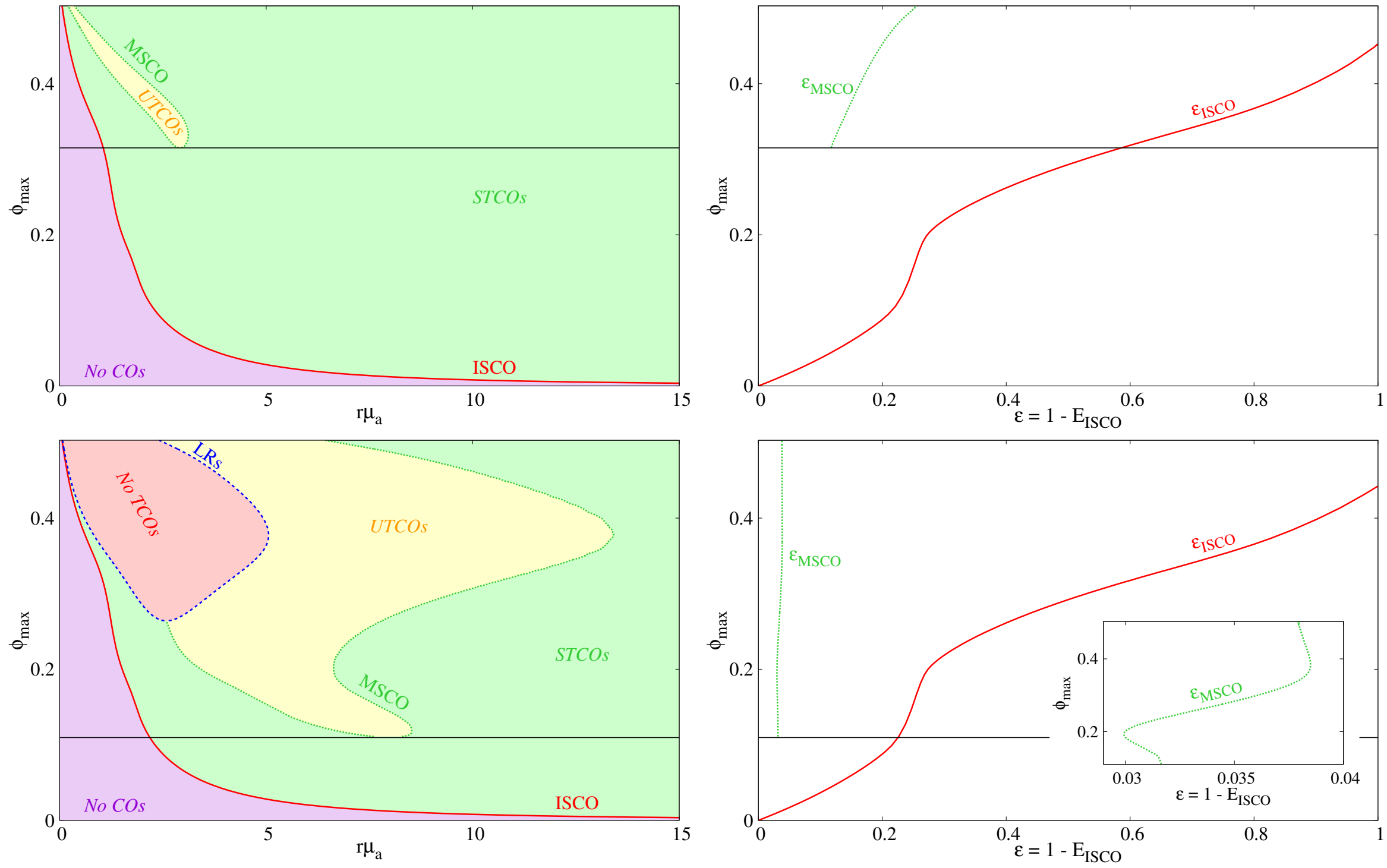

**Figure 7.6:** Same as Figure 7.4 but for axion boson stars with $f_a = 0.05$.

### 7.4.1.2 *Adding scalar field self-interaction: Axion boson stars*

Axion boson stars are solutions of the (complex)-Einstein-Klein-Gordon theory where a massive complex scalar field $\Psi$ with self-interactions is minimally coupled to Einstein's gravity. A detailed study about them was provided in chapter 4. The action is given by Equation 4.1 where we introduced the scalar self-interaction potential $V(\phi)$ based on the QCD axion potential [194] described in Equation 4.7.

The first study about these stars was presented in [190], for the spherical case. A spinning generalisation was presented later in [78]. Here, to probe the effect of the self-interactions on the structure of TCOs and $\epsilon_{\pm}$, we will consider two small values of the decay constant, $f_a = \{0.03, 0.05\}$. The corresponding results are presented in chapter 4 and [78].

In Figure 7.6 we consider $f_a = 0.05$ – a more detailed analysis of the structure of TCOs is done in chapter 4 and [78]. We can see that both the structure of TCOs and LRs, as well as the efficiency follow similar patterns to the previous cases. In particular, $\epsilon_{\text{ISCO}}$ grows monotonically towards unity, moving towards the strong gravity regime. For prograde orbits, at the first solution when MSCO ≠ ISCO, $\epsilon_{\text{ISCO}} \sim 59\%$ and $\epsilon_{\text{MSCO}} \sim 12\%$. After this solution, $\epsilon_{\text{MSCO}}$ increases monotonically until $\sim 26\%$. For retrograde orbits, at the first solution when MSCO ≠ ISCO, $\epsilon_{\text{ISCO}} \sim 23\%$ and $\epsilon_{\text{MSCO}} \sim 3.3\%$. The latter remains approximately constant for other solutions, with a local minimum of $\epsilon_{\text{MSCO}} \sim 3\%$ at $\phi_{\text{max}} \approx 0.193$ and $\epsilon_{\text{MSCO}} \sim 3.8\%$ for the largest values of $\phi_{\text{max}}$.

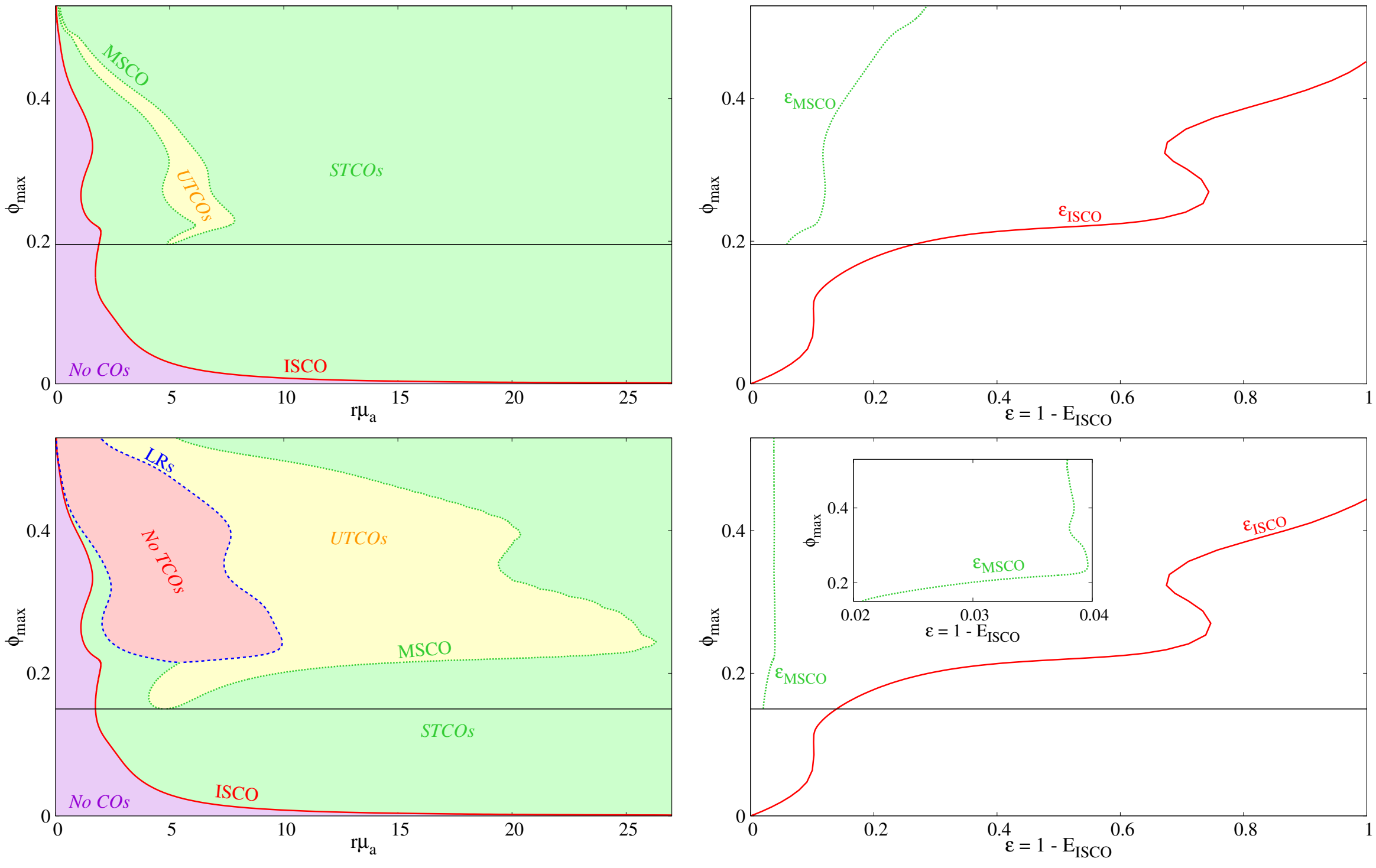

**Figure 7.7:** Same as Figure 7.4 but for axion boson stars with $f_a = 0.03$.

Further decreasing $f_a$ (*i.e.* increasing the self-interactions) introduces more convoluted features - Figure 7.7. This family of axionic stars with $f_a = 0.03$ has the striking feature that $\epsilon_{\text{ISCO}}$ is not longer a monotonically increasing function of $\phi_{\text{max}}$. In fact there are now solutions with degenerated efficiencies, say with $\phi_{\text{max}} = [0.229, 0.376]$. Apart from this novelty, $\epsilon_{\text{ISCO}}$ still approaches unity for large $\phi_{\text{max}}$.

Concerning $\epsilon_{\text{MSCO}}$, in the prograde case, it emerges when $\epsilon_{\text{ISCO}} \sim 26\%$ and is $\epsilon_{\text{MSCO}} \sim 5.9\%$. Then, it varies non-monotonically: at $\phi_{\text{max}} \approx 0.274$, $\epsilon_{\text{MSCO}} \sim 12\%$ (local maximum); at $\phi_{\text{max}} \approx 0.321$, $\epsilon_{\text{MSCO}} \sim 11\%$ (local minimum); for the larger $\phi_{\text{max}}$, $\epsilon_{\text{MSCO}} \sim 32\%$ (global maximum). For retrograde orbits, $\epsilon_{\text{MSCO}}$ is more constant. It emerges when $\epsilon_{\text{ISCO}} \sim 12\%$, with $\epsilon_{\text{MSCO}} \sim 1.2\%$. Then, for $\phi_{\text{ max}} \approx 0.247$, it reaches a local maximum, $\epsilon_{\text{MSCO}} \sim 4\%$. Going further into the strong gravity regime, the efficiency is approximately constant.

### *7.4.1.3 Gauged Boson Stars*

Gauged boson stars can be thought as electrically charged mini-boson stars. They are solutions of the (complex-)Einstein-Klein-Gordon-Maxwell theory,

$$\mathcal{S} = \int d^4x\sqrt{-g}\left[\frac{R}{16\pi} - \frac{1}{4}F_{\mu\nu}F^{\mu\nu} - g^{\mu\nu}D_\mu\Psi^* D_\nu\Psi - \mu^2\Psi^*\Psi\right] , \tag{7.41}$$

where the electromagnetic tensor $F_{\mu\nu} \equiv \partial_\mu A_\nu - \partial_\nu A_\mu$ is defined through the electromagnetic potential $A_\mu$, and $D_\mu \equiv \partial_\mu + iq_E A_\mu$. There is a minimal coupling between the electromagnetic sector and the scalar field through the (gauge) covariant derivative $D_\mu$, which introduces the gauge coupling constant $q_E$.

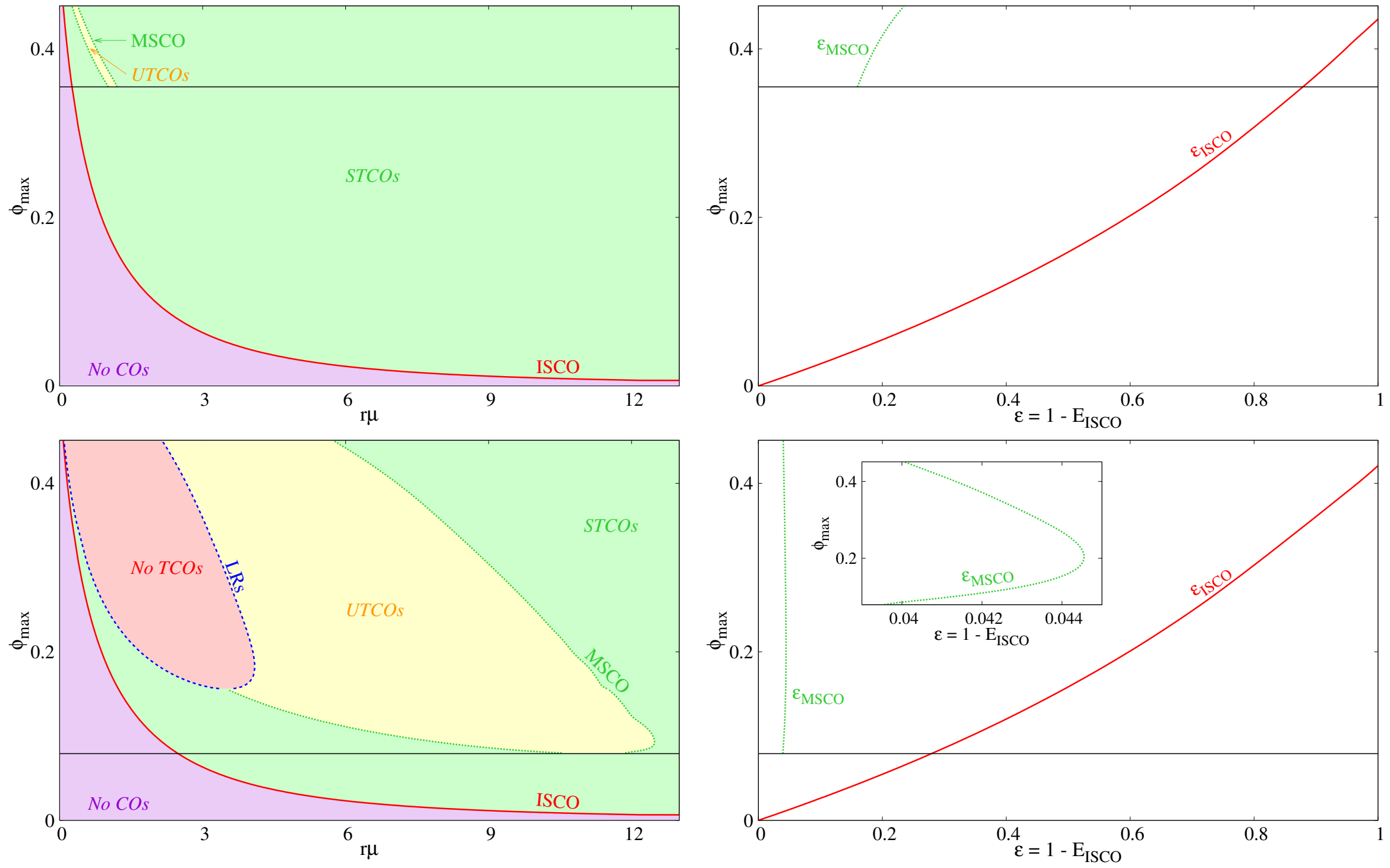

**Figure 7.8:** Same as Figure 7.4 but for gauged boson stars with $q_E = 0.6$.

The first work on gauged boson stars was develop by Jetzer and van de Bij[256] where they obtained spherically symmetric solutions – see also [257]. The rotating generalisation was constructed later in a more general context in [95], (see also [204], [258] for results in a model with a self-interacting scalar field).

The latter are found by using the same ansatz as in Equation 1.16, together with the $U(1)$ form $A = A_\varphi d\varphi + A_t dt$ [95]. This implies that, as before, we have to specify the azimuthal harmonic index $m$. Furthermore, we also need to specify the gauge coupling constant $q_E$. In this chapter we will only consider gauged solution with $m = 1$ and $q_E = 0.6$; the latter choice illustrates the generic features we have seen analysing also other values of $q_E$. Such results are shown in Figure 7.8. Overall we observe that the description for mini-boson stars with $m = 1$ still apply for this case. Being more specific, for the prograde case, $\epsilon_{\rm ISCO} \sim 16\%$ (in contrast to $\sim 25\%$ for mini-boson stars) for the first solution for which $\epsilon_{\rm ISCO} \neq \epsilon_{\rm MSCO}$; then $\epsilon_{\rm MSCO}$ increases gradually until $\sim 24\%$ (in contrast to $\sim 30\%$ for mini-boson stars). For the retrograde case, however, the efficiency difference between gauged and ungauged boson stars is unnoticeable.

The discussion made here prompts the conclusion that, at least for the gauged boson stars reported in [95], the presence of an electric charge does not influence significantly either the structure of TCOs or the efficiency.

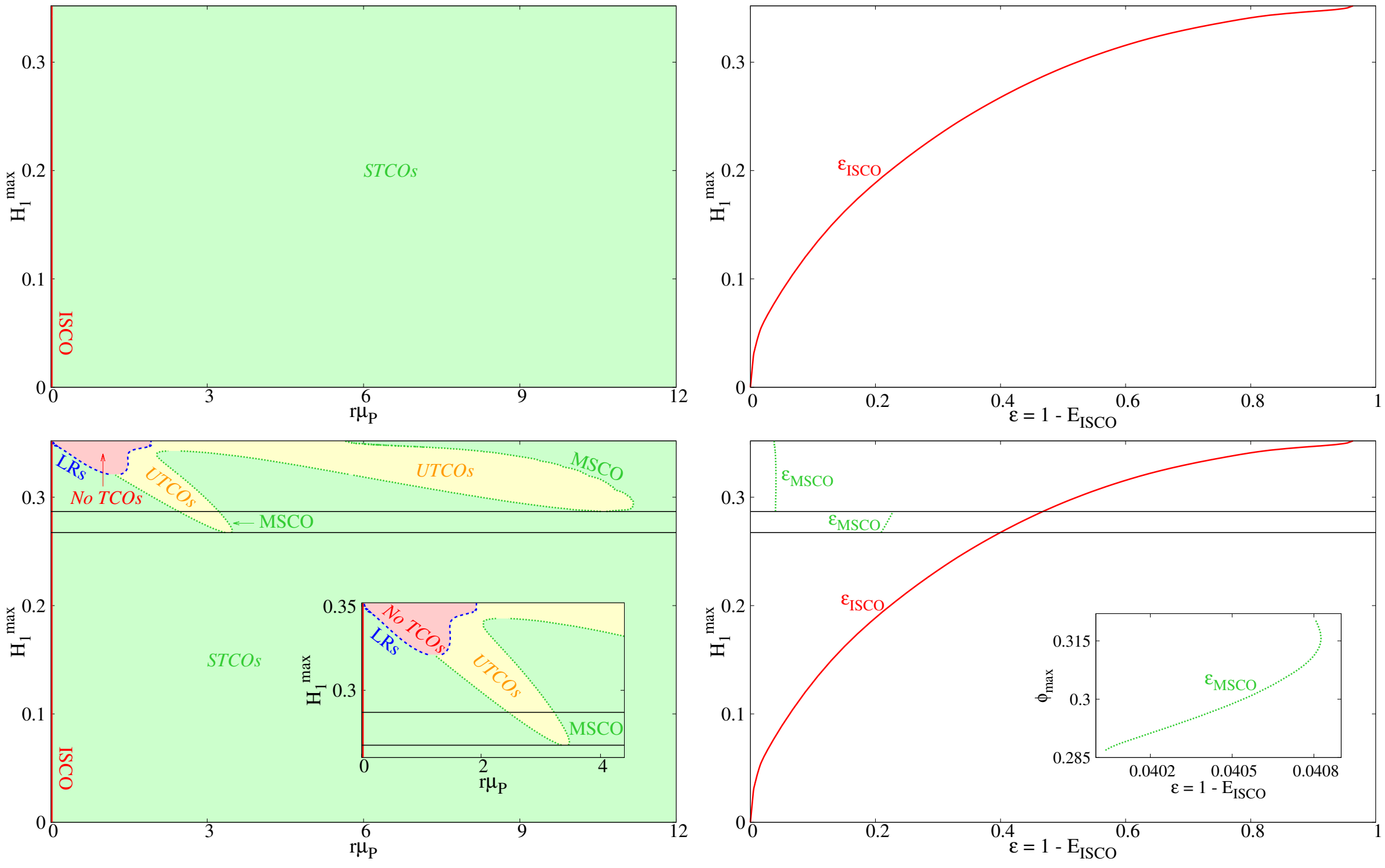

**Figure 7.9:** Same as Figure 7.4 but for Proca stars.

### 7.4.1.4 *Proca Stars*

We now consider vector boson stars, known as Proca stars. They are the horizonless and regular everywhere solutions of the (complex)-Einstein-Proca theory, with the following action,

$$\mathcal{S} = \int d^4x\sqrt{-g}\left[\frac{R}{16\pi} - \frac{1}{4}F_{\mu\nu}\bar{F}^{\mu\nu} - \frac{1}{2}\mu_P^2 A_\mu \bar{A}^\mu\right] , \tag{7.42}$$

where $F_{\mu\nu} = \partial_\mu A_\nu - \partial_\nu A_\mu$ is the field strength written in terms of the 4-potential $A_\mu$. The bar over the field strength and 4-potential, $\bar{F}_{\mu\nu}$ and $\bar{A}_\mu$, corresponds to the complex conjugate, while $\mu_P$ is the vector field mass. These stars were first studied in [156], where both static and rotating numerical solutions where discussed, together with their physical properties and stability. The spinning solutions therein, however, were excited states. The fundamental spinning solutions were discussed in [155], [208].

The Proca potential *ansatz* is,

$$A = \left(iVdt + \frac{H_1}{r}dr + H_2 d\theta + iH_3 \sin\theta d\varphi\right) e^{i(m\varphi-\omega t)} , \tag{7.43}$$

where $V, H_1, H_2$ and $H_3$ are functions that only depend on the $(r,\theta)$ coordinates, and $m \in \mathbb{Z}^+$ is the usual azimuthal harmonic index. Given that there is an infinite number of families of Proca stars with different values of $m$, in this chapter we will only consider the family of $m = 1$ Proca stars (and also nodeless in the radial direction, *i.e.* the fundamental solutions).

Since our bosonic field is now a vector, we can no longer use $\phi_{max}$ to label solutions. This is replaced by the maximal value of the $H_1$ function in the vector ansatz Equation 7.43. As for $\phi_{\max}$ for the previous stars, the $H_1$ function also increases monotonically moving from the dilute regime until the strong gravity regime along the domain of existence of Proca stars. Hence, each individual Proca solution has a different value of $H_1^{\max}$. Therefore, in Figure 7.9, the structure of TCOs and LRs is shown in a $H_1^{\max}$ *vs.* $r\mu_P$ plot (and similarly for the efficiency $\epsilon$).

Figure 7.9 exhibits clear differences between the vector and scalar stars. A notorious one is the possibility of having stable TCOs all the way to the center of the star. Thus, ISCO has $r = 0$. Then $\epsilon_{\rm ISCO}$ amounts to known the energy of the particle sitting at $r = 0$. We see that such $\epsilon_{\rm ISCO}$ goes from zero until close to unity as more compact stars are considered, similarly to the scalar stars.

Concerning $\epsilon_{\rm MSCO}$, the sense of rotation plays a role. For prograde orbits, there is no region of unstable TCOs. Thus $\epsilon_{\rm ISCO} = \epsilon_{\rm MSCO}$. For retrograde orbits, however, unstable TCOs can appear. In fact, there can even be several disconnected regions of such orbits. Thus $\epsilon_{\rm MSCO}$ presents more than one discontinuity. The first discontinuity appears when $\epsilon_{\rm ISCO} \sim 40\%$, and $\epsilon_{\rm MSCO} \sim 21\%$. Then, $\epsilon_{\rm MSCO}$ increases slightly up to $\epsilon_{\rm MSCO} \sim 23\%$ where the second discontinuity appears, dropping further to $\epsilon_{\rm MSCO} \sim 4\%$. For the remaining solutions on the strong gravity regime, it decreases slowly down to $\epsilon_{\rm MSCO} \sim 3.8\%$.

### 7.4.2 BHs

Now we consider (non-Kerr) BH examples. Our illustrations, again, are numerical. The line element considered in this chapter, which is common for the two families of BHs discussed below, is the same used in previous chapters were we obtained several different hairy BHs – Equation 1.19.

For the case of BHs, we choose to show the efficiency as a function of the dimensionless spin, $j = J/M^2$ in the two plots for each family of solutions. The left (right) plot corresponds to prograde orbits (retrograde orbits).

#### *7.4.2.1 BHs with Synchronised Axionic Hair*

To illustrate the structure of TCOs and LRs for a non-Kerr family of BHs that can exhibit large phenomenological deviations from Kerr we consider BHs with synchronised hair. We will consider the axionic model [79], which contains in a particular limit the free scalar field model [37], [86]. Some results for the latter, concerning the efficiency, were recently presented in [65].

As seen and discussed in chapter 5, KBHsASH are stationary, axisymmetric, regular everywhere on and outside the event horizon, asymptotically flat solutions of the (complex-)Einstein-Klein-Gordon theory – *cf.* Equation 4.1. They can be considered as the natural BH generalisation of the axion boson stars studied previously, thus the self-interaction potential $V(\phi)$ follows the QCD axion potential, similar as for stars case – *cf.* Equation 4.7. In this chapter we will only consider the BH generalisation of the axion boson stars with $f_a = 0.05$.

BHs with synchronised axionic hair are composed of a BH horizon surrounded by an axionic scalar field whose angular frequency is synchronised with the angular rotation of the horizon of the BHs. Such synchronisation can be written as $\omega = m\Omega_H$, where $\Omega_H$ is the angular velocity of the horizon. If this synchronisation is not met, the scalar field cannot be in equilibrium with the BH.

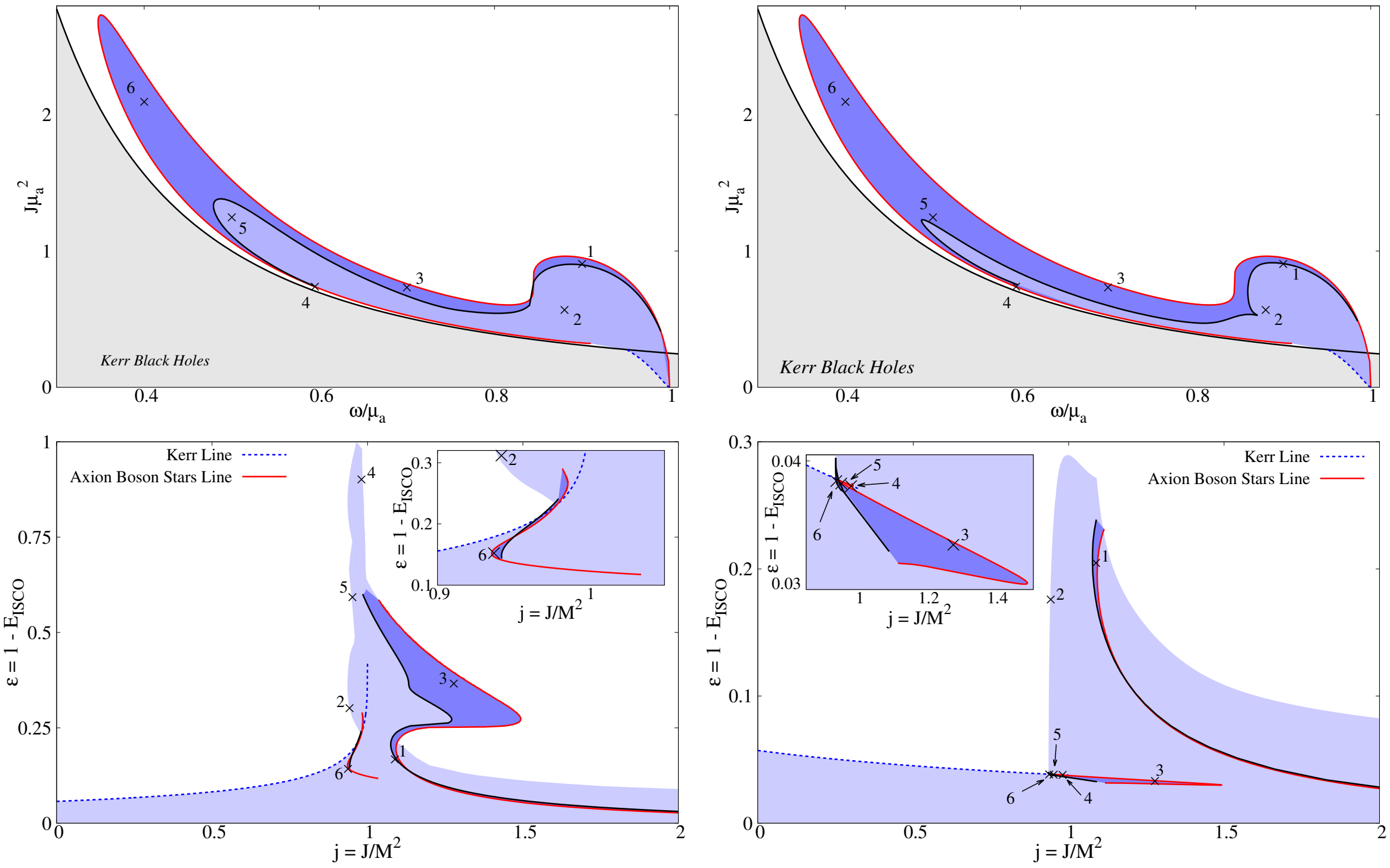

**Figure 7.10:** Domain of existence in an angular momentum $J\mu_a^2$ *vs.* angular frequency of the scalar field $\omega/\mu_a$ plane (*top row*) and efficiency as a function of the dimensionless spin $j = J/M^2$ (*bottom row*) of BHs with synchronised axionic hair with $f_a = 0.05$. Prograde (retrograde) orbits are presented in the *left column* (*right column*). The light blue regions correspond to solutions in which the efficiency is $\epsilon_{\rm ISCO}$; the dark blue regions correspond to solutions for which the efficiency is $\epsilon_{\rm MSCO}$. The inset in both efficiency plots sheds more light onto the regions that are more challenging to analyse on the main plots. Six solutions are highlighted.

This new family of BHs was obtained and discussed in chapter 5 and [79]. For solutions with a small amount of hair, the structure is similar to the Kerr one – *cf.* see the region of lower values of $\phi_{\rm max}$ in Figure 5.6 and Figure 5.7. There is only one unstable LR. Between the event horizon and the LR, there are no TCOs. Between the LR and the ISCO, there are unstable TCOs; and above the ISCO, there are stable TCOs. Thus, for these solutions with a small amount of hair, $\epsilon_{\rm ISCO} = \epsilon_{\rm MSCO}$. The results for solutions in this class (Kerr-like) are represented in light blue for all plots in Figure 7.10. On the other hand, very "hairy" solutions present a more convoluted structure of TCOs – *cf.* see the region of higher values of $\phi_{\rm max}$ in Figure 5.6 and Figure 5.7. There are new disconnected regions of unstable TCOs and forbidden for TCOs, where the latter appear when the scalar field is compact enough to develop extra LRs. Hence, we can have $\epsilon_{\rm ISCO} \neq \epsilon_{\rm MSCO}$. The results for solutions in this class (non-Kerr like) are represented in dark blue color for all plots in Figure 7.10.

In Figure 7.10 we show $\epsilon_{\rm ISCO}$ for BHs with synchronised axionic hair for prograde (*left*) and retrograde (*right*) orbits. Both panels also include an inset plot showing the domain of existence of these BHs in an angular momentum $J\mu_a^2$ *vs.* $\omega/\mu_a$ diagram. Both in the main panels and insets, there are two additional lines. The first one corresponds to no horizon limit: the set of axion boson stars with $f_a = 0.05$. This (red solid) line is known as the *Axion Boson Stars line*. The second (blue dashed) line corresponds to the no hair limit or Kerr limit of the hairy BHs - the *Kerr line*. We have highlighted six particular solutions, numbered 1 to 6, to allow an easier mapping between the domain of existence and the efficiency plot.

Figure 7.10 (left panel) shows there are prograde efficiencies $\epsilon_{\rm ISCO}$ arbitrarily close to the unity, exceeding greatly the maximal efficiency of $\sim 42\%$ (of the Kerr limit). The solutions with the largest efficiencies correspond to solutions in the strong gravity regime, where the ISCO occurs for smaller radii, leading to larger $\epsilon_{\rm ISCO}$. For non-Kerr like solutions – dark blue region – $\epsilon_{\rm MSCO}$ can be as high as $\sim 60\%$, again, larger than the maximal efficiency for Kerr BHs. For very non-Kerr like solutions, $\epsilon_{\rm MSCO}$ drops to around $\sim 20\%$ because a new region of unstable TCOs develops, pushing MSCO outwards.

Figure 7.10 (right panel) addresses the retrograde case. Efficiencies are rather smaller than in the prograde one, since both ISCO and MSCO occur at larger radii. Additionally, a new disconnected regions of (no or unstable) TCOs develop for solutions with far less hair than for the previous case, pushing MSCO outwards. Nevertheless, retrograde efficiencies of $\epsilon_{\rm ISCO} \sim 30\%$ are possible, far larger than those for the retrograde case in Kerr BHs.

#### 7.4.2.2 *Einstein-scalar-Gauss-Bonnet BHs*

The goal of our final example is to stress that, in fact, many models of non-Kerr BHs have small phenomenological differences with respect to the Kerr model. In particular this applies to the efficiencies we have been discussing. We will discuss Einstein-scalar-Gauss-Bonnet BHs, obtained in chapter 6, which are asymptotically flat, regular everywhere outside and at the event horizon, axisymmetric and stationary solutions of the Horndeski shift-symmetric theory. This is a scalar-tensor theory, within the generic class of Einstein-scalar-Gauss-Bonnet models given by Equation 6.2 with $f(\phi) = \phi$.

BHs solution within this theory were first obtained by Sotiriou and Zhou [27], [28]. They obtained analytically, static perturbative solutions (small values of $\alpha$), as well as numerical static solutions (large values of $\alpha$). A similar work for the spinning generalisation of these solutions was reported in chapter 6 and in [31].

The structure of TCOs (not shown here) is always Kerr-like. Thus, similarly to the solutions with small amounts of hair in the previous family of hairy BHs, $\epsilon_{\rm ISCO} = \epsilon_{\rm MSCO}$.

In Figure 7.11 we show the efficiency for Einstein-scalar-Gauss-Bonnet BHs for prograde (*left*) and retrograde (*right*) orbits. We also include insets showing the domain of existence of these BHs in a dimensionless spin, $j = J/M^2$ *vs* $\alpha/M^2$ plot. Four additional lines are exhibited, which are the same reported in Figure 6.3. The first (green dotted) line is known as the *critical line* and corresponds to the limit beyond which the (repulsive) Gauss-Bonnet term prevents the existence of a horizon. The second (black dashed) line corresponds to the set of extremal hairy solutions with a vanishing Hawking temperature - the *extremal line*. The third (blue dashed) line corresponds to non-rotating BHs - the *static line*. Finally, a fourth (solid red) line corresponds to Kerr BHs, in which $\alpha/M^2 = 0$ - the *Kerr line*. Five particular solutions, numbered 1 to 5, are also highlighted, to map their location in the domain of existence and in the efficiency plot.

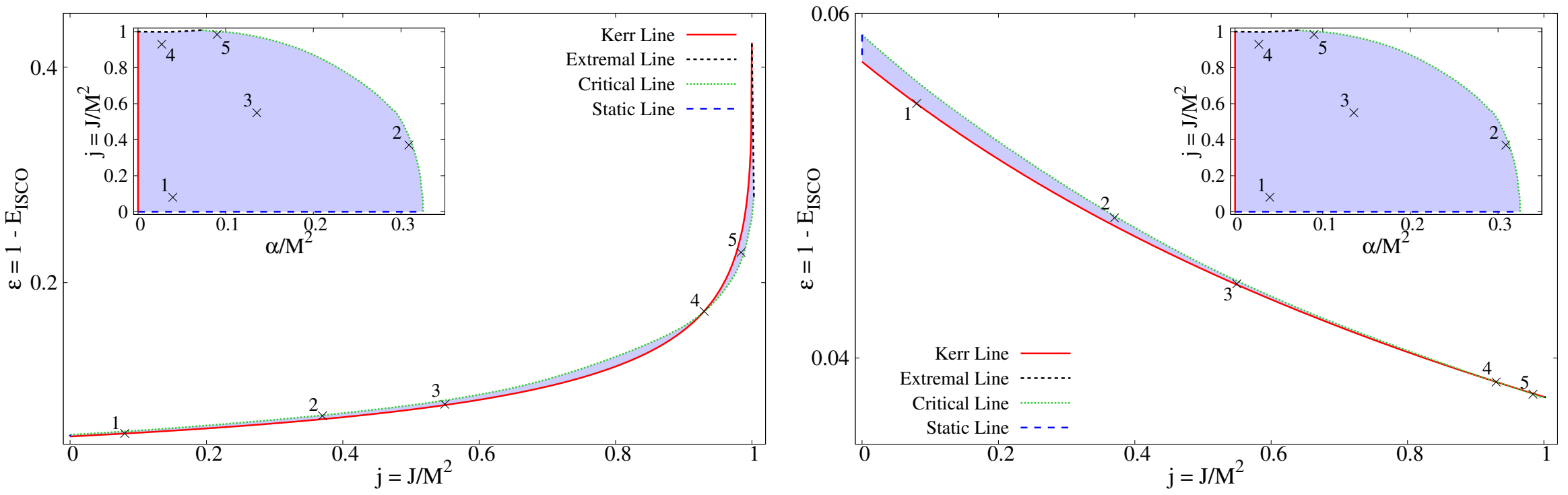

**Figure 7.11:** Efficiency as a function of the dimensionless spin $j = J/M^2$ for Einstein-scalar-Gauss-Bonnet BHs. The *left plot* (*right plot*) exhibits prograde (retrograde) orbits. The insets exhibit the domain of existence of Einstein-scalar-Gauss-Bonnet BHs in a dimensionless spin $j = J/M^2$ *vs.* $\alpha/M^2$ plane. Five solutions are highlighted.

Figure 7.11 (left panel) shows that the efficiency of Einstein-scalar-Gauss-Bonnet BHs is very similar to the efficiency of Kerr BHs, for the same dimensionless spin, $j$. The largest difference is (only) around $\sim 4\%$. For small $j$, the Einstein-scalar-Gauss-Bonnet BHs have a larger efficiency; but, for sufficiently large $j$, the reverse happens. This sort of transition was already discussed in the previous chapter and in [31] (albeit not for the efficiency). The right panel in Figure 7.11 exhibits a similar picture for retrograde orbits. The largest difference is now (only) around $\sim 3\%$ and occurs in the static limit, $j \to 0$. Increasing the spin, this difference monotonically decreases. For large spins, there is almost no difference between hairy and Kerr BHs in terms of efficiency. This result is consistent with the discussion in the previous chapter and in [31].

## 7.5 Discussion and final remarks

In this chapter, we have shown that for stationary, axisymmetric and asymptotically flat compact objects with a $\mathbb{Z}_2$ symmetry, the existence of equatorial LRs leads to a specific structure for equatorial TCOs, independently of the direction of rotation. Such structure is entirely determined by the stability of the LR: for an unstable LR, the region radially immediately above (below) the LR has unstable TCOs (no TCOs) – *cf.* Figure 7.2 (top panel); for a stable LR, the region radially immediately above (below) the LR has no TCOs (has stable TCOs) – *cf.* Figure 7.2 (bottom panel).

As a corollary of this result, for a horizonless object that possesses one unstable LR and another stable LR at a smaller radius than the first, for either sense of rotation, the region between the LRs has no TCOs – *cf.* Figure 7.3 (top panel). Radially immediately above (below) the unstable (stable) LR, there are unstable (stable) TCOs. This implies that it is possible to have stable TCOs closer to the object itself than the LR; thus, a potential ISCO may occur at a smaller radius than the LR. However, one needs to clarify if the motion on such region is perturbatively stable in a direction perpendicular to the equatorial plane.

As another corollary, for asymptotically flat equilibrium BHs, which generically have an unstable LR for either rotation sense [251], the region between the event horizon and the unstable LR contains no TCOs – *cf.* Figure 7.3 (bottom panel). Since the LR is unstable, the region radially immediately above has unstable TCOs; thus, for a BH, the ISCO will always occur at a larger radius than this unstable LR.

In the second part of this chapter, we have studied the efficiency associated to the process of converting gravitational energy into radiation by a material particle falling under an adiabatic sequence of TCOs, for several stars and BHs, namely, three different families of bosonic scalar stars (mini, gauged and axion boson stars), one family of bosonic vector (Proca) stars and two different families of hairy BHs (BHs with synchronised axionic hair and Einstein-scalar-Gauss-Bonnet BHs).

Regarding the several families of bosonic scalar stars, we found that the structure of TCOs is quite similar between them. Moreover, their structure is also similar to that found for some naked singularities – see Refs. [259]–[262]. The efficiency $\epsilon_{\rm ISCO}$ computed at the ISCO, can grow arbitrarily close to unity, both for prograde and retrograde orbits. Also, the efficiency $\epsilon_{\rm MSCO}$, at the MSCO, has the largest values for both stars without self-interactions (mini and gauged boson stars) and prograde orbits.

The family of bosonic vector stars presents a structure of TCOs quite different from their bosonic scalar cousins. For prograde orbits, stable TCOs can exist all the way until $r = 0$. Thus $\epsilon_{\rm ISCO}$ is computed at the origin and it increases monotonically towards values close to 100% for stars in the strong gravity regime. For retrograde orbits, more compact stars develop regions of unstable and no TCOs; thus, the efficiency $\epsilon_{\rm MSCO}$ drops to small values, around $\sim 4\%$.

For BHs with synchronised axionic hair, we found that new disconnected regions of unstable and no TCOs (beside the ones that exist already for Kerr BHs) develop. Thus, the efficiency $\epsilon_{\rm MSCO}$ can drop; nevertheless, it is possible to have solutions in which this efficiency for prograde orbits is much larger than the one for Kerr BHs and even close to the unity. In the case of retrograde orbits, the efficiency can not be as high, but can, nonetheless, be higher than that for the (retrograde) Kerr case.

Finally, concerning the family of Einstein-scalar-Gauss-Bonnet BHs we found that the higher-order correction to Einstein's gravity which arise from the linear coupling between the Gauss-Bonnet term and the scalar field has no strong influence on the efficiency. For prograde orbits, the efficiency is only slightly larger (smaller) than that of Kerr BHs for the same $j$, when $j$ is small (large). For retrograde orbits, the efficiency of Einstein-scalar-Gauss-Bonnet BHs is larger than their Kerr counterpart, but the difference decreases almost to zero as $j$ increases.

CHAPTER 8

# Closing Remarks

In this thesis, we have obtained and studied several properties of various ultracompact objects, including both BHs and stars, in different gravitational models. Such models were constructed by simply adding fields to canonical GR or considering higher curvature corrections to GR. For the former, we add a massive and complex scalar field to GR to avoid Bekenstein's theorem, with different self-interacting potentials. In particular, we obtained BHs with higher azimuthal harmonic index in the case of a massive and complex scalar field without any self-interactions – chapter 3 –, as well as stars and BHs with a massive and complex scalar field with an axionic-like self-interaction potential – chapter 4 and chapter 5. For the latter, we consider the addition of the Gauss-Bonnet (GB) quadratic curvature invariant. Such a term by itself does not influence the theory in four-dimensional gravity, but by coupling it non-minimally to a real scalar field, the GB term becomes dynamical. In chapter 6, we consider a linear coupling between the scalar field and the GB term, giving rise to BHs with a non-trivial scalar field and geometry with higher curvature corrections.

In all cases mentioned so far, and in a large set of different models, ultracompact objects can only be obtained numerically. The numerical procedure for each one was fairly similar, boiling down to defining an ansatz for the metric and fields and the appropriate boundary conditions on the corresponding limits, writing a suitable combination of the equations of motion, and discretising them on a finite 2D grid with $N_r \times N_\theta$ points, where $N_r$ and $N_\theta$ are the number of discrete points on the radial $r$ and angular $\theta$ directions. To solve the equations of motion, we used the professional package FIDISOL/CADSOL which uses a backwards finite difference method together with the Newton-Raphson method – Appendix A. This solver provides the user with numerical error estimates, and, for all solutions studied in this thesis, the maximal numerical error found was of the order of $10^{-3}$. Different numerical tests can also be performed using well-known physical relations, such as, *e.g.* the Smarr relation. Using such physical tests yield error estimates of the same order.

A systematic construction of several tens of thousand of numerical solutions for each model studied in this thesis was performed. With such large catalogue of solutions, we interpolated them into the continuum to analyse firstly the domain where one can find them and secondly their physical properties.

After performing the above analysis, it is possible to find that some solutions may have very different properties compared to a Kerr BH, but others do not deviate much from Kerr. A prime example of the former is the family of Kerr BHs with axionic scalar hair – *cf.* chapter 5. Solutions of this family possess a large variety of properties that can be quite different from the ones found for Kerr. A shortlist of those properties are: the dimensionless spin, $j = J/M^2$, can be larger than unity, violating the Kerr bound, $j \leq 1$; the existence of ergo-Saturn, composed of an ergo-sphere together with an ergo-torus; a more complex structure of circular orbits on the equatorial plane, with the presence of more LRs as well as several new regions of unstable timelike circular orbits. However, the properties in question can only be observed for solutions that comprise a large quantity of scalar hair. For almost bald solutions (almost no scalar hair), none of the aforementioned properties are observed and they are essentially equal to Kerr BHs. For the latter, a good example are the spinning BHs in shift-symmetric Horndeski theory – *cf.* chapter 6. In this family, almost all BHs have the same properties as the Kerr BHs. Only a tiny number of hairy BHs can violate the Kerr bound, but such violation is only by a few percentage ($\sim 1.3\%$). Furthermore, despite the existence of hairy BHs that can violate the Kerr bound, for all properties studied, they all have similar properties to the ones found for Kerr BHs.

In addition to the analysis carried out on the acquisition and study of the aforementioned numerical solutions, we also study some phenomenology properties of those numerical solutions and other solutions in different theories reported by various authors. In particular, we investigated the horizon geometry, where we analysed the possibility of embedding the event horizon of a given BH in Euclidean 3-space $\mathbb{E}^3$ and the relation between the sphericity and linear velocity of the horizon with its embedding – chapter 2. We also studied the structure of equatorial timelike circular orbits around several BHs and stars and the efficiency associated with the energy conversion of the rest mass into radiation of a timelike particle as it falls from infinity until it reaches the innermost stable circular orbit – chapter 7.

From these phenomenology studies, we saw that analogously to what was discussed in previous paragraphs, some families of ultracompact objects can have large phenomenology deviations from the one known for Kerr BHs, while others do not. As a short example, we saw that all families of ultracompact objects that possess a massive complex scalar field (and even with a massive complex vector field) could have much larger efficiencies than the maximal efficiency found for an extreme Kerr BH. Whereas the family of spinning BHs in shift-symmetric Horndeski theory, their efficiencies only differ a few percent frwe have a region in which no (timelike, null, or spacelike) circular orbits existom the value found for Kerr BHs and do not exceed the maximal efficiency for an extremal Kerr BH.

With the numerical solutions reported in this thesis, one can further explore their phenomenology by, *e.g.*, performing numerical time evolutions to study the stability of the solutions and to perform collision to analyse the gravitational waves generated during such a process. Those gravitational waves can be compared to the experimental data already reported by the LIGO/Virgo/KAGRA collaboration to determine if the numerical solutions could generate the measured data and, if not, to impose restrictions on the theoretical model from which the solutions were obtained. Similarly, one can also perform simulations to obtain the shadow and the lensing of light around the solutions to compare it to the data reported by the EHT collaboration.

An extensive catalogue of numerical solutions can also open the door to the use of Deep Learning. Such an approach may help us obtain solutions in regions where the numerical package FIDISOL/CADSOL can not find solutions with reasonable accuracies, such as the region where the BS line and the extremal line spiral in the case of KBHsSH – *cf.* Figure 1.1. A different exciting application of Deep Learning may be its use to obtain numerical solutions as a numerical solver. Such an approach has the advantage of constructing a region of the domain of existence in one go. However, the solutions obtained may not have as good accuracy as those obtained by the numerical package mentioned above. Nevertheless, a numerical approach using both Deep Learning and FIDISOL/CADSOL may help us construct an even more extensive catalogue of solutions with less time and effort.

The following years shall bring new experimental data from several collaborations that will help answer the Kerr hypothesis. With all the experimental data that we have access to now (and that we will have in the future), we live in the perfect time to construct systematically large catalogues of spinning BHs and compact horizonless objects in several different models within or beyond GR and to study their physical properties. This will let us analyse their phenomenology, compare it to the experimental data, and march towards the true answer for the Kerr hypothesis.

# APPENDIX A

# Numerical Techniques: the FIDISOL/CADSOL package

One of the biggest problems when studying black holes solutions in different theories of gravity (by either adding fields to Einstein's gravity or considering high order corrections to the curvature) is finding a consistent way to obtain the solutions. Such problems arise from the fact that Einstein's equations are, in general, a set of coupled nonlinear second-order partial differential equations (PDEs), which makes extremely hard the quest of finding analytic solutions (*i.e* in closed form) such as the ones found, *e.g.*, in pure General Relativity – the Schwarzschild and Kerr solutions. In such cases, the task of finding an analytical solution is simplified by the vanishing of the energy-momentum tensor and the presence of several symmetries. However, in more complex cases, this task becomes increasingly difficult. As such, one may take the pragmatic approach of seeking numerical solutions, when the analytic approach fails.

Within the large set of numerical techniques that exist in the literature that can solve a variety of different classes of PDEs, we are interested in numerical techniques that provide us with stationary solutions from a given theory of gravity. Hence, we need a numerical approach that can solve a set of coupled *elliptic* PDEs. Such numerical approach is provided by the professional package FIDISOL/CADSOL (FInite DIfference SOLver/Cartesian Arbitrary Domain SOLver)[44]–[46]. This software, written in Fortran 90, was first developed by Willi Schönauer and Eric Schnepf in 1987 and further improved some years later, and it solves nonlinear systems of two and three dimensional elliptic and parabolic PDEs, subjected to arbitrary boundary conditions on a rectangular domain (or on any domain that can be analytically transformed to a rectangular domain). It implements a backwards finite difference method [263] together with a root finding method – the Newton-Raphson method [264] –, with self-adaptative grid and consistency order. It can also provide error estimates for the computed solution, allowing the user to judge on its quality.

The fact that this solver uses a Newton-Raphson method implies that it finds the roots of the equation(s). For the type of problems solved in this thesis, the differential equation(s) are written in the generic form

$$P(x, y; u; u_x, u_y; u_{xy}, u_{xx}, u_{yy}) = 0 \,, \tag{A.1}$$

where $u$ is the set of functions we want to compute and $u_x$, $u_y$ and $u_{xy}, u_{xx}, u_{yy}$ are the first, and, respectively, the second derivatives of the function(s) $u$ with respect to the generic coordinates $x$ and $y$ (note that for the problems we solve, $x, y$ are basically the spherical coordinates $r, \theta$). Also, we have to compute the Jacobian for all the functions present in the equations, which is found by simple differentiation of each equation $P$ with respect to $u; u_x, u_y; u_{xy}$ and $u_{xx}, u_{yy}$. One has also to supply an initial guess for each function $u$, together with the boundary conditions for $u$, and also a given mesh in $x, y$ with $N_x \times N_y$ points. If one already has all these requirements, one can implement the corresponding problem in the solver.

The numerical procedure works as follows [44]–[46] (note that the approach here is generic for the Newton-Raphson method): for an approximate initial solution $u^{(1)}$, $P(u^{(1)})$ does not vanish. In the next step one considers an improved solution

$$u^{(2)} = u^{(1)} + s\Delta u, \tag{A.2}$$

supposing that $P(u^{(1)} + s\Delta u) = 0$ (with $s$ a relaxation factor, which is usually chosen as $s = 1$). The expansion in the small parameter $\Delta u$ gives to first order

$$0 = P(u^{(1)} + \Delta u) \approx P(u^{(1)}) + \frac{\partial P}{\partial u}(u^{(1)})\Delta u \ + \ldots \,. \tag{A.3}$$

This algebraic equation is used to determine the correction $\Delta u^{(1)} = \Delta u$. Then one repeats the calculation iteratively ($e.g.$ $u^{(3)} = u^{(2)} + \Delta u$), such that the approximate solutions will converge (provided the initial guess was close enough to the true solution).

In each step, a linear system of algebraic equations is solved, and the residual $||P(u^{(i)})||$ decreases by a factor of approximately 10. As a generic feature of this approach, the iteration stops after $N$ steps, when the Newton residual $P(u^{(N)})$ is smaller than a prescribed tolerance (which is an input parameter). Clearly, it is essential to have a good first guess, when starting the iteration procedure. Such dependence of the convergence on the initial guess is the major disadvantage of this method (note that without a good enough initial guess, the solver will not converge to the true solution, giving us some random wrong answer). However, if we have a good understanding of the problem we plan to solve, we can find a reasonable initial guess for the functions, overcoming such a disadvantage.

The package FIDISOL/CADSOL provides also error estimates for each function, which provides us with a criteria to judge the quality of the computed solution. The errors are computed by the solver, on the "consistency level", namely, the discretized Newton residual, and as discretization error terms in $x, y, z$ directions. The discretization error is estimated through the difference of difference quotients. For example, in Equation A.3, the derivative of the solution $u$ and of the correction function $\Delta u$ are discretized by a difference method with arbitrary consistency orders. Also, the derivatives are replaced, for example, in the form $u_{xx} \Leftarrow u_{xx,d} + d_{xx}$, $\Delta u_{xx} \Leftarrow \Delta u_{xx,d}$ (where the index $d$ means "$discretized$"). In this relation $d_{xx}$ denotes the estimate for the discretization (or truncation) error of $u_{xx}$, defined as $d_{xx} = u_{xx,d,next} - u_{xx,d}$; the index "$next$" stays for the next higher member of the family of backward difference formulas. For convergent problems, the discretized Newton residual decreases with the number of Newton-Raphson iterations. Also, the discretization error terms depend on the grid size and the used consistency order, $i.e.$ on the order of the discretisation of derivatives. Note that for the numerical results reported in this thesis, the order of the discretisation was typically six.

Apart from the error estimates provided by FIDISOL/CADSOL, we use other specific criteria to judge on the quality of the numerical results, as provided by the problem one solves ($e.g.$ the Smarr law and/or the 1st law of thermodynamics when dealing with new classes of BH solutions). Examples of such tests are discussed throughout the thesis.

Overall, this package is an excellent tool for the kind of problems when one wants to find stationary solutions within a given theory of gravity, yielding good quality solutions and corresponding error estimates within a reasonable amount of computational work.

To give a better understanding of how this package works, we shall numerically reconstruction the Kerr solution. As the latter is known analytically, one can compare the numerical and closed form solution to gain insight on the overall quality of the method.

## A.1 Kerr Solution

A good way to get a better intuition and understanding of the solver is by using it to reconstruct known solutions, such as the Kerr one. In order to obtain the Kerr solution, the first step is to define an *ansatz* for the metric, in which we introduce several *ansatz* functions. With this in mind, we take the following *ansatz* for the metric (which is the same used to obtain numerical BHs in the different theory mentioned throughout this thesis),

$$ds^2 = -e^{2F_0} N dt^2 + e^{2F_1}\left(\frac{dr^2}{N} + r^2 d\theta^2\right) + e^{2F_2} r^2 \sin^2\theta (d\varphi - W dt)^2 \,, \tag{A.4}$$

where $\{F_i, W\}_{i=0,1,2}$ are functions that only depend on $r$ and $\theta$, and $N \equiv 1 - r_H/r$, in which $r_H$ is the event horizon radius. This form for the Kerr metric corresponds to a non-standard coordinate system that one does not find in a normal textbook. However, the coordinate transformation to the standard BL coordinates is quite simple, corresponding solely to a shift of the radial coordinate, $r = R - a^2/R_H$, where $R$ and $R_H = M + \sqrt{M^2 - a^2}$ are the radial coordinate and horizon radius in BL coordinates, respectively, and $a \equiv J/M$ is the reduced angular momentum. In this non-standard coordinate system, the *ansatz* functions can be written as [86],

$$e^{2F_1} = \left(1 - \frac{c_t}{r}\right)^2 + c_t(c_t - r_H)\frac{\cos^2\theta}{r^2} \tag{A.5}$$

$$e^{2F_2} = e^{-2F_1}\left\{\left[\left(1 - \frac{c_t}{r}\right)^2 + \frac{c_t(c_t - r_H)}{r^2}\right]^2 + c_t(r_H - c_t)\left(1 - \frac{r_H}{r}\right)\frac{\sin^2\theta}{r^2}\right\} \tag{A.6}$$

$$F_0 = -F_2\ , \quad W = \frac{e^{-2(F_1+F_2)}}{r^3}\sqrt{c_t(c_t - r_H)}(r_H - 2c_t)\left(1 - \frac{c_t}{r}\right) \tag{A.7}$$

where $c_t < 0$ is a constant that does not have a very transparent meaning. However, it is related to the non-staticity of the solution, since $c_t = 0$ correspond to the Schwarzschild metric. With this functions, we can compute several physical quantities of interest, such as the ADM mass, $M$, and angular momentum $J$; the horizon area, $A_H$, temperature, $T_H$, and velocity, $\Omega_H$, [86]

$$M = \tfrac{1}{2}(r_H - 2c_t)\ , \quad J = \tfrac{1}{2}\sqrt{c_t(c_t - r_H)}(r_H - 2c_t)\ , \tag{A.8}$$

$$A_H = 4\pi(r_H - c_t)(r_H - 2c_t)\ , \ T_H = \tfrac{r_H}{4\pi(r_H - c_t)(r_H - 2c_t)}\ , \ \Omega_H = \tfrac{\sqrt{c_t(c_t - r_H)}}{(r_H - c_t)(r_H - 2c_t)} \tag{A.9}$$

Now that we have chosen the metric *ansatz*, we have completed the first step. The second step resides on the imposition of the boundary conditions for all the metric functions. Since the solutions we want to obtain are regular outside of the horizon, stationary, axisymmetric and asymptotically flat, the boundary conditions will be imposed according to these requirements. In this way we impose the following boundary conditions [86],

(i) *Asymptotic boundary conditions*. Asymptotically flatness implies that the solution must approach a Minkowski spacetime at spatial infinity, $r \longrightarrow \infty$, thus,

$$\lim_{r\longrightarrow\infty} F_i = \lim_{r\longrightarrow\infty} W = 0\ . \tag{A.10}$$

where $i = 0, 1, 2$.

(ii) *Axis boundary conditions*. At the poles, *i.e.* at $\theta = 0, \pi$, axial symmetry and regularity imply,

$$\partial_\theta F_i = \partial_\theta W = 0\ . \tag{A.11}$$

Moreover, the absence of conical singularities implies also that $F_1 = F_2$ on the symmetry axis, however, this condition is not impose as a boundary condition but is used as a test for the numerics. Also, due to the symmetry with the equatorial plane we only have to compute the several *ansatz* function on one of the hemisphere and further impose the following boundary condition on the equatorial plane,

$$\partial_\theta F_i|_{\theta=\pi/2} = \partial_\theta W|_{\theta=\pi/2} = 0\ . \tag{A.12}$$

(iii) *Event horizon boundary conditions*. A coordinate transformation can be made in the form, $x = \sqrt{r^2 - r_H^2}$, to simplify the numerical treatment, and with such new coordinate one can compute a power series expansion near the horizon, $x \longrightarrow 0$, of the metric functions,

$$F_i = F_i^{(0)}(\theta) + x^2 F_i^{(2)}(\theta) + \mathcal{O}(x^4)\ , \tag{A.13}$$

$$W = \Omega_H + \mathcal{O}(x^2)\ . \tag{A.14}$$

With this result it is natural to impose the following boundary condition,

$$\partial_x F_i|_{r=r_H} = 0 \ , \qquad W|_{r=r_H} = \Omega_H \ . \tag{A.15}$$

The constant $\Omega_H$ can be immediately interpreted as the angular velocity of the horizon.

The next step is to impose the vacuum Einstein equations,

$$R_{\mu\nu} - \frac{1}{2} g_{\mu\nu} R = 0 \tag{A.16}$$

to get the differential equations for the metric functions. Since the Einstein equations are non-linear and really tricky to compute by hand, we used the computational software MATHEMATICA to help us obtaining such equations. They are quite involved and not particularly enlightening. For our purposes, it is useful to organize these equations so that each equation has second derivatives of a single function. This is achieved by taking the following combinations of the Einstein equations: [86],

$$E_r^r + E_\theta^\theta - E_\varphi^\varphi - E_t^t = 0 \ , \tag{A.17}$$
$$E_r^r + E_\theta^\theta - E_\varphi^\varphi + E_t^t + 2W E_\varphi^t = 0 \ , \tag{A.18}$$
$$E_r^r + E_\theta^\theta + E_\varphi^\varphi - E_t^t - 2W E_\varphi^t = 0 \ , \tag{A.19}$$
$$E_\varphi^t = 0 \ , \tag{A.20}$$

where $E_\nu^\mu \equiv R_\nu^\mu - 1/2 \delta_\nu^\mu R = 0$ are the Einstein equations. Although these 4 combinations give us four equations with second derivatives of a single of the four metric functions, another two combinations can be made to obtain two "constraint" equations that can be used to test the numerical accuracy of the solutions. Such constraint combinations are [86],

$$E_r^r - E_\theta^\theta = 0 \ , \tag{A.21}$$
$$E_r^\theta = 0 \ . \tag{A.22}$$

With the equations of motion written, we introduce a new radial variable $\bar{x} = x/(1+x)$ with the intention of mapping the semi-infinite region $[0, \infty[$ to the finite region $[0, 1]$. With such transformation we can rewrite the equations of motion with the following substitutions [86],

$$x \, \partial_x \mathcal{F} \longrightarrow (1-\bar{x}) \partial_{\bar{x}} \mathcal{F} \ , \qquad x^2 \, \partial_{xx} \mathcal{F} \longrightarrow (1-\bar{x}^2) \partial_{\bar{x}\bar{x}} \mathcal{F} - 2(1-\bar{x}) \partial_{\bar{x}} \mathcal{F} \ , \tag{A.23}$$

where $\mathcal{F} = \{F_0, F_1, F_2, W\}$ are the metric functions and $x = \sqrt{r^2 - r_H^2}$, as before.

Finally, we implemented these equations, together with the boundary conditions, on the FIDISOL/CADSOL solver and used the BLAFIS supercomputer to run the solver. The solver gives us, as an output, discrete values of the metric functions on a chosen grid, composed by discrete points within integration region $0 \leq \bar{x} \leq 1$ and $0 \leq \theta \leq \pi/2$ (the values for $\pi/2 \leq \theta \leq \pi$ are obtained due to the reflexion symmetry along the equatorial plane). Typical grids use 250 points in the $x-$direction and 30 points in the $\theta-$one.

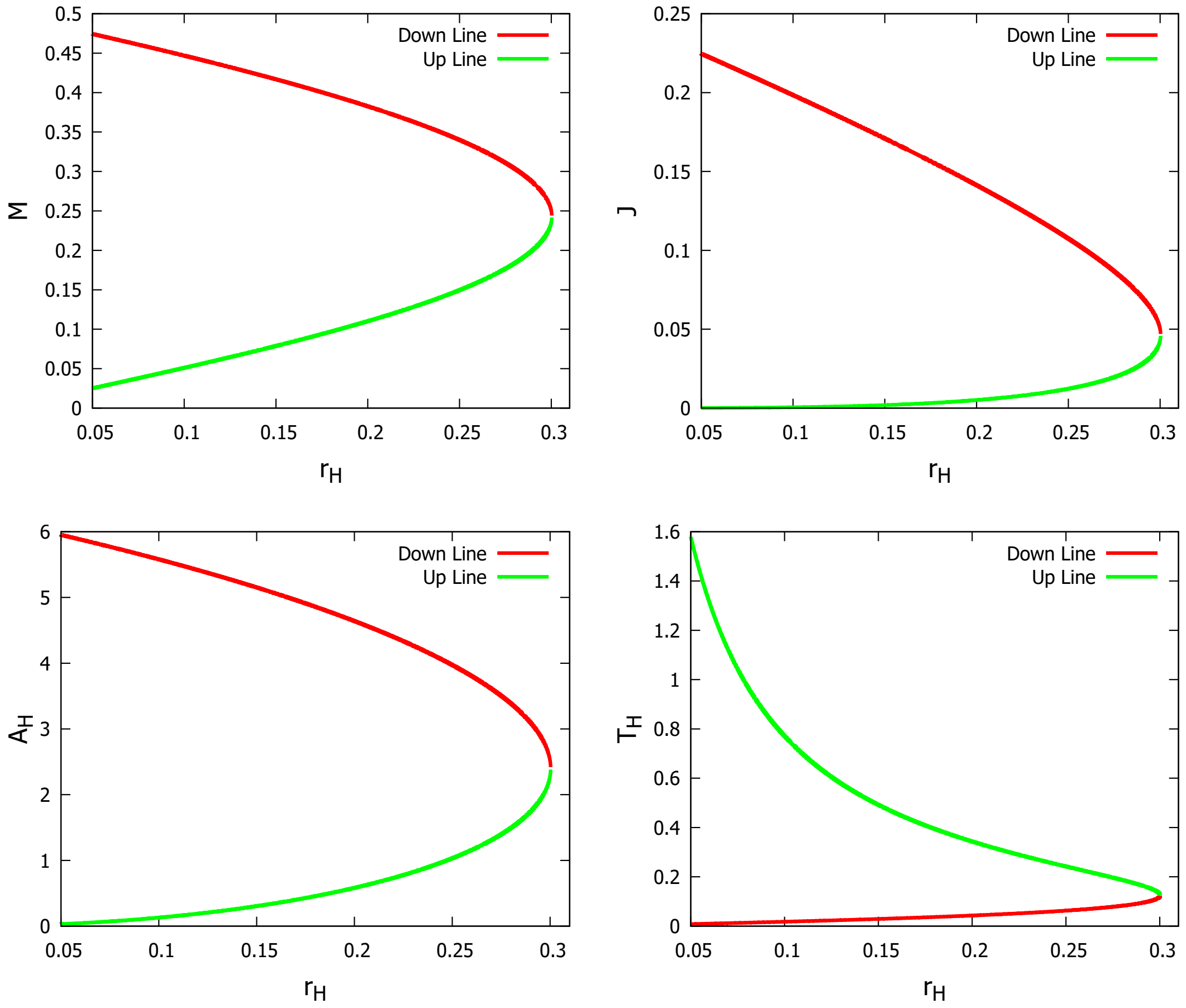

**Figure A.1:** Physical Quantities as a function of the event horizon radius, $r_H$, for a *theoretical* Kerr solution with a fixed value of the horizon angular velocity, $\Omega_H = 1$. One can observe the existence of two branches of solutions.

The best way to compare the analytically known Kerr BH metric, henceforth dubbed *theoretical* solution, with a numerical one, obtained by the above strategy, is to compare the physical quantities of both. In the case of the theoretical Kerr BH, the physical quantities were computed in Equation A.8 and, in Figure A.1, we present the plots of the theoretical lines of the physical quantities as a function of the event horizon radius, $r_H$, for a Kerr solution with $\Omega_H = 1$. In this figure we see the existence of two different branches. The first, down branch (green line) starts from a static Schwarzschild configuration (note that we can only approach the Schwarzschild configuration as a limit, since we fixed the horizon velocity) and evolves until a back bending point, where the second, up branch (red line) start. The latter then evolves until we reach the extremal Kerr solution.

On the other hand, the physical quantities of a numerical BH are encoded in terms of the metric functions, and can be obtained, either at the horizon or at infinity. Considering first the horizon quantities, we can introduce the Hawking temperature [265], $T_H = \kappa/(2\pi)$, where $\kappa$ is the surface gravity, and the entropy, $S = A_H/4$ [196], where,

$$T_H = \frac{1}{4\pi r_H} e^{(F_0 - F_1)|_{r_H}} \,, \qquad A_H = 2\pi r_H^2 \int_0^\pi d\theta \sin\theta \, e^{(F_1+F_2)|_{r_H}} \,. \tag{A.24}$$

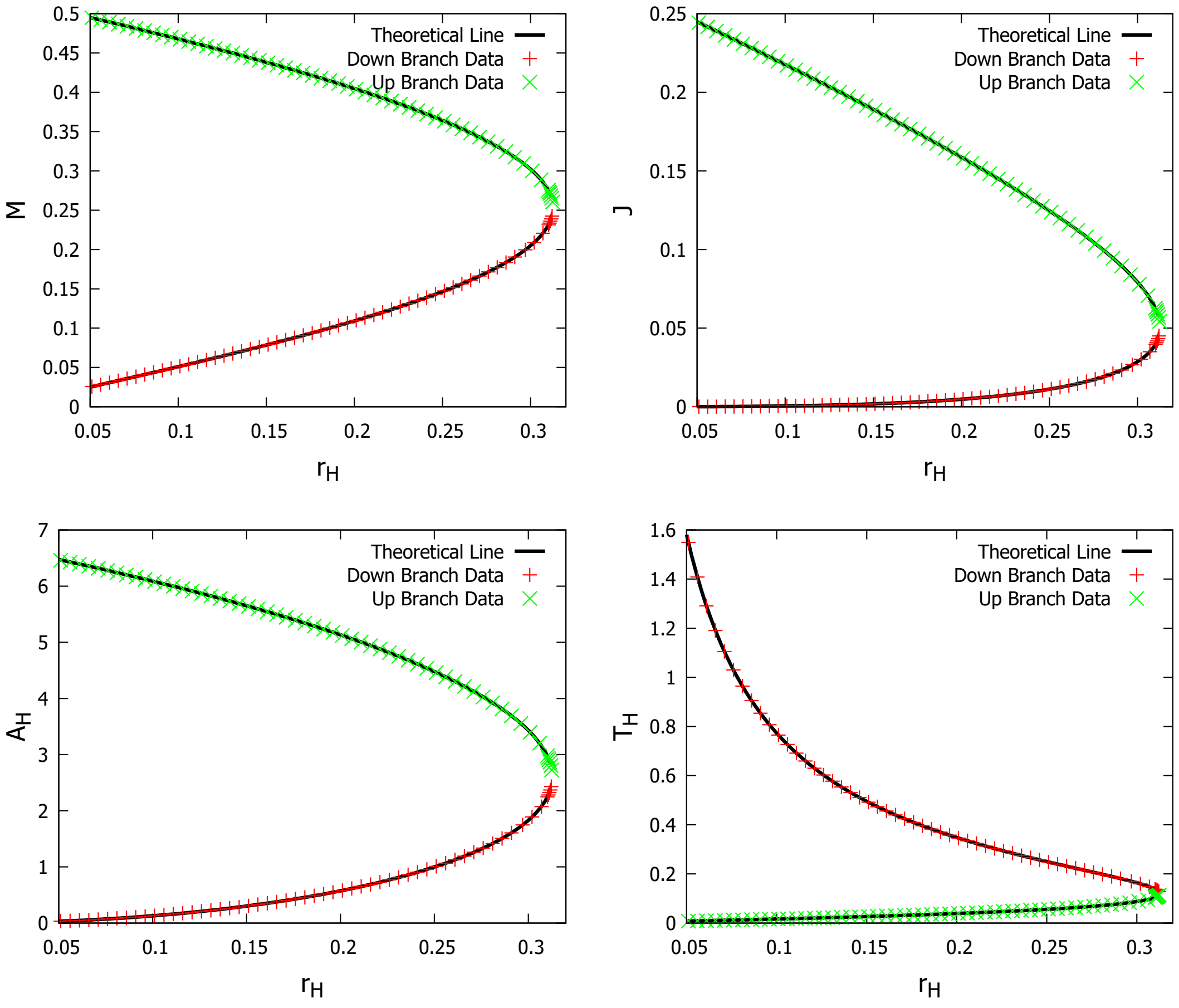

**Figure A.2:** The numerical data and theoretical line of the physical quantities associated to a Kerr solution with a fixed value of the horizon angular velocity, $\Omega_H = 0.96$. It is possible to see a very good agreement between the numerical and the theoretical data.

Also, the horizon angular velocity, $\Omega_H$, is fixed by the horizon value of the metric function $W$,

$$\Omega_H = -\frac{t^2}{t \cdot \varphi} = -\left.\frac{g_{tt}}{g_{t\varphi}}\right|_{r=r_H} = W|_{r=r_H} \ , \tag{A.25}$$

but this quantity will not be computed since it is, together with the horizon radius, an input variable for the solver. Moving to the spatial infinity quantities, we have the ADM mass $M$ and angular momentum $J$. Both can be read from the asymptotic behaviour of the metric functions,

$$g_{tt} = -1 + \frac{2M}{r} + \dots \ , \qquad g_{t\varphi} = \frac{2J}{r}\sin^2\theta + \dots \ . \tag{A.26}$$

In Figure A.2 we have the numerical points and the theoretical line of the four physical quantities mentioned before. Here we choose to vary only the horizon radius, $r_H$, fixing the horizon angular velocity, $\Omega_H$. As one can see, we have two branches, similar as since before, associated to two different cases when one takes the limit $r_H \to 0$: In the down branch, we arrive to a static Schwarzschild configuration; in the up branch, we arrive to an extremal Kerr configuration. Such structure of branches makes clear that for the same horizon radius (and horizon angular velocity) we have, in general, two BHs with different physical quantities, $M, J$. One can also see that the numerical points agree very well with the theoretical line.

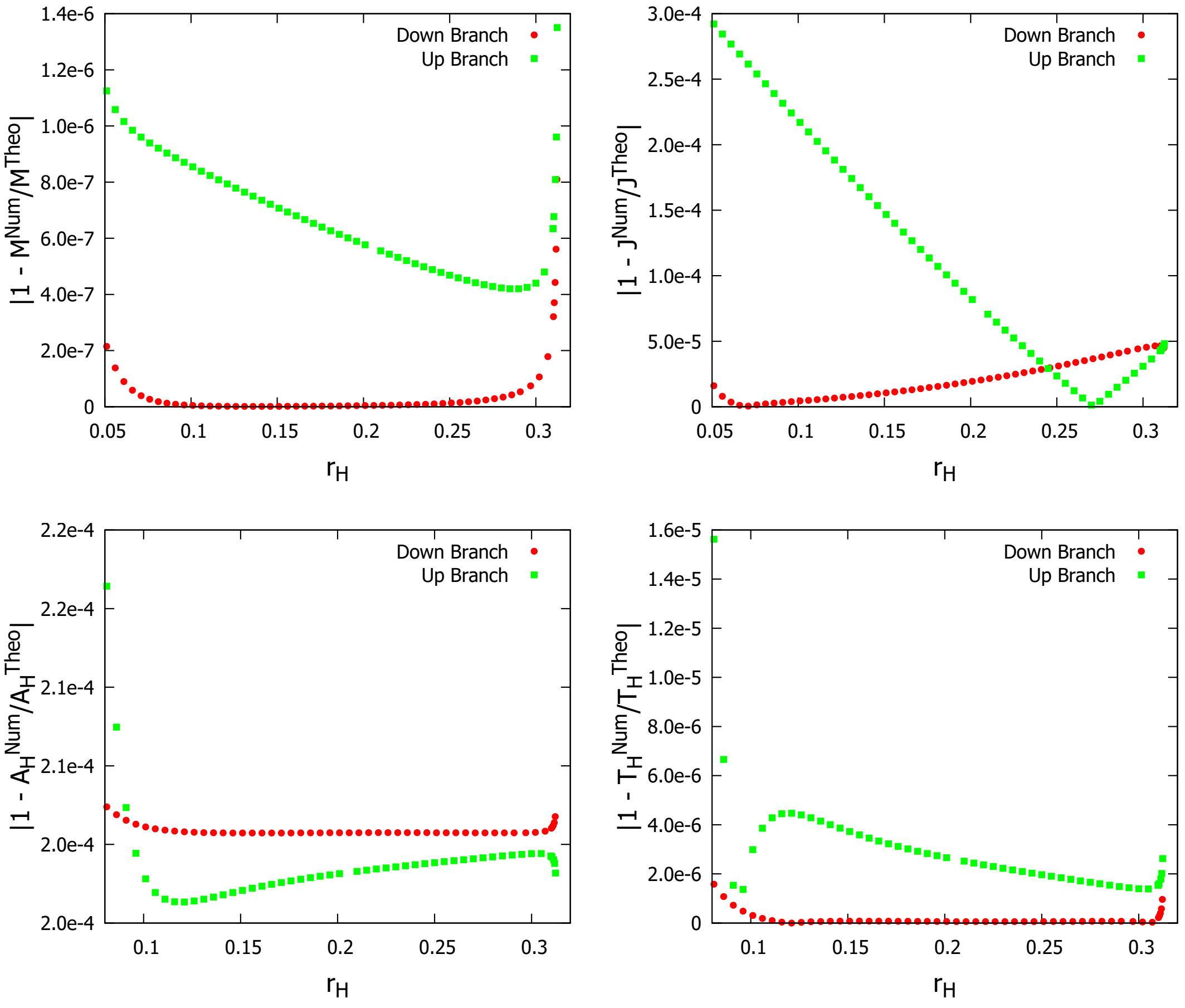

**Figure A.3:** Relative error of the numerical data for the several physical quantities. One sees that the errors vary from $10^{-4}$ and $10^{-7}$ and that the errors tend to increase when we approach both limits of the horizon radius range.

To get a better insight about this, we display in Figure A.3 the relative error of the numerical data. In these figures we can see that the errors vary between $10^{-4}$ and $10^{-7}$, depending on the physical quantity, which is quite acceptable, and provides a good test for this solver. We can also see a general behaviour: the error always increases when we approach both limits of the horizon radius range: $r_H \to 0$ and $r_H \to r_H^{(max)}$, for both branches. In the former case, such behaviour happens because we are approaching a region where the numerics starts to become more difficult to perform, especially in the up branch where we approach extremality. In the latter case, we are in a region where two solutions are very close to each other with almost the same physical quantities, which tends to decrease the accuracy of the solutions found by the solver.

Two hands-on lectures about this solver can be found in [266], [267]. Both lectures were given during the Workshop on Compact Objects, Gravitational Waves and Deep Learning (`https://indico.cern.ch/event/951466/`).

APPENDIX B

# Additional Solutions in Shift-Symmetric Horndeski Theory

## B.1 Perturbative solutions

### B.1.1 Spherically symmetric black holes

The Schwarzschild BH is not a solution of the model given by Equation 6.2 with Equation 6.6, since $R^2_{\rm GB} \neq 0$. Nonetheless, one can construct a perturbative solution around it, as a power series in $\beta$ defined in Equation 6.30. Therefore, we consider a generic expansion[1]

$$N(r) = \left(1 - \frac{r_H}{r}\right) \sum_{k\geqslant 0} \beta^k h_k(r) \, , \qquad \sigma(r) = \sum_{k\geqslant 0} \beta^k \sigma_k(r) \, , \qquad \phi(r) = \sum_{k\geqslant 1} \beta^k \phi_k(r) \, . \tag{B.1}$$

The horizon is still located at $r = r_H$. Then, one solves the EsGB equations order by order in $\beta$.

The choice Equation B.1 leads to a particularly simple structure of the equations for the functions $\{h_k(r), \sigma_k(r), \phi_k(r)\}$, which can easily be solved to an arbitrary order. We have done it up to $k = 12$. These functions are polynomials in $x = r_H/r$, the expression of the first few terms being

$$\begin{aligned}
&h_0(r) = 1, \;\; h_1(r) = 0, \;\; h_2(r) = -\frac{49}{5}x - \frac{29}{5}x^2 - \frac{19}{5}x^3 + \frac{203}{15}x^4 + \frac{218}{15}x^5 + \frac{46}{3}x^6, \\
&\sigma_1(r) = 0, \;\; \sigma_2(r) = -\left(2x^2 + \frac{8x^3}{3} + 7x^4 + \frac{32x^5}{5} + 6x^6\right), \\
&\phi_1(r) = 4x + 2x^2 + \frac{4x^3}{3}, \;\; \phi_2(r) = 0, \\
&\phi_3(r) = -\frac{4}{15}x + \frac{292}{15}x^2 + \frac{1052}{45}x^3 + \frac{22}{5}x^4 - \frac{476}{75}x^5 - \frac{656}{45}x^6 + \frac{20}{3}x^7 + \frac{58}{5}x^8 + \frac{424}{27}x^9 \, .
\end{aligned} \tag{B.2}$$

Unfortunately, no general pattern can be found and the coefficients of the terms in the polynomial expressions of $\{h_k(r), \sigma_k(r), \phi_k(r)\}$ become increasingly complicated, with higher powers of $x = r_H/r$.

[1] Note that $\phi_1(r)$ is a nodeless function, corresponding to the solution of the scalar field Equation 6.12 in a fixed Schwarzschild background. Moreover, one can show analytically that the scalar field remains nodeless even with a non-perturbative approach.

The corresponding expression of the mass function $m(r)$ follows directly from Equation B.1 (we recall that $N = 1-2m(r)/r$). While $m(r)$ is strictly positive, its derivative becomes negative in a region close to the event horizon, the lowest order term being

$$m' = \left(x^2 + x^3 + 13x^4 + x^5 + x^6 - 23x^7\right) 2\beta^2 + \dots \, . \tag{B.3}$$

Thus, for any $\beta$, one finds $m' < 0$ for $r_H \leqslant r \leqslant 1.1049 r_H$. This implies the existence of *negative effective energy densities* ($i.e.$ $\rho^{(\text{eff})} = -T_t^t < 0$) in the model, a feature confirmed by numerics.

We also display the expression of the first few terms for several quantities of interest

$$\begin{aligned}
M &= M^{(0)}\left(1 + \frac{49}{5}\beta^2 + \frac{408253}{3850}\beta^4 + \frac{75242913669527}{26533757250}\beta^6\right) + \dots, \\
T_H &= T_H^{(0)}\left(1 - \frac{1}{15}\beta^2 - \frac{118549}{4950}\beta^4 - \frac{35399108806973}{26533757250}\beta^6\right) + \dots, \\
S &= S^{(0)}\left(1 + \frac{88}{3}\beta^2 + \frac{162064}{675}\beta^4 + \frac{955514545484}{156080925}\beta^6\right) + \dots, \\
Q_s &= r_H\left(4\beta - \frac{4}{15}\beta^3 - \frac{237098}{2475}\beta^5 - \frac{70798217613946}{13266878625}\beta^7\right) + \dots \, ,
\end{aligned} \tag{B.4}$$

with $M^{(0)} = \frac{r_H}{2}$, $S^{(0)} = \pi r_H^2$, $T_H^{(0)} = \frac{1}{4\pi r_H}$ the corresponding quantities for the Schwarzschild solution. One can easily verify that the perturbative expansion satisfies, order by order, the Smarr relation and the 1st law.

### B.1.2 Slowly rotating black holes

The equations of motion possess a simple solution for the case of slowly rotating BH solutions. The latter have been investigated in other gravity theories (see $e.g.$ [268]–[270]) and usually give an idea about some properties of the non-perturbative (in the spin parameter) configurations.

To consider slowly rotating BHs we assume a metric of the following form

$$ds^2 = -N(r)\sigma^2(r)dt^2 + \frac{dr^2}{N(r)} + r^2\left[d\theta^2 + \sin^2\theta(d\varphi - W(r)dt)^2\right] \, , \tag{B.5}$$

with a small $W(r)$, such that, to first order in $W$, the above line element takes the (more) familiar form

$$ds^2 = -N(r)\sigma^2(r)dt^2 + \frac{dr^2}{N(r)} + r^2\left(d\theta^2 + \sin^2\theta d\varphi^2\right) - 2r^2\sin^2\theta W(r) d\varphi dt \, . \tag{B.6}$$

The limit $W(r) = 0$ corresponds to the static EsGB BHs discussed above. Then it is straightforward to prove that, for small rotation, the EsGB equations possess the following first integral

$$\left\{ r^3\left[r - 4\alpha N(r)\phi'(r)\right]\frac{W'}{\sigma}\right\}' = 0 \, . \tag{B.7}$$

The constant of integration is proportional to $J$ – the angular momentum. In the absence of a closed form general expression for the EsGB BHs, the best one can do it to replace in Equation B.7 the corresponding form of the perturbative solution in $\beta = \alpha/r_H^2$ derived above, and integrate for $W(r)$. Then a general expression of the form

$$W(r) = \frac{2J}{r^3} \sum_{k \geqslant 0} w_k(r) \beta^{2k} \, , \tag{B.8}$$

emerges. All functions $w_k(r)$ above can be expressed as polynomials in $x = r_H/r$; here we display the first two functions functions only,

$$w_0(r) = 1 \, , \qquad w_1(r) = -\left( \frac{6}{5}x^2 + \frac{28}{3}x^3 + 3x^4 + \frac{12}{5}x^5 - \frac{10}{3}x^6 \right) \, . \tag{B.9}$$

This approach holds for the first order in $W$, thus for an infinitesimally small angular momentum. Then, physical quantities such as the mass and event horizon area do not change as compared to the static case. On the other hand, the BH acquires a non-trivial angular momentum horizon angular velocity, with leading terms

$$\Omega_H = \frac{2J}{r_H^3} \left( 1 - \frac{63}{5}\beta^2 - \frac{206249189}{1351350}\beta^4 \right) \, . \tag{B.10}$$

The corresponding expression of the reduced horizon angular velocity $\omega_H$ is also of interest, with

$$\omega_H = \Omega_H M = \frac{J}{r_H^2} \left( 1 - \frac{14}{5}\beta^2 - \frac{16415506}{96525}\beta^4 \right) \, , \tag{B.11}$$

a relation which can also be expressed in terms of the dimensionless parameters $j = J/M^2$ and $\alpha/M^2$ as

$$\omega_H = \frac{j}{4} \left[ 1 + \frac{21}{20} \left( \frac{\alpha}{M^2} \right)^2 + \frac{11390263}{3931200} \left( \frac{\alpha}{M^2} \right)^4 \right] \, . \tag{B.12}$$

Therefore, to these orders in perturbation theory, the reduced horizon angular velocity *increases* as compared to a similar (slowly rotating) Kerr BH with the same mass $M$ and angular momentum $J$, a prediction which agrees with our numerical results (see also Figure 6.7).

## B.2 The attractors and the issue of extremal solutions

The numerical results suggest that, unlike the extremal Kerr solution, the extremal EsGB solutions are not regular. Evidence for this conjecture is obtained as follows. Instead of solving the full bulk equations searching for extremal solutions, one tackles the construction of the corresponding near-horizon configurations. In this case, one has to solve a co-dimension one problem (the radial dependence being factorized), whose solutions are easier to study.

Since this problem was already considered in a more general context [271] (see also the corresponding Einstein-dilaton-GB computation in [84]), in what follows we shall review the basic results only. The idea is to consider a construction of the near-horizon limit of the extremal rotating BH as a power series in $\alpha$. The background solution is taken to be the vacuum near horizon extremal Kerr (NHEK) solution in pure Einstein gravity [242]. As we shall see, the $\alpha^2$-corrections to this solution are singular and destroy its smoothness.

Following the usual ansatz in the literature (see *e.g.* [272]) we consider the following line element

$$ds^2 = v_1(\theta)\left(-r^2dt^2 + \frac{dr^2}{r^2} + \beta^2 d\theta^2\right) + v_2(\theta)\sin^2\theta\left(d\phi + Krdt\right)^2 \ , \tag{B.13}$$

where $0 \leqslant r < \infty$, $0 \leqslant \theta \leqslant \pi$, and $\beta,\ K$ are real parameters, while the scalar field depends on $\theta$ only,

$$\phi = \phi(\theta)\ . \tag{B.14}$$

Also, it is convenient to define

$$\cos\theta = u\ , \tag{B.15}$$

such that the line element (Equation B.13) becomes

$$ds^2 = v_1(u)\left(-r^2dt^2 + \frac{dr^2}{r^2} + \beta^2\frac{du^2}{1-u^2}\right) + v_2(u)(1-u^2)\left(d\phi + Krdt\right)^2\ . \tag{B.16}$$

The functions $v_1(u)$, $v_2(u)$ together with the constants $K, \beta$ satisfy a complicated set of ordinary differential equations which result from the EsGB equations. These equations (with $\alpha \neq 0$) appear to possess no analytical solutions. An approximate solution can be constructed, however, by considering an expansion[2] in $\alpha$ around the Einstein gravity solution, with

$$v_1(u) = v_{10}(u) + \alpha^2 v_{12}(u) + \ldots,\ \ v_2(u) = v_{20}(u) + \alpha^2 v_{22}(u) + \ldots,\ \ \phi(u) = \phi_0 + \alpha\phi_1(u) + \ldots, \tag{B.17}$$

and

$$K = K_0 + \alpha^2 K_2 + \ldots,\ \ \beta = \beta_0 + \alpha^2\beta_2 + \ldots\ . \tag{B.18}$$

The lowest order terms in the above expansion corresponds to the Einstein gravity solution [242]

$$K_0 = \beta_0 = 1\ , \qquad v_{10}(u) = \frac{J}{16\pi}(1+u^2)\ , \qquad v_{20}(u) = \frac{J}{4\pi}\frac{1}{(1+u^2)}\ , \tag{B.19}$$

while $\phi_0$ can be set to zero without any loss of generality.

In the next step, we find the expression of $\phi_1(u)$ by solving Equation 6.12 in the NHEK background (Equation B.19)

$$\frac{d}{du}\left((1-u^2)\frac{d\phi_1(u)}{du}\right) + \frac{J(1+u^2)}{16\pi}L_{GB}^{(NHEK)} = 0\ , \tag{B.20}$$

$L_{GB}^{(NHEK)}$ being the GB invariant evaluated for the NHEK geometry,

$$L_{GB}^{(NHEK)} = -\frac{12288\pi^2(-1+15u^2-15u^4+u^6)}{J^2(1+u^2)^6}\ . \tag{B.21}$$

[2]More rigorously, this expansion is in the dimensionless parameter $\alpha/J$.

The general solution of the Equation B.20 reads

$$\phi_1(u) = s_0 + s_1 \log\left(\frac{1+u}{1-u}\right) + \frac{32\pi}{J}\left[\frac{2(1-4u^2-u^4)}{(1+u^2)^3} + \log\left(\frac{1+u^2}{1-u^2}\right)\right], \tag{B.22}$$

with $s_0$, $s_1$ arbitrary constants. One can set $s_0 = 0$ without any loss of generality. For any choice of $s_1$ the function $\phi_1(u)$ necessarily diverges at $u = 1$ and/or $u = -1$. In our approach, we take

$$s_1 = \frac{32\pi}{J}, \tag{B.23}$$

such that $\phi_1(u)$ is divergent at $u = 1$ only.

In the next step, we solve for the corrections to the geometry as encoded in the functions $v_{12}(u)$ and $v_{22}(u)$. Since the equations for these functions are sourced by a divergent scalar field $\phi_1(u)$, one expects them to be divergent as well. This is indeed confirmed by our results, and one finds

$$\begin{aligned} v_{12}(u) = \frac{\pi}{J}\Bigg(&\mathcal{F}_1(u) - 32(4+u(3+4u))\arctan u - 64(1-u^2)\log(1+u^2) \\ &-\beta_2\frac{J^2}{24\pi^2}(2+2u^2+3u\sqrt{1-u^2}\arccos u) - 128(1-u)^2\log(1-u) - 64\sqrt{2}u\sqrt{1-u^2} \\ &+\frac{969}{\sqrt{2}}u\sqrt{1-u^2}\arctan\frac{\sqrt{2}u}{\sqrt{1-u^2}} - 192\sqrt{2}u\sqrt{1-u^2}\,\mathrm{arctanh}\,\frac{\sqrt{1-u^2}}{\sqrt{2}} \\ &+uz_1 + (1+(u-4)u)z_2 - u\sqrt{1-u^2}z_3\Bigg), \end{aligned} \tag{B.24}$$

where $z_1$, $z_2$, $z_3$ are arbitrary constants and we define

$$\begin{aligned} \mathcal{F}_1(u) = \frac{1}{105(1+u^2)^5}\Big(&88054 + 26880u + 759219u^2 + 161280u^3 + 1133035u^4 + 376320u^5 \\ &+1617566u^6 + 430080u^7 + 1109548u^8 + 241920u^9 + 377967u^{10} + 53760u^{11} + 49331u^{12}\Big). \end{aligned}$$

A very similar expression is found for $v_{22}(u)$, with

$$\begin{aligned} v_{22}(u) = \frac{8\pi}{J(1+u^2)^2}\Bigg(&\mathcal{F}_2(u) - 16(4-u(3-4u))\arctan u - 32(1+u)^2\log(1+u^2) \\ &+\beta_2\frac{J^2}{48\pi^2}(-5(1+u^2) + \frac{6u\arccos u}{\sqrt{1-u^2}}) - 64(1-u)^2\log(1-u) + 64\sqrt{2}\frac{u}{\sqrt{1-u^2}} \\ &+\frac{969}{\sqrt{2}}\frac{u}{\sqrt{1-u^2}}\arctan\frac{\sqrt{2}u}{\sqrt{1-u^2}} + \frac{192\sqrt{2}u}{\sqrt{1-u^2}}\,\mathrm{arctanh}\,\frac{\sqrt{1-u^2}}{\sqrt{2}} \\ &-\frac{1}{2}uz_1 + \frac{1}{2}(1+4u+u^2)z_2 - \frac{u}{\sqrt{1-u^2}}z_3\Bigg), \end{aligned} \tag{B.25}$$

with

$$\begin{aligned} \mathcal{F}_2(u) = \frac{1}{105(1+u^2)^5}\Big(&40667 - 20160u + 4707u^2 - 114240u^3 + 515365u^4 - 255360u^5 \\ &+474470u^6 - 282240u^7 + 372733u^8 - 154560u^9 + 174351u^{10} - 33600u^{11} + 35035u^{12}\Big). \end{aligned}$$

Then, with the above expressions, one can prove the existence of a singularity at the poles of the horizon, with the Ricci scalar diverging at $\theta = 0$ ($i.e.$ $u = 1$)

$$R = -\frac{32768\pi^3\alpha^2}{J^3(u-1)} + \frac{49152\pi^3\alpha^2}{J^3} + \mathcal{O}(u-1) \,. \tag{B.26}$$

Finally, let us mention that the above perturbative result does not exclude the existence of regular solutions for $\alpha$ large enough. Thus we have also attempted to solve the field equations of the model within a nonperturbative approach, by solving a boundary value problem. The imposed boundary conditions assure the regularity of the configurations at $u = \pm 1$. However, no such solutions could be found.

APPENDIX C

# Circular Motion of Spacelike Geodesics

Consider the same assumptions and symmetries discussed at the beginning of section 7.2. In such spacetime, the effective Lagrangian of a spacelike test particle can be written as,

$$2\mathcal{L} = g_{\mu\nu}\bar{x}^{\mu}\bar{x}^{\nu} = 1 \; , \tag{C.1}$$

where the bar denotes the derivative with respect to arc length. Assuming that the motion occurs on the equatorial plane, $\theta = \pi/2$, we can write the Lagrangian as,

$$2\mathcal{L} = g_{tt}(r, \theta = \pi/2)\bar{t}^2 + 2g_{t\varphi}(r, \theta = \pi/2)\bar{t}\bar{\varphi} + g_{rr}(r, \theta = \pi/2)\bar{r}^2 + g_{\varphi\varphi}(r, \theta = \pi/2)\bar{\varphi}^2 = 1 \; . \tag{C.2}$$

Hereafter we will drop the radial dependence of the metric functions to simplify the notation. Due to stationarity and axial-symmetry, we can introduce the energy and angular momentum of the spacelike particle,

$$-E \equiv g_{t\mu}\bar{x}^{\mu} = g_{tt}\bar{t} + g_{t\varphi}\bar{\varphi} \; , \quad L \equiv g_{\varphi\mu}\bar{x}^{\mu} = g_{t\varphi}\bar{t} + g_{\varphi\varphi}\bar{\varphi} \; . \tag{C.3}$$

Rewriting the Lagrangian with these new quantities,

$$2\mathcal{L} = -\frac{A(r, E, L)}{B(r)} + g_{rr}\bar{r}^2 = 1 \; , \tag{C.4}$$

where, similar as before, $A(r, E, L) = g_{\varphi\varphi}E^2 + 2g_{t\varphi}EL + g_{tt}L^2$ and $B(r) = g_{t\varphi}^2 - g_{tt}g_{\varphi\varphi}$. We can now introduce the potential $V_1(r)$ as,

$$V_1(r) \equiv g_{rr}\bar{r}^2 = 1 + \frac{A(r, E, L)}{B(r)} \; . \tag{C.5}$$

To have a particle following a circular orbit at $r = r^{\text{cir}}$, both the potential and its radial derivative must be null, hence,

$$V_1(r^{\text{cir}}) = 0 \quad \Leftrightarrow \quad A(r^{\text{cir}}, E, L) = -B(r^{\text{cir}}) \; , \tag{C.6}$$

and

$$V_1'(r^{\text{cir}}) = 0 \quad \Leftrightarrow \quad A'(r^{\text{cir}}, E, L) = -B'(r^{\text{cir}}) \, . \tag{C.7}$$

Along such circular orbit, the angular velocity of the particle (measured by an observer at infinity) is,

$$\Omega = \frac{d\varphi}{dt} = \frac{\bar{\varphi}}{\bar{t}} = -\frac{E g_{t\varphi} + L g_{tt}}{E g_{\varphi\varphi} + L g_{t\varphi}} \, . \tag{C.8}$$

Solving the equation $V_1(r^{\text{cir}}) = 0$ together with the equation for the angular velocity, we can write the energy and angular momentum of the spacelike particle,

$$E_\pm = -\left. \frac{g_{tt} + g_{t\varphi}\Omega_\pm}{\sqrt{-\beta_\pm}} \right|_{r^{\text{cir}}} \, , \quad L_\pm = \left. \frac{g_{t\varphi} + g_{\varphi\varphi}\Omega_\pm}{\sqrt{-\beta_\pm}} \right|_{r^{\text{cir}}} \, , \tag{C.9}$$

where $\beta_\pm \equiv (-g_{tt} - 2g_{t\varphi}\Omega_\pm - g_{\varphi\varphi}\Omega_\pm^2)|_{r^{\text{cir}}} = -A(r^{\text{cir}}, \Omega, \Omega)$ is the same function defined for the timelike particle case, Equation 7.21.

Solving the second equation, $V_1'(r^{\text{cir}}) = 0$, together with the previous results, we can compute the angular velocity of the spacelike particle,

$$\Omega_\pm = \left[ \frac{-g_{t\varphi}' \pm \sqrt{C(r)}}{g_{\varphi\varphi}'} \right]_{r^{\text{cir}}} \, . \tag{C.10}$$

This is the same expression for the angular velocity as we saw for timelike particles, Equation 7.22.

From these results we can conclude that when circular orbits are possible, *i.e.* $C(r) \geqslant 0$, the only difference between the circular motion of timelike and spacelike particles resides on the energy and angular momentum, or more precisely, on their dependency with the $\beta_\pm$ function. When $\beta_\pm > 0$, it is possible to have timelike circular orbits (TCOs) since both the energy and angular momentum of the timelike particle are well defined, but one can not have spacelike circular orbits, since the energy and angular momentum of the spacelike particle are not well defined. Likewise, when $\beta_\pm < 0$ the opposite occurs: it is not possible to have TCOs, but it is possible to have spacelike circular orbits.

It is also possible to conclude that the transition of $\beta_\pm$ from positive to negative values, and vice-versa, is entirely continuous, providing that we can have circular orbits, *i.e.* $C(r) \geqslant 0$.

# APPENDIX D

# List of Publications

This thesis is based on the following scientific works,

- Jorge F. M. Delgado, Carlos A. R. Herdeiro, Eugen Radu, "*Horizon geometry for Kerr black holes with synchronized hair*", **Phys.Rev.D** 97 (2018) 12, 124012, *e-Print:* 1804.04910 [gr-qc]

- Jorge F. M. Delgado, Carlos A. R. Herdeiro, Eugen Radu, "*Kerr black holes with synchronised scalar hair and higher azimuthal harmonic index*", **Phys.Lett.B** 792 (2019) 436-444, [*e-Print:* 1903.01488 [gr-qc]]

- Jorge F. M. Delgado, Carlos A. R. Herdeiro, Eugen Radu, "*Spinning black holes in shift-symmetric Horndeski theory*", **JHEP** 04 (2020) 180, [*e-Print:* 2002.05012 [gr-qc]]

- Jorge F. M. Delgado, Carlos A. R. Herdeiro, Eugen Radu, "*Rotating Axion Boson Stars*", **JCAP** 06 (2020) 037, [*e-Print:* 2005.05982 [gr-qc]]

- Jorge F. M. Delgado, Carlos A. R. Herdeiro, Eugen Radu, "*Kerr black holes with synchronized axionic hair*", **Phys.Rev.D** 103 (2021) 10, 104029, [*e-Print:* 2012.03952 [gr-qc]]

- Jorge F. M. Delgado, Carlos A. R. Herdeiro, Eugen Radu, "*Equatorial timelike circular orbits around generic ultracompact objects*", **Phys.Rev.D** 105 (2022) 6, 064026, [*e-Print:* 2107.03404 [gr-qc]]